\documentclass[10pt,openright,final,a4paper]{memoir}

\usepackage[a4paper]{geometry} 
\let\STARTCODE\relax
\let\STOPCODE\relax

\STARTCODE

\usepackage{color,calc,graphicx,soul,url}
\usepackage{epstopdf,epsfig}
\usepackage{amsmath,amsfonts,mathtools}
\usepackage[utf8]{inputenc}
\usepackage[T1]{fontenc}
\usepackage[final]{pdfpages}
\usepackage{appendix,etoolbox}
\usepackage{amsmath}
\usepackage{pstricks}
\usepackage{pst-all}
\usepackage{pstricks-add}
\usepackage{pst-3d}
\usepackage{pst-func}
\usepackage{pst-circ}
\usepackage{pst-math}
\usepackage{tikz}
\usepackage{titletoc}
\usepackage{watermark}
\usepackage[intoc]{nomencl}
\makenomenclature
\usepackage{eso-pic,everyshi,calc,ifthen,wallpaper}
\usepackage{pgfplots}

\AtBeginEnvironment{subappendices}{%
\chapter*{Appendix}
\addcontentsline{toc}{chapter}{Appendices}
\counterwithin{figure}{chapter}
\counterwithin{table}{chapter}
}

\usetikzlibrary{arrows,calc,intersections,positioning}
\usepgfplotslibrary{groupplots}
\usepgfplotslibrary{patchplots}
\usetikzlibrary{pgfplots.colormaps}
\pgfplotsset{
colormap={Oranje}{color(0cm)=(white); color(0.5cm)=(orange!50); color(2cm)=(red!75)}
}

\usepackage{mathpazo}

\DeclareMathOperator{\tr}{Tr}

\DeclareMathOperator{\atanh}{atanh}
\DeclareMathOperator{\Th}{th}
\DeclareMathOperator{\VAR}{var}
\newcommand{\ud}{\mathrm{d}}
\newcommand{\He}[1]{\mathrm{He}_{#1}}
\DeclareMathOperator{\sgn}{sgn}
\newcommand{\Flip}[1]{\text{$\text{F}_{\text{#1}}$}}

\newcommand*{\plogo}{\includegraphics[width=1\textheight,angle=90]{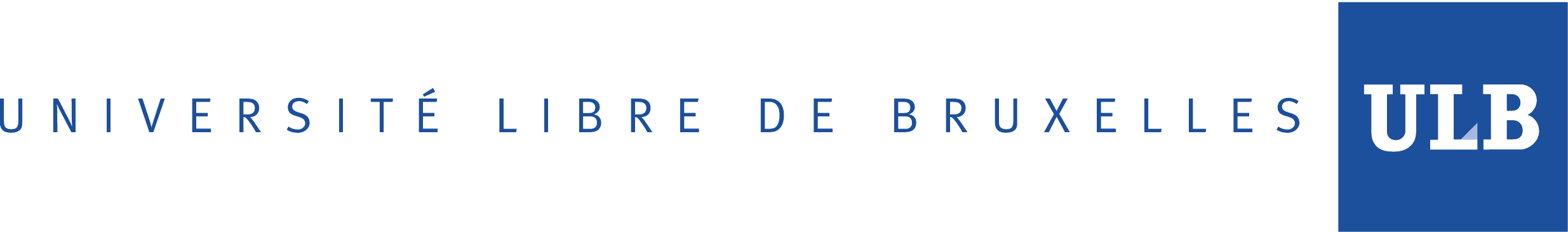}} 
\newcommand*{\sbsem}{\includegraphics[width=0.4\textwidth]{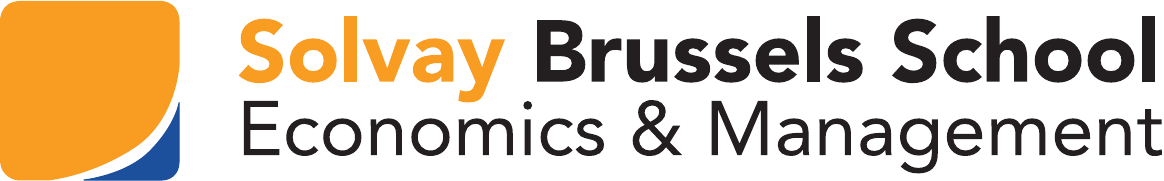}} 
\newcommand*{\thesistitle}{Collective behaviours in the stock market}
\newcommand*{\thesisSubtitle}{A maximum entropy approach}
\newrgbcolor{grr}{0.5 0.5 0.5}
\newrgbcolor{bleuf}{0.2 0.2 1}
\newrgbcolor{bleuff}{0 0 0.5}
\newrgbcolor{brun}{0.5 0.2 0.02}
\definecolor{mpurple}{RGB}{147,112,219}
\newrgbcolor{vert}{0.28 0.6 0.28}
\newrgbcolor{rougef}{0.5 0 0}
\definecolor{rouch}{rgb}{0.85,0,0.1}
\newrgbcolor{vertf}{0 0.5 0.25}
\definecolor{bleuf}{rgb}{0.2,0.2,1}
\definecolor{marron}{rgb}{0.4,0.2,0.15}
\definecolor{orange}{RGB}{255,69,0}
\definecolor{orange2}{RGB}{255,140,0}
\definecolor{LSblue}{RGB}{106,90,205}
\definecolor{ulblue}{RGB}{0,76,147}

\definecolor{nicered}{RGB}{150,0,0}
\usepackage{hyperref}
\hypersetup{
    bookmarks=true,         
    unicode=false,          
    pdftoolbar=true,        
    pdfmenubar=true,        
    pdffitwindow=false,     
    pdfstartview={FitH},    
    pdfnewwindow=true,      
    colorlinks=true,       
    linkcolor=blue!97.5,          
    citecolor=bleuf ,        
    filecolor=magenta,      
    urlcolor=bleuf ,          
    }

\usepackage{breakurl}

\usepackage{caption}
\DeclareCaptionFont{nicered}{\color{nicered!95}}
\setsubsecheadstyle{\normalfont\bfseries\color{nicered!95}}

\makeatletter
\newlength{\numberheight}
\makechapterstyle{TomTom}{%
\cleardoublepage
\setlength{\beforechapskip}{2cm}
\setlength{\midchapskip}{-0.5cm}
\setlength{\afterchapskip}{1.75cm}

\setlength{\numberheight}{20mm}
}
\makeatother
\chapterstyle{TomTom}
\setsecheadstyle{\bfseries\color{nicered!95}}
\STOPCODE

\openright  

\makeatletter
\newenvironment{summary}
               {\begin{Spacing}{0.5}
               \begin{flushleft}\textbf{Summary}\hspace{5pt}\end{flushleft}
               \vspace{-12.5pt}
               \hrulefill\hspace{2cm}\vspace{-5pt}
                 \list{}{\listparindent 0cm%
                        \setlength{\leftmargin}{2cm}
                        \itemindent\listparindent
                        \rightmargin\leftmargin
                        \parsep\z@ \@plus\p@}%
                \item\relax}
               {\endlist\vspace{-10pt}\hrulefill\hspace{2cm}\end{Spacing}\newpage}

\SpecialCoor

\usepackage[
    sorting=none,
    url=false,
    natbib=true,
    doi=true,
    eprint=true,
    firstinits=true,
    backend=bibtex,
    defernumbers=true,
    backref=true
]{biblatex} 
\DeclareNameAlias{sortname}{last-first}

\pgfplotscreateplotcyclelist
{fillcol}{%
color = red!75!blue\\%
color = red!60!blue\\%
color = red!45!blue\\%
color = red!30!blue\\%
color = red!15!blue\\%
color = red!05!blue\\%
}

\makepagestyle{myruled}
\makerunningfootwidth{myruled}{1.15\textwidth}
\makeheadrule {myruled}{\textwidth}{\normalrulethickness}
\makeevenhead {myruled}{\small\textsc{\leftmark}}{} {}
\makeoddhead {myruled}{}{}{\small\rightmark}
\makeevenfoot{myruled}{
\rule{2cm}{0.25pt}\\
\rule{0.25em}{0pt}\textcolor{nicered!95}{\Large\thepage}
}{} {}
\makeoddfoot{myruled}{}{} {
\rule{2cm}{0.25pt}\rule{0.30em}{0pt}\\
\hfill\textcolor{nicered!95}{\Large\thepage}
}
\makeatletter 
\makepsmarks {myruled}{
\nouppercaseheads
\createmark {chapter} {both} {shownumber}{}{. \ }
}
\makeatother
\begin{document}

\frontmatter

\thispagestyle{empty}

\thispagestyle{empty}
\hbox{ 

\begin{tikzpicture}[remember picture, overlay]
    \node[] at (current page.center) {%
        \includegraphics[width=\paperwidth]{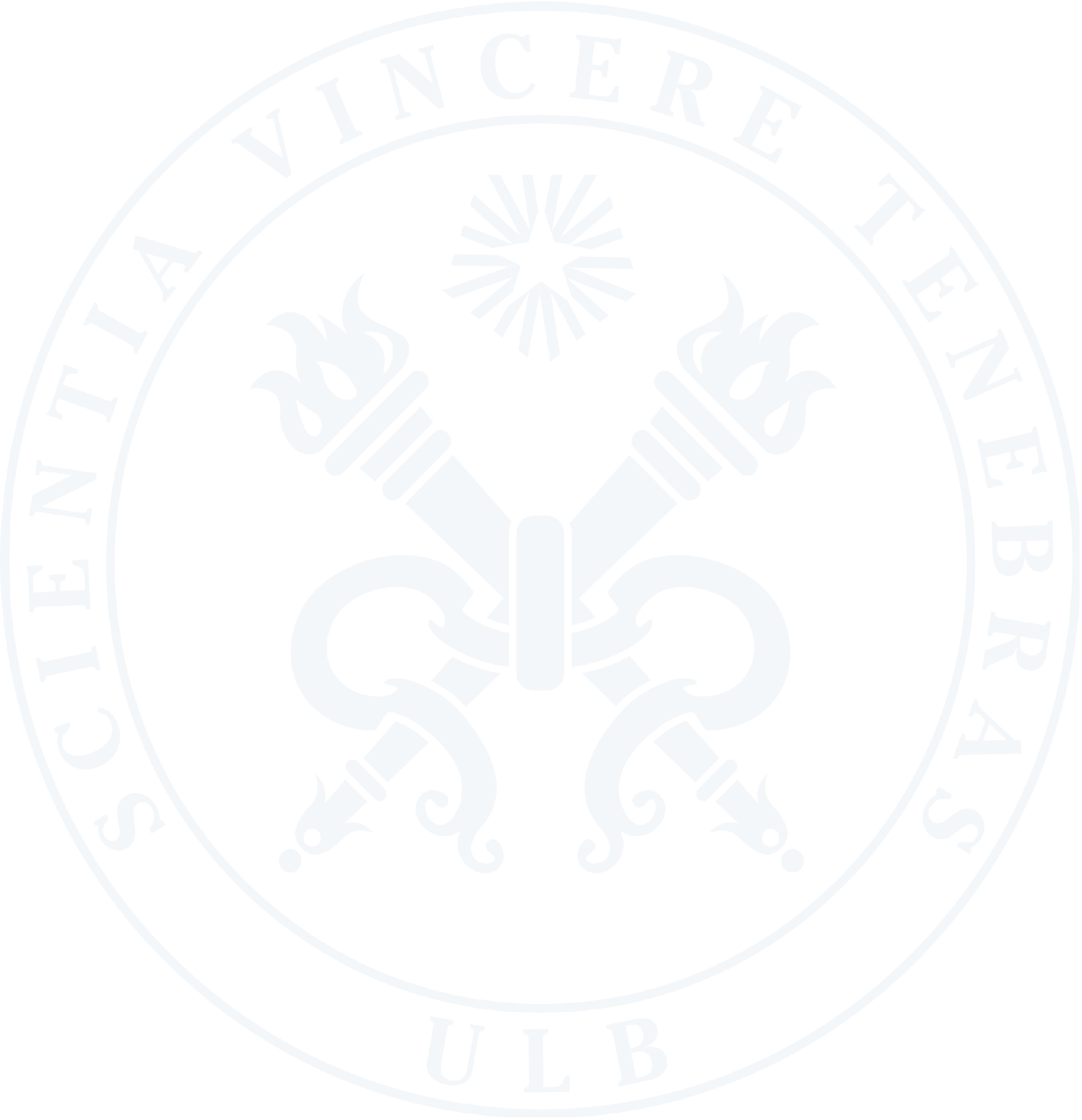}%
    };%
\end{tikzpicture}


\begin{tikzpicture}[remember picture, overlay]
    \node[] at (0.1\marginparwidth,0.5\textheight) {%
        \plogo%
    };%
\end{tikzpicture}


\hspace*{0.2\textwidth} 
\rule{0.5pt}{\textheight} 
\hspace*{0\textwidth} 
\parbox[b]{0.80\textwidth}{ 
\begin{flushright}
{\noindent\Huge\textcolor{black}{\thesistitle}}\\[1\baselineskip] 
{\noindent\huge \textcolor{black}{\thesisSubtitle}}\\[5\baselineskip] 
{\noindent\large Ph.D. Dissertation}\\[5\baselineskip]


{\noindent
\begin{flushright}
\begin{tabular}{r r}
   \textcolor{ulblue}{Advisor:}    & Prof. Philippe Emplit\\
   \textcolor{ulblue}{Co-advisor:} & Prof. Bram De Rock \\
 \end{tabular}
\end{flushright}
}

\vspace{0.2\textheight}

{\noindent\LARGE \textsc{Thomas Bury}}\\ 
{\noindent\large \textcolor{black}{SBS-EM, February 2014}}\\
{\noindent\rule{0.4\textwidth}{0.1pt}}\\
{\hbox{}
\null
\vspace*{\fill}
\hbox{}
\null
\vspace*{\fill}
\hbox{}
\sbsem
}
\end{flushright}
}}

\newpage
\thispagestyle{empty}
\mbox{}
\clearpage


%
%
%



\thispagestyle{empty} 

\null\vfill 

\begin{flushright}
\emph{À mon grand-père, parrain Jean}
\end{flushright}

\vfill\vfill\vfill\vfill\vfill\vfill\null 

\clearpage 

\newpage
\thispagestyle{empty}
\mbox{}
\clearpage

\thispagestyle{empty} 

\null\vfill 

\textit{“Strength does not come from winning. Your struggles develop your strengths. When you go through hardships and decide not to surrender, that is strength.”}

\begin{flushright}
Arnold Schwarzenegger
\end{flushright}

\vfill\vfill\vfill\vfill\vfill\vfill\null 

\cleardoublepage 


\thispagestyle{empty}

I, Thomas Bury, declare that this thesis titled, '\thesistitle' and the work presented in it are my own. I confirm that:

\begin{itemize}
\item[\tiny{$\bullet$}] This work was done wholly or mainly to fulfill the requirements for a doctor's degree at the Université libre de Bruxelles.
\item[\tiny{$\bullet$}] Where any part of this thesis has previously been submitted for a degree or any other qualification at this University or any other institution, this has been clearly stated.
\item[\tiny{$\bullet$}] Where I have consulted the published work of others, this is always clearly mentioned.
\item[\tiny{$\bullet$}] Where I have quoted from the work of others, the source is always given. With the exception of such quotations, this thesis is entirely my own work.
\item[\tiny{$\bullet$}] I have acknowledged all main sources of help.
\item[\tiny{$\bullet$}] Where the thesis is based on work done by myself jointly with others, I have made clear exactly what was done by others and what I have contributed myself.\\
\end{itemize}

\begin{center}
Signed: \tikz[baseline]\node (b){\includegraphics[width=3.5cm]{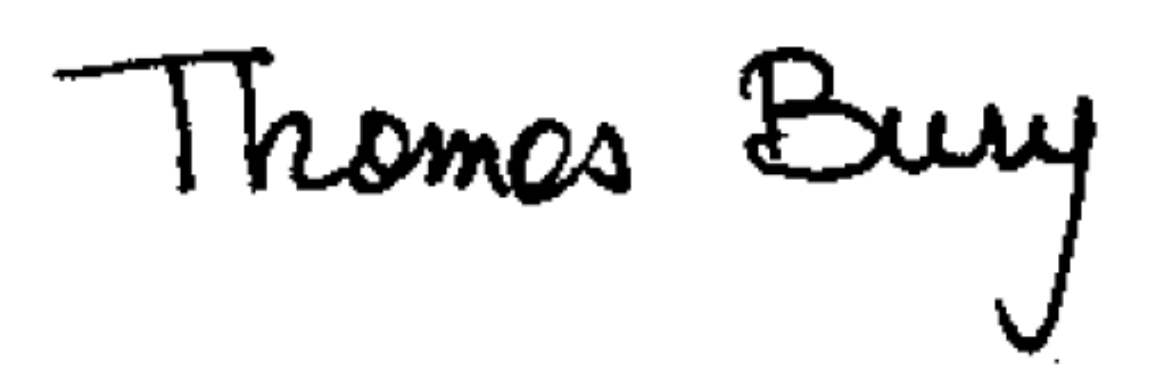}};\tikz\node [coordinate] (e) {};\\
\tikz[anchor=base, baseline,overlay]\draw[line width=0.5pt] (-1,1\baselineskip)-- (3,1\baselineskip);

Date:\qquad February, 2014\\
\tikz[baseline,overlay]\draw[line width=0.5pt] (-1,.5\baselineskip)-- (3,.5\baselineskip);
\end{center}
%

\vspace{2.5cm}

The examining committee will be composed by

\vspace{0.5cm}

\begin{tabular}{llll}
  -Philippe & Emplit      & Advisor                              & Universit\'e libre de Bruxelles \\
  -Bram     & De Rock     & co-Advisor                           & Universit\'e libre de Bruxelles \\
  -Estelle  & Cantillon   & Président du Comité d'accompagnement & Universit\'e libre de Bruxelles \\
  -Davy     & Paindaveine & Member                               & Universit\'e libre de Bruxelles \\
  -David    & Veredas     & Member                               & Universit\'e libre de Bruxelles \\
  -Yann     & Frignac     & Member                               & Télécom SudParis \\
  -Alan     & Kirman      & Member                               & Aix-Marseille Université \\
\end{tabular}

\vspace{3cm}

\textbf{This version is the revised version following comments done during the private assessment.}

\vfill

\clearpage 

\openright

\chapter*[Abstract]{Abstract}
\addcontentsline{toc}{chapter}{Abstract}
\thispagestyle{empty}
Scale invariance, collective behaviours and structural reorganization are crucial for portfolio management (portfolio composition, hedging, alternative definition of risk, etc.). This lack of any characteristic scale and such elaborated behaviours find their origin in the theory of complex systems. There are several mechanisms which generate scale invariance but maximum entropy models are able to explain both scale invariance and collective behaviours.
The study of the structure and collective modes of financial markets attracts more and more attention. It has been shown that some agent based models are able to reproduce some stylized facts. Despite their partial success, there is still the problem of rules design. In this work, we used a statistical inverse approach to model the structure and co-movements in financial markets. Inverse models restrict the number of assumptions. We found that a pairwise maximum entropy model is consistent with the data and is able to describe the complex structure of financial systems. We considered the existence of a critical state which is linked to how the market processes information, how it responds to exogenous inputs and how its structure changes. The considered data sets did not reveal a persistent critical state but rather oscillations between order and disorder.
In this framework, we also showed that the collective modes are mostly dominated by pairwise co-movements and that univariate models are not good candidates to model crashes. The analysis also suggests a genuine adaptive process since both the maximum variance of the log-likelihood and the accuracy of the predictive scheme vary through time. This approach may provide some clue to crash precursors and may provide highlights on how a shock spreads in a financial network and if it will lead to a crash. The natural continuation of the present work could be the study of such a mechanism.


\chapter*[Acknowledgments]{Acknowledgments}
\addcontentsline{toc}{chapter}{Acknowledgments}
\thispagestyle{empty}
\epigraph{\footnotesize \emph{When I was a kid, my grandfather used to say to me that a fellow's life wasn't worth mentioning if he hadn't shared it with some folks along the way.}}{\footnotesize MacGyver}

\emph{Hey kid, what are you gonna do when you grow up?} When I grow up, I would like to be a physicist. Up to now it was my main goal, it still is. Even if the path is not the one I had pictured and despite several difficulties and mistakes, I had the great opportunity to spend a good time in studying physics and sciences. It was fun, motivating and challenging. I am deeply grateful to my advisor Philippe Emplit (aka Flup) for giving me this chance, for his clever advice, his support and his kindness. I am also indebted to Bram De Rock, my co-advisor, and to Estelle Cantillon. They joined the project at an early stage and their help was essential. I enjoyed your different opinions and your expertise.

I thank the jury, Philippe, Bram, Estelle, Davy, David, Yann and Alan, for agreeing to review this thesis and for their useful comments.

Une moitié de ses sept années a été consacrée à l'enseignement. J'ai là encore beaucoup appris. Je remercie les professeurs Philippe Emplit, Jean-Claude Dehaes, Philippe Kinet et Marc Haelterman d'avoir partagé leurs connaissances et leurs visions, chacune différente, de la physique et des sciences en général. Mes collègues et amis: Cyril l'homme polyvalent, Jon le magicien des slides, Ourouk le berceau de la civilisation, Lorentz le dynamique, Mehdi la star, Olivier le sage, Charles l'homme d'affaires, Bertrand l'artiste, Stéphane le coordinateur, Christophe et Fikri les chimistes. J'ai une pensée amicale pour les étudiants, en espérant que les quelques minutes (per capita) en séance d'exercices et de laboratoires n'ont pas été trop pénibles. Pour ma part, l'enseignement a lui aussi été une source d'inspiration grâce à une nécessaire et constante remise en question des acquis, méthodes et matériels liés aux cours.

Je remercie mes partenaires et adversaires de jeux, je veux dire collègues et amis du labo pour l'accueil chaleureux du non-opticien (pire, du non-expérimentaliste) que je suis. Je peux dire sans me tromper que l'ambiance si particulière a fortement contribué à garder ma détermination intacte. Arrivé en tant que rookie lors du Mercato kicker d'hiver 2006, j'ai essayé de faire bonne figure devant les joueurs confirmés que sont Sisse, Adrien, et plus tard Bernard, Laurent. J'attends d'ailleurs encore vos tests d'urine. Concernant les darts, je dois là aussi m'avouer vaincu devant le tandem de choc \emph{Sissadrien} et Bernardhood en solo. Je n'oublierai pas non plus les soirées poker auxquelles Jim nous rejoignait et certaines mains mémorables.
J'ai par la suite profité des discussions sérieuses et moins sérieuses à la cuisine du labo avec François le Français (MacGyver a son fidèle couteau, le labo a son fidèle Français), Maïté (qui sait désormais que le gaz de ville n'est pas comme l'eau courante), Pascal, Sim-Pi, Ibtissame, Alexandra, Mika, Quentin, Yvan, Evdokia, Serena, Piotr et Antoine. Ma gratitude à Serge, Anteo et Toon pour leurs relectures de mes manuscrits et leurs conseils. Les pro-Barça, Akram et Sébastien et le pro-Real, Jassem (Cricri) et leurs dictons ("si la passe est belle, c'est normal de marquer un but de chance", "ce sont les buts qu'on ne marque pas qu'on regrette"). Olivier et Steph qui, après m'avoir laissé une longueur d'avance pour tromper l'ennemi m'en ont mis quelques unes par la suite. Personne n'oubliera les lay-up d'Olivier, l'adresse légendaire de Steph La Capuche et son excursion taminoise prénuptiale (5060 represents) ni la prestation d'Anthoni à son marriage. Enfin, Tchoum avec qui j'ai partagé le bureau mais aussi trop de mal-bouffes, un nombre incalculable de cinés et de craquages en tout genre (nos nombreux posters et autres achats compulsifs parlent d'eux-mêmes).

Mais aussi mes collègues du service SMN: Julio, Xavier, Alain, Nicolas, Pierre-Etienne, Artem, Yvan, Julien, Pierre, Laëtitia.

Je remercie aussi tous ceux pas encore cités avec qui j'ai partagé (et partagerai encore j'espère) de bons moments. Patti la globe-trotteuse pour qui une longue absence n'est jamais synonyme de silence gêné, Diako et sa capacité à encaisser des vannes qui n'a d'égale que celle de Rocky à encaisser les coups, Toon pour son subtil mélange "mr le professeur" le jour et "Rodriguez de la Vega" la nuit, Marina pour les discussions surréalistes et rafraîchissantes, Laurent et Steph pour leur style tout en retenue et leur bonne humeur, Amé et Chris dont les repas ont toujours débouché sur de franches rigolades, Méla dans son Luxembourg lointain, toute la fine équipe du foot dominical et les ami(e)s des guides que j'ai peu à peu perdus de vue (my fault).

Je pense aussi aux Louvanistes de naissance ou par adoption, Antho et Fleur pour les nombreuses soirées jeux, Sarah la Ninjette presque jamais à court d'énergie, Nico (Ramses) et Rosalie pour les barbecs du dimanche, Kev l'original Ninja-Kiwi, Miche, la famille de Jean-Phi pour leur accueil sans faille, tous les potes du foot plus ou moins improvisé au Blocry.

Je ne saurais oublier mes proches qui m'ont soutenu et ce bien avant cette aventure, en particulier ma maman Dominique, ma grand-mère Monique, mon grand-père Jean et ma s\oe{}ur Mahé mais aussi le reste de ma famille à qui je n'accorde que trop peu de temps, j'en suis conscient. Enfin un merci spécial à mon \emph{outlaw-brother}, Jean-Phi, j'assimile notre tandem à Tango et Cash (les roles dépendant de la situation) différents mais complémentaires.


\chapter*[Author's contribution]{Author's contribution}
\addcontentsline{toc}{chapter}{Author's contribution}
\thispagestyle{empty}
\begin{refsection}
Some of the results presented in this thesis have been published or submitted to scientific journals with peer review.


\nocite{moi3,moi2,moi1,moi4,moi5}
\printbibliography[
    heading=subbibliography,
    title={Published papers},
    keyword=MyPub,
]

The first paper concerns some technical developments, part of the results are exposed in sections \ref{sec3:eq} and \ref{sec3:MC}. The second paper sets the basis of the pairwise entropy models and the results are detailed in chapter \ref{chap:SPIMSM}. The third paper explores the market structure, preliminary results are exposed in chapter \ref{chap:SPIMSM} and main results are reported in chapter \ref{chap:marketStruct}. The fourth paper is a statistical study of the criticality hypothesis and is detailed in chapter \ref{chap:crit}.
The fifth paper highlights and characterizes the collective co-movements in financial markets, the results are reported in chapter \ref{chap:flip}.


\end{refsection}

\clearpage
\newpage



\tableofcontents* 

\newpage
\listoffigures 
\newpage
\listoftables 
\newpage
\printnomenclature
\newpage


\nomenclature{$\Pr[R>r]$}{The probability that the random variable $R$ is larger than the value $r$.}%
\nomenclature{$p(x)$}{The probability mass (density) function of the discrete (continuous) random variable $X$.}%
\nomenclature{$f(x;\boldsymbol{\theta})$}{A parameterized probability density function.}%
\nomenclature{$\mathbf{s}$}{A configuration (vector of sign of returns).}%
\nomenclature{$\mathcal{U}(\mathbf{s})$}{A utility function as a function of the configurations.}%
\nomenclature{$U$}{A particular value of the utility function $\mathcal{U}(\mathbf{s})=U$.}%
\nomenclature{$S[p(x)]$ or $S[X]$ }{The entropy of the random variable $X$.}%
\nomenclature{$\mathcal{Z}$}{Partition function or exponential of the cumulant generating function.}%
\nomenclature{$\textbf{J}$}{Mutual influence matrix.}%
\nomenclature{$\textbf{C}$}{Covariance matrix.}%
\nomenclature{$\mathcal{H}(\mathbf{s})$}{Opposite of the utility function $\mathcal{H}(\mathbf{s})\equiv -\mathcal{U}(\mathbf{s})$.}%
\nomenclature{$\theta(\cdot)$}{Heaviside function.}%
\nomenclature{$D_{\mathrm{KL}}(\cdot)$}{The Kullback-Leibler divergence (KLD) between the distribution $p$ and $q$.}%
\nomenclature{$\mathrm{E}_{q}[\cdot]$}{Mathematical expectation with respect to the distribution $q$.}%
\nomenclature{$\mathrm{var}[\cdot]$}{The variance.}%
\nomenclature{$\sum_{\{\mathbf{s}\}}$}{The sum over the $2^N$ binary configurations.}%
\nomenclature{$I(X,Y)$}{The mutual information between the random variables $X$ and $Y$.}%
\nomenclature{$p_{2}(\textbf{s})$}{The maximum entropy distribution with first and second order mutual influences.}%
\nomenclature{$s_{i,t}$}{The sign of the return of the $i$th stock at period $t$.}%
\nomenclature{$a_{n}\asymp b_{n}$}{Meaning: $\lim_{n\rightarrow\infty}n^{-1}a_{n}= \lim_{n\rightarrow\infty}n^{-1}b_{n}$.}%
\nomenclature{$L(\boldsymbol{\theta})$}{Likelihood function.}%
\nomenclature{$\ell(\boldsymbol{\theta})$}{Log-likelihood function.}%
\nomenclature{$\mathrm{PL}(\boldsymbol\theta)$}{Pseudo-likelihood function.}%
\nomenclature{$\mathrm{pl}(\boldsymbol\theta)$}{Log pseudo-likelihood function.}%
\nomenclature{$\mathcal{H}_{t}^{T}$}{History of the process from period $t$ to period $t-T$.}%
\nomenclature{$\Pr(A_{n}\in \ud a)$}{A shortcut for $\Pr(A_{n}\in [a,a+\ud a])$.}%
\nomenclature{$\lim_{x\rightarrow\infty}$}{Limit for $x$ going to infinity.}%
\nomenclature{$\sup_{x\in \mathcal{D}}f(x)$}{Supremum of the function $f(x)$ for arguments belonging to the set $\mathcal{D}$.}%
\nomenclature{$T$}{Stochasticity level. The larger $T$, the larger the stochasticity.}%
\nomenclature{$\beta$}{Inverse stochasticity level ($\beta\equiv T^{-1}$). The larger $\beta$, the smaller the stochasticity.}%
\nomenclature{$\lambda$}{Eigenvalue.}%
\nomenclature{DJ}{Dow Jones.}%
\nomenclature{$\log(\cdot)$}{The logarithm to base $10$.}%
\nomenclature{$\ln(\cdot)$}{The natural logarithm (to base $e$).}%
\nomenclature{rPML}{Regularized pseudo-maximum likelihood.}%
\nomenclature{PMLE}{Pseudo-maximum likelihood estimator.}%
\nomenclature{MLE}{Maximum likelihood estimator.}%
\nomenclature{MS}{Mantegna-Sornette.}%
\nomenclature{BD}{Brock-Durlauf.}%
\nomenclature{KS}{Kolmogorov-Smirnov.}%
\nomenclature{CDF}{Cumulative distribution function.}%
\nomenclature{PDF}{Probability density function.}%
\nomenclature{RMS}{Root mean square.}%
\nomenclature{RMSE}{Root mean square error.}%
\nomenclature{MCMC}{Monte Carlo Markov chain.}%
\nomenclature{MCS}{Monte Carlo steps.}%
\nomenclature{Maxent or ME}{Maximum entropy.}%
\nomenclature{MEP}{Maximum entropy principle.}%
\nomenclature{LDP}{Large deviation principle.}%
\nomenclature{$\sgn$}{The sign operator.}%
\nomenclature{$\Flip{i}$}{Flipping operator (changes the sign of the $i$th asset).}%
\nomenclature{$\Th(\cdot)$ or $\tanh(\cdot)$}{Hyperbolic tangent.}%
\nomenclature{$\atanh(\cdot)$}{Arc hyperbolic tangent.}%
\nomenclature{$R_{\mathcal{U}}$}{Response function of the average utility function.}%
\nomenclature{$R_{m}$}{Response function of the average market orientation.}%
\nomenclature{$\delta_{\mathbf{s}_{t},\mathbf{s}}$}{Kronecker delta, equal to one if $\mathbf{s}_{t}=\mathbf{s}$ and to zero otherwise.}
\nomenclature{IID}{Independent and identically distributed.}%
\mainmatter

\clearpage

\openright

\chapter*[Foreword]{Foreword}
\addcontentsline{toc}{chapter}{Foreword}
\thispagestyle{empty}

\epigraph{\footnotesize \emph{The sciences do not try to explain, they hardly even try to interpret, they mainly make models. By a model is meant a mathematical construct which, with the addition of certain verbal interpretations, describes observed phenomena. The justification of such a mathematical construct is solely and precisely that it is expected to work - that is correctly to describe phenomena from a reasonably wide area. Furthermore, it must satisfy certain esthetic criteria - that is, in relation to how much it describes, it must be rather simple.}}{\footnotesize John von Neumann}


Through the years, each time I've answered the question "\emph{what is your thesis about?}", people were mostly surprised that physics and economics could in some way be related. So, even before introducing my work, I must give some element of answer to this tricky question.

The first two years, I got the opportunity to study economics in a nutshell at the ECARES center. After a while, similarities clearly emerged from the different lectures. Economics scales as physics does. Microeconomics deals with fundamental entities trying to describe the behaviour at the relevant \emph{micro} scale and when the number of entities is too large, macroeconomics takes over at a larger scale trying to describe the average behaviour. This pattern is also used in physics (chemistry, biology, finance, signal processing, etc.). The tools used to link both scales are also partially common to these fields.

The most interesting feature of this conception is that a system (composed by many entities) is more than the sum of its parts. Indeed most of time when the pattern "\emph{many entities} $+$ \emph{interactions}" is met, some very special features emerge at the macro scale which are not anticipated at the micro scale. Lets take the simple example of ants. What can a single or a couple of ants do? Not much since they are relatively simple beings programmed to perform a couple of tasks. What can colonies of ants do? They can solve complex problems, build energy efficient habitat, develop agriculture among other things. Human beings are (roughly) like ants. An ordinary individual can not do much but many of us can do wonderful (or sadly, awful) things.

These complex systems are met in many disciplines and relevant tools have been developed to study them. Progressively these tools have been applied to a priori unrelated disciplines. Nowadays this approach seems pretty natural since the pattern "\emph{many entities} $+$ \emph{interactions}" implies a certain degree of independence to the nature of constituting entities. The most common example is that the same class of models describes neural networks and magnetic materials whereas neurons and atoms have nothing in common at the microscopic scale.

One may ask: \emph{what are the applications?} (implied in everyday life) I believe that before thinking of applications, the very nature of the system should be studied and understood. These tasks are (most of time) spread over several decades. It became clearer that episodic crises are actually a structural issue. A better understanding of crises formation involves necessarily a better understanding of the market structure.

Last, in an interdisciplinary approach, one should take the differences between the different fields into account. The modelling task in economics is considerably harder than in physics. There are two main reasons to this. First, in economics there is no such thing as equation of motion. People are not like Newton's apple (obviously), they do not behave the same way, they learn, they change. Secondly, the three pillars of natural sciences are: theory, simulations and experiments (in the order you like). The last pillar is amputated in economics, you simply can not duplicate the market in thousand of copies and study the ensemble properties. This is why economics is conceptually harder than physics and also why analogies are somewhat dangerous. A way to tackle the problem is to rely heavily on statistics. Some methods developed in physics can be applied to any field dealing with complex systems. However physics is not called upon to replace the fundamental statements of economics but can instead provide a different vantage point from which selected topics can be studied.  For instance, a natural choice is the study of the structure of financial systems (the \emph{spatial} axis). A better understanding of the market structure may also help in tricky tasks as portfolio and fund managing. Several strategies coming from recent developments are already used.


\chapter*[Introduction]{Introduction}
\addcontentsline{toc}{chapter}{Introduction}
\thispagestyle{empty}

\epigraph{\footnotesize \emph{Clouds are not spheres, mountains are not cones, coastlines are not circles, and bark is not smooth, nor does lightning travel in a straight line.}}{\footnotesize Benoît Mandelbrot}

\begin{figure}[!ht]
\begin{center}
\includegraphics[scale=1]{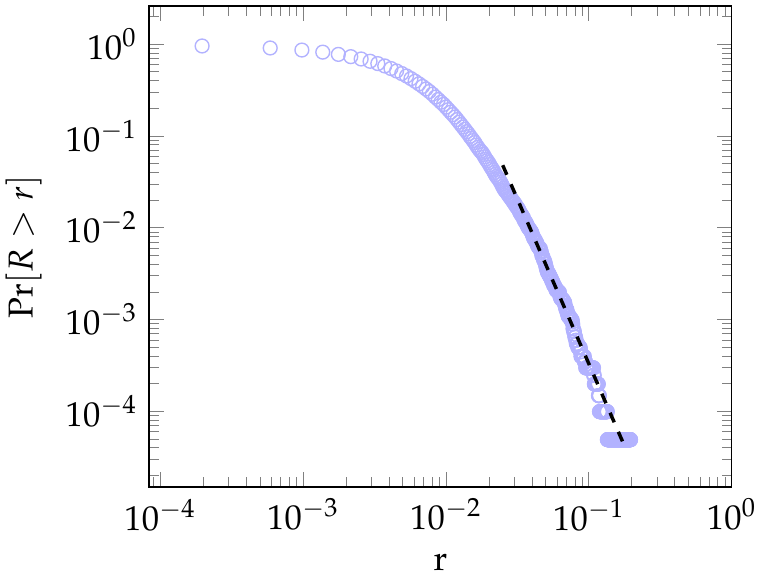}
\end{center}
\caption[Cumulative distribution of the log-returns]{\label{fig:DJretruns} Cumulative distribution of the log-returns for the Dow Jones for the period 1928-2009. The estimation of the exponent $\alpha$ of the tail $x^{-\alpha}$ is equal to $3.5\pm0.1$ (one standard deviation).}
\end{figure}

A hundred years ago, the Italian economist Pareto introduced the notion of power-law describing the wealth distribution. It is a major concept related to the notion of scale invariance which is widely used in finance and economics (fractional Brownian motion, detrended fluctuations analysis, volatility modelling, etc.). This lack of any characteristic scale is surprising at first glance but finds its foundation in the theory of complex systems. Scale invariance is crucial in finance because large absolute returns are power-law distributed as illustrated in Fig-\ref{fig:DJretruns} for the Dow Jones.


Dramatic events in Nature and in human driven systems are most of the time \emph{not} outliers since they are not isolated events, they are part of all the possible occurrences as described by their distribution. It seems rather sensible to find such a feature in financial markets since Nature by itself is also driven by a broad class of events from the rainy summer afternoon to major earthquakes. Markets are like the crust of the Earth, they will undergo quakes almost surely in a long enough observation range; their magnitude $S$ being distributed following a power-law: the larger the magnitude, the lower (but significant) the occurrence probability $\Pr[S>s] \sim s^{-\alpha}$.

Physics is used to deal with scale invariance and power-laws emerging from the pattern "\emph{many entities} $+$ \emph{interactions}". It is therefore not surprising that famous economists (X. Gabaix, T. Lux, etc.) and physicists (H. Stanley, J.P. Bouchaud, etc.) worked together and wrote highly cited papers in prestigious journals (see \cite{Gabaix,Lux} for instance).

Major concerns for which physics may help are the characterization of economic fluctuations \cite{Gabaix,stanley-gabaix2} and their emergence from agent based models \cite{Samanidou,Zhou}.
Despite the breakthroughs made by those models some issues need to be considered. For agent based models, the design of the rules is not an easy task and different sets of rules can lead to the same stylized facts which raises the question \emph{how to test and asses rigorously the rules?} Especially when the rules evolve through time since more and more robot traders with changing strategies are appearing.

Another important question raised by economists and physicists is a possible \emph{critical} state of financial markets. Analogies between stylized facts and critical physical systems are striking (power-law, scaling, structural reorganization, clustering, data aggregation). These features have been studied with a wide variety of models ranging from economics \cite{Gabaix,Mike,Lux} to physics \cite{Sorn} and stochastic processes (multifractal processes) \cite{Muzy}. However the criticality statement is most of the time phenomenological or, worse, an artifact induced by the choice on a particular conditioning variable \cite{Plerou,Potters} rather than rigourously established as in \cite{Kiyono}.

Is there any way to simplify the reality and get a qualitative model describing consistently the market structure and able to reproduce observed collective phenomena leaving aside such particular kinds of rules as is the case in neuroscience \cite{Schneidman}? It is the question that I have tried to answer through this thesis.
In order to avoid rules design, I have followed the inverse formulation: start from the data and try to infer a model and then compare the model to the empirical facts. I used a further simplification, I only consider the sign of returns. The sign is believed to contain information about the structure (based on the experience in magnetic materials, neuronal networks, complex systems, choice theory, etc.).
For that purpose, maximum entropy models are well suited. The use of pairwise maximum entropy (maxent) models has led to a fruitful description of complex systems, particulary in phase transition and magnetic materials (Ising models and spin glasses) \cite{Fischer,Stanley}, but also in neuroscience \cite{Schneidman} and agent based models \cite{Zhou}.  They are related to graphical models, Boltzmann machines, error correcting codes, logistic regression, etc. \cite{Opper}. Maxent models are much more than models recovering moments from data, they are powerful effective models describing collective behaviour and more appropriate than correlations in structure characterization. Moreover their ability to capture statistical dependencies can be tested with the multi-information criterion \cite{Schneid_Multi}. They also allow a comparison with existing behaviours as noise dressed correlation matrix \cite{Laloux} or structural reorganization \cite{Mantegna2,OnnelaPRE}. The remaining results about criticality can also be compared to the rigourous empirical tests derived from statistical physics \cite{Mora}. Furthermore, the market collective dynamics can be highlighted using such a simple model showing clearly periods of larger cross-covariance.

Throughout this work, I have shown that the structure of the financial network is well described by a maxent model, that it provides a framework well suited to study structural reorganizations and clustering features. A maxent model is also suited to study the criticality which is important for understanding how the market processes information and how a shock spreads in the financial network. Last, it is a good candidate to perform spatial guesses of trend reversals using instantaneous market information. Therefore, this inverse approach provides an attractive unified framework to study structural issues and yields interesting perspectives.

Collective behaviour (such as criticality) and structural reorganization are crucial for portfolio management (portfolio composition, hedging, alternative definition of risk, etc.). This approach may provide clues about crash precursors and may be able to cast lights on how a shock spreads and if it will lead to a crash. The natural continuation of the present work could be the study of such a mechanism.

\newpage

The chapters are organized as follows:

\paragraph{Chapter \ref{chap:theory}} introduces concepts and methods used throughout this thesis from a statistical point of view avoiding as much as possible \emph{analogies} to physics. First of all, we introduce the statistical entropy and related concepts such as the Kullback-Leibler divergence (KLD) and the multi-information. The equivalence of the KLD minimization and the likelihood maximization is recalled and the statistical modelling using the maximum entropy principle is sketched. Secondly, one deduces variational methods from the KLD minimization (likelihood maximization). These variational methods are useful for simulations and for the inverse problem (inferring the probability distribution from the data) which is briefly presented. Last, a test of the power-law hypothesis is presented and one introduces the Mantegna-Sornette distance which is used to study the market structure. A mathematical motivation of the maximum entropy principle is given in appendix.

Chapters \ref{chap:SPIMSM} to \ref{chap:flip} are organized in the logical order illustrated by the following thought-line (each step induced the next piece of research)

\begin{figure}[!ht]
\begin{center}
\includegraphics[scale=0.8]{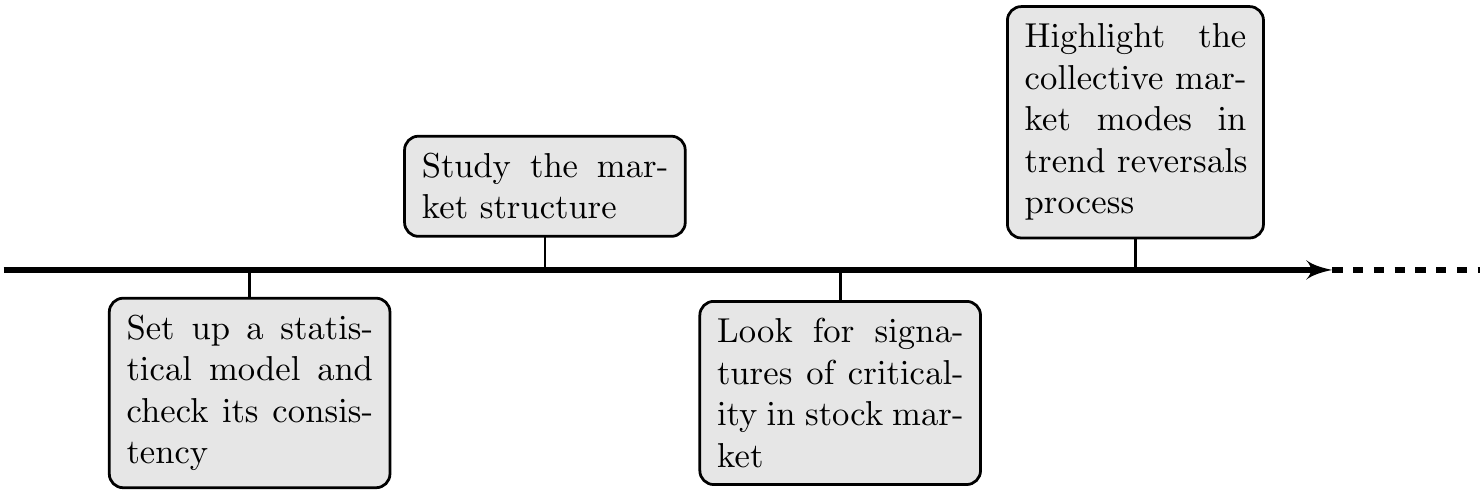}
\end{center}
\caption[Thought-line]{The logical ordering of chapters \ref{chap:SPIMSM} to \ref{chap:flip} respectively.}
\end{figure}

\paragraph{Chapter \ref{chap:SPIMSM}} sets up the pairwise maximum entropy model of stock market. We show that it is a statistically consistent model since pairwise co-movements explain almost all statistical dependencies. We detail the differences with the existing pairwise models in other disciplines.

\paragraph{Chapter \ref{chap:marketStruct}} gives an additional study of the consistency of the pairwise maxent model. We show that the entropy decreases during periods where the absolute market orientation is the largest (during crises and large bullish movements). We explain how Lagrange parameters are related to the market state. In particular, the influence matrix is used to highlight the structural reorganization of a stock market (indirect evidence of order-disorder transition). Last, we make the link to the graph-theoretic approach and we build asset trees on the influence matrix rather than on the correlations. We compare the clustering properties of both approaches.

\paragraph{Chapter \ref{chap:crit}} concerns the study of criticality in financial markets. We perform tests of criticality inspired by statistical physics and we check if signatures of criticality are present in the corresponding maxent pairwise model. In particular, we show that financial markets are closer to the criticality before a crisis. Last, we discuss the interpretation of the criticality in financial markets.

\paragraph{Chapter \ref{chap:flip}} highlights the role of collective market modes in trend reversals prediction. We show that the ensemble's instantaneous state is the most important part for the data studied. This finding reveals the strength of the collective dynamics underlying the trend reversals.

\paragraph{Chapter \ref{chap:social}} proposes a formulation of socio-economics models in term of maximum entropy models. As becomes evident in these models, co-movements are also a fundamental part of the underlying optimization process (utility maximization). Deriving the utility function including a social component on economic considerations requires several assumptions. The application of the maximum entropy principles provides a useful statistical \emph{inverse} formulation which can be interpreted and linked to the optimization process with a minimal set of assumptions. Furthermore, the maxent formulation also provides a convenient framework to discuss the equilibria and their stability. We introduce these notions using methods derived in previous chapters. Finally, we draw the final conclusion and we give some perspectives.

The contents of this thesis is illustrated by the following tag cloud (the size represents the number of times that word has been used.)

\begin{figure}[!ht]
\begin{center}
\includegraphics[scale=1.15]{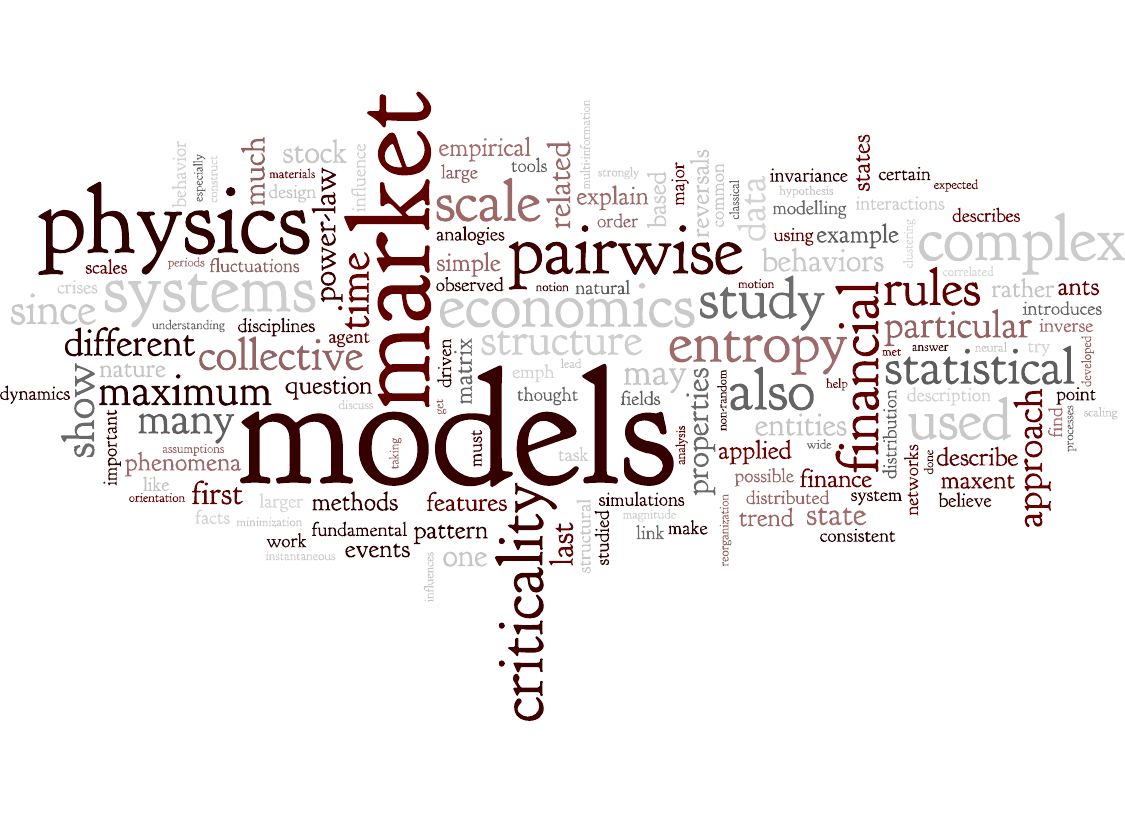}
\caption[Tag cloud]{A tag cloud generated from the content of this thesis using the free application Wordle (\url{http://www.wordle.net/}). The size represents the number of times that word has been used.}
\end{center}
\end{figure}

\chapter{Theory and methods}\label{chap:theory}
\thispagestyle{empty}

  \epigraph{\footnotesize \emph{...You should call it entropy, for two reasons. In the first place your uncertainty function has been used in statistical mechanics under that name, so it already has a name. In the second place, and more important, no one really knows what entropy really is, so in a debate you will always have the advantage.}}{\footnotesize John von Neumann (to Claude Shannon) \cite{EIT}}

\begin{summary}
This chapter is dedicated to theoretical developments and methods used throughout this thesis. The key concepts of entropy, variational methods and maximum entropy principle are briefly introduced from a statistical point of view. A simulation scheme and a power-law hypothesis test are detailed. Last, the link between the correlations and the market structure is explained.
\end{summary}
%

\section{Introduction}
\label{sec3:intro}
The entropy will be a main tool throughout this thesis. It seems therefore necessary to spend some time to explain concepts based on this quantity. In the following, we recall its statistical meaning and define some common tools related to entropy. We explain how it can be used to infer statistical models consistent with data and to the observed phenomena. It should be noted that the maximum entropy principle is equivalent to the maximization of the likelihood of a distribution $p$ closest to the uniform distribution without range restriction and can be used to find the most likely state(s) of a complex system without any a priori restriction, see (\ref{3-DKLlik}) and (\ref{3-LDTmaxent}).
We explain how variational methods derive from the minimization of the KullBack-Leibler discrepancy and their link with the maximum likelihood approach. We present several approximations based on these methods and how simulations can be performed.

We also present the inverse problem consisting in inferring a statistical model from the data. We only present some of the most efficient inference methods. We develop a slightly modified version of a statistical test of the power-law hypothesis.

Last, we sketch how the market topology can be studied through the statistical covariances. Results found with the so-called Sornette-Mantegna distance will be compared to those of the pairwise maximum entropy model in Chap-\ref{chap:marketStruct}.

The chapter is organized as follows. In section \ref{sec3:inv}, the inverse problem is sketched. In section \ref{sec3:data}, a brief description of the data is given. In section \ref{sec3:MEP}, we present the maximum entropy principle. In section \ref{sec3:Relation}, the relation between maximum entropy and graphical models is made, motivating a topological approach of financial networks. In section \ref{sec3:entropy}, we define the entropy and the Kullback-Leibler divergence. In section \ref{sec3:Var}, variational methods are derived from the former concepts. In section \ref{sec3:prob}, we briefly describe how to find the most probable utility and small fluctuations around this state. In section \ref{sec3:TestMEP}, a test of the leading order of a maximum entropy model is depicted. In section \ref{sec3:eq}, self-consistent equations for stationary networks are derived. In section \ref{sec3:MC}, Monte Carlo simulations are detailed. In section \ref{sec3:IsingInv}, some inference methods for the inverse problem are detailed. In section \ref{sec3:zipf}, the consequences of the linearity of the entropy and its relation to Zipf's law are sketched. In section \ref{sec3:DPL}, a statistical test of the power-law hypothesis is discussed. In section \ref{sec3:MSdist}, the topological approach of financial networks is introduced.

\section{Inverse approach}
\label{sec3:inv}
\begin{figure}[!ht]
\begin{center}
\includegraphics[width=0.75\textwidth]{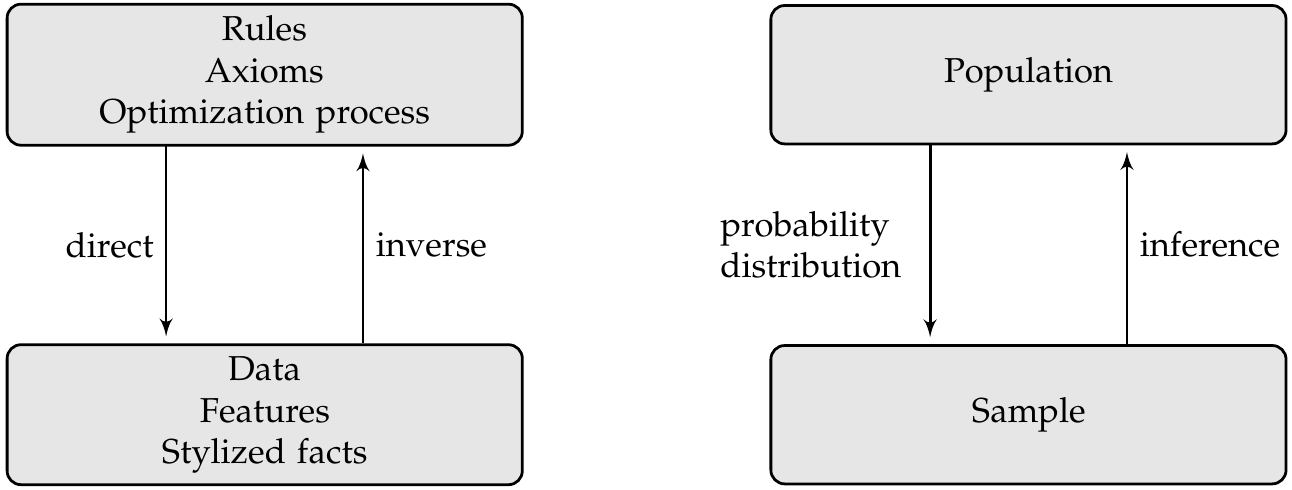}
\end{center}
\caption[Direct and inverse approaches]{\label{fig:DirInv} Schematic representation of the deductive and inductive approaches and the statistical counterpart.}
\end{figure}

First of all, we recall the inverse approach, its aims and concepts. Consider a system: a set of many fundamental entities (like economic agents, ants, neurons, etc.) and the interactions between entities. Generally, one observes features at macro-scale (characteristic scale of the system, eg stock exchange) which are unexpected from the observation at micro-scale (characteristic scale of elementary entities, eg individual economic agent). There are basically two ways to model such a complex system. The direct (deductive) and inverse approaches (inductive), illustrated in Fig-\ref{fig:DirInv}. We note a similarity with the statistical counterpart: the perfect knowledge of a population allows the characterization of a sample (deductive reasoning) and try to infer information based on a partial knowledge as a sample, for instance (inductive reasoning).

In economics, the deductive approach is used in agent based models (ABM). One starts to enunciate a set of rules and one tries to demonstrate some particular properties. This is particularly tricky because there are no universal \emph{laws} like in natural sciences for instance. Furthermore, different sets of rules can eventually lead to the same aggregate (collective) behaviours. Indeed, in complex systems the relevant feature which drives the emergent behaviours is not the nature of elementary entities but rather the kind of interactions (meaning their range, their order and the underlying topology). One can observe similar aggregate features in magnetic materials and in neural networks even if they are very different at individual scale, for instance \cite{Fischer}.
The second approach is to start from the data (one or many samples) assuming that one knows nothing about the system. Specialized mathematical tools have been created especially for this task (eg: the maximum entropy principle).

These two approaches are complementary and models should go back and forth from inductive and deductive approaches until reaching some kind of robust consensus.

\begin{figure}[!ht]
\begin{center}
\includegraphics[width=0.75\textwidth]{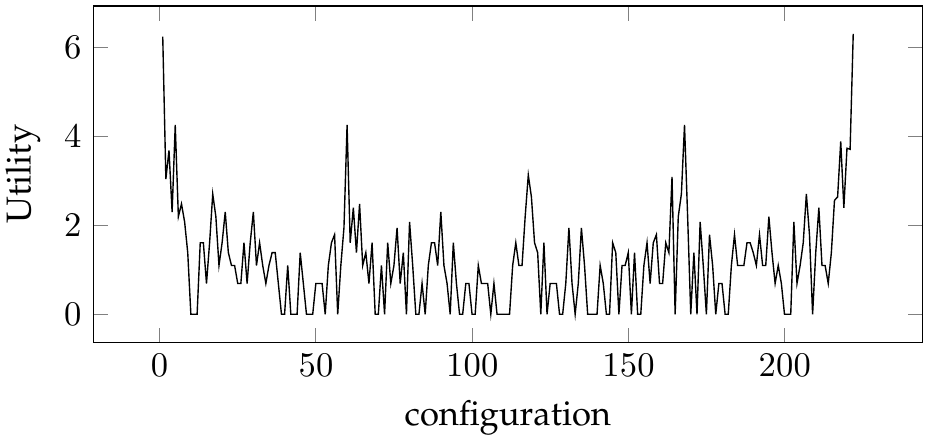}
\end{center}
\caption[Utility function as log-likelihood]{\label{fig:Uloglik} The utility function $\mathcal{U}(\mathbf{s})\simeq \mathrm{Cste}+\ln \hat{p}_{\mathbf{s}}$ for a set of 8 European indices ($\beta$ is set to $1$ for real data). Almost all configurations are properly sampled (see chap-\ref{chap:crit} for details).}
\end{figure}

Can one truly learn anything from the inverse approach? In fact, one can if the relevant variables are well sampled. Let's say that the data are generated by an optimization of an unknown function $\mathcal{U}(\mathbf{s})$ (eg, the social planner's utility) over a certain number of variables $\mathbf{s}$ (eg, agents binary choices). Assume, without loss of generality, that the configurations (vector of all relevant variables) $\mathbf{s}$ are drawn with probability $p(\mathbf{s})=\mathcal{Z}(\beta)^{-1}\exp \beta\mathcal{U}(\mathbf{s})$. When the configuration are properly sampled (says $M$ configurations are recorded), the empirical distribution $\hat{p}_{\mathbf{s}}\equiv M^{-1}\sum_{t=1}^{M}\delta_{\mathbf{s}_{t},\mathbf{s}}$ provides information about the optimization process and the utility function since $\mathcal{U}(\mathbf{s})\simeq \mathrm{Cst}+\beta^{-1}\ln \hat{p}_{\mathbf{s}}$. For a complete discussion with known and unknown relevant variables, see \cite{MMR}. It follows that global maxima of the function $\mathcal{U}(\mathbf{s})$ correspond to the most probable states of the underlying optimization process since $\ln(\cdot)$ is a continuous, strictly increasing function. Local maxima correspond to frequently visited configurations. An example of such utility function is illustrated in Fig-\ref{fig:Uloglik}. In this simple example, there are five main configurations. This is an example of a limited number of configurations with noisy movements driving the system from one to another.

As the utility landscape is rather complex for only $8$ entities, we may expect an even greater complexity for larger systems. Such a complex utility landscape induces several questions: how to identify the most probable states? If the system evolves spontaneously towards its equilibria, are they reached in a reasonable time? can we characterize fluctuations around these states? These questions are approached in the next sections and next chapters.

\section{Data}\label{sec3:data}
In this work, we consider only the sign of returns and instantaneous information (within the defined time bin). The timeseries should therefore be synchronous. The stock exchange closing days, pre-market and after hours trading exchanges are removed. If a time bin is missing for a particular asset, the same time bin should be deleted from the database. The latter case is marginal since we consider indices and highly capitalized companies.

We also consider Stock market indices: Dow Jones, Aex, Bel20, Cac40, Dax, Eurostoxx 50, Ftse, Ibex, Mib. The Dow Jones is the oldest stock market index and is a price-weighted index. The Dow Jones is proportional to the sum of the prices of its 30 components. So, for each US dollar variation, the Dow Jones varies of a given quantity (expressed in \emph{basis point}).  Higher-priced stocks are thus the dominant ones although they may correspond to small capitalizations.
The other indices are calculated on the market capitalization of components: number of shares available for public trading (the so-called float) multiplied by the current price.

\section{Maximum entropy principle}\label{sec3:MEP}
The above discussion immediately raises the question of the right mathematical tool to use to extract information from the data. The method used through this thesis is the maximum entropy principle (MEP) which is a powerful tool for that purpose. It allows to derive the less structured model consistent with some knowledge of the system. It selects the distribution which leaves us with the largest remaining uncertainty consistent with our knowledge (constraints) of the system. In this way, we do not introduce any additional assumptions. Or quoting Edwin Jaynes (the father of the MEP) the MEP is "maximally non-committal with regard to the missing information" \cite{Jaynes}.
Maximizing the entropy can be viewed as a maximization of the likelihood of the distribution $p$ closest to the uniform distribution $U$ without range restriction since $D_{\mathrm{KL}}(p||U)=-S[p]$ up to a constant (see next sections). This method of probability mass function estimation is used in many fields: neuroscience \cite{Schneidman}, econometrics \cite{WuEcon}, etc.

The general MEP is written as a functional maximization ($X$ is a random variable)

\begin{eqnarray}\label{3-MEP}
   & &\max_{\substack{\{p(x)\}}} S[p(x)]= \max_{\substack{\{p(x)\}}} \left\{-\sum_{\{x\}}p(x) \,\ln p(x)\right\}  \\ \nonumber
   &\mathrm{s.t}& p(x)\geq0, \;  \mathrm{E}_{p}[1]=1,\; \mathrm{E}_{p}[f_{i}(x)]=\mu_{i}\quad\mathrm{for}\,i=1,\cdots,m\nonumber
\end{eqnarray}
where our knowledge of the system is encoded by the constraints $\mathrm{E}_{p}[f_{i}(x)]=\mu_{i}$: we know the expected value of some functions of the random variable. The associated Lagrangian is

\begin{equation}\label{Lag}
  L(\{f_{i}\})=-\sum_{\{x\}}p(x) \,\ln p(x)+\lambda_{0}\left(\mathrm{E}_{p}[1]-1\right)+\sum_{i=1}^{m}\lambda_{i}\left(\mathrm{E}_{p}[f_{i}(x)]-\mu_{i}\right)
\end{equation}
The first order condition (functional differentiation with respect to $p(x)$) gives \cite{Jaynes}

\begin{equation}\label{MaxentDist}
  p(x)=\mathcal{Z}^{-1}\exp\left(\sum_{i=1}^{m}\lambda_{i}f_{i}(x)\right)
\end{equation}
where $\mathcal{Z}=\exp(1-\lambda_{0})$. Knowing the expected values of some functions of the state of the system, we are able to derive the probability distribution of the explanatory variable.

The most used distributions are maxent distributions \footnote{a maxent distribution is a distribution with maximum entropy which satisfies the constraints reflecting our knowledge of the system.} as the Gaussian distribution for instance. Assume that our knowledge is restricted to the empirical mean $\mu$ and variance $\sigma^{2}$ of the random variable $X$. The constraints are $p(x)\geq0$, $\mathrm{E}_{p}[1]=1$, $\mathrm{E}_{p}[X]=\mu$ and $\mathrm{E}_{p}[\left(X-\mu\right)^{2}]=\sigma^{2}$. Therefore, we have  $p(x)=\mathcal{Z}^{-1}\exp\left(\lambda_{1}x+\lambda_{2}(x-\mu)^{2}\right)$. The constraint $\mathrm{E}_{p}[X]=\mu$ implies $\lambda_{1}=0$, the constraint $\mathrm{E}_{p}[\left(X-\mathrm{E}_{p}[X]\right)^{2}]=\sigma^{2}$ implies $\lambda_{2}=-(2\sigma^{2})^{-1}$ and the normalization $\mathrm{E}_{p}[1]=1$ implies $\mathcal{Z}=\sqrt{2\pi}\sigma$. We emphasize that maxent models are much more that distributions consistent with some moments. They can reproduce complex structures and capture collective behaviours encountered in complex systems \cite{Fischer,Schneidman}, optimization and probability \cite{Talagrand} but also error-correcting codes \cite{Nish}, etc.
Last, the MEP finds its foundation in the large deviation theory \cite{Ellis,TouchetteLDT}, indeed (\ref{3-Pairmaxent}) is derived only using statistical considerations and gives the most likely way to describe unlikely events given a certain knowledge of the system (see appendix \ref{maxentLD}).

Throughout this work, we use the MEP to find the distribution of the configurations $\mathbf{s}\equiv(s_{1},\ldots,s_{N})$ where each $s_{i}$ is a binary variable. If we observe only the first and second moments, the MEP reads:

\begin{eqnarray}\label{3-Pairmaxent}
  & & \max_{\substack{\{p(x)\}}}   S[p(\textbf{s})]= \max_{\substack{\{p(x)\}}}  \left\{-\sum_{\{\textbf{s}\}}p(\textbf{s}) \,\ln p(\textbf{s}) \right\} \\ \nonumber
   &\mathrm{s.t}&  \sum_{\{\textbf{s}\}}p(\textbf{s})=1,\quad \sum_{\{\textbf{s}\}}p(\textbf{s})s_{i}=q_{i},
   \quad \sum_{\{\textbf{s}\}}p(\textbf{s})s_{i}s_{j}=q_{ij} \nonumber
\end{eqnarray}
Using Lagrange multipliers method, the resulting  two-agent distribution $p_{2}(\textbf{s})$ is given by

\begin{equation}\label{3-PairmaxentDist}
p_{2}(\textbf{s})=\mathcal{Z}^{-1}\exp\left(\frac{1}{2}\sum_{i, j}^{N}J_{ij}s_{i}s_{j}+\sum_{i=1}^{N}h_{i}s_{i}\right)\equiv\frac {e^{- \mathcal{H}(\textbf{s})}}{\mathcal{Z}}
\end{equation}
where $J_{ij}$ and $h_{i}$ are the Lagrange multipliers.

As we will see, an interpretation of such models can be given a posteriori. Moreover such models bring information about collective behaviours, structural reorganization which is crucial for portfolio management, the possible observation of crash precursors and may provide highlights on how a shock spreads and if it will lead to a crash. The natural continuation of the present work can be the study of such a mechanism.

\begin{figure}[!ht]
\begin{center}
\resizebox{0.75\textwidth}{!}{%
 \includegraphics{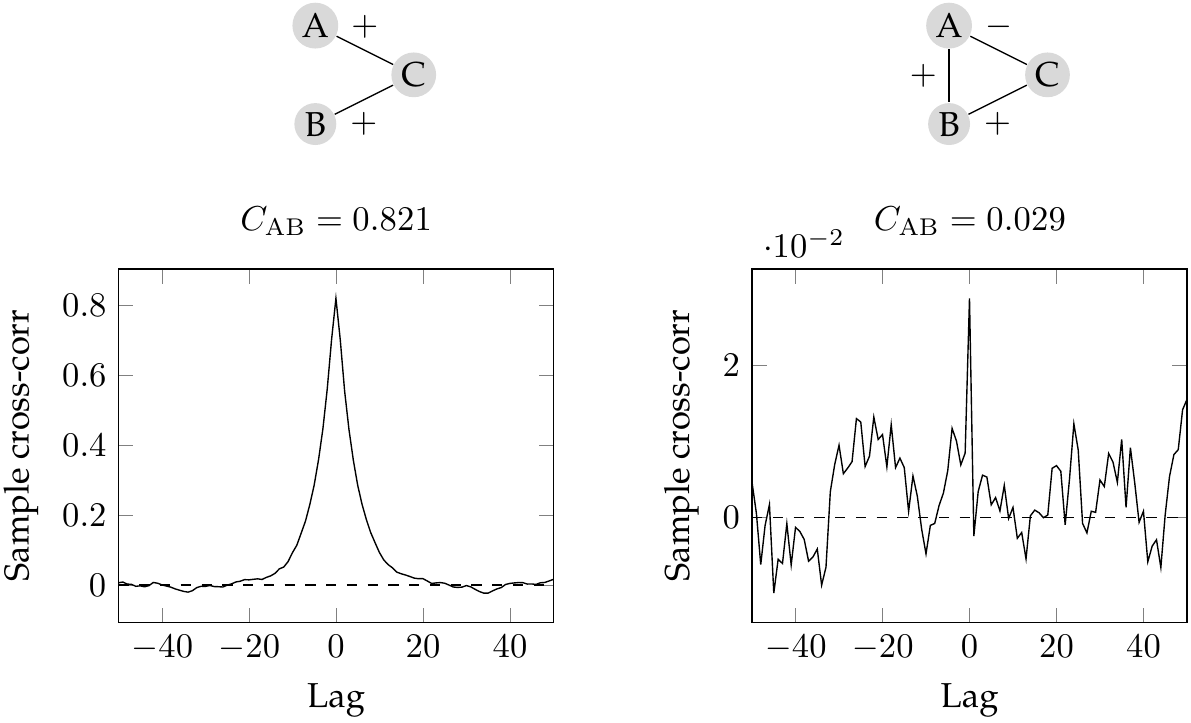}
}
\end{center}
\caption{Correlations induced by common influences.}
\label{fig:Jcorr}
\end{figure}

The parameters appearing in a maxent model provide an attractive alternative to the correlation coefficients which are extensively used in many fields. In portfolio theory for instance, an investor seeks to maximize the expected return for a given level of risk. The variance of the portfolio\footnote{A portfolio is a linear combination of assets.} is a function of the correlation coefficients between the considered assets. However, correlation coefficients are a measure of linear (or monotonic) statistical dependencies, have a significant noisy part \cite{Laloux}, are not appropriate when some assets are \emph{conditionally} independent (partial correlations are better than correlations but entropy is an even more appropriate measure of statistical dependencies).
As an illustration of these features, we give a basic example. Let's consider a financial network of three assets $\{A,B,C\}$. In the first case, $A$ and $B$ influence each other only via the third asset $C$ (one says that $A$ and $B$ are conditionally independent). The influences between the pairs $A-C$ and $B-C$ are both positive. Even if $A$ and $B$ are conditionally independent, we observe a large coefficient correlation, see Fig-\ref{fig:Jcorr}. In the second case, there is a true statistical dependency between all the pairs $A-B$, $A-C$, $B-C$ but one of them is negative. It results a very low cross-correlation.

To see if a pairwise maxent model performs better than simple covariances, we simulate a binary timeseries of three assets with the true mutual influence matrix:

$$
\textbf{J}_{\mathrm{true}}=
\begin{pmatrix*}
    0 & 0 & 2 \\
    0 & 0 & 2 \\
    2 & 2 & 0
\end{pmatrix*}
$$
The simulation returns ($1\times 10^{5}$ Monte Carlo steps, see sec \ref{sec3:MC})  empirical correlations and mutual influence matrices:

$$
\textbf{C}_{\mathrm{emp}}=
\begin{pmatrix*}
    1.00 & 0.94 & 0.97 \\
    0.94 & 1.00 & 0.96 \\
    0.97 & 0.96 & 1.00
\end{pmatrix*}
\qquad
\textbf{J}_{\mathrm{emp}}=
\begin{pmatrix*}
    0.00 & 0.00 & 2.00 \\
    0.02 & 0.00 & 2.00 \\
    2.00 & 2.00 & 0.00
\end{pmatrix*}
$$
The $\textbf{J}_{\mathrm{emp}}$ is estimated by a direct minimization of the entropy (see (\ref{3-Pairmaxent})), therefore is supposed to be "exact" at the given numeric precision and finite size sample. For the second case the simulations are performed with

$$
\textbf{J}_{\mathrm{true}}=
  \begin{pmatrix*}
    0  & 1 &-2 \\
    1  & 0 & 1 \\
    -2 & 1 & 0
  \end{pmatrix*}
$$
then it comes

$$
\textbf{C}_{\mathrm{emp}}=
  \begin{pmatrix*}
    1.00 & 0.06 & -0.87 \\
    0.06 & 1.00 & 0.06 \\
    -0.87 & 0.06 & 1.00
  \end{pmatrix*}
\qquad
\textbf{J}_{\mathrm{emp}}=
  \begin{pmatrix*}
    0.00  & 0.99 & -2.01 \\
    0.99  & 0.00 & 1.00 \\
    -2.01 & 1.00 & 0.00
  \end{pmatrix*}
$$

The parameters of a maxent models are a better measure of statistical dependencies than the correlation coefficients, or even partial correlations because the entropy also captures non-monotonic statistical dependencies.

These features could be interesting in assets selection and hedging (positions taken to offset losses). We will see in chap-\ref{chap:marketStruct} that the pairwise maxent model allows the identification of financial sectors (as clusters in a financial network).

\section{Relation of maxent models to other approaches}\label{sec3:Relation}

Maxent models are related to other models encountered in many disciplines. Among them, graphical models and network theory are interesting related approaches. Suppose that one considers a complicated probabilistic system which is modelled by a Markov random field (a set of random variables on an undirected graph having a kind of conditional independence). We label the state of node $i$ by $s_{i}$. The probability of a configuration $(s_{1}, \cdots, s_{N})$ is written without loss of generality

\begin{equation}\label{GM}
  p(s_{1}, \cdots, s_{N})=\frac {e^{- \mathcal{H}(s_{1}, \cdots, s_{N})}}{\mathcal{Z}}
\end{equation}

\begin{figure}[!ht]
\begin{center}
\includegraphics[width=\textwidth]{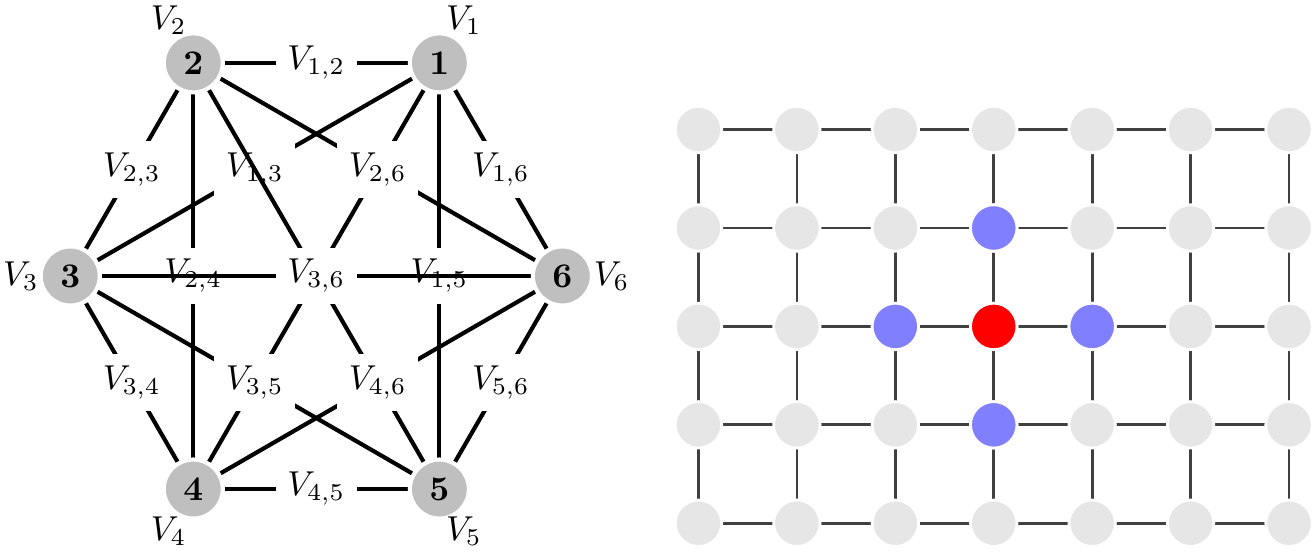}
\end{center}
\caption[Markov networks]{\label{fig:PMRF} \textbf{Left}: a complete graph approximated by unary and binary potentials. \textbf{Right}: a truly pairwise Markov network, the size of maximal clique (subset of nodes in which every node is connected to every other node) is equal to 2. One calls such a network a \emph{Markov} network because given the blue vertices, the red vertex is independent of all other nodes (there is no path from the red vertex to a grey vertex avoiding blue vertices). }
\end{figure}

If random variables are believed to be mutually dependent, but only through the combination of successive local interactions, we can factorize this distribution (\ref{GM}) (up to second order) as

\begin{equation}\label{GM2}
  p(s_{1}, \cdots, s_{N})=\frac {1}{\mathcal{Z}}
  \prod_{i}\phi_{i}({s_{i}})\prod_{(i,j)}\phi_{ij}(s_{i},s_{j})=\frac {1}{\mathcal{Z}}
  e^{- \sum_{i}V_{i}(s_{i})-\sum_{i<j}V_{ij}(s_{i},s_{j})}
\end{equation}
where $(i,j)$ denotes pair of vertices. Graphically speaking, it means that the network (social, financial, neural, etc.) is approximated by independent nodes ($V_{i}(s_{i})$) and pairwise dependence ($V_{ij}(s_{i},s_{j})$) as illustrated in Fig-\ref{fig:PMRF}. This approach is used in belief propagation, image denoising, magnetic materials, disease spreading, etc.

One recovers the pairwise maxent distribution (\ref{3-PairmaxentDist}) if we set $V_{i}(s_{i})=h_{i}s_{i}$ and $V_{ij}(s_{i},s_{j})=J_{ij}s_{i}s_{j}$. Therefore, maxent models provide a way to characterize the underlying financial network. This analogy can be useful especially in migration model like the Schelling segregation Model \cite{Schelling}. The relation between maximum entropy and graphical models motivates a topological approach of financial networks

\section{Entropy}\label{sec3:entropy}
\subsection{Statistical meaning}
One encounters the entropy in almost any fields relying on statistics and probability theory (computer science, physics, neuroscience, communication, finance, economics, etc.). An extensive discussion about entropy and economic modeling can be found in \cite{Aoki}.
To avoid any misunderstanding, we will always refer to its mathematical definition. The entropy is a functional of probability mass function which is intended to be a measure of the average uncertainty or average surprise/likelihood. Formally, the entropy of a discrete random variable $X$ with a probability mass function $p(x)$ (noted $S[p(x)]$ or $S(X)$)  is

\begin{equation}\label{entDef}
  S[p(x)]=-\sum_{x}p(x)\ln p(x)
\end{equation}
We can rewrite the entropy as the average self-information on the random variable $X$: $\mathrm{E}[\ln(1/p(x))]$ where $\ln(1/p(x))$ is the measure of self-information (satisfying additivity, monotonicity and positiveness). A consequence of this definition is that entropy is maximal for uniform distributions. Moreover, we have $S[p(x)]\geq0$. The bivariate expression is

\begin{equation}\label{entMVdef}
  S[X,Y]=-\sum_{x,y}p(x,y)\ln p(x,y)
\end{equation}

\begin{figure}[!ht]
\begin{center}
\includegraphics[width=0.75\textwidth]{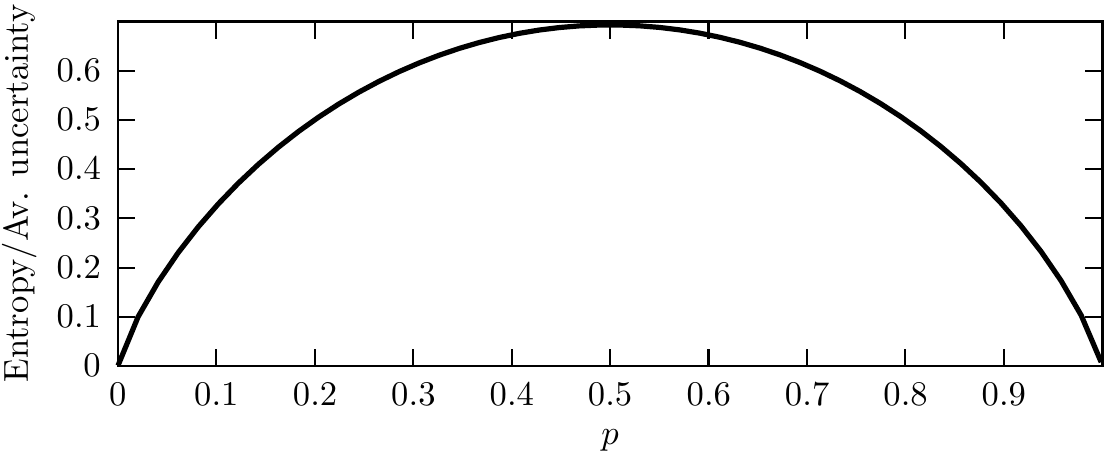}
\end{center}
\caption[Entropy of a coin toss]{\label{fig:entToss} Entropy or average uncertainty of a coin toss for each possible value of head probability.}
\end{figure}

To illustrate  these features, let's consider a simple example. Suppose that you bet ($1:1$ odds) on a coin toss. If the coin is fair, the probability to get head is $p=0.5$. In this situation, intuitively the uncertainty about the toss result is maximal. You have no propensity to bet on head rather than on tail. However if the coin is not fair, say that the probability of head event is $p=0.8$, you may want to play forever and always bet on head to make some money on long term. The uncertainty about the toss result is lower than in the fair toss. It is what entropy measures, uncertainty about the result of a random experiment. Indeed uncertainty about an event is null if this event is sure ($p=1$) and maximal when all events can occur with the same probability. The entropy is defined to capture these properties. We illustrate the entropy of a coin toss in Fig-\ref{fig:entToss} for all possible values of $p$\footnote{By convention $0\ln 0\equiv0$}.

\subsection{Combinatoric meaning}\label{sec3:combinatoric}

The entropy can also be thought as a counting function \footnote{Historically, the entropy was defined as a counting function. Although useful, this approach is somewhat old fashioned and the statistical definition is more versatile and more powerful.} and therefore as a measure of diversity \cite{Aoki}. Indeed, the maximum number of outcomes in a combinatoric problem is equivalent to the Shannon entropy. Consider $N$ repeats of a random experiment with $K$ outcomes of probability $p_{k}$ where $k=1,\ldots,K$. The number of outcomes is $\Omega= \frac{N!}{(Np_{1})!\cdots (Np_{K})!}$. Using Stirling's approximation $\ln N!\simeq N\ln N-N$, one gets $\ln\Omega\simeq-N\sum_{i=1}^{K}p_{i}\ln p_{i}$. In particular, one can count the number of microstates corresponding to a given value $U$ of the utility function $\mathcal{U}(\mathbf{s})$ (where $\mathbf{s}$ is a configuration, eg: a vector of choices): $\Omega=\Omega(U)$. If the utility is a continuous variable, one uses the cumulative counting function: $\Sigma(U)=\sum_{\{\mathbf{s}\}}\theta (\mathcal{U}(\mathbf{s})-U)$ (the number of configurations having a utility smaller or equal to a given value $U$) and the associate density $\rho(U)=\sum_{\{\mathbf{s}\}}\delta (\mathcal{U}(\mathbf{s})-U)$. In this case $S(U)\simeq \ln \rho(U)$. The "function" $\theta(\cdot)$ is the Heaviside function (unit step function) and $\delta(\cdot)$ is the Dirac delta, the derivative of the Heaviside function (as defined in the theory of distributions).

To picture the interpretation of such a counting function, assume that we observe an economic system composed of distinguishable agents, say: $A$, $B$, $C$, $D$. They must make a choice between two resources: $\$$ or $\pounds$. There are $2^4$ configurations, each of them being a \emph{microstate} corresponding to one of the $5$ macrostates (all agents choose $\$$, three agents choose $\$$, etc.). It is the classical occupancy problem. The $5$ macrostates and the corresponding microstates are depicted in the following table:

\begin{center}
\begin{tabular}{cccc}
\toprule
  resource $\$$ & resource $\pounds$ & number of microstates & probability \\ \midrule
  $ABCD$ &  & $1$ & $1/16$ \\ \hline
  $ABC$ & $D$ &  &  \\
  $ABD$ & $C$ & $4$ & $4/16$ \\
  $ACD$ & $B$ &  &  \\
  $BCD$ & $A$ &  &  \\  \hline
  $AB$ & $CD$ & &  \\
  $AC$ & $BD$ & &  \\
  $AD$ & $BC$ & $6$ & $6/16$  \\
  $BC$ & $AD$ & &  \\
  $BD$ & $AC$ & &  \\
  $CD$ & $AB$ & &  \\  \hline
  $A$ & $BCD$ &  &  \\
  $B$ & $ACD$ & $4$& $4/16$ \\
  $C$ & $ABD$ &  &  \\
  $D$ & $ABC$ &  &  \\ \hline
   & $ABCD$ & $1$ & $1/16$ \\
\bottomrule
\end{tabular}
\end{center}

If the configurations are uniformly distributed, the most probable state of such a system is $2$ agents choose the first resource and the $2$ other agents choose the second resource. It is the most diversified state since they are $6$ ways to get it.

The maximum entropy can thus be thought as a maximization of the economic diversity subjected to some constraints (resource allocation, capital repartition, etc.). For a complete discussion of the combinatoric approach, see \cite{Aoki}.
\subsection{Kullback-Leibler divergence}
Another related useful quantity is the Kullback-Leibler divergence (KLD), also called relative entropy, between two distributions $p(x)$ and $q(x)$:

\begin{equation}\label{KLDef}
  D_{\mathrm{KL}}(p||q)=\sum_{x}p(x)\ln \frac{p(x)}{q(x)}
\end{equation}
This quantity can be thought as a measure of distance in the functional space of distribution (even if it is not a \emph{metric}). The KL-divergence is encountered in any field using inference method. Indeed, the KL-divergence between a continuous parameterized candidate $f(x;\boldsymbol{\theta})$  and empirical density $p_{\mathrm{emp}}(x)$ ($N$ IID samples and regularity conditions assumed) is

\begin{eqnarray}
  D_{\mathrm{KL}}(p_{\mathrm{emp}}||f(x;\boldsymbol{\theta})) &=& \int_{-\infty}^{\infty} \sum_{i=1}^{N}\frac{\delta(x-x_{i})}{N}
  \ln \left(\frac{N^{-1}\sum_{i=1}^{N}\delta(x-x_{i})}{f(x;\boldsymbol{\theta})}\right) \mathrm{d}x\\
   &=& \frac{1}{N}\sum_{i=1}^{N}\ln\left(\frac{N^{-1}}{f(x_{i};\boldsymbol{\theta})}\right) \\
   &=& -\ln N-\frac{1}{N}\ln L(\boldsymbol{\theta})\label{3-DKLlik}
\end{eqnarray}
So minimize the KL-divergence is equivalent to maximize the likelihood $ L(\boldsymbol{\theta})$ and maximize the entropy is equivalent to maximize the likelihood of the distribution $p$ closest to the uniform distribution $U$ without range restriction since $D_{\mathrm{KL}}(p||U)=-S[p]$ up to a constant. Furthermore, the Fisher information metric is the Hessian matrix of the KL-divergence, measuring the curvature of the log-likelihood and thus the information content of the distribution about the parameter $\theta$. Furthermore, it is possible to show that the Fisher information matrix is the Hessian matrix of the KLD \cite{Cover}.

There are several interpretations of the KLD (geometric, statistical, etc). For our purpose, we can interpret it as the measure of expected difference between the true and approximated utility functions. Let's say that the data are generated by the true distribution (and the true utility function) $p(\mathbf{s})=\mathcal{Z}(\beta)^{-1}\exp \beta\mathcal{U}(\mathbf{s})$ and that we infer a model from the data using the MEP $p_{\mathrm{ME}}(\mathbf{s})=\mathcal{Z}_{\mathrm{ME}}(\beta)^{-1}\exp \beta\mathcal{U}_{\mathrm{ME}}(\mathbf{s})$. The KLD is rewritten up to a constant as $\mathrm{E}\left[\mathcal{U}-\mathcal{U}_{\mathrm{ME}}\right]$ where $\mathrm{E}\left[\cdot\right]$ is the expectation with respect to the true distribution. The minimization of the KLD will therefore provide the less misspecified model given a partial knowledge of the system.

\begin{figure}[!ht]
\begin{center}
\includegraphics[width=0.75\textwidth]{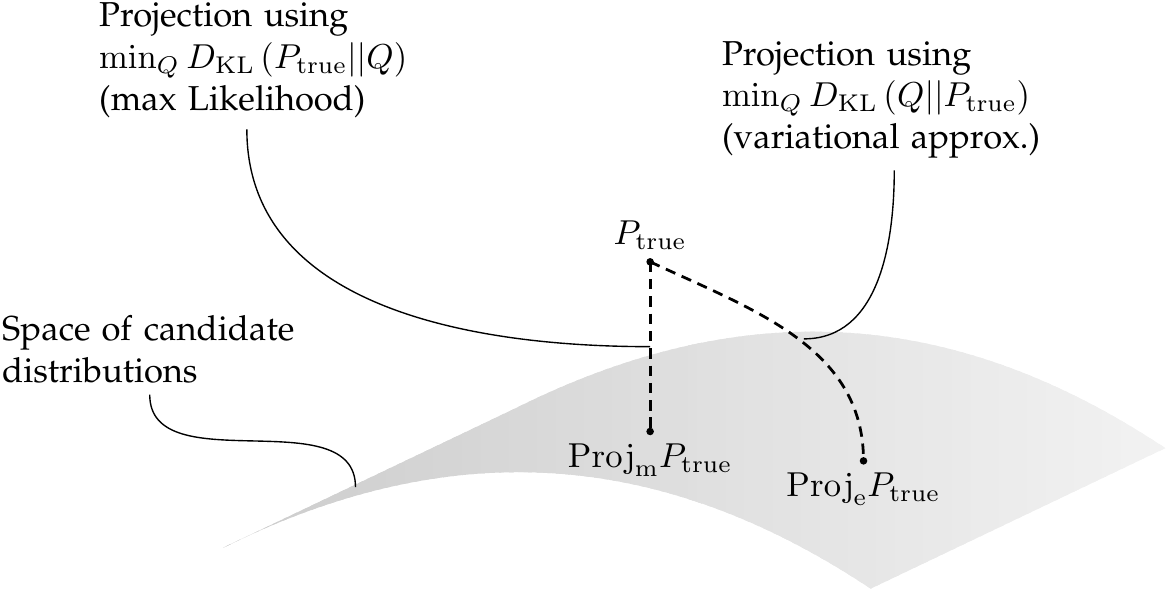}
\end{center}
\caption[Projection and KLD]{\label{fig:projKLD} Geometrical view of the KLD minimization. The maximum likelihood method provides the so-called mixture projection ($\mathrm{Proj}_{\mathrm{m}}P_{\mathrm{true}}$). The minimization of the KLD with distribution transposed gives the so-called exponential projection ($\mathrm{Proj}_{\mathrm{e}}P_{\mathrm{true}}$).}
\end{figure}

For practical use, the maximum likelihood method is not always feasible and it is sometimes easier to deal with $D_{\mathrm{KL}}(f(x;\boldsymbol{\theta})||p_{\mathrm{true}})$ than with $D_{\mathrm{KL}}(p_{\mathrm{true}}||f(x;\boldsymbol{\theta}))$. The minimization of the former KLD will provide a tractable approximated distribution. Even if the maximum likelihood is the best projection, it is not always tractable due to the exponential number of configurations $2^{N}$. As the KLD is a non-symmetric measure of dissimilarity, there are two kinds of projections on the space of candidate distributions (a parametric family). This idea is illustrated in Fig-\ref{fig:projKLD}, see \cite{AmariBook} for details.

\section{Variational methods}\label{sec3:Var}
\subsection{Lower bound}
We have seen that the Kullback-Leibler divergence is a useful tool for statistical inference. Here, we explain that this measure of dissimilarity can also be used to set up variational methods deriving from the minimization between the true (but sometimes intractable) distribution $p$ and an approximated tractable distribution $q$. For a Gibbs distribution $p(\textbf{s})=\mathcal{Z}_{p}^{-1}\exp\left(-\mathcal{H}(\textbf{s})\right)$, KL-divergence is written as

\begin{equation}\label{VarPrinc}
D_{\mathrm{KL}}(q||p)=\mathrm{E}_{q}[\mathcal{H}(\textbf{s})]-S[q(\textbf{s})]+\ln \mathcal{Z}_{p} \geq 0
\end{equation}
where $\mathrm{E}_{q}[\cdot]$ is the expectation with respect to the distribution $q$.

The variational minimization of the KL-divergence is equivalent to the minimization of the following functional $\mathcal{F}[q(\textbf{s})]=\mathrm{E}_{q}[\mathcal{H}(\textbf{s})]-S[q(\textbf{s})]$ since $\ln \mathcal{Z}_{p}$ is a normalizing constant depending only on the true distribution $p$. Moreover as the KL-divergence is positive (or equal to zero) one gets the variational bound $\mathcal{F}[q]\geq-\ln \mathcal{Z}_{p}$.
The statistical foundation of variational methods allows their applications in any fields, not only in physics (where $\mathcal{F}[q]$ is called the \emph{variational free energy}). The $\mathcal{F}$-functional is particulary important because any cumulant, denoted by $\langle\cdot\rangle_{\mathrm{c}}$, of the pairwise maxent model (\ref{Lagrange}) is given by

\begin{equation}\label{CumAve}
\langle s_{i_{1}}\ldots s_{i_{N}}\rangle_{\mathrm{c}}=\partial^{N}\ln \mathcal{Z}/\partial h_{i_{1}}\ldots \partial h_{i_{N}}
\end{equation}
The exact $\mathcal{F}$-functional (or $\ln\mathcal{Z}$ since $\mathcal{F}[p]=-\ln\mathcal{Z}$) is thus the cumulant-generating function.

Another useful formulation \cite{Barb} consists to rewrite the approximated distribution (without loss of generality) as $q(\textbf{s})=\mathcal{Z}^{-1}_{q}\exp\left(-\mathcal{H}_{q}(\textbf{s})\right)$. The lower bound of the minimization (\ref{VarPrinc}) becomes $-\ln \mathcal{Z}_{q}+\mathrm{E}_{q}[\mathcal{H}(\textbf{s})-\mathcal{H}_{q}(\textbf{s})]\geq -\ln \mathcal{Z}_{p}$ or equivalently $\mathrm{E}_{q}[\mathcal{H}_{q}(\textbf{s})]-S[q(\textbf{s})]+\mathrm{E}_{q}[\mathcal{H}(\textbf{s})-\mathcal{H}_{q}(\textbf{s})]\geq -\ln \mathcal{Z}_{p}$. It is equivalent to approximate the exact (but untractable) partition function or the exact $\mathcal{F}$-functional.

%
%

\subsection{Variational approximation}

Unfortunately, most of time $\ln\mathcal{Z}$ can not be exactly computed except when entities are independent (no mutual influences). A possible way to get a tractable approximation is to expand $\mathcal{F}=-\ln\mathcal{Z}$ "around" the independent model \cite{Georges,Plef}. The trick is to weight the magnitude order of the co-movement strengths by considering $p(\textbf{s};\alpha)=\mathcal{Z}(\alpha)^{-1}\exp\left(\alpha\sum_{ij}J_{ij}s_{i}s_{j}+\sum_{i}h_{i}s_{i}\right)$ (the true distribution $p(\textbf{s})$ is recovered when $\alpha=1$) and then minimize the KL-divergence with defined values for the first moment $\mathrm{E}_{\alpha}[s_{i}]=q_{i}$. The independent variables $q_{i}$ are explicitly introduced by inverting  $\mathrm{E}_{\alpha}[s_{i}]=q_{i}$ or equivalently by taking the Legendre transform of the $\mathcal{F}$-functional. One expands the functional $\mathcal{G}(\alpha,\{q_{i}\})=-\ln\mathcal{Z}+\sum_{i}h_{i}(\alpha)q_{i} $ in the domain of almost independent entities where $\alpha$ is close to zero:

\begin{equation}
\mathcal{G}(\alpha)=\mathcal{G}(0)+\frac{\partial \mathcal{G}}{\partial\alpha}\Big|_{\alpha=0}\alpha
+\frac{\partial^{2}\mathcal{G}}{\partial\alpha^{2}}\Big|_{\alpha=0}\frac{\alpha^{2}}{2!}
+\mathcal{O}(\alpha^{3})
\end{equation}

Noting that $q_{i}=\frac{\partial\ln\mathcal{Z}}{\partial h_{i}}$ and $h_{i}=\frac{\partial\mathcal{G}}{\partial q_{i}}$, we have:

\begin{eqnarray}
  \mathcal{G}(0) &=& \frac{1}{2}\sum_{i}\left[(1+q_{i})\ln(\frac{1+q_{i}}{2})+
    (1-q_{i})\ln(\frac{1-q_{i}}{2})\right]\\
  \frac{\partial \mathcal{G}}{\partial\alpha}\Big|_{\alpha=0} &=&
  -\frac{1}{2}\sum_{i,j}J_{ij}q_{i}q_{j} \\
  \frac{\partial^{2}\mathcal{G}}{\partial\alpha^{2}}\Big|_{\alpha=0} &=&
  -\frac{1}{2}\sum_{i,j}J_{ij}^{2}(1-q_{i}^{2})(1-q_{j}^{2})
\end{eqnarray}

If the $J_{ij}$ are drawn from the Gaussian distribution, one can show that terms beyond the second order can be neglected when the size $N$ tends to infinity \cite{Plef}. So minimize the KL-divergence between the tractable and true distributions is equivalent to minimize the following functional (up to second order)

\begin{equation}
\begin{split}\label{GTAP}
    \mathcal{G}_{\mathrm{2nd}}&=\frac{1}{2}\sum_{i}\left[(1+q_{i})\ln(\frac{1+q_{i}}{2})+
    (1-q_{i})\ln(\frac{1-q_{i}}{2})\right]\\
    &-\frac{1}{2}\sum_{i,j}J_{ij}q_{i}q_{j}
    -\frac{1}{4}\sum_{i,j}J_{ij}^{2}(1-q_{i}^{2})(1-q_{j}^{2})
    +\mathcal{O}(\alpha^{3})
\end{split}
\end{equation}

The third order reads

\begin{flalign}
 \frac{\partial^{3}\mathcal{G}}{\partial\alpha^{3}}\Big|_{\alpha=0} =
  -2\sum_{ij}J_{ij}^{3}q_{i}q_{j}(1-q_{i}^{2}) (1-q_{j}^{2})
  -\sum_{i,j,k}J_{ij}J_{jk}J_{ki}(1-q_{i}^{2}) (1-q_{j}^{2}) (1-q_{k}^{2})
\end{flalign}
This expansion can be continued to arbitrary high order, see \cite{Georges} for explicit derivation of higher orders and a diagrammatic formulation. This functional is particulary important for the inverse problem (finding the parameters from data) but also in approximated description of equilibria.

This expansion is valid if the radius of convergence $\rho$ is such that $\rho>1$ \cite{Plef}. To study $\rho$, we use the property stating that $\partial\mathcal{G}/\partial\alpha$ has the same  radius of convergence of $\mathcal{G}(\alpha)$ which is useful since one knows an exact relation for the first derivative

\begin{equation}
\frac{\partial \mathcal{G}}{\partial\alpha} = -\frac{1}{2}\sum_{i,j}J_{ij}q_{i}q_{j}-\frac{1}{2\beta}\sum_{i,j}J_{ij}\chi_{ij}(\alpha)
\end{equation}
where the covariances are given by $\chi_{ij}(\alpha)=\left[\langle s_{i}s_{j}\rangle_{\alpha}-q_{i}q_{j}\right]$ and the expectation $\langle B\rangle_{\alpha}$ stands for \newline $\mathcal{Z}(\alpha)^{-1}\tr \left(B\exp(\alpha\sum_{ij}J_{ij}s_{i}s_{j}+\sum_{i}h_{i}s_{i})\right)$.

The radius of convergence is equal to the distance between $\alpha=0$ and the closest singularity of $\partial\mathcal{G}/\partial\alpha$. It is then equivalent to study the singular eigenvalues of $\boldsymbol{\chi}$. Following \cite{Plef}, one can use the resolvent $R(z,\alpha)=(\boldsymbol{\chi}^{-1}-zI)^{-1}$. The singularities of $R(z,\alpha)$ correspond to the eigenvalues of $\boldsymbol{\chi}^{-1}$. We are only interested by the null eigenvalues of $\boldsymbol{\chi}^{-1}$ ($z=0$ in the resolvent). If we take a circle $\gamma_{0}$ centered (counter clockwise) in $z=0$ but excluding any other singularity, the functional expansion is valid for any value of $\alpha$ if and only if

\begin{equation}
\frac{1}{2\pi i}\oint_{\gamma_{0}}z^{k}\,R(z,\alpha) \ud z=0 \qquad \text{for all}\quad k=0,1,\ldots,N-1
\end{equation}
because if the resolvent is a holomorphic function in the domain $\Gamma_{0}$ bordered by $\gamma_{0}$, this integral is equal to zero but the converse is not true. Moreover, we must check this feature for all the possible value of the multiplicity of the null eigenvalue. The Laurent series may include negative powers equal to the multiplicity $m$. As $\gamma_{0}$ does not encircle any other singularity but $z=0$, $z^{m}R(z,\alpha)$ is a holomorphic function in $\Gamma_{0}$ but $z^{m-1}R(z,\alpha)$ is not.

The expansion of $\mathcal{G}$ makes sense if this feature is true for any $\alpha \leq 1$. The validity is thus linked to the eigenvalues of the Hessian matrix of the KL-divergence. Last, we note that $\mathcal{G}$ is a convex functional of $q_{i}$ as the covariance matrix $\boldsymbol{\chi}$ is a positive semidefinite matrix and is equal to the inverse of the Hessian matrix $(\textbf{H}(\mathcal{G}))_{ij}=\partial^{2} \mathcal{G}/\partial q_{i}\partial q_{j}$.

\section{Most probable state and fluctuations}\label{sec3:prob}

Another useful maximization principle is the derivation of the most probable state, or the most probable utility. Up to now, one has the configuration distribution not the utility distribution. To derive this state, the partition function is rewritten

\begin{eqnarray}
  \mathcal{Z}&=& \sum_{\{\mathbf{s}\}}\exp\left(\mathcal{U}(\mathbf{s})\right) \\
             &=& \int\limits_{-\infty}^{+\infty}\ud U \left(\sum_{\{\mathbf{s}\}}\delta(U-\mathcal{U}(\mathbf{s}))\right)e^{U}\\
             &=& \int\limits_{-\infty}^{+\infty}\ud U \,\rho(U)\exp(U) \\
             &=& \int\limits_{-\infty}^{+\infty}\ud U \,e^{S(U)+U} \\
\end{eqnarray}
where $S(U)$ is the entropy at fixed utility \footnote{$\ln \rho(U)=S(U)$, see Sec-\ref{sec3:combinatoric}.} (or the \emph{incertitude}) $\ln \rho(U)$ and $\rho(U)$ is the state density. By analogy with the previous discussion the quantity $S(U)+U$ is called the $\mathcal{F}$-density. Taking quantities per agent $Ns(u)\equiv S(U)$ and $Nu=U$, the partition function becomes

\begin{equation}
  \mathcal{Z}=N\int\limits_{-\infty}^{+\infty}\ud u\, e^{N(s(u)+u)}
\end{equation}
and thus the probability density function of the utility is

\begin{equation}\label{FreeMin}
  p(u)=\frac{e^{N(s(u)+u)}}{\mathcal{Z}}=\rho(u)\frac{e^{N u}}{\mathcal{Z}}
\end{equation}
The most probable utility is the utility maximizing $f(u)\equiv s(u)+u$: a maximal incertitude $s(u)$ providing the higher utility $u$. The most probable state will correspond to the mean utility if the utility distribution is sharply peaked. Moreover for a Gibbs distribution, the spontaneous fluctuations are linked to the response function to a shock in the stochasticity level $R_{\mathcal{U}}(T)\equiv-\partial \langle\mathcal{U}\rangle/\partial T=T^{-2}\VAR{[\mathcal{U}]}$. It results that the fluctuations around the mean utility value are given by

\begin{equation}\label{flu}
  \frac{\sqrt{\VAR{[\mathcal{U}]}}}{\langle\mathcal{U}\rangle}=\frac{\sqrt{T^{2}R_{\mathcal{U}}(T)}}{\langle\mathcal{U}\rangle}
\end{equation}
Generally, this quantity scales as $N^{-1/2}$ and fluctuations around the mean utility are negligible for large systems. However, the latter statement is invalid in the vicinity of a particular value of the stochasticity level  where the response function per capita $R_{\mathcal{U}}(T)/N$ admits a vertical asymptote when $N\rightarrow\infty$ \cite{Fischer}. If the system is large and, loosely speaking, if the system is strongly ordered (or disordered), one can take the quadratic approximation of $f(u)$ around the most probable state $\bar{u}$. It comes $f(u)\simeq f(\bar{u})-2^{-1}T^{-2}r_{u}^{-1}(u-\bar{u})^{2}$ where $r_{u}=N^{-1} R_{\mathcal{U}}$. The utility pdf is rewritten

\begin{equation}\label{FreeMinApprox}
  p(u)=p(\bar{u})\exp\left(\frac{-N(u-\bar{u})^{2}}{2T^{2}r_{u}}\right)
\end{equation}
which is a Gaussian pdf of mean $\bar{u}$ and variance $N^{-1}T^{2}r_{u}$.

\section{Testing the order of maxent models}\label{sec3:TestMEP}

An important issue is to determine which order we should keep in statistical modeling. A possible test is the multi-information criterion \cite{Schneid_Multi}. In the following, we sketch this method.

\begin{figure}[!ht]
\begin{center}
\includegraphics[width=0.35\textwidth]{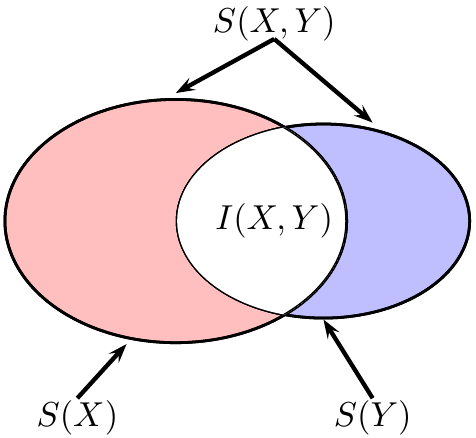}
\end{center}
\caption[Mutual information]{\label{fig:MutInf} Schematic representation of individual (marginal), joint entropies and mutual information.}
\end{figure}

Under assumptions of additivity, continuity and monotonicity, one can show \cite{Cover} that the measure of statistical interdependence of two random variables ($X$ and $Y$) is the so-called mutual information $I(X,Y)$ which reads

\begin{equation}\label{MutInf}
  I(X,Y)=D_{\mathrm{KL}}\left(p(x,y)||p(x)p(y)\right)=\mathrm{E}_{p(x,y)}\left[\ln\frac{p(X,Y)}{p(X)p(Y)}\right]
\end{equation}
The mutual information measures the distance between the joint distribution and the product of marginals (statistical independence), in other words it measures the amount of information that one random variable embeds about another one (reduction in the uncertainty due to the knowledge of the other random variable).  If $X$ and $Y$ are independent then $I(X,Y)=0$ because the knowledge of $X$ does not tell anything about $Y$. Indeed, one can rewrite $I(X,Y)=S(X)+S(Y)-S(X,Y)$ which is schematically illustrated in Fig-\ref{fig:MutInf}.

To illustrate its power as a measure of statistical dependency, let's consider a classical example. Let $X$ and $Y= X^2$ be two random variables. Obviously, $Y$ is a function of $X$ and only $X$. A good measure of statistical dependencies should return "$1$" (perfect dependency between the two random variables). However, the correlation coefficient $\mathrm{corr}(X,Y)$ is theoretically equal to zero\footnote{Taking the geometrical interpretation, $\mathrm{corr}(X,Y)$ is the inner product between two vectors. Therefore its is a measure of the angle between the random variables. The symmetry of the dependence induces the result.}. The joint entropy $S(X,Y)$ measure the information that $X$ and $Y$ share and should therefore be zero in this example or the redundancy $R(X,Y)\equiv \left[S(X)+S(Y)\right]^{-1}I(X,Y)$ should be equal to one.

\begin{figure}[!ht]
\begin{center}
\includegraphics[width=0.5\textwidth]{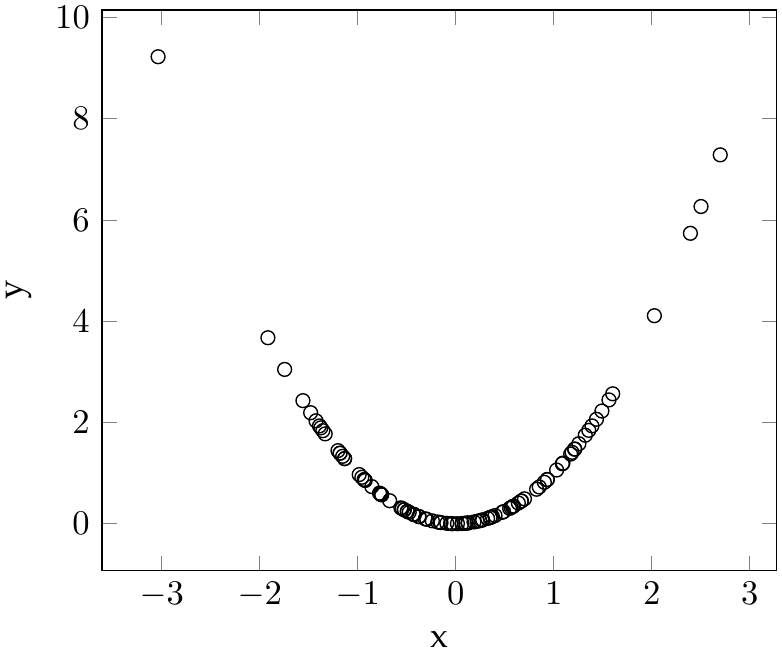}
\end{center}
\caption[Statistical dependencies]{\label{fig:MIcorrRV} Typical statistical dependency not detected by the correlation coefficient. The correlation coefficient is equal to $0.11$ (very small correlation) and the redundancy is equal to $0.83$ (large fraction of shared information).}
\end{figure}

For $100$ realizations of a Gaussian random variable $X$ and taking $Y=X^2$, it returns
$\mathrm{corr}(X,Y)=0.11$ and $R(X,Y)=0.83$. Therefore the knowledge of $X$ (or $Y$) is equivalent to the knowledge of $Y$ (or $X$) as illustrated in Fig-\ref{fig:MIcorrRV}.

More generally, the multi-variate definition of the multi-information is

\begin{equation}\label{MultiInf}
  I(\{X_{i}\})=D_{\mathrm{KL}}(p(x_{i},\cdots,x_{N})||p(x_{1})\cdots p(x_{N}))=\sum_{i=1}^{N}S(X_{i})-S(X_{1},\cdots,X_{N})
\end{equation}
We note that $I(\{X_{i}\})$ is greater than or equal to zero since the presence of correlations decreases the entropy.
The main idea of the multi-information criterion is to decompose the multi-information (MI) into a sum of the entropies of successive marginals:

\begin{equation}\label{MultiInfCon}
  I(\{X_{i}\})=\sum_{i=1}^{N}S(X_{i})-S(X_{1},\cdots,X_{N})=\sum_{k=2}^{N}I_{\mathrm{C}}^{k}(\{X_{i}\})
\end{equation}
where the connected information of order $k$ is the difference between entropies of the $k-1$th and $k$th order marginals

\begin{equation}\label{ConnectedInf}
I_{\mathrm{C}}^{k}(\{X_{i}\})=S[p_{k-1}]-S[p_{k}]
\end{equation}
The connected information represents the amount by which the maximal possible value of the entropy decreases when we include the $k$th order marginal in the description \cite{Schneid_Multi}.
Using these quantities, one can build a test to know which order one should include in a maxent model.
To test if (say) a pairwise correlation model explains satisfactorily data statistics, one evaluates the ratio between $S(p_{1})-S(p_{2})$ and the Kullback-Leibler discrepancy $I_{N}\equiv D_{\mathrm{KL}}\left(p_{N}||p_{1}\right)$, where $S(p_{2})$ is the entropy of the pairwise model. If this ratio is close to 1, the pairwise correlations explain most of available information. As we saw, the multi-information $I_{N}=S(P_{1})-S(P_{N})$ measures the total amount of statistical dependencies in the system.

Last, we build a benchmark. We compute the multi-information criterion (MIC) for a truly binary pairwise maxent model

\begin{equation}
p_{2}(\textbf{s})=\mathcal{Z}^{-1}\exp\left(\frac{1}{2}\sum_{i, j}^{N}J_{ij}s_{i}s_{j}+\sum_{i=1}^{N}h_{i}s_{i}\right)\equiv\frac {e^{- \mathcal{H}(\textbf{s})}}{\mathcal{Z}}\label{Lagrange}
\end{equation}
where $J_{ij}$ and $h_{i}$ are Lagrange multipliers, $\mathcal{Z}$ a normalizing constant (the partition function) equal to $ \tr\exp(\sum_{ij}J_{ij}s_{i}s_{j}+\sum_{i}h_{i}s_{i})$\footnote{The trace operator $\tr$ denotes the sum over all configurations $\sum_{\{\textbf{s}\}}$} and $s_{i}=\pm1$ for all $i=1,\cdots,N$. These binary variable may be thought as buy/sell, bullish/bearish, etc. We simulate (see section \ref{sec3:MC}) samples of length $T$ with $J_{ij}=1$ if $i$ and $j$ are nearest neighbour on a square lattice and $J_{ij}=0$ otherwise (we set $h_{i}=0$). A configuration $\textbf{s}=(s_{1},\cdots,s_{N})$ is recorded each $N$ steps after an equilibration period of $1\times10^4$ rounds. We compute empirical relative frequencies of configurations, we estimate the independent and pairwise maxent distribution using a regularized pseudo-maximum likelihood method to infer Lagrange parameters (see section \ref{sec3:IsingInv}). The results are reported in Table-\ref{tab3:MICbench}.

\begin{table}[!ht]
\caption[Multi-information criterion]{Second order multi-information criterion $I_{2}/I_{N}$ computed for truly pairwise generated samples.}
\label{tab3:MICbench}
\begin{center}
\begin{tabular}{lcr}
\hline
N             & sample length ($T$) & MIC                                 \\ \hline
9             &  $5\times 10^{3  }$ & 0.9881                               \\
9             &  $1\times 10^{5  }$ & 0.9918                              \\
16            &  $5\times 10^{3  }$ & 0.9918                                \\
16            &  $1\times 10^{5  }$ & 0.9926                               \\
36            &  $1\times 10^{5  }$ & 0.9922                                \\
\hline
\end{tabular}
\end{center}
\end{table}

\section{Equilibrium}\label{sec3:eq}

If the constraints $\mathrm{E}_{p}[f_{i}(x)]=\mu_{i}$ in (\ref{3-MEP}) do not depend on time (stationary condition), Lagrange parameters should be constant. Therefore the Gibbs distribution (\ref{Lagrange}) is the equilibrium distribution. For a binary state, say yes/no choice, the mean consensus of agent $i$ is

\begin{equation}\label{EOM}
  \langle s_{i}\rangle=\Big\langle \tanh(\sum_{j}J_{ij}s_{j}+h_{i})\Big\rangle
\end{equation}
the right hand side (RHS) is in general untractable \cite{Baxter}. Several approximation schemes have been derived, some of which are explained hereafter.

\subsection{Variational approximation}

One can derive an approximation introducing $\mathcal{G}$ in (\ref{CumAve}). At the first order, we get

\begin{equation}\label{3-MF}
  \langle s_{i}\rangle=\tanh\left(\sum_{j}J_{ij}\langle s_{j}\rangle+h_{i}\right)
\end{equation}
at second order, it comes

\begin{equation}\label{3-TAP}
  \langle s_{i}\rangle=\tanh\left(\sum_{j}J_{ij}\langle s_{j}\rangle-
  \sum_{j}J_{ij}^{2}\langle s_{i}\rangle (1-\langle s_{j}\rangle^{2})
  +h_{i}\right)
\end{equation}
For Gaussian influences $J_{ij}$ and for large enough networks, the second order is the leading order \cite{Plef}.

\subsection{Expansion in terms of power of cumulants}

We propose a cumulant expansion of the averaged hyperbolic tangent \cite{moi3}. It will lead to interesting results. Indeed if we expand the averaged hyperbolic tangent up to third order (in what follows, we use the notation $X_{i}$ in place of $h_{i}^{\mathrm{eff}}=\sum_{j}J_{ij}s_{j}+h_{i}$  and $\langle\cdot\rangle_{\mathrm{c}}$ stands for the cumulant average), we get

\begin{equation}\label{expansion}
\begin{split}
  \langle \Th(X_{i})\rangle\simeq &\Th(\langle X_{i}\rangle)+\frac{1}{2}\Th^{\prime\prime}
    (\langle X_{i}\rangle)\langle(X_{i}-\langle X_{i}\rangle)^{2}\rangle\\
    &+\frac{1}{6}\Th^{\prime\prime\prime}(\langle X_{i}\rangle)\langle(X_{i}-\langle X_{i}\rangle)^{3}\rangle
\end{split}
\end{equation}
where the prime stands for the derivative with respect to $X_{i}$,
$\langle(X_{i}-\langle X_{i}\rangle)^{2}\rangle=\langle X_{i}^{2}\rangle_{\mathrm{c}}$ and $\langle(X_{i}-\langle X_{i}\rangle)^{3}\rangle=\langle X_{i}^{3}\rangle_{\mathrm{c}}$.

The last two terms in the right hand side (RHS) are respectively the variance of $h_{i}^{\mathrm{eff}}$ and a term proportional to the skewness of the distribution. First of all, we note that third and higher order cumulants are relevant only if the distribution is significantly different from the Gaussian one. Indeed the normal distribution is the only one with all its cumulants equal to zero excepted the two first ones \cite{Lukacs}.
For a general distribution simplifications can occur for the third and fourth cumulants. If the  distribution of $X_{i}$ is symmetric, the skewness will be zero. The third cumulant is related to the skewness $\gamma_{1}$ of the distribution by the relation $\gamma_{1}=\kappa_{3}/\kappa_{2}^{3/2}$. Another useful feature is the kurtosis $\beta_{2}=\mu_{4}/\kappa_{2}^{2}$. It quantifies the peakedness of the distribution. For comparison with the normal distribution peakedness, we consider the excess kurtosis $\gamma_{2}=\kappa_{4}/\kappa_{2}^{2}$. If the effective fields distribution has a peakedness similar to the gaussian one, the fourth centered moment will be $\mu_{4}=3\kappa_{2}^{2}$.

However for the interesting case where the system is close to an order-disorder transition (the net mean orientation reaches a bifurcation point like illustrated in Fig-\ref{fig:Ising}), in a first approach all the cumulants should be considered. The equilibria are given by, up to second order for simplicity,

\begin{equation}\label{MMFE}
\begin{split}
  q_{i}(t)=&\Th\langle h_{i}^{\mathrm{eff}}\rangle
    \left(1-[\langle {h_{i}^{\mathrm{eff}}}^{2}\rangle-\langle h_{i}^{\mathrm{eff}}\rangle^2]\right)\\
    &+\Th^{3}\langle h_{i}^{\mathrm{eff}}\rangle
    \left(\langle {h_{i}^{\mathrm{eff}}}^{2}\rangle-\langle h_{i}^{\mathrm{eff}}\rangle^2\right)
    \end{split}
\end{equation}
where we defined $q_{i}=\langle s_{i}\rangle$.

The variance of $h_{i}^{\mathrm{eff}}$ appears explicitly in this relation. It takes into account the heterogeneity of the $h_{i}^{\mathrm{eff}}$, including higher order cumulants would amount to include a deviation from the normal distribution. If the variance of $h_{i}^{\mathrm{eff}}$ is negligible, the zeroth order approximation

\begin{equation}\label{MFE}
    q_{i}=\Th\langle h_{i}^{\mathrm{eff}}\rangle
\end{equation}
makes sense.

Through the term  $\langle{h_{i}^{\mathrm{eff}}}^{2}\rangle$, equilibria depend on the 2-agents correlations $q_{ij}\equiv\langle s_{i}s_{j}\rangle$. Indeed if the external inputs $h_{i}$ are zero, (\ref{MMFE}) can be rewritten as


\begin{equation}\label{MMFE-2}
\begin{split}
m_{i}=&\Th\langle h_{i}^{\mathrm{eff}}\rangle
    +\left[-\Th\langle h_{i}^{\mathrm{eff}}\rangle+\Th^{3}\langle h_{i}^{\mathrm{eff}}\rangle\right]\times\\
    &\left(\sum_{k, j}J_{ij}J_{ik}(q_{jk}-q_{j}q_{k})\right)
\end{split}
\end{equation}

The covariances $q_{jk}-q_{j}q_{k}=\langle s_{j}s_{k}\rangle-\langle s_{j}\rangle\langle s_{k}\rangle$ quantify the difference between interacting pairs and independent ones.
If agents $j$ and $k$ are independent, then $\langle s_{j}s_{k}\rangle=\langle s_{j}\rangle\langle s_{k}\rangle$ and the associated covariance will be zero. We need the equation for the $k$-agents correlations but each of these equations involves a higher $(k+1)$-agents correlations. So we have a system of equations up to arbitrary order. We could truncate this system at the $k$th order, take the corresponding zeroth order approximation and replace the solution in the lower order.
We can also neglect the off-diagonal terms, \emph{ie} take into account only the one agent fluctuations. The we get $\langle s_{i}s_{i}\rangle_{\mathrm{c}}\approx (1-q_{i}^{2})$.  The three agents cumulant is approximatively given by $2(q_{i}^{3}-q_{i})$, and so on. Regrouping terms in cumulants power,  we get

\begin{equation}
\begin{split}\label{3-static}
  \langle \Th(X_{i})\rangle=&\Th(\langle X_{i}\rangle)+\frac{1}{2}\Th^{\prime\prime}
    (\langle X_{i}\rangle)\kappa_{2}\\
    &+\frac{1}{6}\Th^{\prime\prime\prime}(\langle X_{i}\rangle)\kappa_{3}\\
    &+\frac{1}{24}\Th^{\prime\prime\prime\prime}(\langle X_{i}\rangle)(\kappa_{4}+3\kappa_{2}^{2})+\cdots
\end{split}
\end{equation}

\subsection{Edgeworth series}

Another way to get a tractable approximation is to expand the probability density function (pdf) in Edgeworth series (an algorithm to compute the different terms of this expansion is given in \cite{Blin}). The Edgeworth series is an asymptotic expansion of the probability density function of a random variable in powers of the second cumulant (the variance is taken as the parameter of the expansion).

Loosely speaking, the Edgeworth series is a reordered Taylor expansion of the logarithm of the characteristic function. Consider the characteristic function $\phi_{\mathrm{un}}(t)$ of a random variable with an unknown distribution $p_{\mathrm{un}}(x)$: $\phi_{\mathrm{un}}(t)=\int_{D}p_{\mathrm{un}}(x)e^{itx}\ud x=\exp\left(\sum_{k}(k!)^{-1}(it)^{k}\kappa_{k}\right)$ and the characteristic function of a reference distribution $\phi_{\mathrm{ref}}(t)=\int_{D}p_{\mathrm{ref}}(x)e^{itx}\ud x$. The Taylor expansion of $\phi_{\mathrm{un}}(t)/\phi_{\mathrm{ref}}(t)$ (with a Gaussian reference random variable $\mathcal{N}(0,1)$, for instance) is

\begin{equation}
  \phi_{\mathrm{un}}(t)=\phi_{\mathrm{ref}}(t)\left[1+c_{1}t+\frac{c_{2}}{2!}t^{2}+\ldots\right]
\end{equation}
using $(-it)^{k}\phi_{\mathrm{ref}}(t)\Leftrightarrow p_{\mathrm{ref}}^{(k)}(x)$, property of the Fourier transform (where $p_{\mathrm{ref}}^{(k)}(x)$ is the $k$th derivative of the reference distribution), we get

\begin{equation}
  p_{\mathrm{un}}(x)=p_{\mathrm{ref}}(x)-c_{1}p_{\mathrm{ref}}^{(1)}(x)+\frac{c_{2}}{2!}p_{\mathrm{ref}}^{(2)}(x)+\ldots
\end{equation}
The coefficients $\{c_{k}\}$ are determined by the cumulants.

Ordering terms by derivatives order leads to the Gram-Charlier series and ordering terms by power of the standard deviation leads to the so-called Edgeworth-Petrov series.

The approximated pdf is given by

\begin{equation}
    \tilde{p}(x)=f(x)\left[1+\frac{\kappa_{3}}{6}\He{3}(x)+\frac{\kappa_{4}}{24}\He{4}(x)+\frac{\kappa_{3}^{2}}{72}\He{6}(x)\right]
\end{equation}
where $\He{n}$ are the modified Hermite polynomials, $\kappa_{n}$ the $n$th cumulants and $f(x)$ the normal pdf $\mathcal{N}(\mu,\sigma)$. The average of the hyperbolic tangent is therefore approximated by

\begin{equation}\label{Edge}
    \langle \Th(x)\rangle\simeq\int_{-\infty}^{\infty}\Th(x)\tilde{p}(x)\ud x
\end{equation}

\subsection{Comparison of methods}

A straightforward benchmark is the homogeneous pairwise Markov network (also called nearest neighbours Ising model)  illustrated in the right panel of Fig-\ref{fig:Lattice} with $J_{ij}=J$ for the 4 neighbours and $h_{i}=0$ for all $i$. The exact mean consensus can be analytically computed \cite{Baxter} and is equal to

\begin{equation}
m(J)=
\begin{cases}
0,&  J<J_{c}\equiv\frac{2}{\ln(1+\sqrt{2})}\\
\left[1-\sinh(2 J)^{-4}\right]^{\frac{1}{8}}, & J>J_{c}
\end{cases}
\label{3-OnsagerSol}
\end{equation}

\begin{figure}[!ht]
\begin{center}
\includegraphics[width=0.75\textwidth]{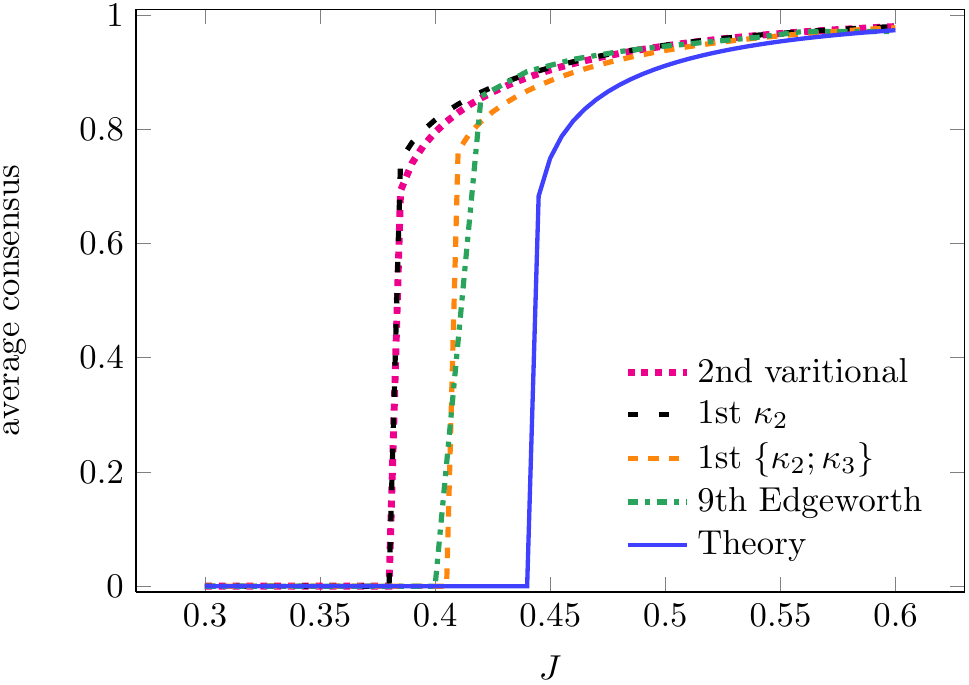}
\end{center}
\caption[Equilibrium approximations]{\label{fig:EqMag} Illustration of the second order variational (\ref{3-TAP}), the first order $\kappa_{2}$, the first order $\{\kappa_{2};\kappa_{3}\}$, the ninth order Edgeworth series approximations and the exact average consensus. }
\end{figure}

The different approximations are illustrated in Fig-\ref{fig:EqMag}.

We note that the first order in single agent variance (first order in $\kappa_{2}$) gives similar results than the second order variational approximation. Including the first order in $\kappa_{3}$ gives better results.

\section{Road to equilibrium and Monte Carlo simulations}\label{sec3:MC}

\subsection{Monte Carlo Markov Chain}

To perform a simulation, we need to describe how the Gibbs distribution can be reached as the equilibrium distribution of a given Markov process. A way to reach a Gibbs distribution

\begin{equation}
p_{2}(\textbf{s})=\mathcal{Z}^{-1}\exp\left(\frac{1}{2}\sum_{i, j}^{N}J_{ij}s_{i}s_{j}+\sum_{i=1}^{N}h_{i}s_{i}\right)\equiv\frac {e^{- \mathcal{H}(\textbf{s})}}{\mathcal{Z}}\label{3-Gibbs}
\end{equation}
is given by the following dynamics (the so-called Glauber dynamics \cite{Glauber}). Namely, one takes a randomly chosen entity $i$ and an attempt to flip the associated binary variable $s_{i}$ is performed with a rate depending on an exponential weight, the other orientations remaining fixed. We define the reversal operator $\Flip{i}$ such that
$\Flip{i}\textbf{s}=\Flip{i}(s_{1},\ldots,s_{i},\ldots,s_{N})=(s_{1},\ldots,-s_{i},\ldots,s_{N})$. This asynchronous updating involves that two consecutive configurations only differ by a single reversal.
To find the exponential rate, we consider the evolution of the probability mass function (PMF) for this dynamics which is given by the master equation

\begin{equation}
\frac{\mathrm{d}}{\mathrm{d}t}p(\textbf{s}; t)=
\sum_{i = 1}^{N} \Big\{\omega(s_{i}|\, -s_{i})\;p(\Flip{i}\textbf{s}; t)-\omega(-s_{i}|\, s_{i})\;p(\textbf{s}; t)\Big\}\label{3-Master}
\end{equation}
where $\omega(s_{i}|\, -s_{i})$ is the transition rate from configuration $\Flip{i}\textbf{s}$ to configuration $\textbf{s}$.
They are derived from the transition probability $\mathrm{P}[s_{i,t+\tau}=-s_{i,t}|s_{i,t},\, \mathbf{s}_{-i,t}] \equiv W(-s_{i}|s_{i},0) = \omega(-s_{i}|\, s_{i})\,\tau+o(\tau)$.

The master equation states that the variation of the PMF is equal to the inward probability flow minus the outward probability flow \cite{Kampen}. At equilibrium, this dynamics should lead to the Gibbs distribution (\ref{3-Gibbs}). A sufficient condition to reach equilibrium is

\begin{equation}\label{3-DBal}
  \omega(s_{i}|\, -s_{i})\;p_{2}(\Flip{i}\textbf{s})-\omega(-s_{i},|\, s_{i})\;p_{2}(\textbf{s})=0
\end{equation}
As we are only interested in the equilibrium PMF and not how one reaches it, we can choose any transition rates satisfying (\ref{3-DBal}). A convenient choice for simulation (discrete time) is to take the transition probability

\begin{equation}\label{3-TP}
   W(-s_{i}|s_{i})= \frac{1}{2}\left[1-s_{i,t}\tanh\left(\sum_{j}J_{ij}s_{j,t}+h_{i}\right)\right]
\end{equation}

Simulations are performed following the scheme\newline

\textbf{Algorithm}
\begin{enumerate}
  \item Choose an entity uniformly at random.
  \item Compute the transition probability (\ref{3-TP}).
  \item Generate a uniform random number $x\in[0,1]$, if $W(-s_{i}|s_{i})>x$, accept the reversal.
  \item Parameterize time such that a Monte Carlo step (MCS) corresponds to $N$ reversal attempts.
  \item Wait for equilibration.
  \item Store the desired statistics.
\end{enumerate}

\begin{figure}[!ht]
\begin{center}
\includegraphics[width=0.5\textwidth]{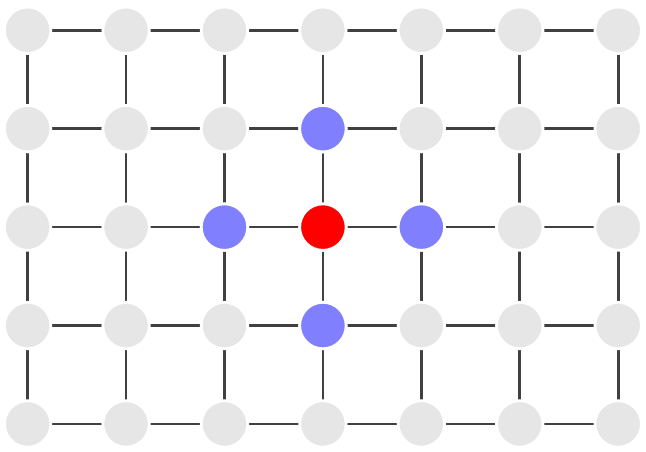}
\end{center}
\caption[Idealized city]{\label{fig:Lattice} Idealized city where each agent has exactly 4 neighbours. Each agent has to make a yes/no choice, interacts positively in the same fashion with his neighbours and has no idiosyncratic preferences.}
\end{figure}

A more detailed discussion (equilibration time, proper definition of statistics, etc.) can be found in \cite{BinderMC}. To fix ideas, we consider an idealized city where each agent has exactly 4 neighbours as illustrated in Fig-\ref{fig:Lattice}. Each agent has to make a yes/no choice described by $s_{i}=\pm 1$, interacts positively in the same fashion with his neighbours and has no idiosyncratic preferences.

\begin{figure}[!ht]
\begin{center}
\resizebox{0.75\textwidth}{!}{%
\includegraphics{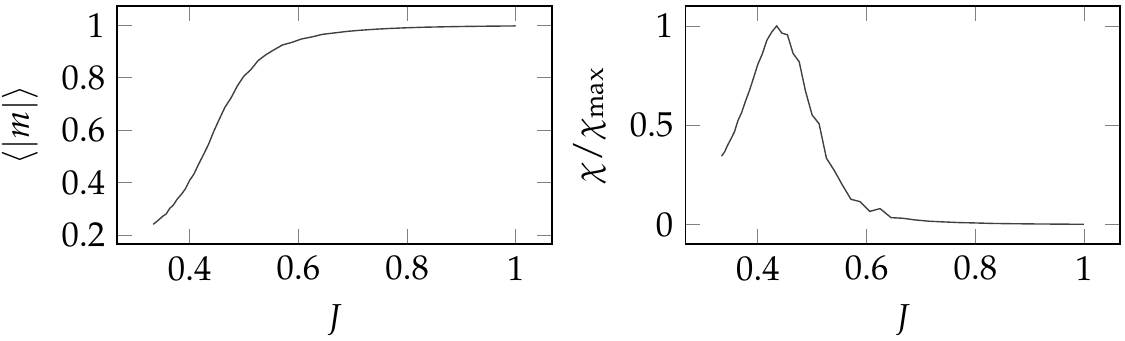}
}
\end{center}
\caption[Monte Carlo estimation of the consensus]{\label{fig:Ising} The mean value $\langle |m|\rangle$ and the variance ($\chi$) of the absolute value of the consensus $|m|$ for the idealized city where each agent has 4 neighbours and where mutual influence $J$ is homogeneous. If the mutual influence is large enough, the consensus takes a non-zero value. We observe that the variance is larger at the particular value at which the consensus goes from 0 to 1.}
\end{figure}

Depending on the strength of the mutual influence $J$ (weight of each of the 4 edges), the mean consensus $\langle m \rangle=\langle N^{-1}\sum_{i}s_{i}\rangle$ can be either equal to zero either non-zero \cite{Stanley}. Moreover if idiosyncratic preferences $h_{i}$ are set to zero, the Gibbs distribution (\ref{3-Gibbs}) is invariant under reversal $s\rightarrow -s$. Thus, to get the mean consensus from simulations, one should measure the mean absolute value of the consensus  $\langle |m|\rangle$. The mean value and the variance ($\chi$) of the absolute value of the consensus are illustrated in Fig-\ref{fig:Ising}

\subsection{Approximated dynamics}

One can derive the exact consensus evolution under the latter dynamical scheme. The exact, but in general untractable, evolution equation is

\begin{equation}\label{exact}
   \frac{\mathrm{d}m_{i}(t)}{\mathrm{d}t}= - m_{i}(t)+\big\langle \Th(h_{i}^{\mathrm{eff}})\big\rangle
\end{equation}
Using our previous approximation of the average hyperbolic tangent, we get

\begin{equation}\label{EOM-cumul2}
\begin{split}
    \frac{\mathrm{d}m_{i}(t)}{\mathrm{d}t}=& - m_{i}(t)+\Th(\langle X_{i}\rangle)+
    \sum_{j=1}^{R(n)}\kappa_{2}^{j}\mathrm{A}^{(n)}_{j}(\langle x_{i}(t)\rangle)\\
    &+\mathcal{O}(\kappa_{2}^{R(n)},\kappa_{3})
\end{split}
\end{equation}
where the cumulants are those of the $X_{i}=h_{i}^{\mathrm{eff}}$.

Both $\mathrm{A}_{j}^{(n)}$ and $R(n)$ depend on the truncation order $n$ of $\langle \Th(X_{i})\rangle$. Furthermore the coefficients $\mathrm{A}_{j}^{(n)}$ depend on powers of $\Th(\langle X_{i}\rangle)$. This approximation can be extended to any arbitrary order. The coefficients $\mathrm{A}_{j}^{(n)}$ are obtained by substituting the centered moments by the cumulants in (\ref{expansion}) using their recurrence relation \cite{Blin}. For example, up to the second order in $\kappa_{2}$ and first order in $\kappa_{3}$, it reads

\begin{equation}
\begin{split}
  \langle \Th(X_{i})\rangle\simeq &\Th(\langle X_{i}\rangle)+\frac{1}{2}\Th^{\prime\prime}
    (\langle X_{i}\rangle)\kappa_{2}\\
    &+\frac{1}{6}\Th^{\prime\prime\prime}(\langle X_{i}\rangle)\kappa_{3}\\
    &+\frac{1}{24}\Th^{\prime\prime\prime\prime}(\langle X_{i}\rangle)3\kappa_{2}^{2}
\end{split}
\end{equation}
where the assumption $\kappa_{4}+3\kappa_{2}^{2}\simeq 3\kappa_{2}^{2}$ was used.
Higher order terms involve higher powers of $\kappa_{2}$ but also products of  $\kappa_{2}$ powers with higher order cumulants.

\begin{figure}[!ht]
\begin{center}
\includegraphics[width=0.75\textwidth]{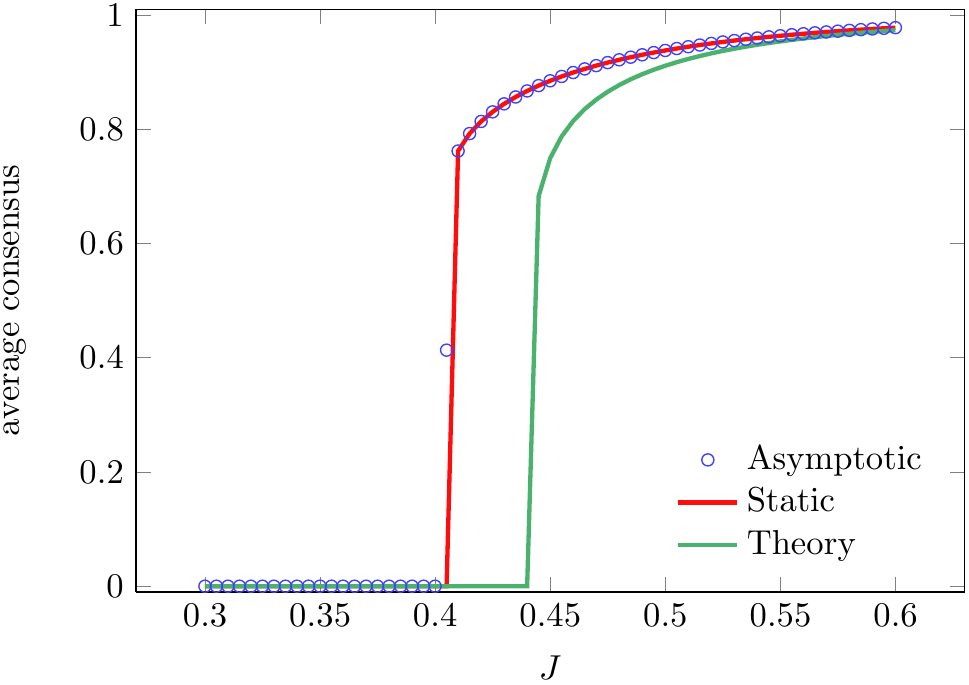}
\end{center}
\caption[Asymptotic and equilibrium solutions]{\label{fig:AsympMag} Illustration of the static solution at the first order $\{\kappa_{2};\kappa_{3}\}$, the asymptotic solution of dynamical evolution at the first order $\{\kappa_{2};\kappa_{3}\}$ and the theoretical equilibrium consensus.}
\end{figure}

The asymptotic value for a homogeneous pairwise Markov network is illustrated in Fig-\ref{fig:AsympMag} and compared to the static solution (\ref{3-static}).

\section{Inverse problem: parameters estimation}\label{sec3:IsingInv}

The network reconstruction (estimation of $\mathbf{J}$ and $\mathbf{h}$) is an important task for applications. A direct estimation (solving the constraints of the maxent problem) is unfeasible for more than 20 entities. To overcome this problem, many schemes were proposed. We present here the estimation methods that we consider the most relevant. Many other schemes exist and the literature is growing due to a renewed interest in this kind of inverse problems.

\subsection{Regularized pseudo-maximum likelihood}\label{sec3:rpml}

The regularized pseudo-maximum likelihood method is powerful for Lagrange parameters estimation of the pairwise maximum entropy model while common maximum likelihood (ML) is untractable (ML involves the computation of $\mathcal{Z}(\mathbf{J},\mathbf{h})$ involving $2^{N}$ terms) \cite{Aurell}. This method can be thought as an autologistic regression to predict binary outcomes (flipping events). The main idea is to factorize spatially the distribution and to consider only conditional probabilities.

Let $\{\mathbf{s}_{t}\}_{t=1}^{T}$  be a sequence of random vectors generated by the distribution $p_{T}(\{\mathbf{s}_{t}\}_{t=1}^{T})$ which can be rewritten as

\begin{equation}\label{fact}
  p_{T}(\{\mathbf{s}_{t}\}_{t=1}^{T})=p_{1}(\mathbf{s}_{1})\prod_{\tau=1}^{T-1}
  \frac{p_{\tau+1}(\{\mathbf{s}_{t}\}_{t=1}^{\tau+1})}{p_{\tau}(\{\mathbf{s}_{t}\}_{t=1}^{\tau})}
  =p_{1}(\mathbf{s}_{1})\prod_{\tau=1}^{T-1}p_{\tau+1}(\mathbf{s}_{\tau+1}|\{\mathbf{s}_{t}\}_{t=1}^{\tau})
\end{equation}
If this distribution is untractable, one can not use directly the maximum likelihood method. However, one can replace $p_{T}(\{\mathbf{s}_{t}\}_{t=1}^{T})$ by an approximated distribution $q_{T}(\{\mathbf{s}_{t}\}_{t=1}^{T};\boldsymbol\theta)$. This function is referred to as the pseudo-likelihood (noted $\mathrm{PL}(\boldsymbol\theta)$ hereafter). Even if the problem is now misspecified, one can estimate the parameters $\boldsymbol\theta$ by minimizing the KL-divergence of the empirical distribution $p_{\text{emp}}$ relative to $q$. Using (\ref{3-DKLlik}), we get

\begin{equation}\label{PMLC}
D_{\mathrm{KL}}(p_{\text{emp}}||q(\boldsymbol\theta))=-\frac{1}{T}\ln \mathrm{PL}(\boldsymbol\theta)-\ln T
\end{equation}
A convenient choice for the misspecified likelihood function $\mathrm{PL}(\boldsymbol\theta)$ is the product of spatial conditionals $P(s_{i,t}|\mathbf{s}_{-i,t};\, \boldsymbol\theta)$. For a N-dimensional sample of length $T$, the objective function to be maximized is

\begin{equation}\label{PML}
  \mathrm{pl}(\boldsymbol\theta)=\frac{1}{T}\ln\mathrm{PL}(\boldsymbol\theta)=\frac{1}{T}\sum_{t=1}^{T}\sum_{i=1}^{N}
  \ln p(s_{i,t}|\mathbf{s}_{-i,t};\, \boldsymbol\theta)
\end{equation}
where conditional probabilities of a memoryless model are

\begin{equation}
p(s_{i,t}|\mathbf{s}_{-i,t};\, \boldsymbol\theta)=\frac{1}{2}
\left[1+s_{i,t}\tanh\left(\sum_{ j\neq i}J_{ij}s_{j,t}+h_{i}\right)\right]
\end{equation}
and

\begin{equation}
p(s_{i,t}|\mathcal{H}_{t}^{T};\, \boldsymbol\theta)=\frac{1}{2}
\left[1+s_{i,t}\tanh\left(\sum_{ j\neq i}J_{ij}s_{j,t}+h_{i}
+\sum_{\tau=1}^{T}\sum_{j}K_{ij}^{\tau}s_{j,t-\tau}\right)\right]
\end{equation}
for a model involving some memory.

The resulting pseudo-maximum likelihood estimator (PMLE) is consistent (converges in probability to the true value $\boldsymbol\theta_{0}$) \cite{Hyv}.

A regularization term is added to the PL function to prevent overfitting which is a negative multiple of the $l_{2}$-norm of parameters to be estimated, for instance. The regularized PL (rPL) objective function is thus $\mathrm{PL}(\boldsymbol\theta)-\lambda\, \|\boldsymbol\theta\|_{2}^{2}$ with $\lambda>0$. If the network is believed to be sparse, a $l_{1}$ regularization term should be used \cite{Aurell} (small values of the parameters are projected on zero).

\subsection{Inversion of self consistent equation} \label{subsec:diagonalTrick}

As one can obtain mean values $q_{i}$ and covariances $C_{ij}$ (also noted $\chi_{ij}$) from recorded data, the self-consistent equations (\ref{3-MF}) and (\ref{3-TAP}) can be inverted. At the first order, one has

\begin{eqnarray}
  q_{i}&=&\Th(\sum_{j}J_{ij}q_{j}+h_{i}) \\
  C_{ij} &=& \frac{\partial\Th(\sum_{k}J_{ik}q_{j}+h_{i})}{\partial h_{j}}=(1-q^{2}_{i})\left[\sum_{k}J_{ik}C_{kj}+\delta_{ij}\right]
\end{eqnarray}
The first order estimators $\tilde{\mathbf{J}}^{\text{\tiny{1st}}}$ and $\tilde{\mathbf{h}}^{\text{\tiny{1st}}}$ are

\begin{eqnarray}\label{1stOrInv}
\tilde{\textbf{J}}^{\text{\tiny{1st}}}&=&\textbf{P}^{-1}-\mathbf{C}^{-1}\\
  \tilde{h}_{i}^{\text{\tiny{1st}}}&=&\Th^{-1}(q_{i})-\sum_{j}\tilde{J}^{\text{\tiny{1st}}}_{ij}q_{j}
\end{eqnarray}
where $P_{ij}=(1-q_{i}^2)\delta_{ij}$ and $\tilde{J}^{\text{\tiny{1st}}}_{ii}=0$ (no self-influence).

At second order, using (\ref{3-TAP}), one has

\begin{eqnarray}\label{3-TAPinv}
\tilde{J}^{\text{\tiny{2nd}}}_{ij}&=&-2(\tilde{J}_{ij}^{\text{\tiny{2nd}}})^{2}q_{i}q_{j}-\tilde{J}_{ij}^{\text{\tiny{2nd}}}\\
  \tilde{h}_{i}^{\text{\tiny{2nd}}}&=&\Th^{-1}(q_{i})-\sum_{j}\tilde{J}^{\text{\tiny{2nd}}}_{ij}q_{j}+q_{i}
\sum_{j}(\tilde{J}_{ij}^{\text{\tiny{2nd}}})^{2}(1-q_{j}^{2})
\end{eqnarray}
where $\tilde{J}^{\text{\tiny{2nd}}}_{ii}=0$.

Finally, to avoid to compute higher orders (which can be tricky or leads to multi-valued solutions) one considers the so-called diagonal trick \cite{Tanaka}. The idea is that diagonals entries $\tilde{J}^{\text{\tiny{1st}}}_{ii}$ are related to the whole second order and a part of third order. Another main improvement of this method is obtained by inverting (\ref{3-MF}) or (\ref{3-TAP}) in each of their basins of attraction \cite{Nguyen}.

\section{Entropy and Zipf's law}\label{sec3:zipf}
We saw in sec \ref{sec3:combinatoric} that the maximum number of outcomes in a combinatoric problem is equivalent to the Shannon entropy. The entropy can be formally expressed as a function of the utility. The expansion of the entropy around the mean utility $U$ is written (where $U$ is the notation for $\langle\mathcal{U}\rangle$)

\begin{equation}
  S(\mathcal{U})\simeq S(U)-\frac{1}{T}(\mathcal{U}-U)+\frac{1}{2T^{2}R_{\mathcal{U}}} (\mathcal{U}-U)^{2}
\end{equation}
For ranks (ordered states) distributed following a power-law, the quadratic and higher order terms are sub-intensive; the entropy should be a linear function of the utility \cite{Step}. Indeed for a Zipf's law $p(r)=A r^{-\alpha}\equiv\mathcal{Z}^{-1}\exp{\mathcal{U}(r)}$ where the utility is $\mathcal{U}(r)=\ln(A\mathcal{Z})-\alpha\ln r$ and $r$ is the rank associated to a state, the entropy is exactly a linear function of the utility. The number of outcomes by units of utility is

\begin{eqnarray}
  \frac{\ud r(\mathcal{U})}{\ud \mathcal{U}} &=& \left(\frac{\ud \mathcal{U}(r)}{\ud r}\right)^{-1}=-\frac{r(\mathcal{U})}{\alpha} \\
   &=& -\frac{(A\mathcal{Z})^{1/\alpha}}{\alpha}e^{-\mathcal{U}/\alpha}
\end{eqnarray}
taking the logarithm, one has

\begin{equation}
  \mathcal{S}(\mathcal{U})=-\frac{\mathcal{U}}{\alpha}+\mathrm{Cst}
\end{equation}

It is a very particular case because the fluctuations are always very large (since $R_{\mathcal{U}}$ is proportional to the variance of the utility, see sec \ref{sec3:prob}). To illustrate the entropy-utility relation, consider the Brock and Durlauf model with an homogeneous and complete social network. With rational expectation (see chap \ref{chap:social} for a detailed presentation), one has $\mathcal{U}(m(\mathbf{s}))=J(2N)^{-1}m(\mathbf{s})^{2}$ where $J$ is the strength of social interactions, $m(\mathbf{s})=N^{-1}\sum_{i}s_{i}$ is the consensus associated to the choice vector $\mathbf{s}$ and $N$ the number of agents (the choice of the $i$th agent is described by $s_{i}=\pm 1$). The entropy $s(m)\equiv S(m)/N$ per capita is

\begin{equation}
s(m)=-\frac{1-m}{2}\ln\left(\frac{1-m}{2}\right)-\frac{1+m}{2}\ln\left(\frac{1+m}{2}\right)
\end{equation}

\begin{figure}[!ht]
\begin{center}
\resizebox{0.5\textwidth}{!}{%
\includegraphics{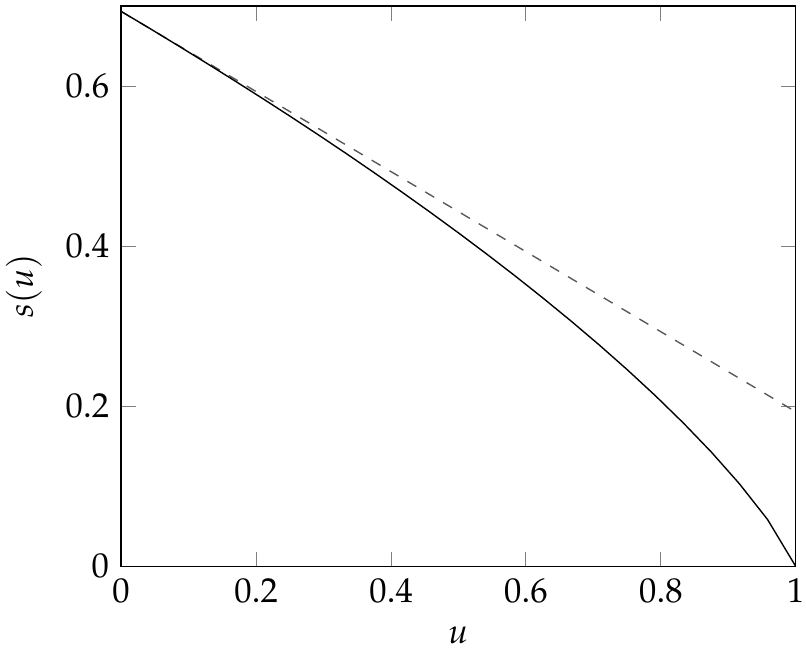}
}
\end{center}
\caption[Entropy-utility relation]{Entropy as a function of utility (bold line) for the Brock-Durlauf binary choice model with an homogeneous complete social network. At first glance, the function is close to a linear relation but the curvature is negative everywhere. For a restricted social network (nearest neighbors), the curvature can be equal to zero for a particular value of the utility. }
\label{fig3:ENtUt}
\end{figure}

Writing the entropy as a function of the reduced utility per capita $u=m^{2}$, we get

\begin{equation}
s(u)=-\frac{1-\sqrt{u}}{2}\ln\left(\frac{1-\sqrt{u}}{2}\right)
-\frac{1+\sqrt{u}}{2}\ln\left(\frac{1+\sqrt{u}}{2}\right)
\end{equation}
This relation is illustrated in Fig-\ref{fig3:ENtUt}. The function is close to a linear relation but the curvature is negative everywhere. We do not expect to observe a Zipf's law in this system if configurations are well sampled.
For a restricted social network (nearest neighbors), the curvature can be equal to zero for a particular value of the utility.

Hereafter, we detail a statistical test of power-laws.

\section{Discrete power-law}\label{sec3:DPL}
\begin{figure}[!ht]
\begin{center}
\includegraphics[width=0.85\textwidth]{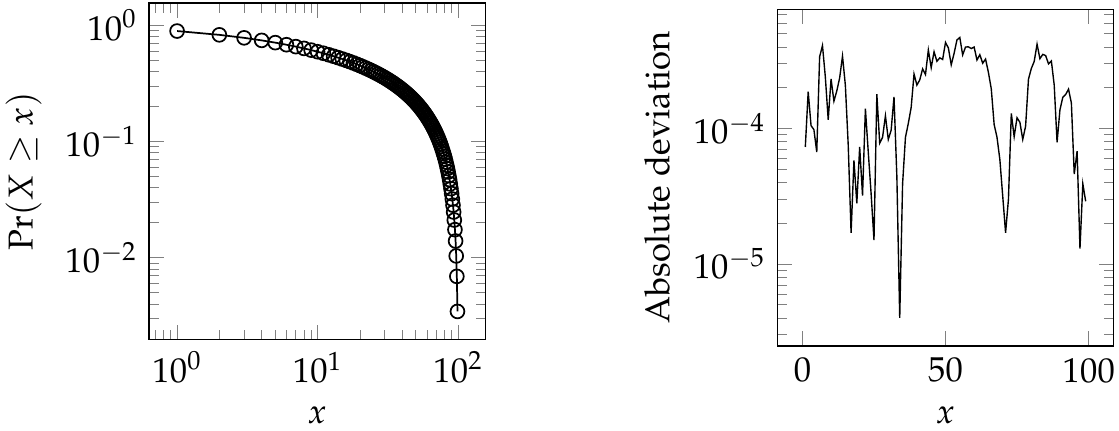}
\end{center}
\caption[Power-law]{\label{fig:PLG} The theoretical CDF (left panel) for $\beta=0.750$ and $x_{\mathrm{max}}=100$. The empirical sample (length $10^6$) was generated with the same parameters. The absolute deviation between theoretical and empirical CDF is illustrated in the right panel.}
\end{figure}

Financial markets are a typical example of complex systems exhibiting collective behaviours and special features such as volatility clustering and power-law. In our application, we will need to test if some distributions are really power-law or not. A statistical test for power-law is given in \cite{Clauset}. We adapt this test to discrete power-law with a natural upper bound. Before considering the discrete case, we note that if the distribution $p(x)\sim x^{-\beta}$ has a finite upper bound $x_{\mathrm{max}}$, then the cumulative distribution function (CDF) will not be a straight line in a log-log plot, see Fig-\ref{fig:PLG}, because

\begin{equation}\label{3-PLCDF}
  \Pr[X\geq x]=\mathrm{Cst} \int_{x}^{x_{\mathrm{max}}}y^{-\beta}\ud y= \frac{\mathrm{Cst}}{1-\beta}
  \left[x_{\mathrm{max}}^{1-\beta}-x^{1-\beta} \right]
\end{equation}
where the constant appears to normalize the distribution to 1 and $\beta>1$. Taking the logarithm of both sides, it comes

\begin{equation}
  \log \Pr[X\geq x]= \log\left(x^{1-\beta}-x_{\mathrm{max}}^{1-\beta}\right)+\log \frac{\mathrm{Cst}}{\beta-1}
\end{equation}
The dependent variable $\log \Pr[X\geq x]$ is a linear function of $\log x$ only when $x_{\mathrm{max}}\rightarrow \infty$


The statistical test proposed in \cite{Clauset} consists in the following scheme

\begin{enumerate}
  \item Determine the best fit of the power-law to the data using maximum-likelihood estimator.
  \item Calculate the Kolmogorov-Smirnov (KS) statistics for the goodness-of-fit. The KS statistics is the maximum absolute value between empirical CDF and the CDF of the estimated power-law.
  \item Generate a large number ($\sim 1000)$ of synthetic data sets.
  \item Calculate the p-value as the fraction of the KS statistics for the synthetic data sets whose value exceeds the KS statistics of the real data.
  \item If the p-value is sufficiently small ($\sim 0.05$), the power-law is ruled out.
\end{enumerate}

The MLE estimator of a discrete power-law with a natural cut-off $x_{\mathrm{max}}$ is derived from the first order condition for the log-likelihood based on $N$ observations

\begin{equation}
  \ell(\beta)=\ln L(\beta)= -\beta \sum_{i=1}^{N}\ln x_{i}-N \ln\left(\sum_{x=1}^{x_{\mathrm{max}}}x^{-\beta}\right)
\end{equation}
taking the derivative with respect to $\beta$ leads to the MLE $\beta_{\mathrm{MLE}}$ satisfying

\begin{equation}\label{PLMLE}
  \frac{1}{N}\sum_{i=1}^{N}\ln x_{i}=  \frac{\sum_{x=1}^{x_{\mathrm{max}}}x^{-\beta_{\mathrm{MLE}}}\ln x_{\mathrm{max}}}{\sum_{x=1}^{x_{\mathrm{max}}}x^{-\beta_{\mathrm{MLE}}}}
\end{equation}

The standard deviation of $\beta_{\mathrm{MLE}}$ is obtained by taking the expansion of the likelihood around $\beta_{\mathrm{MLE}}$

\begin{equation}
  \ell(\beta) = \ell(\beta_{\mathrm{MLE}})+\frac{1}{2!}\frac{\partial^{2}\ell(\beta)}{\partial\beta^{2}}\Big|_{\beta_{\mathrm{MLE}}}
  (\beta-\beta_{\mathrm{MLE}})^{2}
\end{equation}
identifying the terms to the Gaussian approximation $-\ln(\sigma \sqrt{2\pi})-\frac{1}{2}\left(\frac{x-\beta}{\sigma}\right)^{2}$, we get

\begin{equation}
  \sigma_{\beta_{\mathrm{MLE}}}=\frac{1}{\sqrt{N\left[\frac{\zeta^{''}(x_{\mathrm{max}},\beta_{\mathrm{MLE}})}
  {\zeta(x_{\mathrm{max}},\beta_{\mathrm{MLE}})}-\left(\frac{\zeta^{'}(x_{\mathrm{max}},\beta_{\mathrm{MLE}})}
  {\zeta(x_{\mathrm{max}},\beta_{\mathrm{MLE}})}\right)^{2}\right]}}
\end{equation}
where $\zeta(x_{\mathrm{max}},\beta)=\sum_{x=1}^{x_{\mathrm{max}}}x^{-\beta}$ and the prime stands for the derivative with respect to $\beta$.

Synthetic data distributed as a discrete power-law with a finite upper bound are generated as follows. One generates a realization $u$ of a uniform random variable $U$ in $[0,1]$, one calculates $\sum_{x=1}^{x_{\mathrm{max}}}x^{-\beta}$ and the cumulative sum $\sum_{y=1}^{x}y^{-\beta}$. The smallest integer $x$ such that $\sum_{y=1}^{x}y^{-\beta}\geq u \, \sum_{x=1}^{x_{\mathrm{max}}}x^{-\beta}$ is stored. This process is repeated to generate a sample of desired length. An example is illustrated in Fig-\ref{fig:PLG}, the absolute deviation between the theoretical and simulated CDF is smaller than $10^{-3}$.

\begin{figure}[!ht]
\begin{center}
\resizebox{\textwidth}{!}{%
\begin{tabular}{c}
    \includegraphics[scale=1]{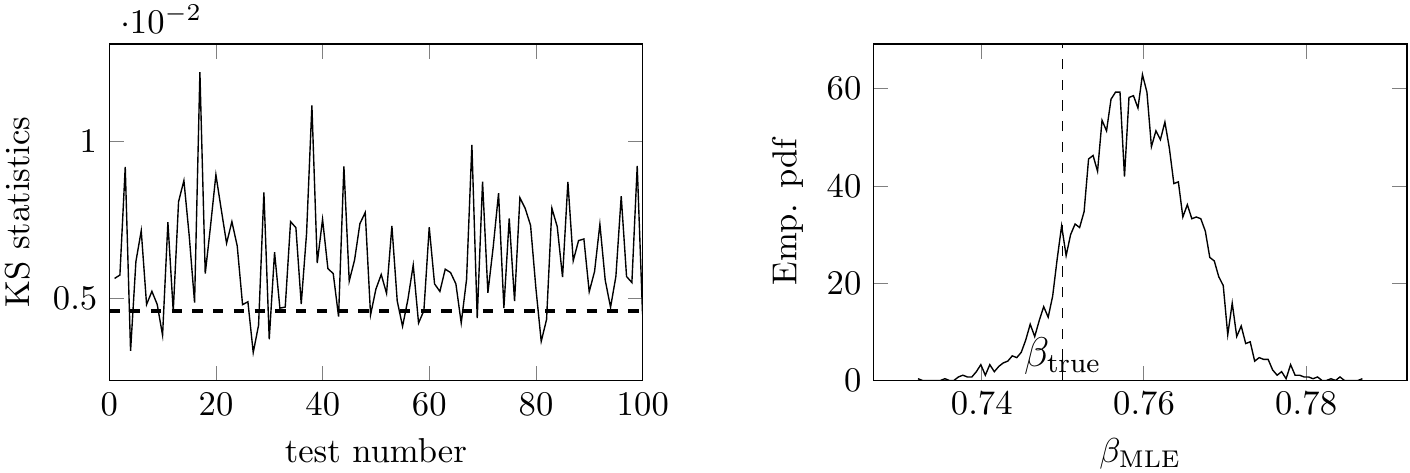} \\
  \includegraphics[scale=1]{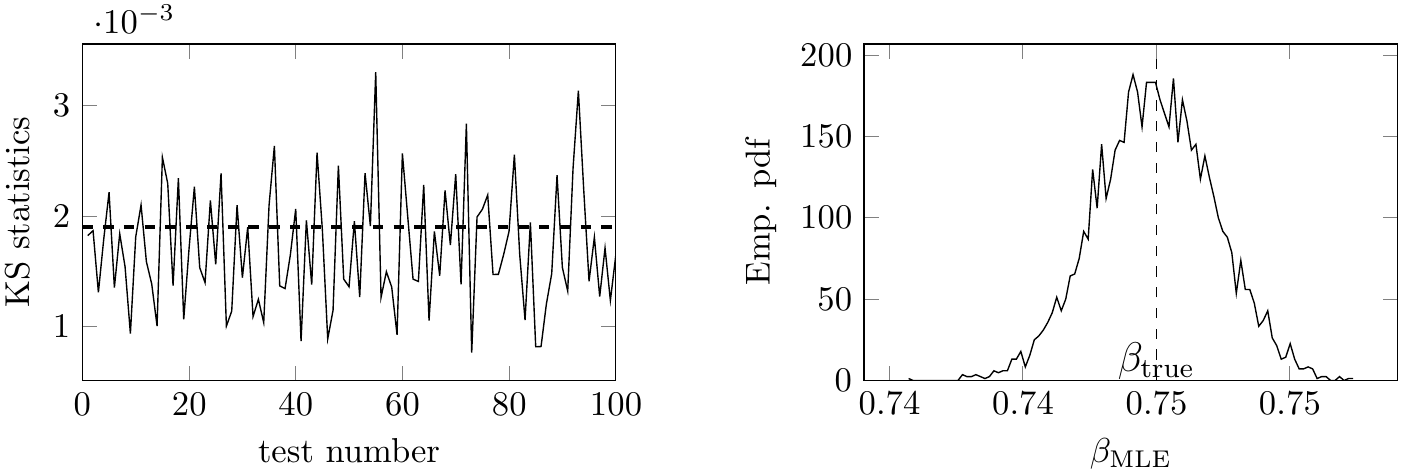} \\
\end{tabular}
}
\end{center}
\caption[Kolmogorov-Smirnov statistics and max-lik estimator]{\label{fig:TPL} KS statistics and MLE distribution for $10^4$ data assumed to be power-law distributed (top panels) and for $10^5$ data (bottom panels). Each of the 1000 samples is generated using ranks between 1 and 100 and $\beta_{\mathrm{true}}=0.75$. The dashed line stands for the true value of the exponent.}
\end{figure}

As an example we run the test on $10^4$ and $10^5$ synthetic data (integers between 1 and 100 simulated with $\beta_{\mathrm{true}}=0.750$). The maximum likelihood estimators are respectively $\beta_{\mathrm{MLE}}=0.758(7)$ and $\beta_{\mathrm{MLE}}=0.750(2)$. We run the test for 1000 synthetic sets that returns p-values $p=0.86$ and $p=0.35$, in both case the power-law is not ruled out. The KS statistics and the distribution of MLE from the 1000 synthetic sets are illustrated in Fig-\ref{fig:TPL}.

\section{Mantegna-Sornette distance and market topology}\label{sec3:MSdist}

The Mantegna-Sornette distance (MS-distance) provides a way to study the market topology, especially using minimal spanning tree (also called asset tree) \cite{Mant}. This distance between two assets is defined as

\begin{equation}\label{MSD}
  d_{ij}=\sqrt{2(1-C_{ij})}
\end{equation}
where $C_{ij}$ are the correlation coefficients of the log-returns. The motivation is the following. Define the log-return as $r_{t}=\ln p_{t}-\ln p_{t-1}$, where $p_{t}$ is the price at time $t$. For $T+1$ observations, one defines the temporal vector return $\mathbf{r}(k)$ of an asset labelled $k$ as $\mathbf{r}(k)=(r_{1}(k),\cdots,r_{T}(k))$. This vector is then normalized as

\begin{equation}\label{MSDnorm}
  \mathbf{w}(k)=\frac{\mathbf{r}(k)-\langle \mathbf{r}(k)\rangle}{\sqrt{\langle \mathbf{r}(k)\mathbf{r'}(k)\rangle-\langle \mathbf{r}(k)\rangle^{2}}}
\end{equation}
where $\langle \cdot\rangle$ is the temporal average over the observation period and the prime stands for the transposition. This definition implies $\|\mathbf{w}(k)\|=1$ and $\langle\mathbf{w}(k)\rangle=0$, thus the correlation coefficients are $C_{ij}=\mathbf{w}(i)\mathbf{w'}(j)$ (an inner product). The distance between normalized returns vectors is then given by $d_{ij}=\|\mathbf{w}(i)-\mathbf{w}(j)\|=\sqrt{2(1-C_{ij})}$.

\begin{figure}[!ht]
\begin{center}
\includegraphics[width=0.75\textwidth]{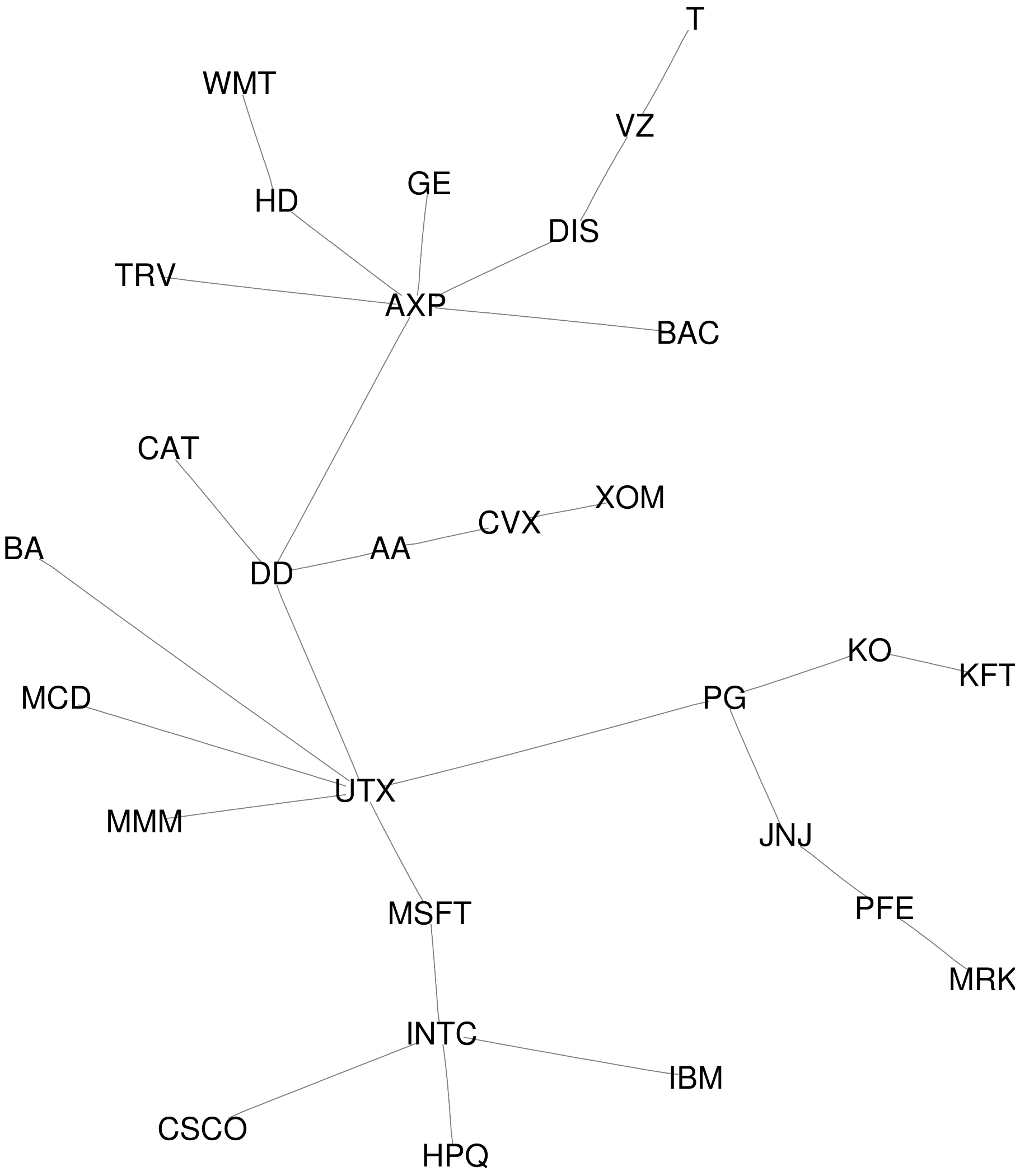}
\end{center}
\caption[Assets tree]{\label{fig:DJmstMant} The minimum spanning tree based on the Mantegna-Sornette distance. The correlation coefficients of  29 large capitalization US companies are computed over 2500 trading days (10 trading years). The edge length is proportional to the distance between stocks. Companies are denoted by their ticks, available on Yahoo Finance for instance.}
\end{figure}

This distance is useful to study the market topology \cite{Onnela,OnnelaPRE}, in particular the asset trees build with the MS-distance are non-random and seem to be scale-free trees (the degree distribution is a power-law) and they exhibit dynamic reorganization. The minimum spanning tree (MST) drawn with those weights is illustrated in Fig-\ref{fig:DJmstMant} for 29 large capitalization US companies. The degree distribution and the length of the MST highlight hierarchical structures and dynamic reorganization \cite{Mantegna}.
The financial network illustrated in Fig-\ref{fig:DJmstMant} clearly shows the existence of \emph{hubs} (highly connected companies like UTX).

\begin{figure}[!ht]
\begin{center}
\includegraphics[width=\textwidth]{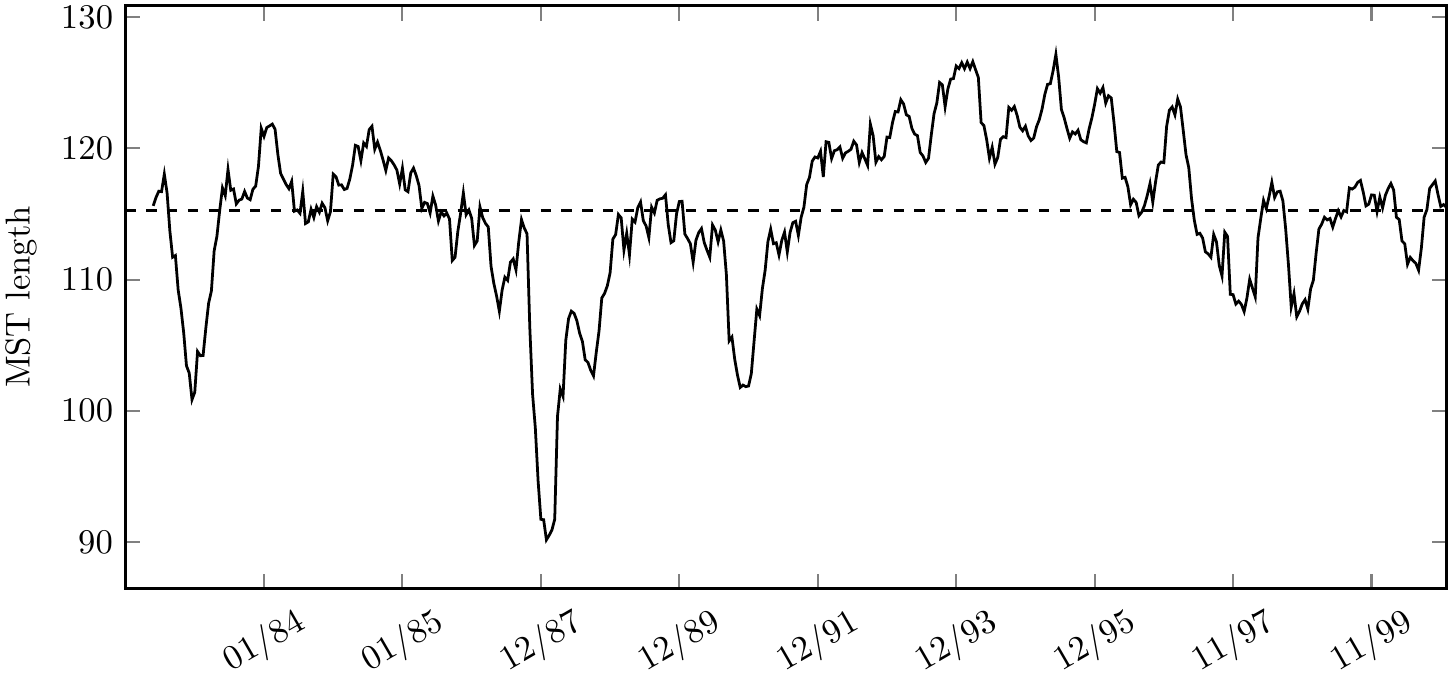}
\end{center}
\caption[Length of a assets tree through time]{\label{fig:MSTlengthMant} The length of the MST through time for  115 large capitalization US companies are computed over 4500 trading days. The distances are computed on a time window of 100 trading days width translated by 5 days each step. The dashed line stands for the length mean value.}
\end{figure}

It was shown that the length (sum of the vertices weights) of the MST decreases during crises. This feature is illustrated in Fig-\ref{fig:MSTlengthMant} where the length decreases in the interval containing the Black Monday (October 19, 1987). Some of these features will be studied within the maximum entropy framework.

\section{Conclusion}

We saw that the entropy is a measure of statistical dependencies. The variational methods have been introduced in a statistical framework using the Kullback-Leibler discrepancy. The minimization of the KLD is equivalent to the maximization of the likelihood and to minimization of the $\mathcal{F}$-functional. It follows that the maximum entropy principle is equivalent to maximize the likelihood of the distribution closest to the uniform distribution without range restriction or to find the most probable state conditionally our knowledge of the system.

Using these concepts, the inverse problem was set up and approximations of untractable maxent distributions were detailed.
Last, we saw the relation between the correlation of stock returns and the market structure.

\begin{subappendices}
\section{Large deviations theory}\label{sec3:LDT}


The maximum entropy principle, the entropy and the KLD find their foundations in the large deviations theory (LDT). The theory of large deviations studies the exponential decay of probabilities in random systems and thus concerns the asymptotic behaviour of tails of sequences of probability distributions. The former definition may seem somewhat vague but intuitively, the most interesting events in a random system (the financial markets in our concern) are the rare or unlikely events like crashes and large returns. The LDT is the natural framework for the characterization of such events in terms of probability (fluctuations around the most probable state).

In fact, we already saw heuristically a result of the LDT in Sec-\ref{sec3:Var} and Sec-\ref{sec3:prob}. Hereafter, we will see a more general version of these statements.

It is possible to show that the LDT is the mathematics of systems of many interacting entities \cite{Ellis,TouchetteLDT}. Following the modern interpretation, one has a nice probabilistic derivation of the main variational principles. Namely, one can identify \cite{TouchetteLDT}:

\begin{table}[!ht]
\caption[Correspondence with the LDT]{Correspondence with the large deviations theory.}
\label{tab3:LDTcorresp}
\begin{center}
\begin{tabular}{p{0.425\textwidth} c p{0.425\textwidth}}
  \hline
  \textbf{Random systems} &  & \textbf{Large deviation theory} \\
  \hline
  Macro-state / observable (utility, etc.)     & $\leftrightarrow$ & Random variable \\
  Entropy          & $\leftrightarrow$ & Rate function (up to a minus sign) \\
  Free "energy" (mean utility plus entropy at fixed utility, see Sec-\ref{sec3:Var} and Sec-\ref{sec3:prob})      & $\leftrightarrow$ & Scaled cumulant generating function (SCGF)\\
  Equilibrium      & $\leftrightarrow$ & Most probable state \\
  Maxent principle & $\leftrightarrow $& Contraction principle (constrained minimization of the unconstrained rate function) \\
  Minimization of the free energy (hereafter the $\mathcal{F}$-functional or cumulant generating function)               & $\leftrightarrow$ &
  Minimization of the rate function with weighted states (non uniform prior distribution) \\
  Legendre-Fenchel transform linking the entropy and the SCGF      & $\leftrightarrow$ &
  Saddle point approximation \\
  \hline
\end{tabular}
\end{center}
\end{table}

It is worth to present (heuristically) these quantities and principles to give a more rigourous justification to the maximum entropy principle, entropy and Kullback-Leibler discrepancy.

\subsection{Example: independent binary variables}
\begin{figure}[!ht]
\begin{center}
\includegraphics[width=0.5\textwidth]{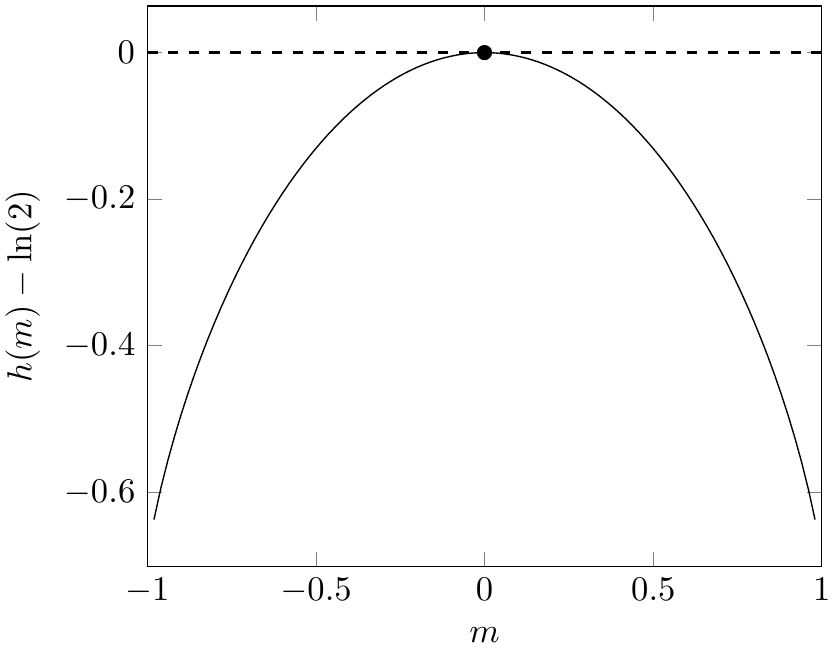}
\end{center}
\caption[Entropy of independent signs]{\label{fig:EntIndSign} The exponential rate characterizing the probability of market net orientation for independent signs.}
\end{figure}

First of all, let's consider an example. As in this thesis we only consider the sign of the returns, a relevant example is a set of $N$ independent (for simplicity) binary variables $s_{i}\in\{-1,1\}$ where $i=1, \ldots, N$. Assume that the market configurations $\mathbf{s}=(s_{1},\ldots,s_{N})$ are uniformly distributed (again for simplicity). The market net orientation  is $M_{n}=N^{-1}\sum_{i=1}^{N}s_{i}=N_{+}-N_{-}$ where $N_{+}$ is the number of positive signs and $N_{-}$ is the number of negative signs. The number of configurations having a given net orientation $M_{n}=m$ is

\begin{equation}
  \Omega(m)=\frac{N!}{N_{+}!N_{-}!}
\end{equation}
where $N_{\pm}=N(1\pm m)/2$. Using the Laplace approximation of the factorial function (Stirling approximation)\footnote{Using the Gamma function $\Gamma(N+1)=N!$ for integers, it comes  $\Gamma(N+1)=\int_{0}^{\infty}x^{N}e^{-x}\ud x=N\int_{0}^{\infty}\exp(N\ln Ny-Ny)\ud y$ where $y=xN^{-1}$. The Laplace approximation gives $\int_{0}^{\infty}\exp(N\ln y-Ny)\ud y\simeq e^{-N}\sqrt{2\pi/N}$. Then $\ln \Gamma(N+1)\simeq N\ln N -N+2^{-1} \ln(2\pi N)=
N\ln N -N+O(\ln N)$. The use of the Laplace approximation plays a major role in the LDT, as we will see hereafter.}, it comes ($m$ is restricted to the range $[-1,1]$)

\begin{equation}
  \Omega(m)\simeq e^{N h(m)} \qquad \mathrm{where} \qquad h(m)=-\frac{1-m}{2}\ln\left(\frac{1-m}{2}\right)-\frac{1+m}{2}\ln\left(\frac{1+m}{2}\right)
\end{equation}
Ones identifies the entropy $h(m)$ of a Bernouilli distribution. Therefore, the probability $p(m)\equiv\Pr(M_{n}=m)$ to observe a market net orientation $m$ with these assumptions is

\begin{equation}
 p(m)=\frac{\#\text{configurations having a net orientation equal to }m}{\text{total number of configurations}}=\frac{\Omega(m)}{2^{N}}\simeq e^{N (h(m)-\ln 2)}
\end{equation}
where $h(m)-\ln 2$ is negative for each possible value of $m$, excepted for $m=0$. The probability to observe a net orientation close to $1$ is small if the signs are independent and not biased by external information. The exponential rate  $h(m)-\ln 2$ is illustrated Fig-\ref{fig:EntIndSign}. We observe in this figure that there is a single point where the probability does not decay exponentially. We will see that this point corresponds to a law of large numbers (LLN).

\subsection{Basic results}

In a nutshell, a large deviation principle (LDP) for a random variable $A_{n}$ is an asymptotic property $\Pr(A_{n}\in \ud a)\asymp \exp(-nI(a))\ud a$ where $A_{n}\in \ud a$ is a shortcut for $A_{n}\in [a,a+\ud a]$ and $\asymp$ is the asymptotic equality\footnote{$\Pr(A_{n}\in \ud a)\asymp \exp(-nI(a))\ud a$ means $\Pr(A_{n}\in \ud a)= \exp\left(-nI(a)+O(\ln n)\right)\ud a$ or in other words $\lim_{n\rightarrow\infty}-n^{-1}\ln \Pr(A_{n}=a)=I(a)$. So "$\asymp$" means that the dominant part of $\Pr(A_{n}\in \ud a)$ is the decaying exponential as $n\rightarrow\infty$.}. The rate function (or minus the entropy) $I(\cdot)$ gives the decreasing exponential rate of the probability density function. There are two main results, known as the Varadhan and Gärtner-Ellis (GE) theorems.

The Varadhan theorem allows to derive the scaled cumulant generating function from the knowledge of the rate function $I(\cdot)$. Without mathematical rigour, if $A_{n}$ satisfies a LDP with a rate function $I(a)$, then the SCGF $\lambda(f)$ of a continuous function $f(\cdot)$ of $A_{n}$ is given the Legendre-Fenchel (LF) transform of the rate function: $\lambda(f)=\sup_{a}\{f(a)-I(a)\}$. In other words, the rate function characterizing the asymptotic behaviour of a macroscopic variable (utility, etc.) is given by the Legendre-Fenchel transform of the SCGF.
The heuristic proof for a linear function $f(a)=ka$ is derived by introducing the LDP in the definition of the SCGF:

\begin{eqnarray}
 \lambda(k)\equiv \lim_{n\rightarrow\infty}\frac{1}{n}\ln\mathrm{E}[e^{nkA_{n}}] &=&
  \lim_{n\rightarrow\infty}\frac{1}{n} \ln\int_{\mathbb{R}}e^{nka}\Pr(A_{n}\in \ud a)\\
    &=& \lim_{n\rightarrow\infty}\frac{1}{n} \ln\int_{\mathbb{R}}e^{n(ka-I(a))} \ud a\\
    &=& \sup_{a}\{ka-I(a)\} \qquad \text{Laplace approximation}
\end{eqnarray}

The GE theorem tells that if the SCGF $\lambda(k)=\lim_{n\rightarrow\infty}\frac{1}{n}\ln\mathrm{E}[e^{nkA_{n}}]$ is differentiable everywhere then $A_{n}$ satisfies the LDP $\Pr(A_{n}\in \ud a)\asymp \exp(-nI(a))\ud a$ with a rate function $I(a)=\sup_{k}\{ka-\lambda(k)\}$.

A third useful result is the contraction principle. It states that one can derive a LDP from another known LDP. If $A_{n}$ satisfies a LDP with a rate function $I_{A}(a)$ and if another random variable $B_{n}$ admits a representation $B_{n}=f(A_{n})$, where $f(\cdot)$ is a continuous mapping (possibly many-to-one) then $B_{n}$ satisfies a LDP with a rate function $I_{B}(b)=\inf_{a:f(a)=b}I_{A}(a)$. We can think to a utility function ($U_{n}$) which could be written as a function of another aggregate variable ($K_n$). Then the contraction principle gives the rate function of the utility $U_{n}(K_{n})$ as a constraint minimization of the unconstraint rate function of $K_{n}$.

\begin{figure}[!ht]
\begin{center}
\includegraphics[width=0.675\textwidth]{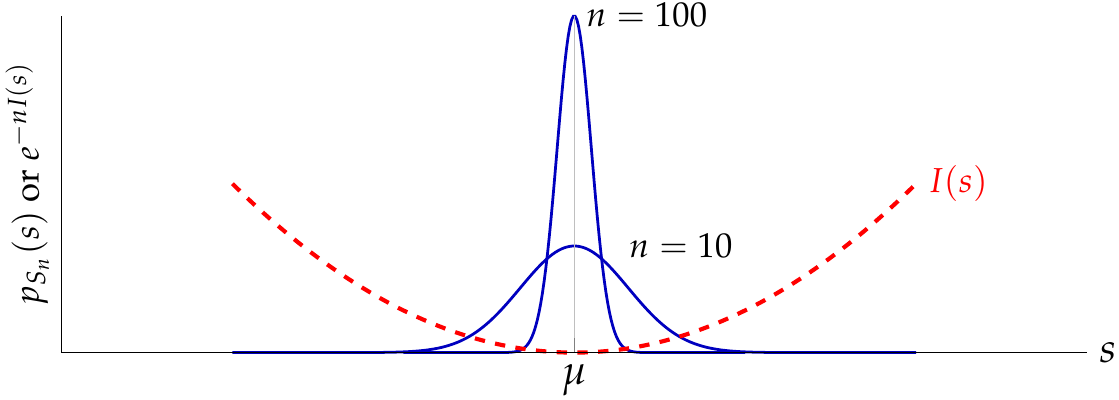}
\end{center}
\caption[Rate function for a Gaussian sample mean]{\label{fig:Ratefunc} The Rate function for a Gaussian sample mean vs the value of a realization of the sample mean.}
\end{figure}

As an example, let $n$ IID Gaussian $\mathcal{N}(\mu,\sigma)$ random variables (RV) and $S_n$  be the sample mean $S_{n}\equiv n^{-1}\sum_{i=1}^{n}X_{i}$. The SCGF is

\begin{eqnarray*}
  \lambda(k) &=&\lim_{n\rightarrow\infty}\frac{1}{n}\ln\mathrm{E}\left[e^{nk(n^{-1}\sum_{i=1}^{n}X_{i})}\right]  \\
             &\stackrel{\mathrm{Ind}}{=}& \lim_{n\rightarrow\infty}\frac{1}{n}\ln\prod_{i=1}^{n}\mathrm{E}\left[e^{kX_{i}}\right]
             =\ln \mathrm{E}\left[e^{ kX}\right]=\mu k+\frac{(k\sigma)^2}{2}
\end{eqnarray*}
which is everywhere differentiable with respect to $k$ so the GE theorem applies. The rate function is therefore $I(s)=\sup_{k\in \mathbb{R}}\{ks-\lambda(k)\}=\sup_{k\in \mathbb{R}}\{ks-\mu k+2^{-1}(k\sigma)^2 \}$ which implies $k_{\mathrm{max}}(s)=(s-\mu)/\sigma$. Finally, the rate function is

\begin{equation*}
  I(s)=\frac{(s-\mu)^2}{2\sigma^2}
\end{equation*}
which is illustrated in Fig-\ref{fig:Ratefunc}. We note that $I(s)$ is quadratic so the fluctuations of the Gaussian sample mean are Gaussian, as expected.

The large deviation theory is an extension of the law of large numbers (LLN) and of the central limit theorem (CLT). The law of large numbers is obtained by minimizing the (strictly convex) rate function and the central limit theorem by taking the second order Taylor expansion of the rate function. If $I(a)$ is derived from the GE theorem then $I(a)$ is strictly convex (the LF transform yields to convex functions and by assumption, the SCGF is differentiable everywhere) and $I(a)\geq0$ (if $I(a)<0$ then $\Pr(A_{n}\in \ud a)\asymp \exp(-nI(a))\ud a$ diverges).

\begin{itemize}
  \item[LLN:] If the rate function has a unique global minimum and is strictly convex, then $k(a_{\mathrm{min}})=0=I'(a_{\mathrm{min}})$ implying $I(a_{\mathrm{min}})=0=k(a_{\mathrm{min}})a_{\mathrm{min}}-\lambda(k(a_{\mathrm{min}}))$. The expansion of $I(a)$ to the zeroth order gives the LLN: $\lim_{n\rightarrow\infty}\Pr(A_{n}\in [a_{\mathrm{min}},a_{\mathrm{min}}+\ud a])=\exp(-nI(a_{\mathrm{min}}))=1$. The probability to deviate from the most probable state tends to zero when the system size tends to infinity.

      To illustrate this result, consider again the sample mean of $n$ $X_i$ IID random variables $\{X_i\}$: $A_{n}=n^{-1}\sum_{i=1}^{n}X_{i}$. The SCGF is $\lambda(k)=\ln\mathrm{E}\left[e^{kX}\right]$. If $\lambda(k)$ is differentiable everywhere then $k(a)$ is the unique root of $\lambda'(k)=a$ and $a(k)$ is the unique root of $I'(a)=k$. At $a_{\mathrm{min}}$ one has $a_{\mathrm{min}}=\lambda'(0)=\lim_{n\rightarrow\infty}\mathrm{E}[A_{n}]=\mu$. Then the previous result gives $\lim_{n\rightarrow\infty}\Pr(n^{-1}\sum_{i=1}^{n}X_{i}\in [\mu-\epsilon,\mu+\epsilon])=1$.

  \item[CLT:] Under the same assumptions, the expansion to the second order around $a_{\mathrm{min}}$ gives $I(a)\simeq I(a_{\mathrm{min}})+I'(a_{\mathrm{min}})(a-a_{\mathrm{min}})+2^{-1} I''(a_{\mathrm{min}})(a-a_{\mathrm{min}})^{2}$ thus $I(a)\simeq 2^{-1} I''(a_{\mathrm{min}})(a-a_{\mathrm{min}})^{2}$. Around the minimum of the rate function $a_{\mathrm{min}}$, we have the approx of Gaussian fluctuations: $\Pr(A_{n}\in [a,a+\ud a])\simeq\exp\left(-nI''(a_{\mathrm{min}})(a-a_{\mathrm{min}})^{2}/2\right)\ud a$. The probability to observe a realization in an interval close to the most probable state is  Gaussian.
\end{itemize}

Let's illustrate these results and their limitations. Consider the Brock and Durlauf binary choice model in presence of social interaction \cite{Brock}. It is possible to show (See chap-\ref{chap:social}) that the deterministic part of the social planer's utility function for an homogeneous complete social network (each agent is influenced by all the others with the same social strength) can be written as

\begin{equation}
 \mathcal{U}(\mathbf{s})=\frac{J}{2n}\left(\sum_{i=1}^{n}s_{i}\right)^{2}+h\sum_{i}s_{i}
\end{equation}
where $J$ is the strength of social interactions, $s_{i}$ is the binary choice of the $i$th agent and $h$ is the idiosyncratic preference. Then the static mean consensus $M_{n}=n^{-1}\sum_{i=1}^{n}s_{i}$ is distributed as

\begin{equation}
p_{M_{n}}(m)=e^{-n\beta J \phi(m)-\ln\mathcal{Z}(\beta)}\quad\text{with}\quad \phi(m)=2^{-1}m^{2}-\beta^{-1}J^{-1}\ln\left(2\cosh \beta(J m+h)\right)
\end{equation}

\begin{figure}[!ht]
\begin{center}
\includegraphics[width=0.75\textwidth]{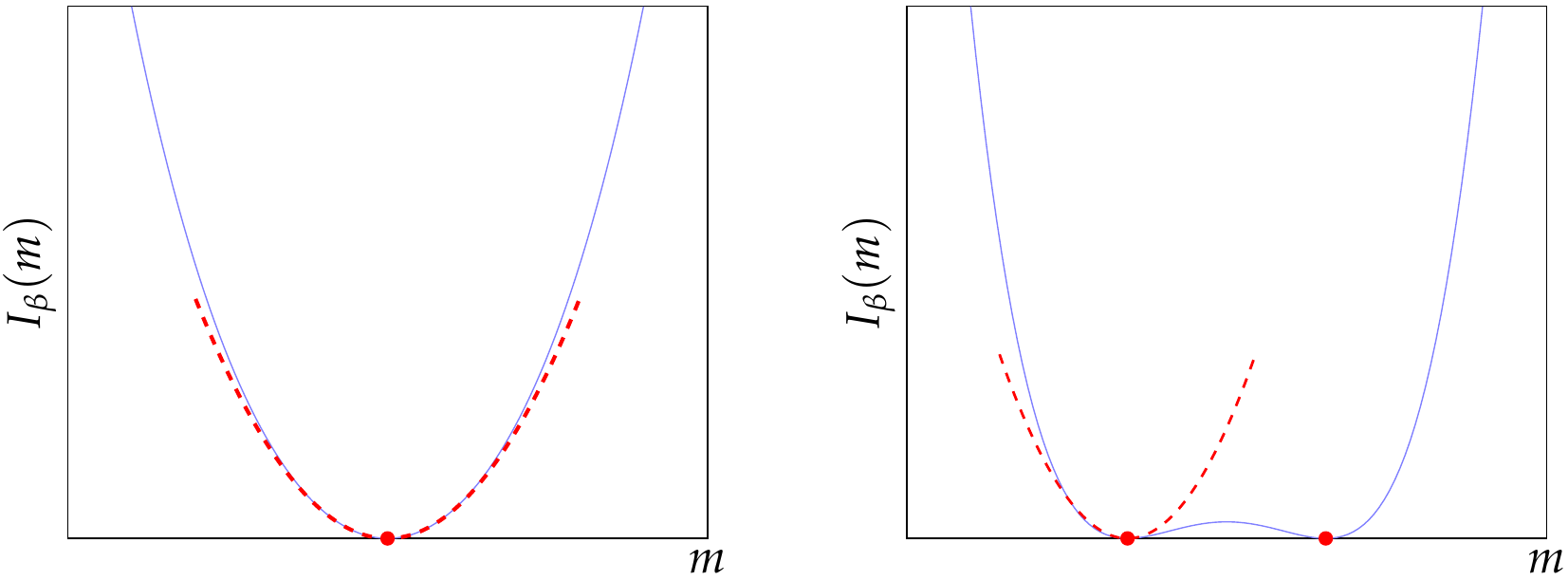}
\end{center}
\caption[LDT, LLN and CLT]{\label{fig:LDTclt} The rate function for the binary choice problem in presence of social interactions. The social network is homogeneous (each agent interacts in the same way with all the others). One makes the assumption of rational expectations. The left panel illustrates the disordered state (zero spontaneous consensus) and the right panel illustrates the ordered states (non-zero spontaneous consensus). The zeros of the rate function are illustrated by dots and the dashed lines stand for the quadratic approximation.}
\end{figure}

The rate function $I_{\beta}(m)\equiv\beta J \phi(m)+n^{-1}\ln\mathcal{Z}$ characterizes the mean consensus distribution meaning that its mathematical properties rule the socio-economic behaviours as illustrated in Fig-\ref{fig:LDTclt}. The minimization of the rate function $I_{\beta}(m)$ gives the self-consistent equation $m=\tanh(\beta J m)$ which is precisely the equilibria derived by Brock and Durlauf in the binary choice model for the homogeneous complete social network \cite{Brock}.

This model is interesting because it illustrates two very different regimes: zero and non-zero spontaneous consensus (respectively disordered and ordered states). In the disordered state, the LLN holds because $I_{\beta}(m)$ has a single zero which means a single accumulation point. More and more probability mass is accumulated at $m=0$ when the number of agents increases. The CLT also holds in the disordered state because $I_{\beta}(m)$ is locally quadratic which means that deviations larger than (says) $3$ standard deviations around the mean value $m=0$ of the consensus are very unlikely.

On the other hand, near the transition zero-non/zero spontaneous consensus, the LLN and the CLT are no longer valid. As described by Brock and Durlauf, this model can have two equilibria (depending on the value of the product $\beta J$). The breakdown of the LLN is a consequence of multiple zeros of the rate function. There are two points where the probability distribution does not decay exponentially in the Brock and Durlauf model when $\beta J>1$ \cite{Brock} as illustrated in the left panel of Fig-\ref{fig:LDTclt}. The CLT is no longer valid because the rate function is not locally quadratic neither around the mean value $m=0$ nor around the accumulation points (dots in the Fig-\ref{fig:LDTclt}). It results to near the transition zero- non zero spontaneous consensus, the fluctuations around the mean value are larger than Gaussian fluctuations. In such models, the consequence of collective behaviours is that "unlikely" or "extreme" events could not be so rare. Moreover, since large deviations are not so rare, the mean value of the consensus could be an irrelevant aggregate variable for a macroscopic description.

\subsection{Entropy and maximum entropy principle}\label{maxentLD}

A heuristical link between the rate function and the entropy is given by the Sanov theorem. Consider a random experiment with a discrete sample space $\Omega=\{q^{-1},\cdots,q^{-1}\}$ (uniform outcomes) repeated $n$ times. The probability to observe a given sequence of $n$ outcomes is $\Pr(\boldsymbol{\omega})=q^{-n}$. The empirical relative frequencies are given by the random vector $\mathbf{L}^{n}=n^{-1}\sum_{j=1}^{n}(\delta_{\omega_{j},1};\cdots;\delta_{\omega_{j},q})$. What is the probability that the empirical relative frequencies are equal to $n^{-1}(k_{1};\cdots;k_{q})\equiv n^{-1}\mathbf{k}$ with $\sum_{i=1}^{q}k_{i}=n$? In this particular case, one can derive straightforwardly the answer:

\begin{equation}\label{PottsFact}
  \Pr(\mathbf{L}^{n}=n^{-1}\mathbf{k})=\frac{1}{q^{n}} \frac{n!}{\prod_{i=1}^{q}k_{i}!}
\end{equation}
because there are $n!/\prod_{i=1}^{q}k_{i}!$ ways to have $\mathbf{L}^{n}=n^{-1}\mathbf{k}$. Using the Laplace approximation of the factorial function (Stirling approximation), one obtains

\begin{equation}\label{PottsStir}
  \Pr(\mathbf{L}^{n}=n^{-1}\mathbf{k})\asymp e^{-n D_{\mathrm{KL}}\left(n^{-1}\mathbf{k}||\Pr(\boldsymbol{\omega})\right)}
\end{equation}
where $D_{\mathrm{KL}}(\mu||\nu)=\sum_{x}\mu(x)\ln \frac{\mu(x)}{\nu(x)}$ is the relative entropy. As $\nu(\boldsymbol{\omega})=q^{-n}$, the relative entropy is equal to the opposite of the entropy $S[\mu(x)]=-\sum_{x}\mu(x)\ln \mu(x)$ up to a constant (see Sec-\ref{sec3:entropy}). The rate function is then nothing but the statistical entropy.

Furthermore, if one uses the contraction principle one obtains the maximum entropy principle \cite{Jaynes}. Namely, the contraction principle allows to derive the probability measure maximizing the unconstrained entropy consistently with a given representation (ie a set of constraints). In words, the most likely way to describe unlikely events is to minimize the rate function (maximize the entropy) with respect to some knowledge about the considered system. If a random variable $B_{n}$ admits a continuous representation $f(A_{n})$ in terms of another random variable $A_{n}$ satisfying a LDP, then one obtains
\begin{eqnarray}
  \Pr(B_{n}\in \ud b) &=& \int\limits_{\{a:f(a)\in \ud b\}}\Pr(A_{n}\in \ud a) \\
   &\asymp& \int\limits_{\{a:f(a)\in \ud b\}}\exp(ns_{A}(a))\ud a  \qquad \text{with}\quad s_{A}(a)\equiv -I_{A}(a)\\
   &\asymp&  \exp\left(\sup_{\{a:f(a)= b\}}\{n s_{A}(a)\}\right)\ud b \label{3-LDTmaxent}
\end{eqnarray}
and thus the rate functions are related (contraction principle) as

\begin{equation}
  s_{B}(b)=\sup_{\{a:f(a)= b\}}\{ s_{A}(a)\}
\end{equation}

We can think to a utility function ($U_{n}$) which could be written as a function of another aggregate variable (says $K_n$). Then the contraction principle gives the rate function of the utility $U_{n}(K_{n})$ as a constraint minimization of the unconstraint rate function of $K_{n}$.

Formally, one can show \cite{Ellis,TouchetteLDT} that if one considers $n$ random variables $\{X_{i}\}$ (with uniform prior) associated to $n$ entities and if the $K$ observed quantities $\{\mathrm{E}[f_{k}(X)]\}$ are functions of $\mathbf{X}\equiv(X_{1};\cdots;X_{n})$ then the rate function is given by the following optimization problem

\begin{eqnarray}\label{3-MEPv1}
  S[\{\mu_{k}\}]=& \sup &S[p(\mathbf{X})]= -\sum_{\{\mathbf{X}=\mathbf{x}\}}p(\mathbf{x}) \,\ln p(\mathbf{x})  \\ \nonumber
   &\mathrm{s.t}& p(\mathbf{X})\geq0, \;  \mathrm{E}_{p}[1]=1,\; \mathrm{E}[f_{k}(X)]=\mu_{k}\quad\mathrm{for}\,k=1,\cdots,K\nonumber
\end{eqnarray}

\subsection{Minimum free energy principle}

Assume that the prior distribution is given by a Gibbs distribution \newline $\mathrm{Pr}_{\beta}(\ud\boldsymbol{\omega})\equiv\mathcal{Z}_{n}^{-1}(\beta)\exp(-\beta H_{n}(\boldsymbol{\omega}))\Pr(\ud \boldsymbol{\omega})$ rather than a uniform one. If a  LDP holds for a random variable $M_{n}$ and if the "energy" (the opposite of a utility function) $H_{n}$ can be restated as a function of $M_{n}$ (eg: the net consensus) then

\begin{eqnarray}
  \mathrm{Pr}_{\beta}(M_{n}\in dm) &=& \mathcal{Z}_{n}^{-1}(\beta)\int\limits_{\{\omega:M_{n}(\boldsymbol{\omega})\in\ud m\}}\exp(-\beta H_{n}(m))\Pr(\ud \boldsymbol{\omega}) \\
  &=& \mathcal{Z}_{n}^{-1}(\beta)\exp(-\beta H_{n}(m))\int\limits_{\{\omega:M_{n}(\boldsymbol{\omega})\in\ud m\}}\Pr(\ud \boldsymbol{\omega}) \\
  &=& \mathcal{Z}_{n}^{-1}(\beta)\exp(-\beta H_{n}(m))\Pr(M_{n}\in\ud m) \\
   &\asymp& \exp\left(-n\left[\frac{\beta H_{n}(m)}{n}-s(m)-\phi(\beta)\right]\right)\ud m
\end{eqnarray}
where $\phi(\beta)$ is the SCGF which can be rewritten as $\phi(\beta)=\lim_{n\rightarrow\infty}-n^{-1}\ln \mathcal{Z}_{n}(\beta)$.

The most probable state is then given by the infimum of the rate function $I_{\beta}(m)=\beta \frac{H_{n}}{n}(m)-s(m)-\phi(\beta)$ which leads to the so-called (for historically reason) minimum free energy principle.

\begin{equation}\label{MinFEP}
  \phi(\beta)=\inf_{m}\{\beta \frac{H_{n}(m)}{n}-s(m)\}
\end{equation}
where $\beta \frac{H_{n}}{n}(m)-s(m)$ is the density of free energy as a function of the realized value $m$ of $M_{n}$. The resulting value of the consensus $m$ has a large utility together with a large economic diversity \cite{Aoki}.

\subsection{Domain of validity}

The Legendre-Fenchel transform $\lambda(k)=\sup_{x}\{kx-f(x)\}$ yields to convex functions, the GE theorem does not allow to calculate non-convex rate functions (the rate function may have several minima, for instance). Assuming that the rate function is non-convex, the double Legendre-Fenchel transform yields to the convex hull of the non-convex function. If the scaled cumulant generating function is differentiable everywhere then the Legendre-Fenchel transform is an involution (is its own inverse). Furthermore if the SCGF is differentiable everywhere and strictly convex, then $\lambda'(k)$ is a monotonically increasing function and $\lambda'(k)=x$ can be inverted (one-to-one matching) $f'(x)=k$ (where the prime stands for the derivative with respect to the independent variable).

If the SCGF has a non-differentiable point (e.g. as $|x|$ at $x=0$), its Legendre-Fenchel transform has a linear part on the interval $[x_{l}\equiv\lambda'(k_{\mathrm{nondiff}}^{-}),x_{r}\equiv\lambda'(k_{\mathrm{nondiff}}^{+})]$.

If $f(x)$ is non-convex then the SCGF $\lambda(k)$ must have a non-differentiable point (the demonstrations can be found in \cite{Rock}).

Another restricting condition is the limit of infinitely large system $n\rightarrow\infty$. This condition is required in the Laplace approximation of the integral, which leads to the Legendre-Fenchel transform. If this condition is not met, the link between the SCGF and the rate function is only a qualitative one.

\section{Laplace approximation}

\begin{figure}[!ht]
\begin{center}
\includegraphics[width=0.75\textwidth]{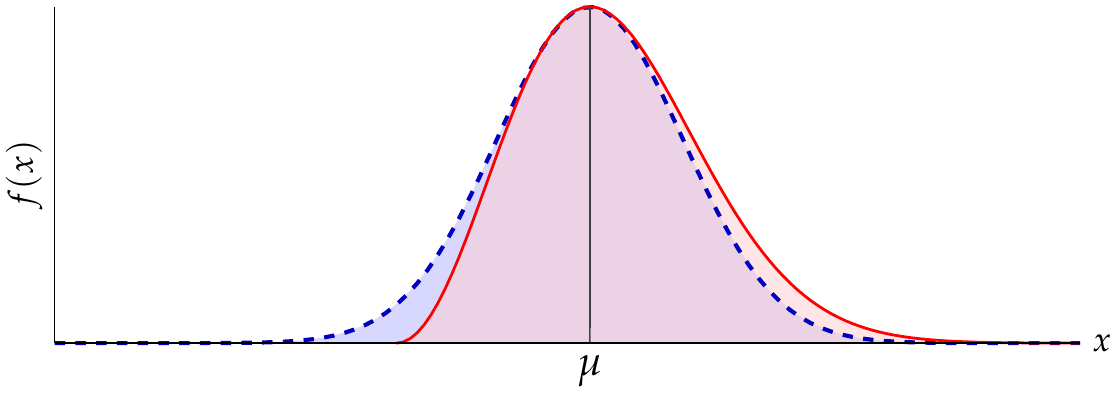}
\end{center}
\caption[Laplace approximation]{\label{fig:Laplace} The Laplace approximation is illustrated by the dashed curve.}
\end{figure}

The Laplace approximation (or saddle point approximation) is a method to approximate integrals which have a large integrand; this method is illustrated in Fig-\ref{fig:Laplace}. Let $f(x)$ a regular enough function with a single maximum at $x_{\mathrm{max}}$ then $f(x)\simeq f(x_{\mathrm{max}})-2^{-1}|f''(x_{\mathrm{max}})| \,(x-x_{\mathrm{max}})^{2}$. The Laplace approximation is
%

\begin{equation*}
  \mathcal{I}(n) \equiv \int_{\mathbb{R}}e^{nf(x)}\ud x \simeq e^{nf(x_{\mathrm{max}})} \int_{\mathbb{R}}e^{-\frac{n |f''(x_{\mathrm{max}})|}{2}(x-x_{\mathrm{max}})^{2}}\ud x
   = e^{nf(x_{\mathrm{max}})} \sqrt{\frac{2\pi}{n |f''(x_{\mathrm{max}})|}}
\end{equation*}
The last equality follows from Gaussian integration. One has: $\ln \mathcal{I}(n)\simeq nf(x_{\mathrm{max}})$ for $n\gg 1$.

\end{subappendices}

\chapter{Statistical pairwise interaction model of the stock market}\label{chap:SPIMSM}
\thispagestyle{empty}
\begin{summary}
Financial markets are a classical example of complex systems as they comprise many interdependent stocks. As such, we can obtain a surprisingly good description of their structure by taking into account only the sign of their variations. Models have been applied and gave some valuable results but at the price of restrictive assumptions on the market dynamics or others are agent-based models with rules designed in order to recover some empirical behaviours (power laws, etc.). Here we show that the pairwise model is actually a statistically consistent model with observed first and second moments of the stocks orientation without making such restrictive assumptions. This is done with an approach based only on empirical data of returns. Our data analysis suggests that the actual interaction structure may be thought as a pairwise maximum entropy model on a complex network with mutual influences scaling as the inverse of system size. This has potentially important implications since many properties of such a model are already known and some techniques can be straightforwardly applied. Typical behaviours, as multiple equilibria or metastable states, different characteristic time scales, spatial patterns, order-disorder, could find an explanation in this picture.
\end{summary}

\begin{figure}[!ht]
\begin{center}
\includegraphics[scale=0.8]{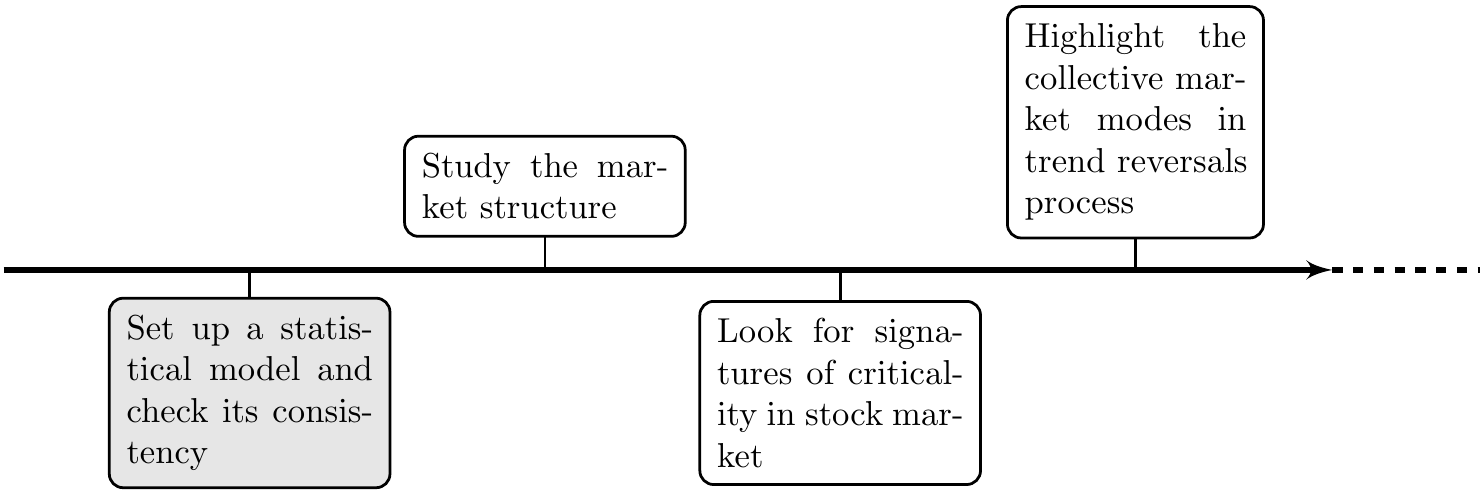}
\end{center}
\end{figure}

\newpage
\section{Introduction}
\label{intro}
A very interesting feature of complex systems is that sometimes the microscopic details of interactions are not necessary to explain the observed macroscopic structures (at least qualitatively). The most famous examples are the Ising and spin glasses models where the interactions are taken as constant or randomly distributed in a given neighbourhood. In these models, the complicated interactions between electrons are simply replaced by pairwise interactions with a dedicated coupling parameter.
It is amazing that the pairwise maximum entropy (maxent) model also describes neural populations \cite{Schneidman}. The authors of this paper showed that the activity time-series of neural populations can be described by such a simple model (however with a major difference with the physical Ising model).

This suggests that the most relevant properties governing the macroscopic behaviours of such complex systems are not the nature of the microscopic entities but are the order of interactions, their range and the topology. Moreover these simple rules are able to explain sophisticated behaviours such as order-disorder transition, memory, clustering and many more collective behaviours.

One also finds collective phenomena in finance \cite{Dal,Jr}, non-random correlations \cite{Laloux} and complex structures \cite{Mant,Onnela}.
The authors of these papers report a certain degree of co-movement with different tools, especially during crisis. Using the Kuramoto model, they recently showed \cite{Dal} that synchronization is observed during crisis. We note that the Kuramoto model is also used in neuroscience and it is related to a kind of pairwise maximum entropy model used in the description of phase transition. Moreover, financial correlation matrices are known to be noise dressed but a \emph{market-mode} (eigenvector with roughly equal components on all stocks) is however observed. Using the eigenvalues and eigenvectors of correlations matrices of some main financial market indices, one can reach the conclusion of a probable existence of a global collective-mode. Last, using tools of graph theory and financial correlations, one can show that markets are strongly re-organized during successive periods of crisis and "normal" operating state \cite{Mant,Onnela}.

Such phenomena can occur in systems composed of many interacting entities (where \emph{interaction} is taken at the larger sense of mutual influence or simply taken as a measure of co-movements).
Moreover, as recently observed \cite{Petra}, the financial and neural networks have topological similarities (modular, hierarchical, small-world organization highlighted by an asset tree based approach). In this view, the pairwise maximum entropy paradigm seems to be an attractive candidate to explain the market structure. A version of this model was already applied to finance but with the restricting assumption that the market dynamics follows the soft-spins Langevin dynamics \cite{Rosenow}. There are also Ising like models which are agent-based models with specific rules such as \emph{"do what your neighbours do"} or more complex dynamical rules \cite{Born,Zhou,Sherr}. The latter approach is thus a different one that Rosenow's (or the present) approach where the elementary entities are stocks and not traders (the most accessible observables are price returns).

The aim of this work is to show that market behaviour can be explained without such hypothetical rules and that the aforementioned collective phenomena result from the mutual influences of underlying constitutive entities, the stocks (in the same spirit of the characterization of collective phenomena in neural networks without using other information than their activity time series \cite{Schneidman}).
We emphasize that this approach is a data-based approach. We do not introduce any rules or dynamical restriction. We only require that the model fits first and second empirical moments. The reason is that underlying microscopic details seem to be unnecessary for the macroscopic description of such phenomena. Indeed macroscopic behaviours in magnetic materials and in neural networks are consistently described by maximum entropy models even though electrons and neurons are undoubtedly completely different elementary entities at the individual scale (as well as their microscopic dynamics).
Furthermore, agent-based models can reveal interesting behavioural patterns but since such different dynamics as the neurons potential activity dynamics and spin dynamics can lead to the same macroscopic patterns, it seems natural to propose a complementary statistical and data-based approach allowing to relax almost any assumption.

Here, we consider stocks as economic entities influencing each other. The interaction process itself is not detailed. Instead, we propose a derivation of the pairwise model only based on the (incomplete) information embedded in data with no restricting assumption on an underlying dynamics. The only (rough) assumption that we made is binarizing prices to interpret daily movement as a positive (or negative) orientation. Such a simplification has already shown its power in neural networks and magnetic materials (at least in structure studies) where the complex interaction process is approximated by a pairwise model and relevant variables (action potential and spin) are binarized.
In this work we provide evidence that a pairwise maxent model on a complex network can accurately describe the stock market. We show that almost all the interaction strengths are Gaussian random variables, that Gaussian influences are compatible with non-Gaussian eigenvalues of the returns correlation matrix and that the mean influence scales as a power close to $-1$ of the system size. Furthermore frustration seems to be a key property since approximately half of the interaction strengths are negative.
We also propose an economic interpretation based on the mutual influence scheme developed in \cite{Brock}. Furthermore the interaction strengths can be thought as incentive since they are related to the Hessian matrix of the utility function \cite{Mas}.

With these features, we conclude that the proposed model may fall into the class of the exact mean field models. We also show that we can reproduce the largest (non-Gaussian) eigenvalue of the returns correlation matrix corresponding to the market-mode \cite{Laloux}, making the link with the random matrix approach. This mapping and the first clue of the reproduction of the market-mode suggest that a link to critical phenomena can be done in this paradigm.
Moreover the topological similarities between the market and neural networks can find their origin in this common statistical model.
Other properties as the existence of hierarchical structures \cite{Mant,Onnela}, possibility of the order-disorder transition and synchronization \cite{Jr,Dal} can potentially be explained by the pairwise paradigm.

The chapter is organized as follows. In section \ref{sec:model}, we present the model, its economic interpretation and the link between the interaction matrix and the moments. In sections \ref{sec:consistency} and \ref{sec:consistency2}, we give evidence that the information embedded in the data is mostly explained by pairwise but no higher-order interactions. In section \ref{sec:dist}, we study the distribution of the influence matrix.

\section{The model}\label{sec:model}
\subsection{Inferred distribution}
Our aim is to set up a model describing the market state and its structure based only on statistical considerations. This requires a way to infer the probability distribution in order to get the observables (here, the associated moments). The model will also allow the study of the market structure. All these quantities will be defined below, see Sec-\ref{sec3:MEP} for details.
We consider a set of $N$ market indices or $N$ stocks with binary states $s_{i}$ ($s_{i}=\pm1$ for all $i=1,\cdots,N$). A system configuration will be described by a vector $\textbf{s}=(s_{1},\cdots,s_{N})$. The binary variables will be equal to $1$ if the associated closing price is larger than (or equal to) the opening one and equal to $-1$ if not. We choose open-to-close rather than close-to-close returns to avoid the over-night effect and the weekend gap (Friday-Monday closings). A configuration $\textbf{s}$ is a binary version of stock returns. Such a simplification of returns is made to study the market structure and will be justified a posteriori if the results are consistent with the data. One knows that this approximation is already useful in the description of neural populations \cite{Schneidman} and that neural networks are similar to financial networks \cite{Petra}. We may thus think that it will also be the case in finance; this will be justified a posteriori as the model gives consistent results. We may also consider this simplification as a study of the return signs. Indeed, stock returns can be rewritten as $r_{t}=\sgn(r_t)|r_{t}|$. Signs of stock returns are sometimes considered as uncorrelated and attract less attention \cite{Cont}. However correlations may appear in complicated (non-linear) fashion as synchronization during crises \cite{Jr}. It seems interesting to study orientation changes.

\begin{figure}[!ht]
\begin{center}
\resizebox{0.9\textwidth}{!}{%
  \includegraphics{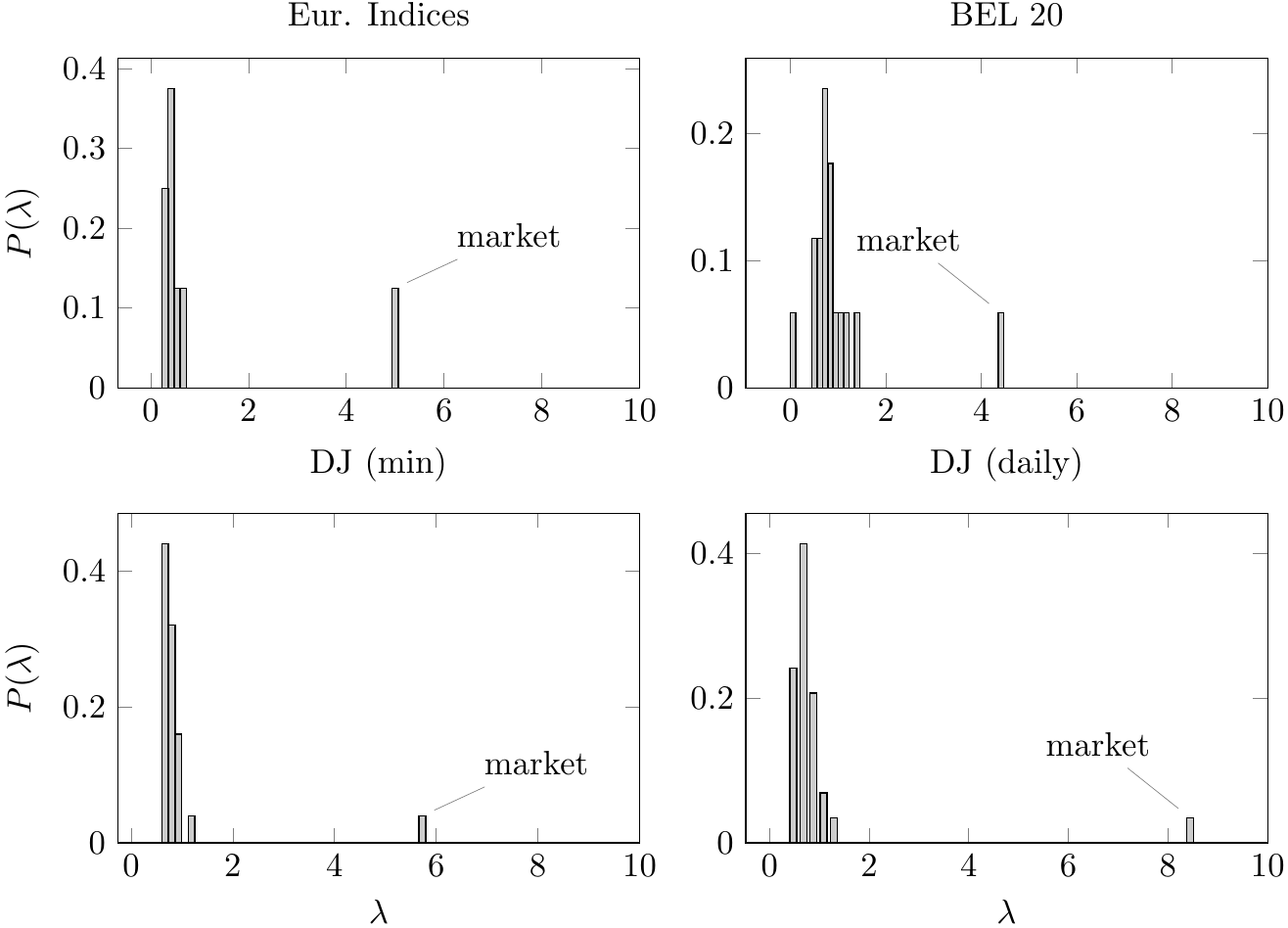}
}
\end{center}
\caption[Indices eigen-mode]{Probability distribution of eigenvalues of the correlation matrix of the binarized returns for 8 European indices (top-left), Bel20 (top-right), Dow Jones at minute sampling (bottom-left) and Dow Jones daily (bottom-right). The market-mode is pinned.}
\label{fig:DJmode}
\end{figure}

A first clue that it is not a too rough an approximation is that it preserves the market eigen-mode (largest eigen-value of the price-returns covariance matrix) \cite{Laloux} as illustrated in Fig-\ref{fig:DJmode}

Another motivation of this approximation is that the resulting binary pairwise model allows collective phenomena which are observed in the market. We will discuss the description of the collective phenomena (structure reorganization, synchronization, etc.) by this pairwise model in another chapter.

We seek to establish the less structured model explaining only the measured mean orientations $q_{i}$ and instantaneous pairwise correlations $q_{kl}$ in terms of theoretical moments $\langle s_{i}\rangle$ and $\langle s_{k}s_{l}\rangle$ without making any further assumption. The brackets $\langle\cdot\rangle$ denote the average with respect to the unknown distribution $p(\textbf{s})$. As the entropy of a distribution measures the randomness or lack of interaction among binary variables, a way to infer such probability distribution knowing the mean orientations and correlations is the maximum entropy principle (MEP). Jaynes showed how to derive the probability distribution using the maximum entropy principle \cite{Jaynes}. It consists in the following constrained maximization (see Sec-\ref{sec3:MEP} for details and Sec-\ref{sec3:LDT} for mathematical motivation)

\begin{eqnarray}\label{maxent}
  & &  \max_{\substack{\{p(\mathbf{s})\}}} S(\textbf{s})= \max_{\substack{\{p(\mathbf{s})\}}}\left\{ -\sum_{\{\textbf{s}\}}p(\textbf{s}) \,\ln p(\textbf{s})\right\}  \\ \nonumber
   &\mathrm{s.t}&  \sum_{\{\textbf{s}\}}p(\textbf{s})=1,\quad \sum_{\{\textbf{s}\}}p(\textbf{s})s_{i}=q_{i},
   \quad \sum_{\{\textbf{s}\}}p(\textbf{s})s_{i}s_{j}=q_{ij} \nonumber
\end{eqnarray}
Using Lagrange multipliers method, the resulting  two-agent distribution $p_{2}(\textbf{s})$ is given by

\begin{equation}
p_{2}(\textbf{s})=\mathcal{Z}^{-1}\exp\left(\frac{1}{2}\sum_{i, j}^{N}J_{ij}s_{i}s_{j}+\sum_{i=1}^{N}h_{i}s_{i}\right)\equiv\frac {e^{- \mathcal{H}(\textbf{s})}}{\mathcal{Z}}\label{04-Lagrange}
\end{equation}
where $J_{ij}$ and $h_{i}$ are the Lagrange multipliers and $\mathcal{Z}$ a normalizing constant (the partition function). They can be expressed in terms of partial derivatives of the entropy as

\begin{equation}
\frac{\partial S(\textbf{s})}{\partial q_{i}} = -h_{i} \qquad
\frac{\partial S(\textbf{s})}{\partial q_{ij}} = -J_{ij}\label{multipliers}
\end{equation}

Thus preferences are conjugated to mean orientations and pairwise influences to pairwise correlations. So the parameters $J_{ij}$ can be thought as a measure of co-movements and $h_{i}$ as a measure of individual movements or as external influences.

Cumulants are obtained from this model and we give their relation to the interaction strengths. As the statistical model (\ref{04-Lagrange}) is expressed as a Gibbs distribution, we have the relations

\begin{equation}\label{moments}
\langle s_{i_{1}}\ldots s_{i_{N}}\rangle_{\mathrm{c}}=\partial^{N}\ln \mathcal{Z}/\partial h_{i_{1}}\ldots \partial h_{i_{N}}
\end{equation}
where $\langle\cdot\rangle_{\mathrm{c}}$ is the cumulant average \cite{Kubo}. This relation gives the link between $\mathbf{J}$ and pairwise correlations. If the partition function $\mathcal{Z}$ cannot be explicitly computed, we can use Plefka series \cite{Plef} or a variational cumulant expansion \cite{Barb} (see Sec-\ref{sec3:Var}).

\begin{figure}[!ht]
\begin{center}
\resizebox{0.85\textwidth}{!}{%
  \includegraphics{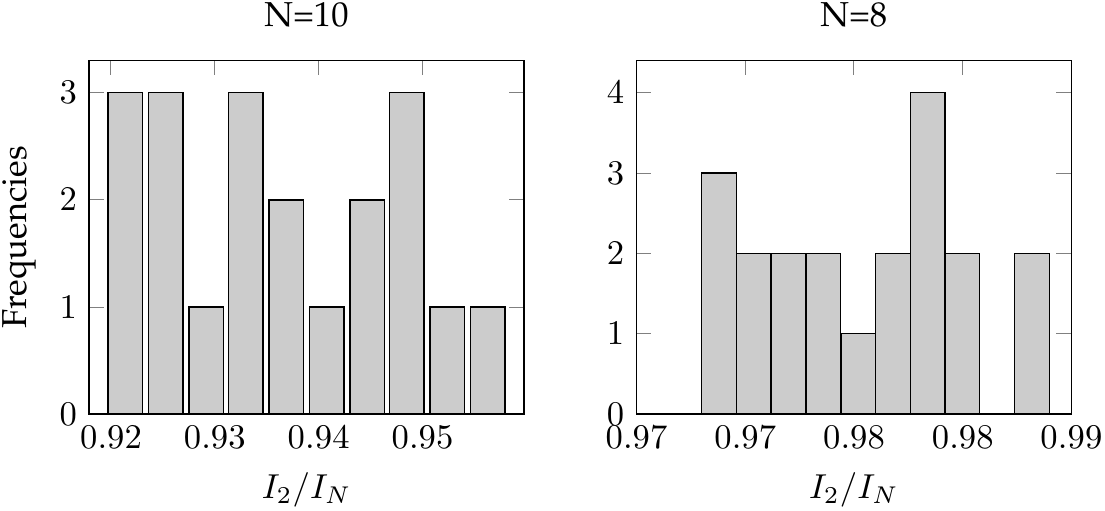}
}
\end{center}
\caption[Multi-information distribution]{Multi-information ratio $I_{2}/I_{N}$ for 20 sets of 8 randomly chosen stocks (right) and for 20 sets of 10 randomly chosen stocks. Sampling time-scale is the minute, the sample length is $3\times 10^{4}$ points and parameters were estimated with a regularized pseudo-maximum likelihood method.}
\label{fig:MI}
\end{figure}

Finally, we test if higher order influences should be ruled out (see Sec-\ref{sec3:TestMEP} for details). We proceed by using the multi-information criterion \cite{Schneid_Multi,Schneidman}. We sketch hereafter the basic idea of this criterion. Considering a financial network of $N$ entities, one can obtain maximum entropy distributions $p_{k}(\textbf{s})$ which are consistent with kth-order moments (for any $k=1,\cdots,N$) like in (\ref{maxent}). The case $k=N$ is an exact description of the financial network. Thus the entropies $S_{k}\equiv S[p_{k}]$ of these distributions decrease with increasing $k$ toward the true entropy $S\equiv S[p_{N}]$ since more correlation reduces the entropy. The multi-information $I_{N}\equiv D_{\mathrm{KL}}\left(p_{N}||p_{1}\right)$ is a measure of the total amount of correlations in the system (where $ D_{\mathrm{KL}}$ is the Kullback-Leibler divergence). Thus if the ratio $I_{2}/I_{N}=(S_{1}-S_{2})/(S_{1}-S_{N})$ is close to $1$ then pairwise correlations provide an effective description of the correlation structure.
For a set of 8 European indices, we obtain $I_{2}/I_{N}=98.2\%$ which means that pairwise correlations represent most of correlations. For the Dow Jones (minute sampling time-scale and $3\times10^{4}$ points), we obtain $I_{2}/I_{N}=95.7\%$ in average. In the latter case we consider 20 sets of 8 randomly chosen stocks and 20 sets of 10 randomly chosen stocks (values for which direct sampling of the distribution gives a good estimate); the results are illustrated in Fig-\ref{fig:MI}.

\begin{figure}[!ht]
\begin{center}
\resizebox{0.6\textwidth}{!}{%
  \includegraphics{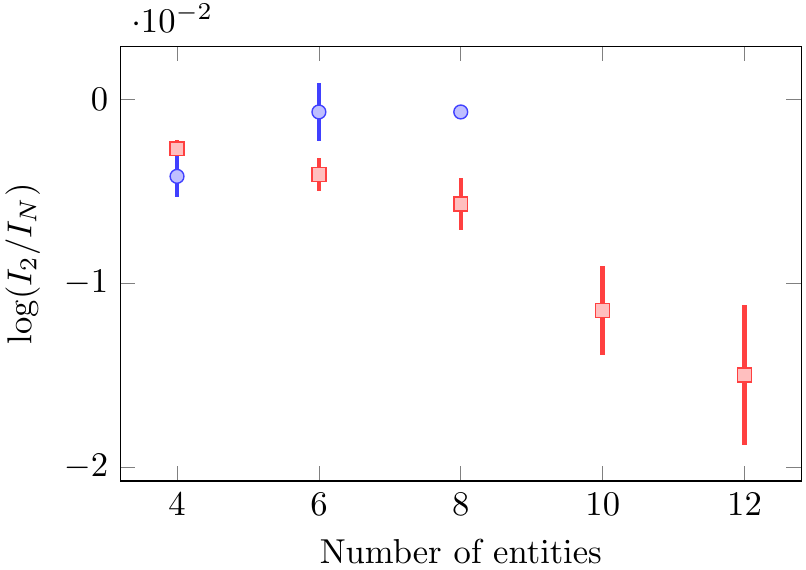}
}
\end{center}
\caption[Multi-information vs the number of stocks]{Multi-information ratio $I_{2}/I_{N}$ for the European indices (dots) and for the Dow Jones at minute time-scale (squares). Each point is the average on 20 randomly chosen groups of size $N$. Error bars represent the standard deviation over those groups.}
\label{fig:MIeurDJ}
\end{figure}

Last we compute the multi-information ratio for groups of different sizes. The values are computed on 20 randomly chosen groups for each considered system size and the $\textbf{J}$ matrix is estimated with the regularized pseudo-maximum likelihood method (see Sec-\ref{sec3:rpml}). The results are illustrated in Fig-\ref{fig:MIeurDJ}.

We observe that the multi-information ratio is close to $1$ whatever the size but it decreases whit increasing size for the Dow Jones while it increases with increasing size for the European indices set. The pairwise model is thus able to explain almost all the correlation structure (at least $95\%$ of it).

\subsection{Interpretation}

The Gibbs distribution (\ref{04-Lagrange}) is similar to those given by Brock and Durlauf in the discrete choice problem \cite{Brock} and in stochastic  models in macroeconomics \cite{Aoki}, but also to the Ising model used in description of magnetic materials and neural networks \cite{Rosenow,Schneidman}. This is also a special case of Markov random fields \cite{Kindermann} (see Sec-\ref{sec3:Relation}). We emphasize that the Gibbs distribution and the concept of information entropy naturally arise from stochastic modelling in economics. This is discussed at length in \cite{Aoki}.
We interpret the objective function $\mathcal{H}(\textbf{s})$ defined by the MEP as follows.
Pairwise interactions between economic entities are modelled by interaction strengths $J_{ij}$ (which describe how $i$ and $j$ influence each other). They can be thought as a measure of the degree of co-movement (coherence) of a time-series pair. As possible underlying causes of those interactions, we may think to the economic background, company management, traders strategies, etc. This should be investigated in an econometrical study. Actually, the causes underlying the interaction process seem to be unnecessary in the description of emergent macroscopic behaviours. Indeed the complicated interactions between magnetic moments or between neurons are efficiently simplified in their maximum entropy description but one still reproduces the main macroscopic features observed in these systems. In this description, the crucial features are the scaling (dependence or independence on the system size) of interaction strengths and the order of interactions.
The interaction matrix $\textbf{J}$ is set to be symmetric in this first approach. There is disagreement or conflict between entities when the weighted product of their orientations $J_{ij}s_{i}s_{j}$ is negative. If two shares are supposed to move together ($J_{ij}>0$), a conflicting situation is the one where they do not have the same orientation (bearish or bullish).

We include idiosyncratic preferences or individual biases of stocks, here the willingness to be bullish or not. These Lagrange multipliers $h_{i}$ can also be interpreted as external influences on entities $i$ induced by the macroeconomic background. By example a company can prosper and make benefits during a crisis period and the associated stock can still fall simultaneously because investors are negatively influenced by the economic background. The stock will have a propensity to fall even if profits are made. If the orientation of the stock satisfies its preference, $h_{i}s_{i}$ will be positive. The total conflict of the system is then given by

\begin{equation}\label{Hfunc}
\mathcal{H}(\textbf{s})=-\frac{1}{2}\sum_{i=1}^{N}\sum_{j=1}^{N}J_{ij}\, s_{i}s_{j}-\sum_{i=1}^{N}h_{i}s_{i}
\end{equation}
We interpret $\mathcal{H}(\textbf{s})$ as the opposite of the so-called utility function $\mathcal{U}(\textbf{s})=-\mathcal{H}(\textbf{s})$ with a pairwise interacting and idiosyncratic parts \cite{Brock}. Consequently interaction strengths can be viewed as incentive complementarities. Indeed we have $\partial^{2} \mathcal{U}/\partial s_{i}\partial s_{j}=J_{ij}$ . The larger $J_{ij}s_{i}s_{j}$, the stronger the strategic interaction between $i$ and $j$.

We emphasize that this pairwise maxent model is forced upon us as the statistically consistent model with measured orientations and correlations. It is not an analogy based on specific hypotheses about the market dynamics and it necessarily implies a multivariate picture of the markets as it should be.

Lastly, we can ask what happens if some of the pairwise influences are negative. This case leads to an interesting situation where the number of possible equilibria may explode. Let's illustrate this phenomenon through a simple example. Assume that we observe 3 stocks $\{a,b,c\}$ such that $J_{ab}=1=J_{bc}$ and $J_{ac}=-1$. There is no way to have no conflict at all and there are several configurations for the minimal (maximum) value of the conflict/utility. Indeed the utility function is $\mathcal{U}(\mathbf{s})=J_{ab}s_{a}s_{b}+
J_{bc}s_{b}s_{c}+J_{ac}s_{a}s_{c}=s_{a}s_{b}+s_{b}s_{c}-s_{a}s_{c}$ and the following configurations (respectively $a$, $b$,$c$) $\uparrow\uparrow\uparrow$, $\downarrow\downarrow\downarrow$, $\downarrow\downarrow\uparrow$, $\uparrow\uparrow\downarrow$, $\uparrow\downarrow\downarrow$, $\downarrow\uparrow\uparrow$ lead to the same total conflict/utility (iso-utility configurations). This is sometimes called \emph{frustration} or more preferably, a \emph{degenerate} utility  value (since many micro-configurations correspond to the same macro-state, here $\mathcal{U}=1)$. With many more entities, the conflict/utility landscape may comprise many valleys as previously illustrated in Fig-\ref{fig:Uloglik}.

\section{Mean field mapping}\label{sec:consistency}

\subsection{Parameters estimation}

The parameters $\{J_{ij},h_{i}\}$ can potentially be exactly computed by performing explicitly the maximization (\ref{maxent}) so that the theoretical moments $\langle s_{i}\rangle$ and $\langle s_{i}s_{j}\rangle$ match the empirical ones $q_{i}$ and $q_{ij}$. This method requires the computation of $2^N$ terms. If this number is too large, the computation is unfeasible and we can benefit from one of the methods described in \cite{Roudi}, see Sec-\ref{sec3:IsingInv} for details. The parameters should be valued such that the constraints are satisfied in (\ref{maxent}). Generally, redrawing the parameters from their distribution will lead to wrong values of the first and second moments. Therefore knowing only the functional form of the distribution is insufficient, we must know their exact values. In this section we use a second order mean-field inversion \cite{Roudi}. Generally this inversion method requires ten or so entities and a sample size $T$ larger than the number of entities $N$. In the following we have $T>20N$ and $N>10$.
This inversion technique, to infer interaction strengths from data, is based on the following relation ($i\neq j$)

\begin{equation}\label{inversion}
    (\mathbf{C}^{-1})_{ij}=-J_{ij}-J_{ij}^{2}\,q_{i}q_{j}
\end{equation}
Given the relation (\ref{inversion}), if the data are noise dressed, the inferred interaction matrix will also be noise dressed. Moreover, as the proposed model is a maximum entropy model, the parameters should be adjusted to satisfy the constraints in (\ref{maxent}). Thus any inversion method will be noise sensitive. Last, we note that the MEP is also sample-dependent since Lagrange multipliers are fitted to reproduce first and second moments. It does not necessarily mean that $J_{ij}$ are time-dependent but it seems intuitive that they are actually time-dependent since a company can die out, be restructured or removed from its index.

\subsection{Mean field}

The previous model can be thought as a pairwise maxent model on a complex network \cite{Dor}. Indeed the objective function of this model can be rewritten as

\begin{equation}
\mathcal{H}(\textbf{s})=-\frac{1}{2}\sum_{i=1}^{N}\sum_{j=1}^{N}J_{ij}A_{ij}\, s_{i}s_{j}-\sum_{i=1}^{N}h_{i}s_{i}
\end{equation}
where $A_{ij}$ are entries of the adjacency matrix, equal to one if the nodes $i$ and $j$ are connected and equal to zero if they are not. Most of time, this kind of models are not exactly solvable. However in a particular case (the so-called mean field model), the model is theoretically tractable (see Sec-\ref{sec3:Var} for details).

For a complete graph ($A_{ij}=1$ for all pairs) the mean-field solution, described by Thouless-Anderson-Palmer (TAP) equations, is exact if: the number of nodes tends to infinity and if the $J_{ij}$ are independent and identically distributed (IID) Gaussian random variables with mean and variance scaling as $N^{-1}$ \cite{Dor,Plef}. One knows that it is not the case for neural networks \cite{Schneidman}. We can check if financial networks can be described by the mean-field solution. In this case, the observed mean orientations should be well approximated by TAP equations

\begin{equation}\label{TAP}
\langle s_{i}\rangle_{\mathrm{c}}=\tanh\left(h_{i}+\sum_{j}J_{ij}\langle s_{j}\rangle_{\mathrm{c}}-\sum_{j}J_{ij}^{2}\langle s_{i}\rangle_{\mathrm{c}}[1-\langle s_{j}\rangle_{\mathrm{c}}^{2}]\right)
\end{equation}

\begin{figure}[!ht]
\begin{center}
\resizebox{0.75\textwidth}{!}{%
  \includegraphics{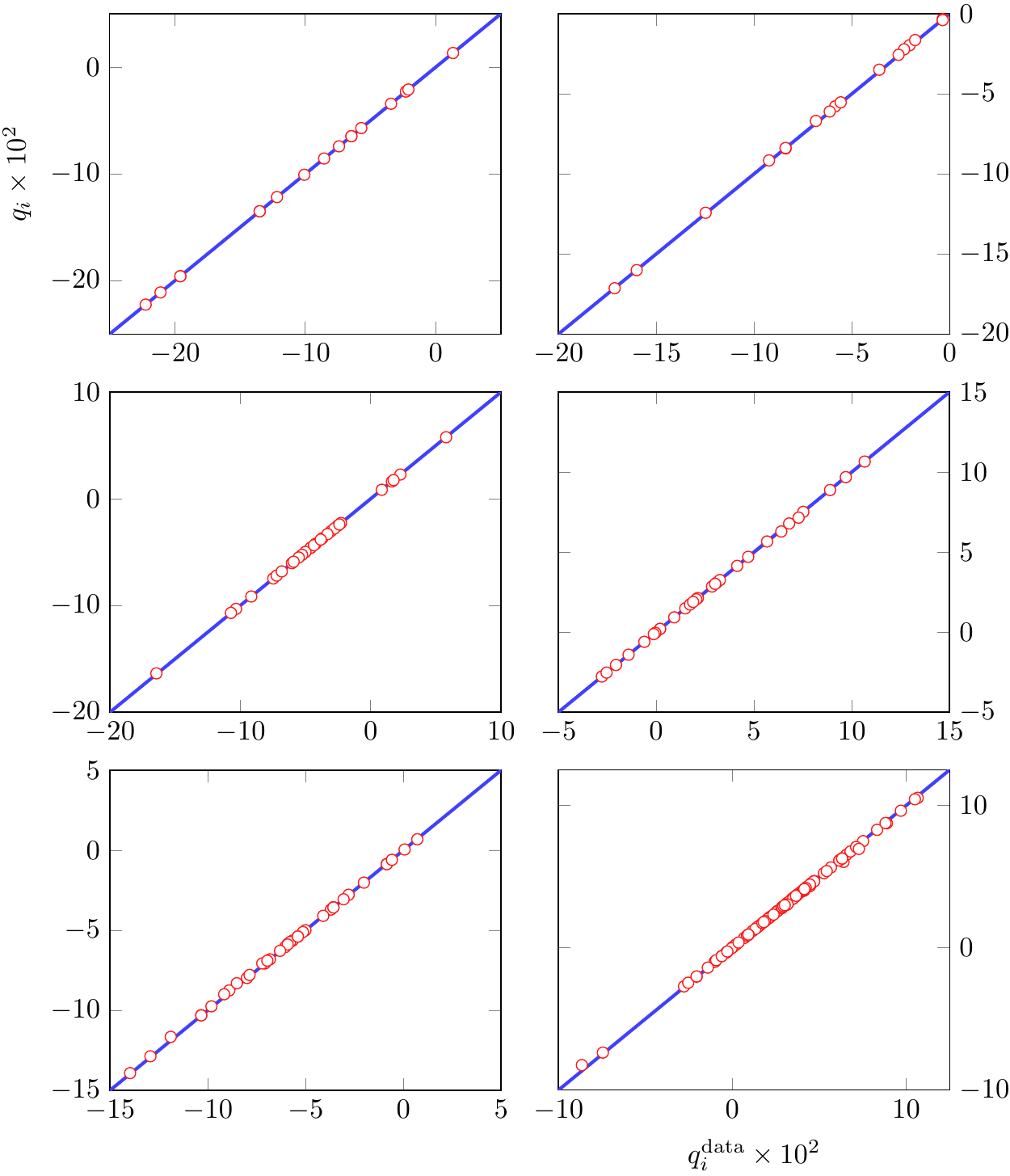}
}
\end{center}
\caption[Empirical and approximated orientations]{Comparison of TAP mean orientations (circles) and empirical ones. The straight line shows equality. Respectively from top left to bottom right (with increasing system size): BEL20, AEX, DAX, DJ, CAC, S$\&$P100.}
\label{fig:choices}
\end{figure}

Below, we show that first and second empirical cumulants are indeed well approximated by TAP mean-field for different market indices and for different system sizes. We consider the $N$ stocks of the BEL20, AEX, DAX, Dow Jones, CAC40 and S$\&$P100 indices respectively during $T=1050$, $T=1400$, $T=1550$, $T=2500$, $T=1550$ and $T=2500$ trading days, such that $T\gg N$ (a trading year is usually about 250 trading days). All these data can be downloaded from the web site Yahoo! Finance \cite{Yahoo}. We compute TAP mean orientations of each stock in this large time window and we compare them with their empirical mean values. The results are illustrated in Fig-\ref{fig:choices}.

TAP mean orientations are indeed a good description of empirical mean orientations, the typical relative deviation is less than $1\%$. As a further test, we also compare empirical variances of orientations to their TAP values. Variances of orientations are $\langle s_{i}^{2}\rangle_{\mathrm{c}}=1-\langle s_{i}\rangle_{\mathrm{c}}^{2}$ inserting the TAP approximation leads to $\langle s_{i}^{2}\rangle_{\mathrm{c}}=1-\tanh^{2}(h_{i}+\sum_{j}J_{ij}\langle s_{j}\rangle_{\mathrm{c}}-\sum_{j}J_{ij}^{2}\langle s_{i}\rangle_{\mathrm{c}}[1-\langle s_{j}\rangle_{\mathrm{c}}^{2}])$.


Variances are also well approximated by TAP variances, the typical relative deviation is about $1\%$. Using error propagation, one can evaluate the error on the estimation of
third order cumulants $\langle s_{i}^{3}\rangle_{\mathrm{c}}=2(\langle s_{i}\rangle_{\mathrm{c}}^{3}-\langle s_{i}\rangle_{\mathrm{c}})$ and higher order cumulants which are expressed in terms of TAP orientations.
The TAP mean field method is exact, in the so-called thermodynamic limit $N\rightarrow\infty$, for the infinite-range interactions provided that the following condition is satisfied \cite{Plef}

\begin{equation}\label{TAP2}
x\equiv 1-(1-2Q_{2}+Q_{4})>0 \quad \mathrm{with} \quad Q_{\nu}=N^{-1}\sum_{i=1}^{N}q_{i}^{\nu}
\end{equation}
We checked that this condition is fulfilled for each of the previous data sets and so our use of TAP equation was justified. We showed that a mean-field version of the maxent model on a complex graph can accurately describe the stock market for different and typical system sizes as TAP equations give results consistent with the data.

\section{Beyond mean field mapping}\label{sec:consistency2}
In the following, we go a step further from the mean field formulation and we perform Monte Carlo simulations (see Sec-\ref{sec3:MC} for details) to compute the different moments (means, covariances and correlations coefficients) and to compare them to the empirical ones.

We apply the pairwise model to a set of six major market indices (AEX, Bel-20, CAC 40, Xetra Dax, Eurostoxx 50, FTSE 100). We selected only European indices because some financial issues are specific to Europe and we consider indices because as they are the driving force of the respective stock markets \cite{Shap}, they will reflect the main properties of the underlying stock set. We observe 2253 configurations from 6/06/2002 to 14/06/2011 \cite{Yahoo}, a nine year long time series including two large crises. Later, we will also analyse the stocks composing the Dow Jones and the S$\&$P100 indices, and another set of 116 stocks. The small number of entities allows a direct computation (potentially \emph{exact}) of Lagrange parameters through the optimization (\ref{maxent}). As mentioned above, higher-order interactions can be involved in the interaction structure. In order to show that pairwise correlations are prevailing, we compute the Kullback-Leibler (KL) divergence, $D_{\scriptstyle{\mathbf{KL}}}(P_{2}\|P_{\scriptstyle{\mathbf{data}}})$ between the two-agent maximum entropy (ME) distribution $P_{2}$ and the empirical one $P_{\scriptstyle{\mathbf{data}}}$. The KL divergence is equal to $2.27\times 10^{-2}$ for the ME distribution inferred from 2253 observations. It must be compared to $D_{\scriptstyle{\mathbf{KL}}}(P_{1}\|P_{\scriptstyle{\mathbf{data}}})=1.48$ for the independent agents model $P_{1}$. The closer to zero this quantity is, the closer $P_{2}$ to $P_{\scriptstyle{\mathbf{data}}}$ is. Moreover, the multi-information $I_{N}=S(P_{1})-S(P_{N})$ in this application is equal to $98.5\%$. The pairwise correlations model is effective since it explains almost all the available information; only $1.5\%$ of information is due to higher-order interactions.

\begin{figure}[!ht]
\begin{center}
\includegraphics[width=0.5\textwidth]{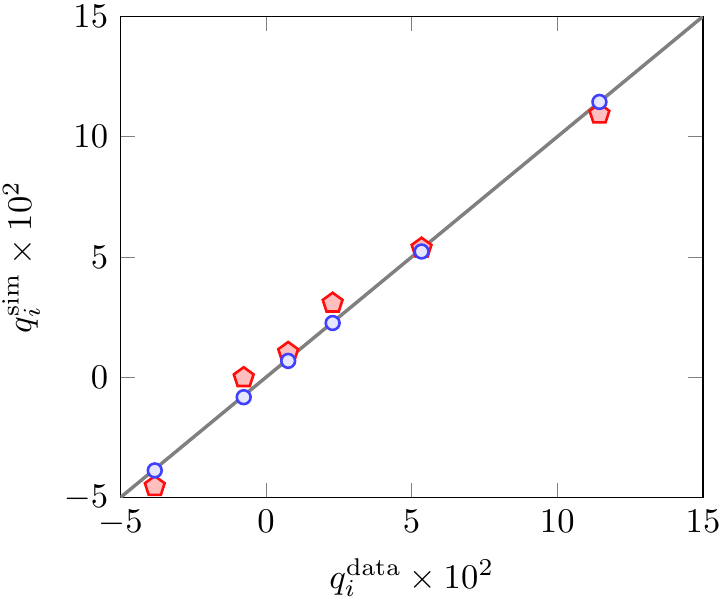}
\end{center}
\caption[Orientations with exact Lagrange parameters]{Comparison of simulated orientations and the actual ones. The straight line shows equality. The circles stand for simulations with exact Lagrange parameters and pentagons stand for approximated Lagrange parameters (\ref{1stOrInv}).}
\label{fig:comp}
\end{figure}

As a further test of the pairwise model consistency, we compare the average index orientations $q_{i}=T^{-1}\sum_{t=1}^{T}s_{i,t}$ obtained by simulation to the real ones. We simulate the process by doing $1\times 10^{5}$ equilibration Monte Carlo time steps\footnote{A Monte Carlo step is a sequence of $N$ iterations, where $N$ is the number of entities in the considered system.} (MCS) and we compute the average on the next $2\times 10^{7}$ MCS in order to reduce the variance of the estimator. The flipping attempts are simulated by the so-called Glauber dynamics (see Sec-\ref{sec3:MC}). Namely, we take an entity $i$ randomly chosen and the attempt to flip the associated binary variable $s_i$ is performed with a rate depending on an exponential weight, the other orientations remaining fixed. We compute the time average for each index from the data and we compare it to the value obtained with the simulation; they are illustrated in Fig-\ref{fig:comp}.

\begin{figure}[!ht]
\begin{center}
\includegraphics[width=.85\textwidth]{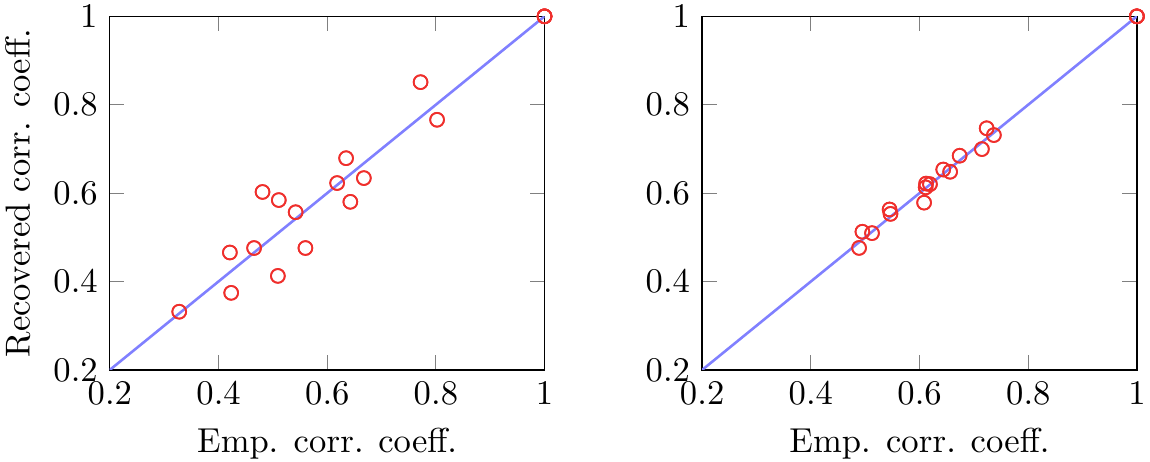}
\end{center}
\caption[Finite size effects in simulations]{Reproduced correlation coefficients from MCMC versus empirical ones. The straight line shows equality. The result based on 130 observations (left) and the result based on 2258 observations (right).}
\label{fig:corr}
\end{figure}

The root mean squared error (RMSE) is equal to $7.0\times 10^{-4}$, which represents $1.5\%$ of the root mean squared (RMS) value of the six arithmetic means (equal to $4.9\times 10^{-2}$). We reproduce quantitatively the average orientation of the six indices on the observation period. Moreover, since we obtained the probability distribution, we can compare the correlation coefficients resulting from the sampling of the proposed probability distribution to the empirical ones. We sample the probability law with exact Lagrange parameters (superscripts ME) $p_{2}(\textbf{s},\mathbf{J}^{\mathrm{\scriptstyle{ME}}},\mathbf{h}^{\mathrm{\scriptstyle{ME}}})$ by a Monte Carlo Markov chain (MCMC). We take $1.2\times 10^{6}$ equilibration steps and $1.2\times 10^{4}$ independent sampling steps between each sample. Fig-\ref{fig:corr} illustrates the reproduced correlation coefficients with the maximum entropy estimation versus the empirical ones. The results for only 130 observations (chosen arbitrarily corresponding to half a year) are conclusive. Indeed the RMSE represents $8.3\%$ of the RMS value and the correlation coefficient of the empirical and simulated values is equal to $0.963$. Including more observations (2258 trading days) allows us to reduce the dispersion in the results (correlation coefficient of the empirical and simulated values equal to $0.997$; the RMSE represents $1.8\%$ of the RMS value). We note that it is effective even with few data.

\begin{figure}[!ht]
\begin{center}
\resizebox{1\textwidth}{!}{%
  \includegraphics{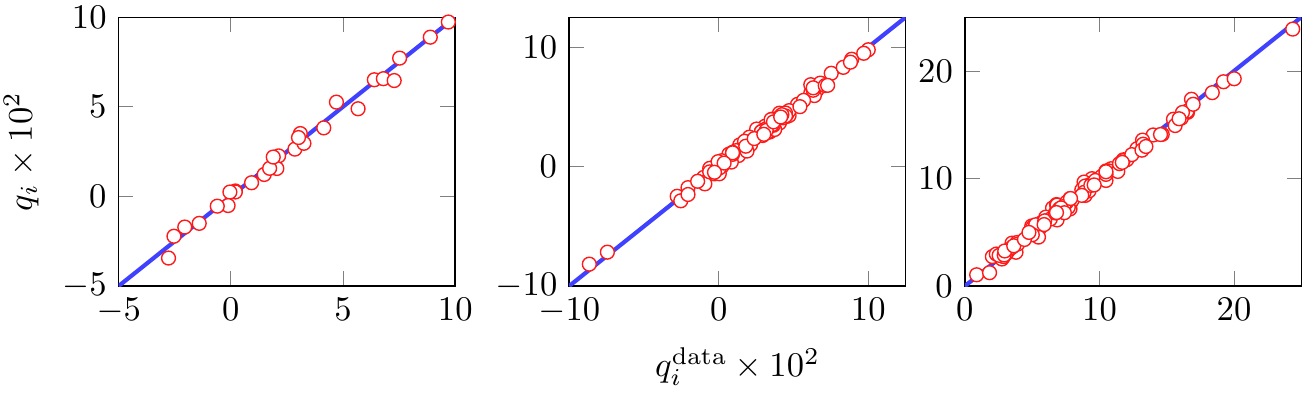}
}
\end{center}
\caption[Dow Jones and S$\&$P100 orientations]{Comparison of simulated orientations and the actual ones. From left to right: DJ, S$\&$P100 and Onnela's set. The straight line shows equality.}
\label{fig:choicesAl}
\end{figure}

\begin{figure}[!ht]
\begin{center}
\resizebox{1\textwidth}{!}{%
  \includegraphics{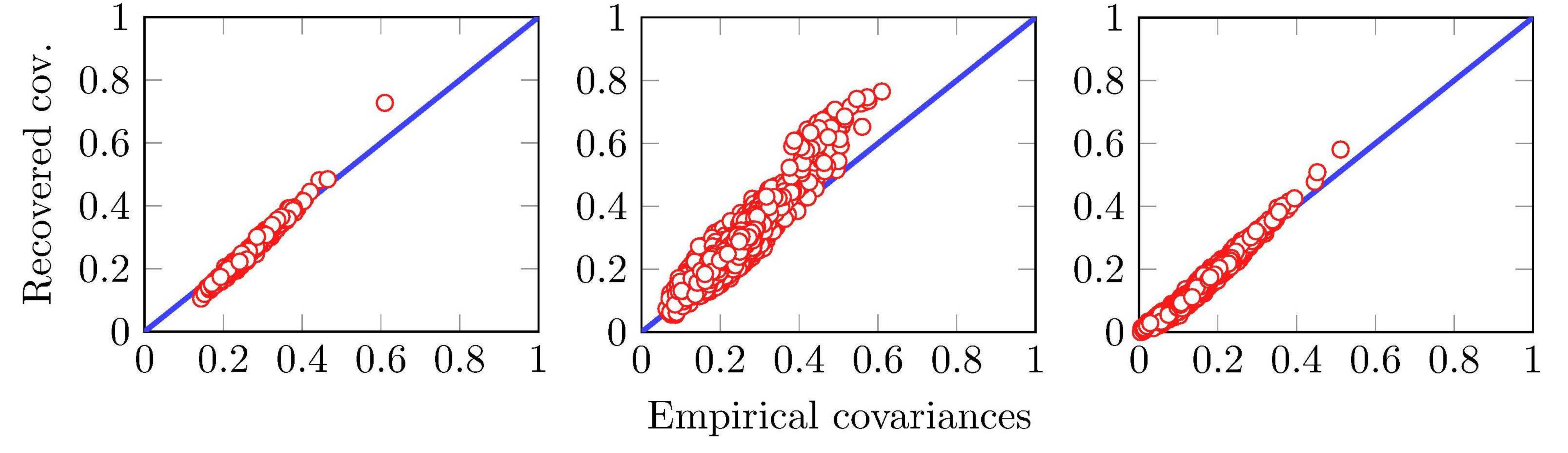}
}
\end{center}
\caption[Simulated covariances]{Reproduced covariances versus empirical ones. From left to right: DJ, S$\&$P100 and Onnela's set. The straight line shows equality.}
\label{fig:covAll}
\end{figure}

We perform the same work for the Dow Jones and the S$\&$P100 indices (2500 configurations observed from 10/10/2001 to 02/08/2011). We also consider 116 stocks from the New York Stock Exchange available on the Onnela's website \footnote{\url{http://jponnela.com/}} extending from the beginning of 1982 to the end of 2000 (4800 trading days). For these larger stock sets, the exact entropy maximization (\ref{maxent}) is not computationally tractable. We use instead an approximated method (in our application the rPLM method performs best, see Sec-\ref{sec3:rpml} for details).
The results for the first and second reproduced moments ($2\times10^{6}$ equilibration MCS, values estimated on $2\times10^{7}$ samples recorded each $N$ iterations) are illustrated in Fig-\ref{fig:choicesAl} and Fig-\ref{fig:covAll}.

The correlation coefficient between the reproduced and empirical values is respectively $0.998$, $0.996$ and $0.997$ for the net orientations illustrated in Fig-\ref{fig:choicesAl} and $0.989$, $0.964$, $0.997$ for the covariances illustrated in Fig-\ref{fig:covAll} which shows the strong linear statistical relation between the empirical and the reproduced values.
The relative deviations between the RMSE and the RMS values are respectively $2\%$, $7\%$ and $6\%$ for the net orientations and $9\%$, $17\%$, $8\%$ for the covariances.

We have seen that, in addition of the multi-information criterion, the net orientations and the covariances are reproduced from this model even with few data. We conclude that the proposed pairwise interaction structure is a trustful one; this means that interactions are believed to be pairwise and symmetric ones and that they cause correlations. However, it is not obvious that all the entries of the influence matrix should be considered as real information and not noise. We study the distribution of the pairwise influence hereafter.

\section{Distribution of the pairwise influences}\label{sec:dist}

The good adequation between empirical and TAP cumulants suggests that the market network should be like a complete graph, with pairwise influences which should be Gaussian ones and scale as the inverse of the system size. However, the real financial network may not actually be a complete graph even if the only null entries of the interaction matrix are the diagonal ones. Indeed one knows that a part of the correlations is noise \cite{Laloux}. Moreover, the finite size of the sample also implies errors in the parameters estimation.
It would be nice if, in addition, the interaction matrix entries $J_{ij}$ were actually gaussian random variables as required in the TAP mean-field approach. This would make the link with the Gaussian spin glass theory \cite{Fischer} used in physics, information theory, optimization, herd behaviour, etc. We want to emphasize that one should not confuse the interaction matrix with the covariance matrix of the returns. The fact that $\textbf{J}$ entries are normally distributed does not mean that there are only noisy movements in the market. The $\textbf{J}$ matrix describes the pairwise interactions, not directly the correlations.

\begin{figure}[!ht]
\begin{center}
\resizebox{\textwidth}{!}{%
  \includegraphics{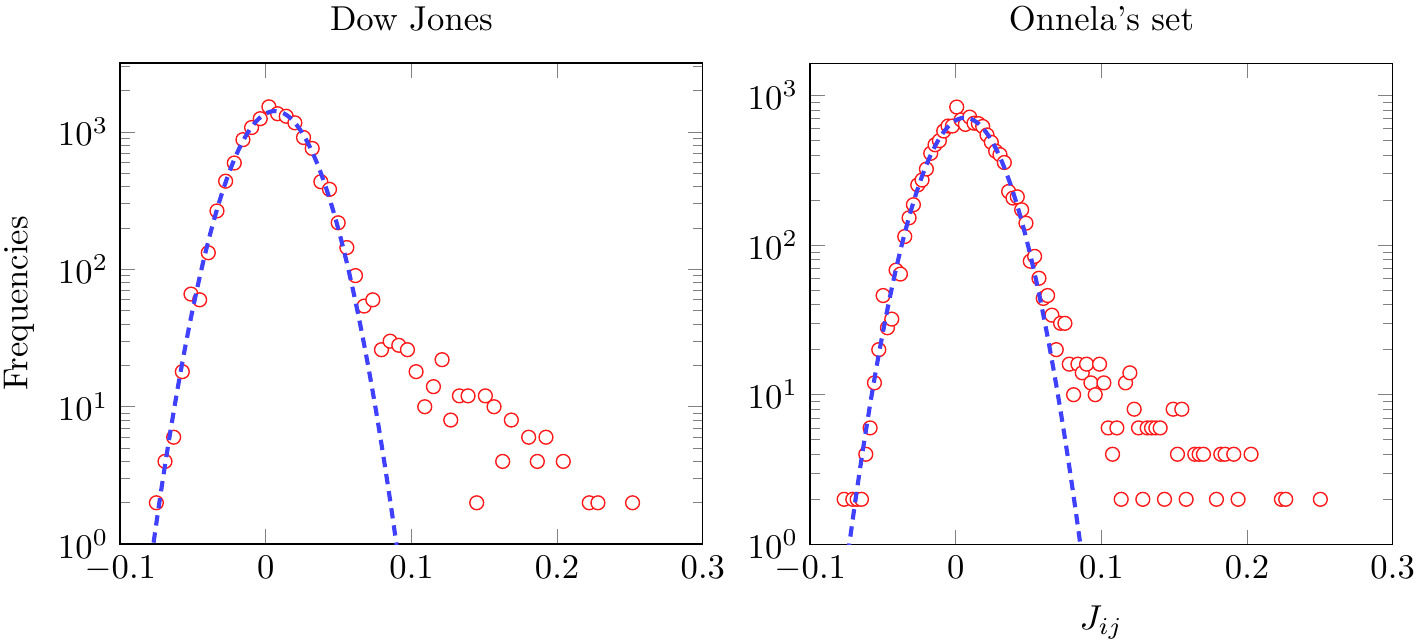}
}
\end{center}
\caption[Empirical frequencies of pairwise influences]{Left: Empirical frequencies of pairwise influences for the DJ (daily time-scale) and right: the Onnela's set . The dashed line is a Gaussian fit of the influences frequencies distribution amputated of its upper tails.}
\label{fig:distr}
\end{figure}

\begin{figure}[!ht]
\begin{center}
\resizebox{\textwidth}{!}{%
  \includegraphics{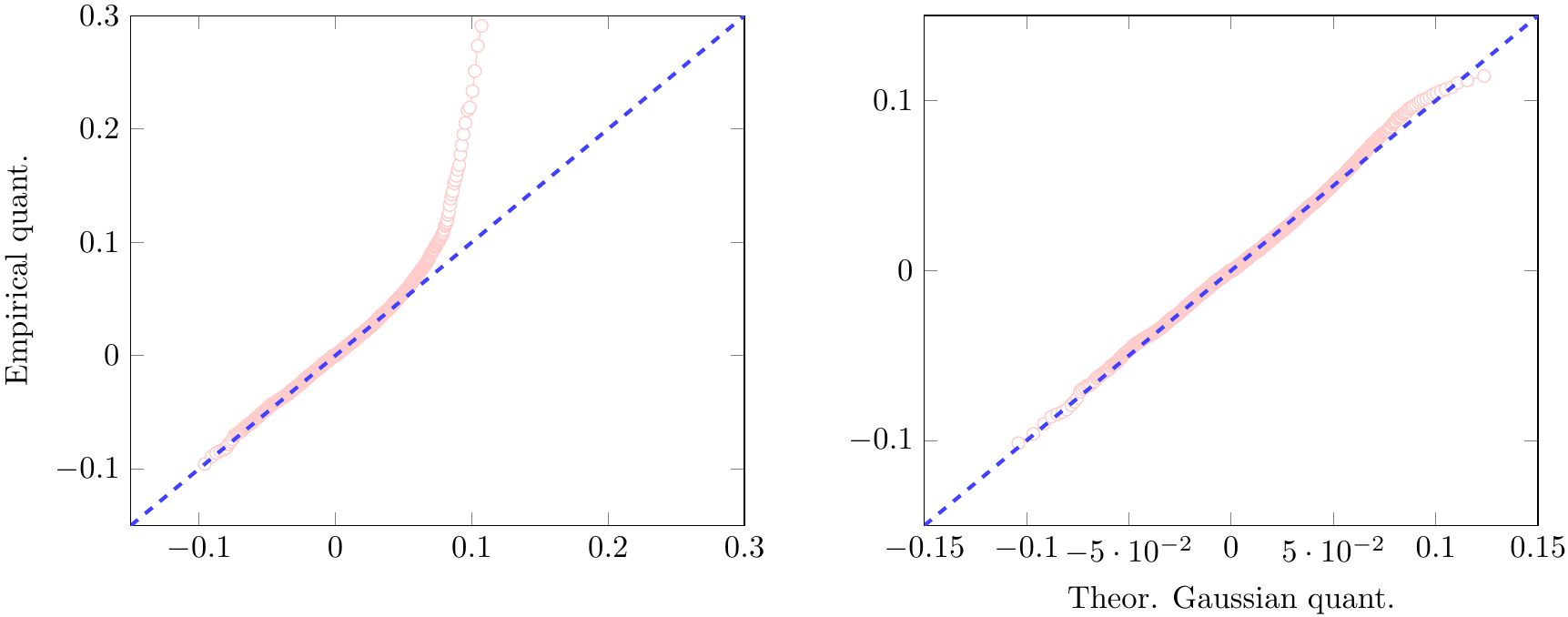}
}
\end{center}
\caption[Q-Q plot of the pairwise influences]{Comparison of S$\&$P100 empirical 1000-quantiles (circles) and theoretical ones. The straight line shows equality. Respectively from left to right: all the $4950$ entries of the $\mathbf{J}$ matrix and the results without the last 200 entries.}
\label{fig:quant}
\end{figure}

We illustrated in Fig-\ref{fig:distr} the empirical frequencies of the estimated mutual influences, estimation performed by inverting a second order mean-field approximation of the self consistent equation as described in Sec-\ref{subsec:diagonalTrick}. We consider the Dow Jones index and a set of 116 NYSE stocks observed during 4800 trading days (available at \url{www.jponnela.com}). The frequencies distribution does not seem to be exactly Gaussian since the upper tail is fatter than in the Gaussian distribution. To formalize this observation, we first use a qualitative normality test. We compare the empirical 1000-quantiles (permilles) with the theoretical 1000-quantiles. If the $J_{ij}$ are Gaussian random variables, we should obtain a linear relation between both these quantities. We illustrated our results in Fig-\ref{fig:quant}.

We tested the normality of the interaction strengths for the previous six market indices. We obtained similar results than those illustrated in Fig-\ref{fig:quant}. The upper tail of the empirical distribution is also found fatter than the Gaussian one but the bulk of the distribution seems to be Gaussian.
Then we use the $\chi^{2}$ and the Jarque-Bera statistical normality tests on the $\mathbf{J}$ upper triangular part amputated of its upper tail. They do not lead to the rejection of the null hypothesis that the bulk of the underlying distribution is a Gaussian one.

Last, to evaluate the importance of the noise in the estimation, we simulate binary time-series (for different sizes and sample lengths) with the maximum entropy conditional flipping probability $p(s_{i,t}=-s_{i,t-1}|\textbf{s}_{-i,t})$ given the state at time $t$ excluding the $i$th entity. The influence matrix was taken homogenous with all entries equal to the empirical mean $\overline{J_{ij}}$ of the considered index in those simulations. We then estimate the influence matrix with those artificial data. Ideally, the standard deviation of estimated artificial influences $\sigma_{\mathrm{noise}}$ should be much smaller than the one of real influences $\sigma_{J}$. The results are reported in Table-\ref{tab:noise}. Depending on the sample length, the noise seems to be significantly but not the dominant part of the estimation excepted for large system size.

\begin{table}[!ht]
\caption[Noise quantification in influences estimation]{Quantification of noisy part of the variance of inferred mutual influences.}
\label{tab:noise}
\begin{center}
\begin{tabular}{lcr}
\hline
Index         & sample length ($T$) & $\sigma_{\mathrm{noise}}/\sigma_{J}$\\ \hline
AEX(daily)    &  $1.4\times 10^{3}$ & 0.22                                \\
DJ(min)       &  $3.0\times 10^{4}$ & 0.24                                \\
DJ(daily)     &  $2.5\times 10^{3}$ & 0.31                                \\
Onnela(daily) &  $4.8\times 10^{3}$ & 0.74                                \\
Cac(daily)    &  $1.5\times 10^{3}$ & 0.75                                \\
\hline
\end{tabular}
\end{center}
\end{table}

\begin{figure}[!ht]
\begin{center}
\resizebox{\textwidth}{!}{%
  \includegraphics{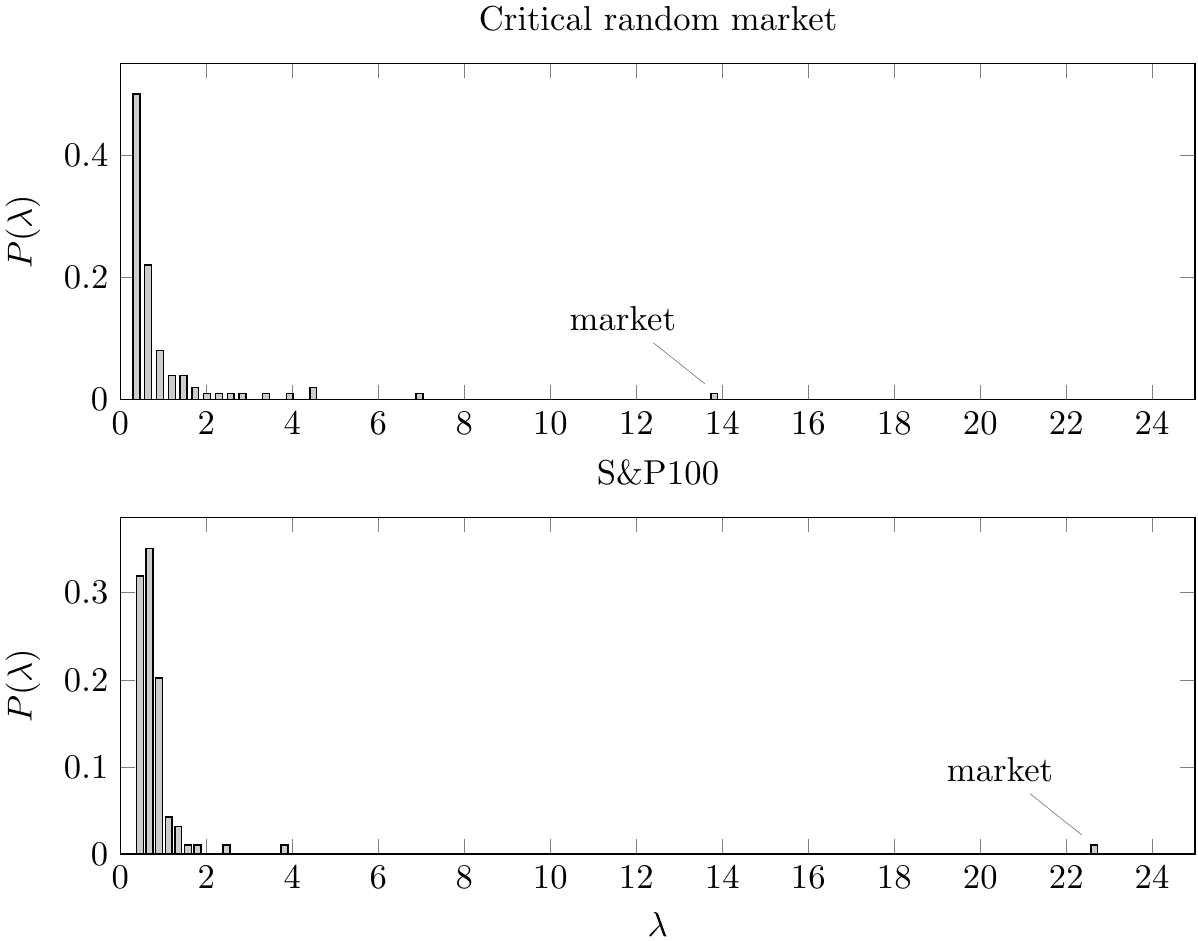}
}
\end{center}
\caption[Critical market mode]{(Top) Typical probability distribution of the eigenvalues of th returns covariance matrix at the transition. A \emph{critical} random market is able to exhibit non-Gaussian covariance matrix. (Bottom) The empirical probability distribution of the eigenvalues of the covariance matrix of the S$\&$P100 index.}
\label{fig:crit}
\end{figure}

However it is not obvious if the upper tail can be neglected or not (one knows that one cannot neglect the non-Gaussian part of the correlation matrix). The non-Gaussian part of the distribution may also be an inference artefact (since less than $10\%$ of the influences are non-Gaussian ones).
We are tempted to let the door open to the case of Gaussian influences. Indeed, in addition to the previous evidence of TAP matching, Gaussian interactions are compatible with the observed market eigen-mode. Consider the simplest situation where $J_{ij}$ are really \emph{IID} Gaussian random variables with zero mean (thus including the frustration since half of the pairwise influences are negative). The largest eigenvalues of the returns covariance matrix are linked to eigenvalues of the $\textbf{J}$ matrix by the relation $[1-J_{\lambda}+J^2]^{-1}$ in the mean field approach, where $J_{\lambda}$ is an eigenvalue of the $\mathbf{J}$ matrix and $J^{2}\equiv N\,\mathrm{VAR}(J_{ij})$ \cite{Fischer}. This quantity is large when $J_{\lambda}$ lies in the vicinity of $1+J^{2}$. When the $J_{ij}=J_{ji}$ are IID Gaussian variables, the largest eigenvalue of the interaction matrix is equal to $2J$. A special case is the one where $J=1$ which corresponds to the transition in the Sherrington-Kirkpatrick model. The largest eigenvalue of the covariance matrix diverges in the limit of infinite number of entities. We illustrated this behaviour for $N=100$ interacting stocks with IID Gaussian interaction strengths in Fig-\ref{fig:crit}.

In the present applications entries of the interaction matrix do not seem to have a common mean and variance; therefore the relation between both kinds of eigenvalues is more complex than the former one. It is then non-obvious to conclude wether the interaction strengths are actually Gaussian or whether the right fat-tail of their distribution is a true deviation to the normal distribution (and not an inference artifact). The possible interpretation of a market behaving as a critical complex system will be investigated in detail in a dedicated chapter.

The possible normality of interactions has another consequence: the $\mathcal{U}(\mathbf{s})$ function defines a Gaussian process. Our model is thus a random utility model and tools of the random matrix theory \cite{Car} can be useful to study the market structure, as they already are in the study of stock return correlations \cite{Laloux}. We also checked that a significant part of interaction strengths are negative ($37.2\%$ for the S$\&$P100 and $24.3\%$ for the Dow Jones). Together with the former observation of a possible market mode even with truly Gaussian $J_{ij}$, we may think that the  frustration is a main feature of the market interaction structure. We may think the frustration as competitive influences between cyclic sectors (more correlated to the global health of the worldwide economy and thus privileged by the investor during a growth period) and the defensive sectors.


%

\begin{figure}[!ht]
\begin{center}
\resizebox{0.7\textwidth}{!}{%
  \includegraphics{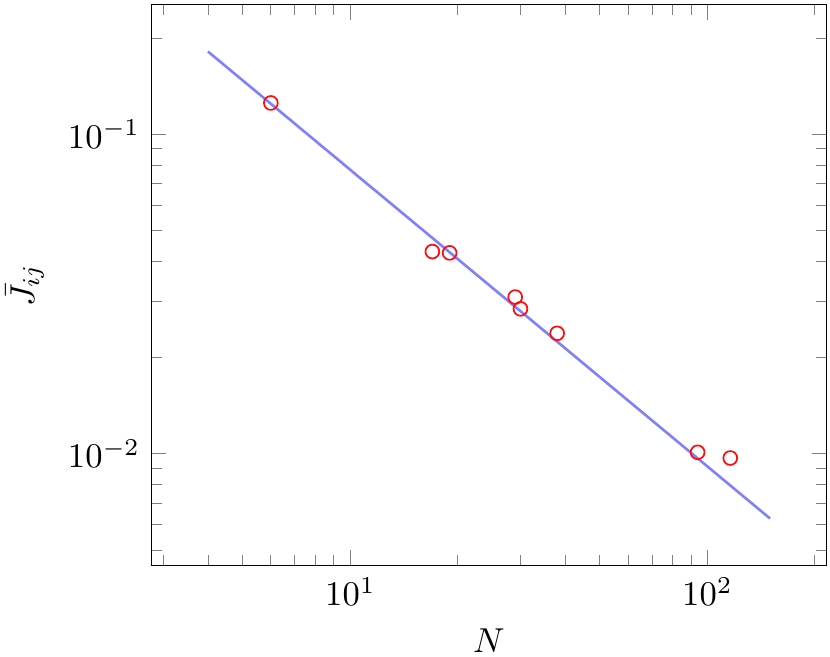}
}
\end{center}
\caption[Scaling of the pairwise influences]{Log-log plot of the mean interaction strengths as a function of the typical system sizes (circles). The straight line is a non-linear fit (power-law).}
\label{fig:MeanJ}
\end{figure}

Another main feature is the scaling of the mean interaction strengths as a function of the system size, as needed in the TAP mean-field approach. To ensure that the $\mathcal{H}$ function (\ref{Hfunc}) is extensive (scaled as $\mathcal{H}\propto N$), the mean strength $\overline{J_{ij}}$ should be scaled as $\overline{J_{ij}}\propto N^{-1}$ \cite{Binder}. Hereafter, we show that mean interaction strengths exhibit indeed these scaling properties for characteristic system sizes encountered in stock markets. We infer interaction strengths on a common time window of 1000 trading days (four years long time series) for the following indices (given in increasing size): BEL20, AEX, DAX, DJ, CAC40, S$\&$P100 and Onnela's set.
We add a supplementary point by computing the interaction strengths between six major European indices (adding another order of magnitude to the typical system size). The results are illustrated in Fig-\ref{fig:MeanJ}

We adjust a power law $aN^{-\alpha}$ to the data (illustrated by a straight line in a log-log graphic). The resulting coefficient of determination $R^{2}=0.997$. The estimation of the slope is $\hat{\alpha}=0.928\pm0.030$ (mean $\pm$ s.d). We conclude from this analysis that the mean strength scales as $\overline{J_{ij}}\propto N^{-\alpha}$ with alpha close to $1$, in the interval of characteristic system sizes encountered in financial markets. This implies that the utility function (\ref{Hfunc}) may be an extensive one (proportional to the size of the system) and thus that the quantities which derive from this function may be correctly scaled.
We note this is not the case for neural networks where the typical interaction strengths seem to be constant for growing $N$. In a complex system this situation is equivalent to lowering the stochasticity (sometimes called the temperature by analogy to physical systems) leading to a \emph{frozen} state. The scaling $\overline{J_{ij}}\propto N^{-1}$ implies on the contrary that financial systems will not freeze and will not have the error-correcting property \cite{Schneidman}, meaning that one can not recover the entire market state by an observation of a small part of it.

\begin{figure}[!ht]
\begin{center}
\resizebox{0.75\textwidth}{!}{%
  \includegraphics{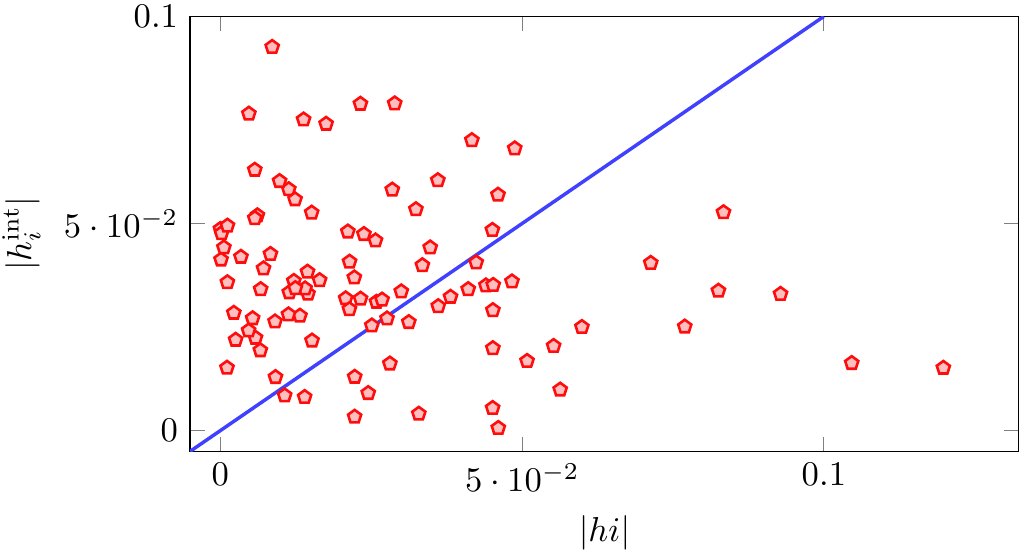}
}
\end{center}
\caption[Collective vs individual biases]{Comparison of the S$\&$P100 index internal bias $h_{i}^{\mathrm{int}}$ experienced by a stock versus its individual bias $h_{i}$. In the upper left triangle, the internal bias dominates the intrinsic bias. Similar results are obtained even for smaller indices (like the BEL20 or AEX).}
\label{fig:bias}
\end{figure}

Since interaction strengths can be weak, we may ask if they have actually a predominant role in the market structure or if the values of interesting quantities are principally determined by individual bias $h_{i}$. From the relation (\ref{04-Lagrange}) we conclude that the orientation of each stock $s_{i}$ is subjected to a total bias $h_{i}+2^{-1}\sum_{j}J_{ij}s_{j}$. Interactions play a key role if the internal bias $h_{i}^{\mathrm{int}}=2^{-1}\sum_{j}J_{ij}s_{j}$ is significant compared to the individual bias $h_{i}$. We checked that they are in average of the same magnitude order. The results for the S$\&$P100 index are illustrated in Fig-\ref{fig:bias}.

Collecting previous results, we gave empirical evidence that the financial market is described by a statistical model equivalent to an infinite range (mean-field) pairwise maxent model. The spin glass theory provides an effective toolbox to study the financial markets structure as a complex system \cite{Rosenow,Binder}.

However, we do not identify this statistical model to the Sherrington-Kirkpatrick model because there is no guarantee that interactions are quenched (static mean and variance) or even drawn from the same distribution. If the parameters are not quenched, their values can possibly change before the equilibration (if there is any) of the system.

\section{Conclusion}

We provided empirical evidence that the financial network is accurately described by a statistical model which can be thought as an pairwise maxent model on a complex (possibly complete) graph with scaled mutual influences. This results lays down the pairwise model as a consistent paradigm in the study of stock market since first and second order influences are the dominant ones.
In particular, we showed that orientations are accurately inferred by the TAP equation (in the stability domain) and reproduced by Monte Carlo simulations. Linked to this result, we checked that almost all the interaction strengths are Gaussian random variables, their average values scale as $N^{-\alpha}$ with $\alpha$ close to $1$. A significant part of the interaction strengths are negative, leading to multiplication of equilibria and metastable states. Moreover, we showed that this model with truly Gaussian and scaled ($N^{-1}$) influences is able to reproduce the market eigen-mode.
Consequently the proposed model may be thought as an exact mean-field one and the market state cannot be deduced by an observation of a small part of it.
Some methods developed in the spin glasses and neural networks theories could be applied in the study of the financial network, but we must pay attention to the specificities of each discipline, like the characteristic system size and the scaling of interactions for instance.
Some of the consequences are the existence of metastable states, the emergence of collective phenomena and spatial patterns, etc.
Furthermore, the processes taking place in the stock market should then occur at different timescales. The finite size of the stock market avoids the thermodynamic limit even as an approximation. Indeed the characteristic index size is about $N=10^2$ or $N=10^3$, much smaller than in physical or biological systems. Even if the relevant variables are correctly scaled, the fluctuations can be significant because at equilibrium they typically scale as $\sqrt{N}$.

Other potentialities could be the clustering analysis, the characterization of the financial network (confirming the small-worldness and scale-freeness within this framework), the study of crises through the interaction matrix and Monte-Carlo simulations. Some of these aspects will be investigated in other chapters of this work.

\chapter{Market structure explained by pairwise interactions}\label{chap:marketStruct}
\thispagestyle{empty}
\begin{summary}
Financial markets are a typical example of complex systems where interactions between constituents lead to many remarkable features. Here, we show that a pairwise maximum entropy model (or auto-logistic model) is able to describe switches between ordered (strongly correlated) and disordered market states. In this framework, the influence matrix may be thought as a dissimilarity measure and we explain how it can be used to study market structure. We make the link with the graph-theoretic description of stock markets reproducing the non-random and scale-free topology, shrinking length during crashes and meaningful clustering features as expected.  The pairwise model provides an alternative method to study financial networks which may be useful for characterization of abnormal market states (crises and bubbles), in capital allocation or for the design of regulation rules.
\end{summary}

\newpage

\begin{figure}[!ht]
\begin{center}
\includegraphics[scale=0.8]{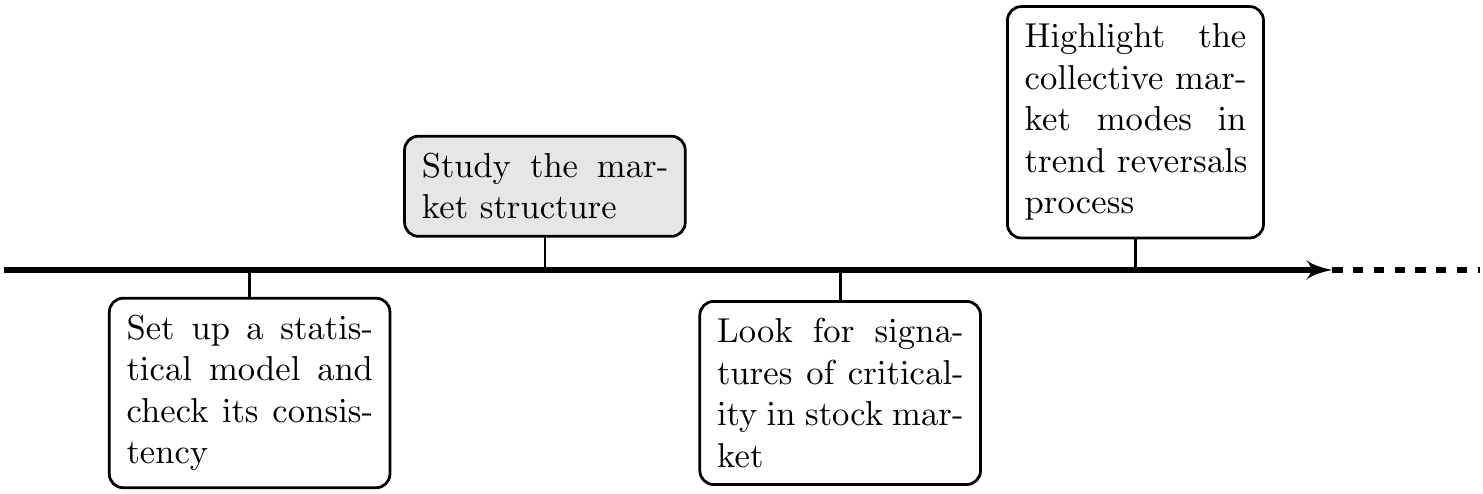}
\end{center}
\end{figure}

\newpage

\section{Introduction}
\label{sec5:intro}
Complex systems are particularly interesting because they exhibit very sophisticated behaviours caused by, a priori, simple rules. Indeed, magnetic materials and neural networks, for instance, have some striking features such as phase transitions, memory, complicated equilibria structures and clustering. It is remarkable that these properties are caused by such simple interactions as pairwise ones.
In the previous chapter, we gave evidence that the markets are driven by such simple rules and that the higher-order interactions encountered in financial systems are the pairwise ones. Typical characteristics of a complex system are numerous entities and interaction rules (with a degree of non-linearity), all leading to the emergence of collective behaviours. These behaviours depend, in general, more on the interactions (e.g their scaling and their order) and their effects than on the intrinsic nature of the elementary constitutive entities taken individually. The market can be viewed as such a system. The entities can be stocks or traders interacting through non-obvious rules. We note that we should interpret \emph{interaction} at the larger sense of mutual or reciprocal influence.

What one knows is that the markets exhibit features such as synchronization \cite{Dal}, structural reorganization \cite{OnnelaPRE,PeronChaos}, power laws \cite{stanley-gabaix1,stanley-gabaix2}, hierarchical and non-randomness \cite{Petra}.
What one does not know is the true market dynamics. Even if trading rules are known, microscopic equations of motion are unknown. This is a fundamental difference between finance and physics/neuroscience.

A natural approach, given the above considerations, is a statistical modelling collecting and using at best the available amount of information and allowing (in a certain sense) the emergence of critical properties. This is exactly the purpose of the maximum entropy modelling in complex systems theory.
Indeed the maximum entropy principle (MEP) allows the selection of the less restricting model on the basis of incomplete information. We choose this data-based approach to avoid the use of any particular microscopic schemes (e.g. trader-agent-based rules, a priori unknown) which are difficult to assess experimentally or to avoid any analogy (even if some of such models are valuable \cite{Rosenow}). The reason is that, even if one does not know the underlying microscopic processes, the macroscopic collective behaviours can still be described by an \emph{effective} model.
One has long experience of this powerful approach in the description of phase transitions and magnetic materials \cite{Fischer}. More recently, it has led to valuable results about the description of real neural networks  \cite{Schneidman}. Moreover, this approach also has counterparts in economics. Indeed, in addition to the  statistical meaning of the entropy, one can interpret it as a measure of the economic activity \cite{Aoki} and it is linked to the central concept of \emph{utility} of many interacting economic entities \cite{Brock,Mas}.

An important outcome of such a modelling is a convenient simplified version of the real interaction structure that is still consistent with the data and observed collective phenomena. In the following, we derive the model from this point of view and we study the structural properties of the resulting complex network. The critical properties will be investigated in another work.

The chapter is organized as follows. In section \ref{sec5:model}, we briefly recall the model. In section \ref{sec5:order}, we show an order-disorder transition through actual data. In section \ref{sec5:int}, we highlight the properties of the interaction matrix and its link to the crises. Finally, in section \ref{sec5:graph} we explain the link with the graph-theoretic approach and the topological evolution of the market network.

\section{The model}\label{sec5:model}
The aim is to set up a statistical model describing the market state. This requires a way to infer the probability distribution in order to get the observables (here, the associated moments). The model will also allow the study of the market structure. We consider a set of $N$ market indices or $N$ stocks with binary states $s_{i}$ ($s_{i}=\pm1$ for all $i=1,\cdots,N$). A system configuration will be described by a vector $\textbf{s}=(s_{1},\cdots,s_{N})$. The binary variable will be equal to 1 if the associated index is bullish and equal to $-1$ if not. A configuration $(s_{1},\cdots,s_{N})$ is a binary version of the index returns.

We seek to establish a less structured model explaining only the measured index mean orientation $q_{i}=\langle s_{i}\rangle$ and instantaneous pairwise correlations $q_{kl}=\langle s_{k}s_{l}\rangle$. The brackets $\langle\cdot\rangle$ denote the average with respect to the unknown distribution $p(\textbf{s})$. As the entropy of a distribution measures the randomness or the lack of interaction among the binary variables, a way to infer such probability distribution knowing the mean orientations and the correlations is the maximum entropy principle. It consists in the following constrained maximization:

\begin{eqnarray}
  & &\max_{\substack{\{p(\mathbf{s})\}}} S(\textbf{s})= \max_{\substack{\{p(\mathbf{s})\}}}\left\{ -\sum_{\{\textbf{s}\}}p(\textbf{s}) \,\ln p(\textbf{s})\right\}  \\ \nonumber
   &\mathrm{s.t}&  \sum_{\{\textbf{s}\}}p(\textbf{s})=1,\quad \sum_{\{\textbf{s}\}}p(\textbf{s})s_{i}=q_{i},
   \quad \sum_{\{\textbf{s}\}}p(\textbf{s})s_{i}s_{j}=q_{ij} \nonumber
\end{eqnarray}
The resulting  two-agent distribution $p_{2}(\textbf{s})$ is the following

\begin{equation}
p_{2}(\textbf{s})=\mathcal{Z}^{-1}\exp\left(\frac{1}{2}\sum_{i, j}^{N}J_{ij}s_{i}s_{j}+\sum_{i=1}^{N}h_{i}s_{i}\right)\equiv\frac {e^{- \mathcal{H}(\textbf{s})}}{\mathcal{Z}}
\end{equation}
where $J_{ij}$ and $h_{i}$ are Lagrange multipliers and $\mathcal{Z}$ a normalizing constant (the partition function), see Sec-\ref{sec3:MEP} for details.

\section{Order-disorder transition}\label{sec5:order}

\begin{figure}[!ht]
\begin{center}
\resizebox{0.85\textwidth}{!}{%
\includegraphics[height=3cm]{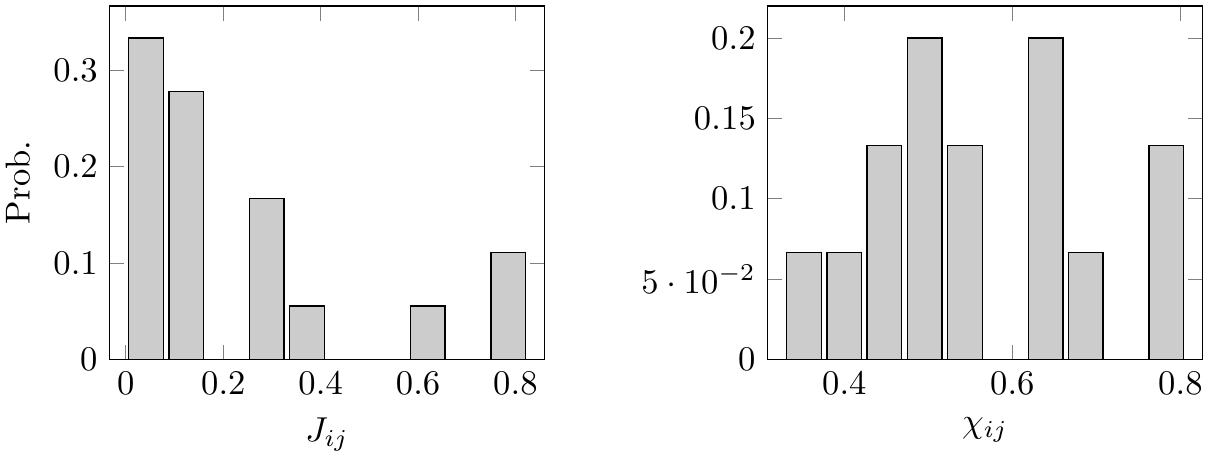}
 }
\end{center}
\caption[Influences and correlations distributions]{\label{fig:prob} Left: maximum entropy distribution of the interaction strengths $\tilde{\textbf{J}}^{\mathrm{\scriptstyle{ME}}}$ and right: empirical distribution of the pairwise correlations obtained from the collected data.}
\end{figure}

One of the most exciting features of the model is the emergence of collective behaviours even if the interactions are weak. The aim is to provide quantitative empirical evidence that the pairwise modelling is a consistent paradigm to describe collective behaviours in markets.
In the following, we apply the pairwise model to a set of six major market indices (AEX, Bel-20, CAC 40, Xetra Dax, Eurostoxx 50, FTSE 100). We selected only European indices because some financial issues are specific to Europe and we consider indices because they are the driving force of the respective market places \cite{Shap}, they will reflect the main properties of the underlying stock set. We observe 2253 configurations from 6/06/2002 to 14/06/2011 \cite{Yahoo}. We take a nine year long time series including two large crises. The daily sampling is enough since we want to study large crises, and the two principal peaks of the Fourier transform are centered on frequencies $f_{1}=6\times10^{-4}\,\mathrm{d}^{-1}$ and  $f_{2}=1.2\times10^{-3}\,\mathrm{d}^{-1}$; the unit \emph{day} stands for trading day. The first frequency $f_{1}$ is the crisis occurrence frequency in our time window, the corresponding period is $T_{1}=1.7\times10^{3}\,\mathrm{d}$ . Later, we will also analyse the stocks composing the Dow Jones and the S$\&$P100 indices, and another set of 116 stocks.
First of all, we give the magnitude order of the interaction strengths and of the empirical pairwise correlations in Fig-\ref{fig:prob}.

The $J_{ij}$ are all positive; we can therefore use the net mean orientation as an order parameter to describe switches between strongly and weakly correlated states. The mean value of $h_{i}$ is about $0.0113$.

\begin{figure}[!ht]
\begin{center}
\resizebox{0.85\textwidth}{!}{%
 \includegraphics{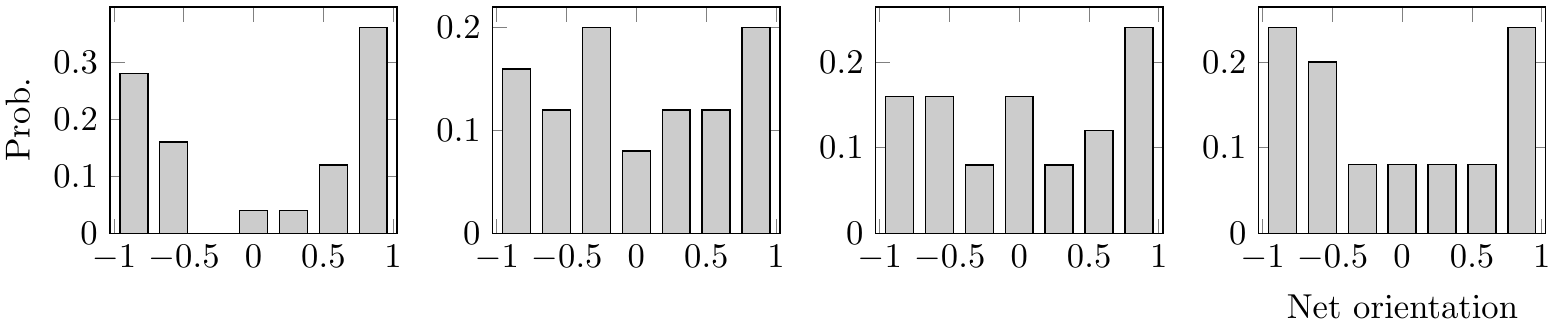}
 }
\end{center}
\caption[Bimodal distribution of the orientation]{Empirical probability distribution of the net orientation on four successive periods, each of 25 trading days. Time goes from left to right. The last time window corresponds to the irregularity induced by the Fukushima nuclear accident.}
\label{fig:EmpTrans}
\end{figure}

As the previous pairwise model describes market indices quantitatively, we expect to observe an order-disorder transition in this system; we give below some empirical evidence that these transitions actually appear.
As the interaction strengths are all positive, the system is ordered if the net orientation distribution has two modes near the extreme values $-1$ and $1$ and disordered if the distribution has a unique mode. Indeed in an ordered situation, each index tends to have the same orientation as the others. Furthermore, in the absence of external influences, both extreme values are equivalent (as a consequence of the symmetry under sign exchange), and the distribution is thus bimodal. One of the extreme values can be favoured following the values taken by the external influences $h_{i}$. It will be a first clue that the system is reorganized if the distribution changes in such a way (having two modes and then a unique one, and reciprocally).
We compute the system net orientation $q(\tau,\Delta t)=(\Delta t\,N)^{-1}\sum_{i}\sum_{t=\tau}^{\tau+\Delta t}s_{i,t}$ on successive periods $\Delta t$ of 25 trading days (without overlapping), and we show that the net orientation probability distribution can be bimodal or not on successive time windows. The resulting empirical distributions for observations from 5 November 2010 to 30 March 2011 are illustrated in Fig-\ref{fig:EmpTrans}.

In Fig-\ref{fig:EmpTrans} we see that the empirical probability distribution has initially two modes at extreme orientation values then has no clear mode, and finally again has two modes. During this period, initially the indices move in an organized way then in a disorganized fashion, and finally the Fukushima nuclear accident caused a large global market fall followed by a large recovery. During this event, the indices were in co-movement. So the system is initially ordered then disordered for two periods and then again ordered.

\begin{figure}[!ht]
\begin{center}
\resizebox{0.7\textwidth}{!}{%
\includegraphics{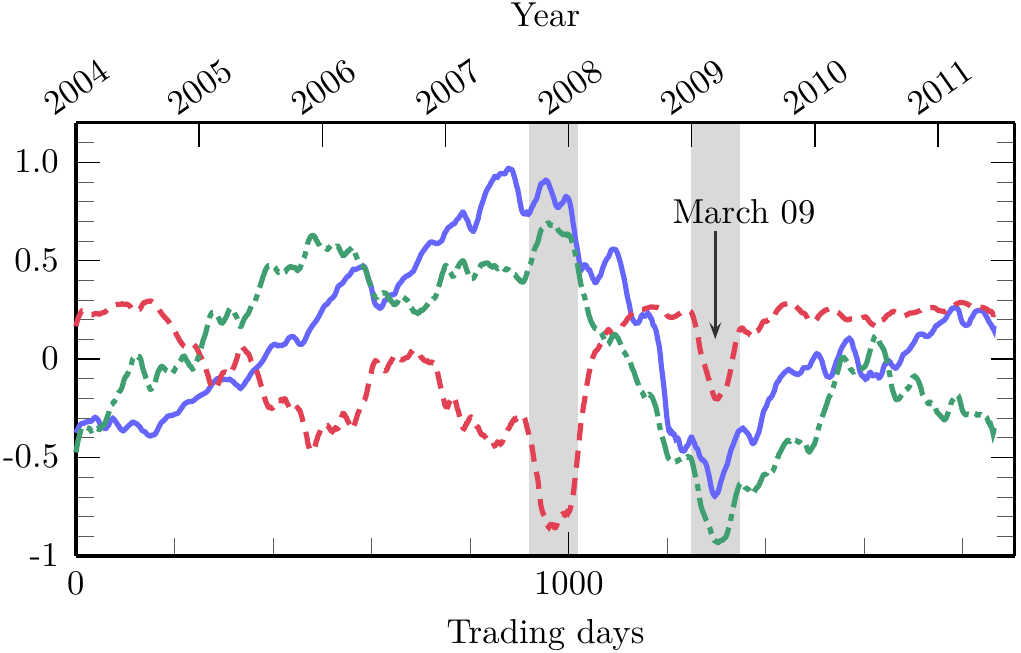}
}
\end{center}
\caption[Entropy during crises]{The normalized sum of indices (full line), the normalized net orientation (dashed-dotted line) and the normalized mean-field entropy (dashed line). The curves have been smoothed. The last major crisis is pointed out by an arrow. The shaded portions show orientation extrema and entropy minima. }
\label{fig:EurEntrSmooth}
\end{figure}

A more accurate way to characterize financial irregularities is to study the entropy $S(\textbf{s})$ on a sliding window (here, 300 trading days shifted by 1 day). Indeed the entropy is a statistical measure of the amount of correlations. The stronger the correlations, the lower the entropy. The main issue is that in general entropy can not be obtained exactly because it requires the computation of $2^{N}$ terms.  We compute the zeroth order approximation (see methods) of the entropy on those time windows (much faster than the exact computation). We saw in Sec-\ref{sec3:Var} that the zeroth order approximation of the entropy \cite{Binder} is

\begin{equation}\label{MFentr}
  S_{0}(\textbf{s})= -\sum_{i=1}^{N}\left\{\frac{1+q_{i}}{2}\ln\left(\frac{1+q_{i}}{2}\right)+
  \frac{1-q_{i}}{2}\ln\left(\frac{1-q_{i}}{2}\right)\right\}
\end{equation}
The entropy is maximal when the average orientations, computed on the corresponding time window, are  equal to zero and is minimal when the indices have the same orientation. During a disordered period, the entropy should be large and during a synchronized (ordered) period the entropy should be low. We should thus observe entropy minima simultaneously to orientation extrema (bubbles or crashes). We check in the results illustrated in Fig-\ref{fig:EurEntrSmooth} that orientation extrema and entropy minima are related to the periods of synchronization described in \cite{Dal}.

We observe large falls of the entropy when the net orientation is much larger than its mean (the mean is set to zero in Fig-\ref{fig:EurEntrSmooth}). The shaded portions show the orientation extrema and entropy minima on this time window. They correspond (chronologically) to the end of the growth period and the end of the collapse. Furthermore the correlation coefficient of the net-orientation and the financial time series is equal to $0.82$ showing a high degree of linear statistical dependency. We conclude that the entropy minima are thus related to financial irregularities (large upward or downward movements).

This is an empirical evidence that order-disorder transitions occur in markets. This interpretation is supported by the recent results obtained in \cite{Dal}, where the authors showed that market irregularities present a high degree of synchronization, meaning an ordered state.
The economic consequence is that the whole market is correlated when such transitions occur. It also means the absence of a characteristic scale for the fluctuations and the emergence of power-laws.

In appendix, we illustrate in Fig-\ref{fig:EurEntr} a larger version of Fig-\ref{fig:EurEntrSmooth}.

\section{Dynamics of interactions}\label{sec5:int}

Linked to the above, such a transition occurs if the stochasticity changes or the interaction strengths change. A possible interpretation of time-varying interaction strengths is that some learning or adaptive process takes place through time. This means that the market adjusts the interactions between its entities in some adaptive processes so the $\{J_{ij},h_{i}\}$ are time dependent. The reason is that the background, namely worldwide economic conditions, changes through time and goes through economic fluctuations with contractions (recessions) and expansions (growths). As the correlations are explained by the pairwise interactions, it also means that the correlations to be do not necessarily match past correlations (non-stationarity).

\begin{figure}[!ht]
\begin{center}
\resizebox{0.8\textwidth}{!}{%
  \includegraphics{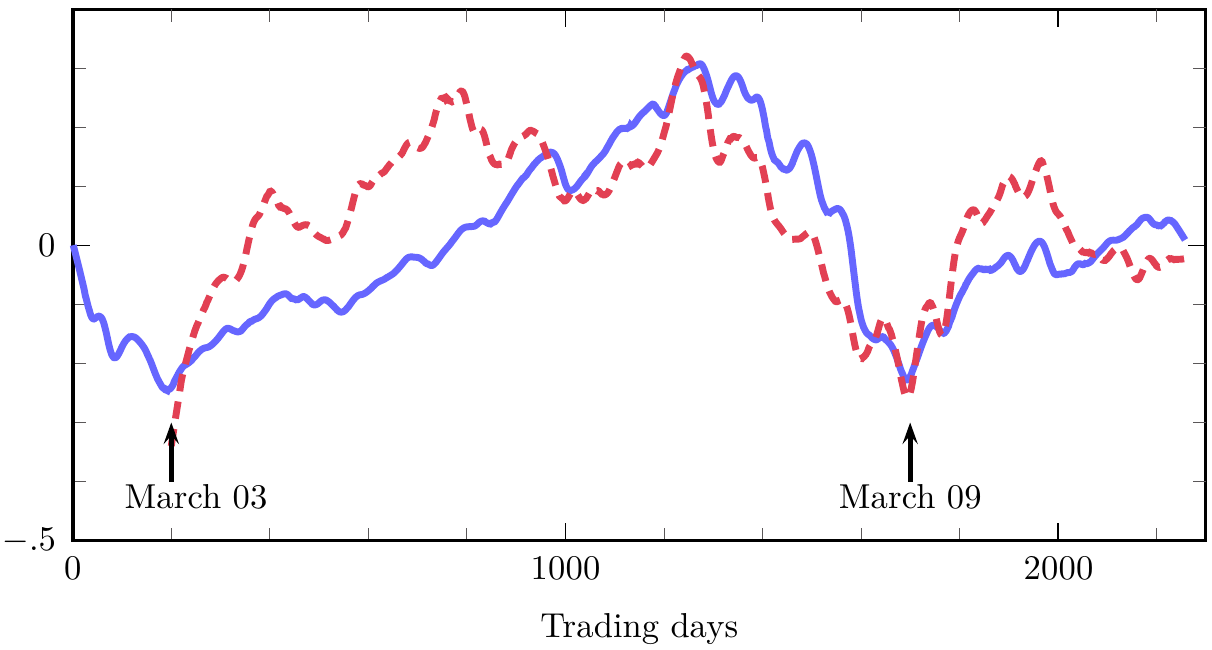}
}
\end{center}
\caption[Individual biases during crises]{The aggregate preference (dashed line) and the normalized sum of indices (full line); both curves have been smoothed. The last two major crises are pointed out by arrows. }
\label{fig:EurPref}
\end{figure}

Following this interpretation, we expect that the temporal behaviours of the interaction strengths and external influences are related to market evolution. This is indeed true, as we will see below. First of all, we study the preference evolution of the six previous indices (reflecting the current state of the European economy) and its link to the crises.
We use a sliding temporal window of width $T=200$ trading days shifted by a constant amount of  $\Delta t=2$ trading days. We show that the aggregate preference $h=\sum_{i}h_{i}$ is negative during a crisis (or during a significant contraction) as illustrated in Fig-\ref{fig:EurPref}.

The first negative incursion corresponds to the 2002-2003 crisis and the second one to the 2008-2009 crisis \cite{Yahoo}.
As expected the external influences are decreasing when the market undergoes a crash.

\begin{figure}[!ht]
\begin{center}
\resizebox{0.8\textwidth}{!}{%
\includegraphics{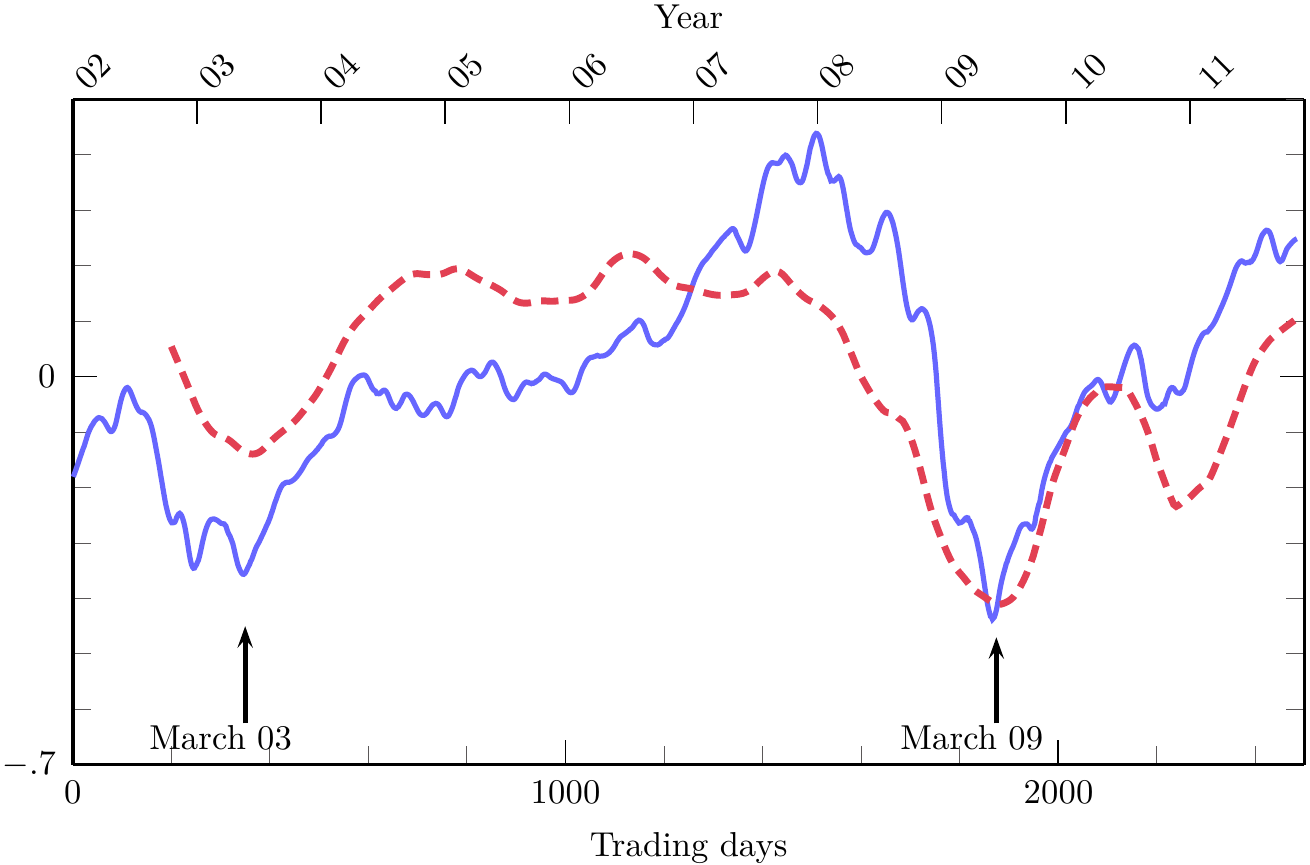}
}
\end{center}
\caption[Diagonal influences]{The normalized Dow Jones index is plotted as a full line; the trace minus its temporal average is the dashed line. We used a sliding temporal window of width equal to 200 trading days shifted each time by 5 trading days.}
\label{fig:DJspectre}
\end{figure}

More interestingly, we will study the spectrum of the interaction matrix. Indeed the spectrum evolution is related to the market evolution. The spectrum of the interaction matrix of a stock set has an interesting feature; we will show it for the Dow Jones index. We collected data for the Dow Jones index from the 10 October 2001 to 1 August 2011 \cite{Yahoo}, and we extract the interaction strengths using the third-order approximation described in \cite{Tanaka}.
The trace of the interaction matrix, the sum of its eigenvalues, has the following interesting property. It decreases during a crisis; specifically, the trace minus its temporal average becomes negative if there is a substantial fall of the index, this feature is illustrated in Fig-\ref{fig:DJspectre}.

The trace of the exact interaction matrix should be zero (without self-interactions) but, with the Tanaka's diagonal trick detailed in Sec-\ref{subsec:diagonalTrick}, the diagonal entries are related to the second-order term and to a part of the third-order of the Plefka series \cite{Tanaka,Plef}. The second-order term of the Plefka series is negative, the sign of the third-order term depends on the product of the interaction strengths. The temporal variation of the trace reflects the temporal variation of these second and third order terms. These terms are particularly important near a transition. This explains why the trace of the interaction matrix is smaller than its mean value during a crisis. Indeed during a crisis all stocks act in similar way: they fall down. They thus have similar mean orientation (down) and the resulting system state is an ordered one. Before the crisis, during a common market growth or steady state, the price of some stocks rises (on average) and some others fall leading to a dispersion of the mean orientations. This is indirect evidence of a transition from one regime to another and of coordination. It is consistent with the results obtained above and in \cite{Dal,Jr}. In appendix, we illustrate in Fig-\ref{fig:DJspecHuge} a larger version of the Fig-\ref{fig:DJspectre}.

\begin{figure}[!ht]
\begin{center}
\resizebox{0.8\textwidth}{!}{%
\includegraphics{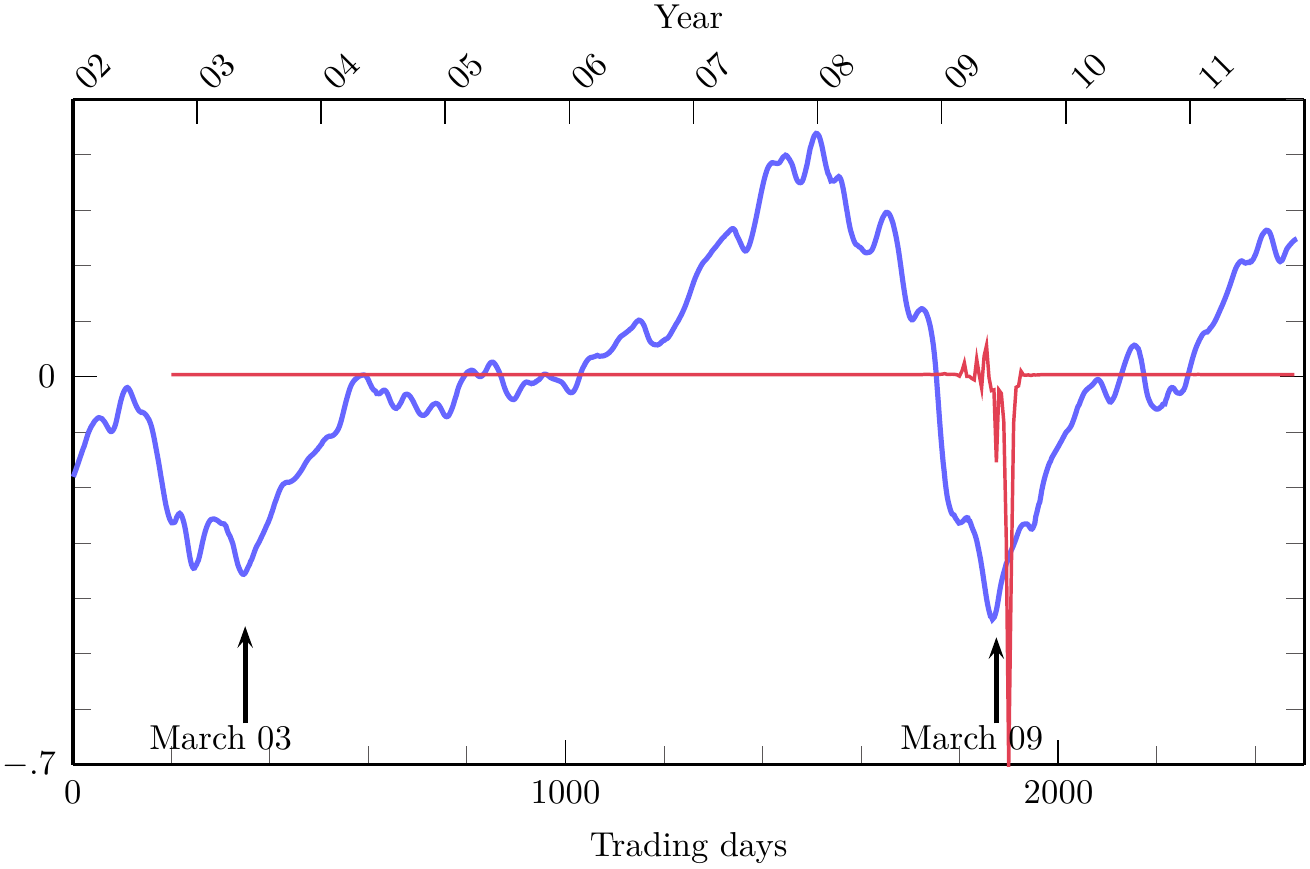}
}
\end{center}
\caption[Determinant of the influence matrix]{The normalized Dow Jones index is plotted as a thick line; the thin line illustrates the determinant minus its temporal average. We used a sliding temporal window of width equal to 200 trading days shifted each time by 5 trading days.}
\label{fig:DJspectreDet}
\end{figure}

Similarly, the determinant of the $\textbf{J}$-matrix undergoes a large variation during the crisis period, this behaviour is illustrated in Fig-\ref{fig:DJspectreDet}.

\section{Link to the graph-theoretic approach}\label{sec5:graph}

\begin{figure}[!ht]
\begin{center}
\resizebox{0.8\textwidth}{!}{%
\includegraphics{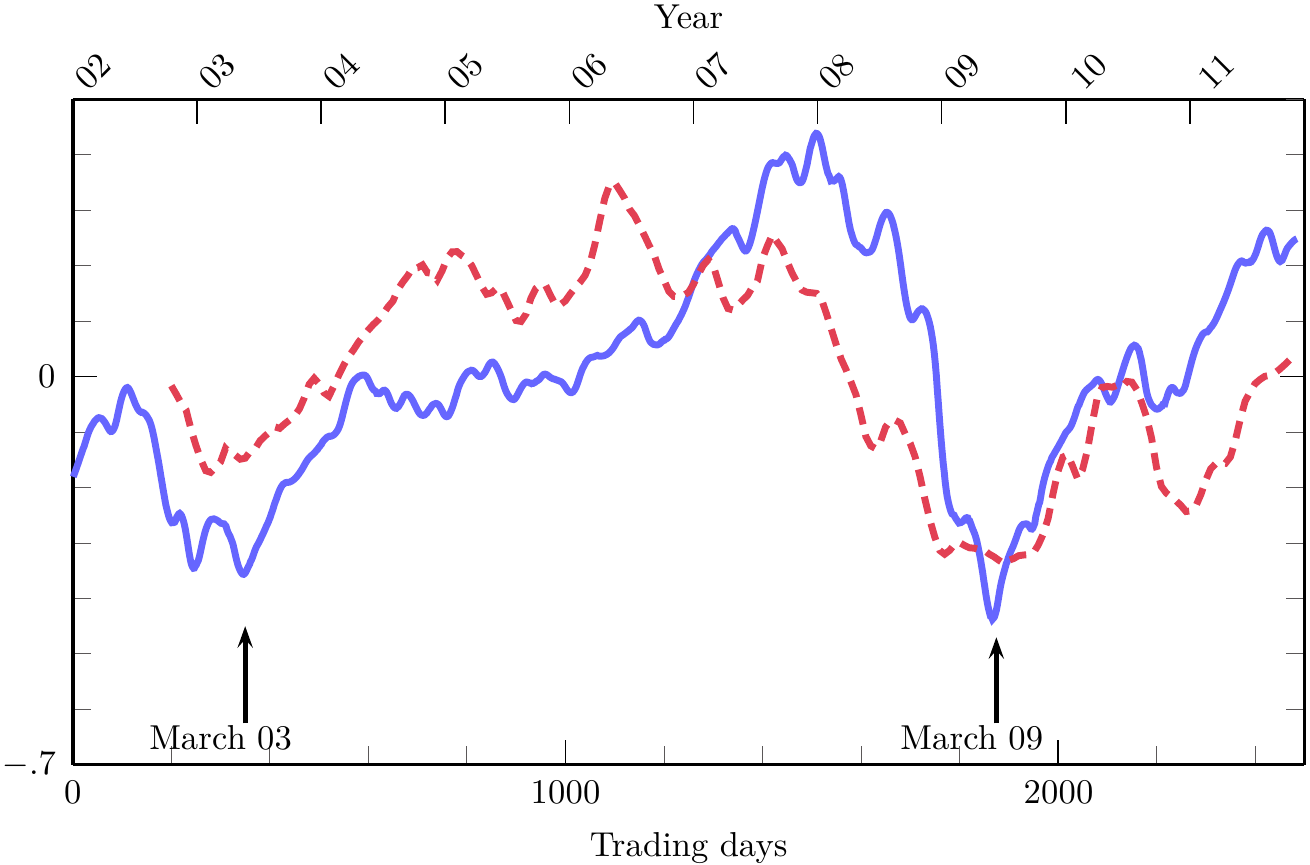}
}
\end{center}
\caption[Length of the Dow Jones assets tree]{The normalized Dow Jones index is plotted in full line, the relative difference to the time average of the length $l(t)=\left[L(t)-\langle L\rangle\right]/\langle L\rangle$ is the dashed line (where the brackets denote the temporal average). We use a sliding window of 100 trading days shifted by 10 trading days each time.}
\label{fig:DJmst}
\end{figure}

Hereafter, we make the link with the previous spectrum feature and the observation that the length of the minimum spanning tree (MST) based on the Sornette-Mantegna distance (see Sec-\ref{sec3:MSdist}) decreases during a crash \cite{Mant,Onnela}, meaning that stocks are highly correlated during these events (as they should be in an order-disorder transition). We will see that we recover this feature with the pairwise model with a distance based on interaction strengths in place of correlation coefficients. Indeed the interaction matrix can be thought of as the weight matrix of an undirected complete graph. Using a modified version of the method proposed in \cite{Ding} and computing the minimum spanning tree length $L(t)$ (the sum of the edges weights of the MST), we also observe that this length decreases during a crash, as expected; the results for the Dow Jones index are illustrated in Fig-\ref{fig:DJmst}.

\begin{figure}[!ht]
\begin{center}
\resizebox{0.9\textwidth}{!}{%
\includegraphics{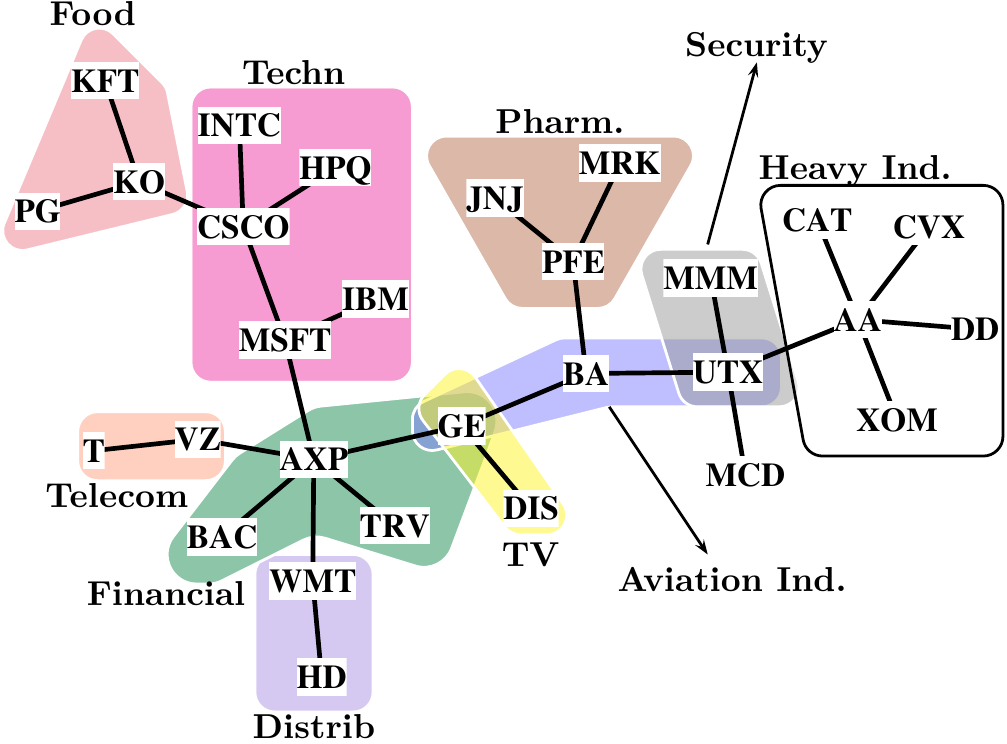}
}
\end{center}
\caption[The Dow Jones assets tree]{The minimum spanning tree based on the interaction matrix $\mathbf{J}$ is estimated on 2500 trading days. The companies are denoted by their ticks; they can be found on any financial website (\emph{Google finance} for instance).  }
\label{fig:DJclusters}
\end{figure}

Moreover, it also allows cluster identification. Indeed, it is known that the asset tree based on the Sornette-Mantegna distance allows regrouping some stocks in clusters following their economic sectors \cite{Onnela}. As correlations are caused by the interactions, it is not surprising that the MST of the network defined by the interaction matrix also allows cluster identification. This approach has the advantage of not being limited to linear or monotonic statistical dependencies. The clustering feature is illustrated in Fig-\ref{fig:DJclusters}.

We note that General Electric (GE) is not the most connected node but it is a cental one in the sense that it appears in three different clusters, as such it is still considered as the \emph{root} of the MST and defines the generational direction. This approach provides a different classification than the one given in \cite{Onnela} or given by Forbes for instance. Indeed, Forbes classification is given by sector then by industry. Disney and Walmart are classified in the same sector, \emph{services}; this category is too vague to be an useful tag. Similarly, General Electric is tagged by Forbes as \emph{industrial goods} and then as \emph{diversified machinery} but this company also provides financial services, aircraft engines, TV channel broadcasting, etc. It is then clear that this company should be classified with more than one tag, as does the proposed method. In this point of view, the internal structure of each company seems to be the crucial information to identify stock clusters.

\begin{figure}[!ht]
\begin{center}
\resizebox{1\textwidth}{!}{%
\begin{tabular}{c c}
 \includegraphics[scale=1]{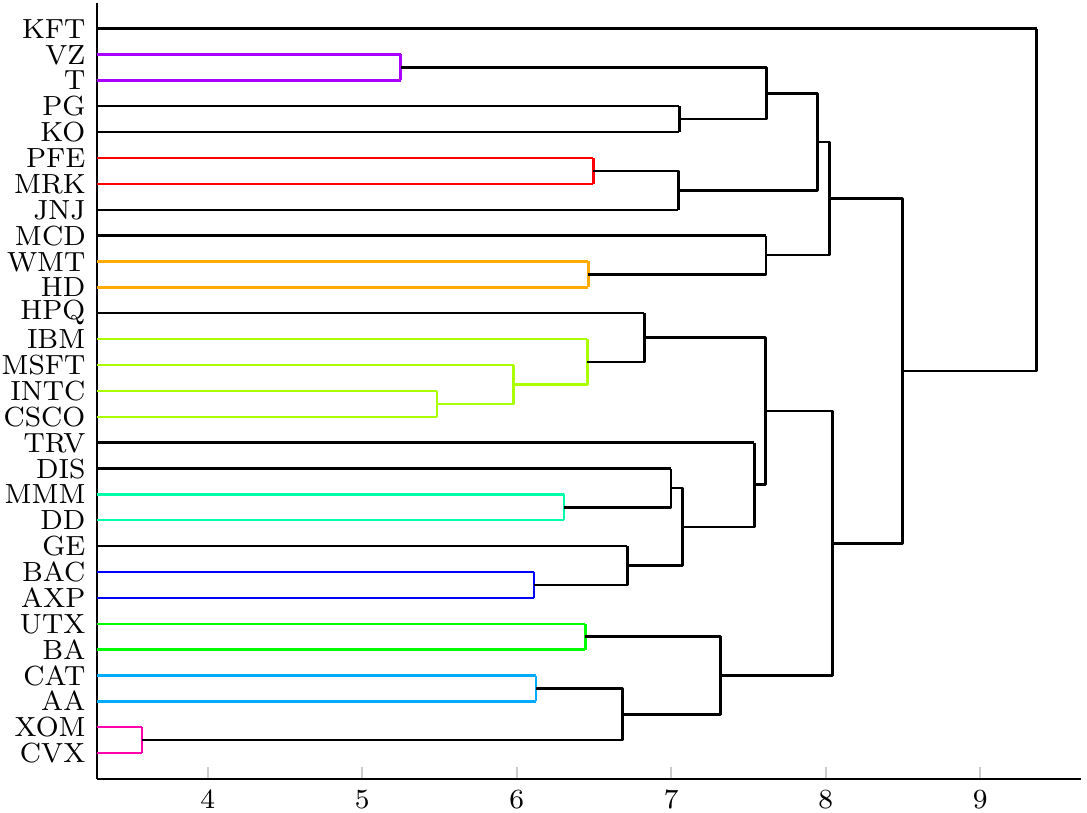}& \includegraphics[scale=1]{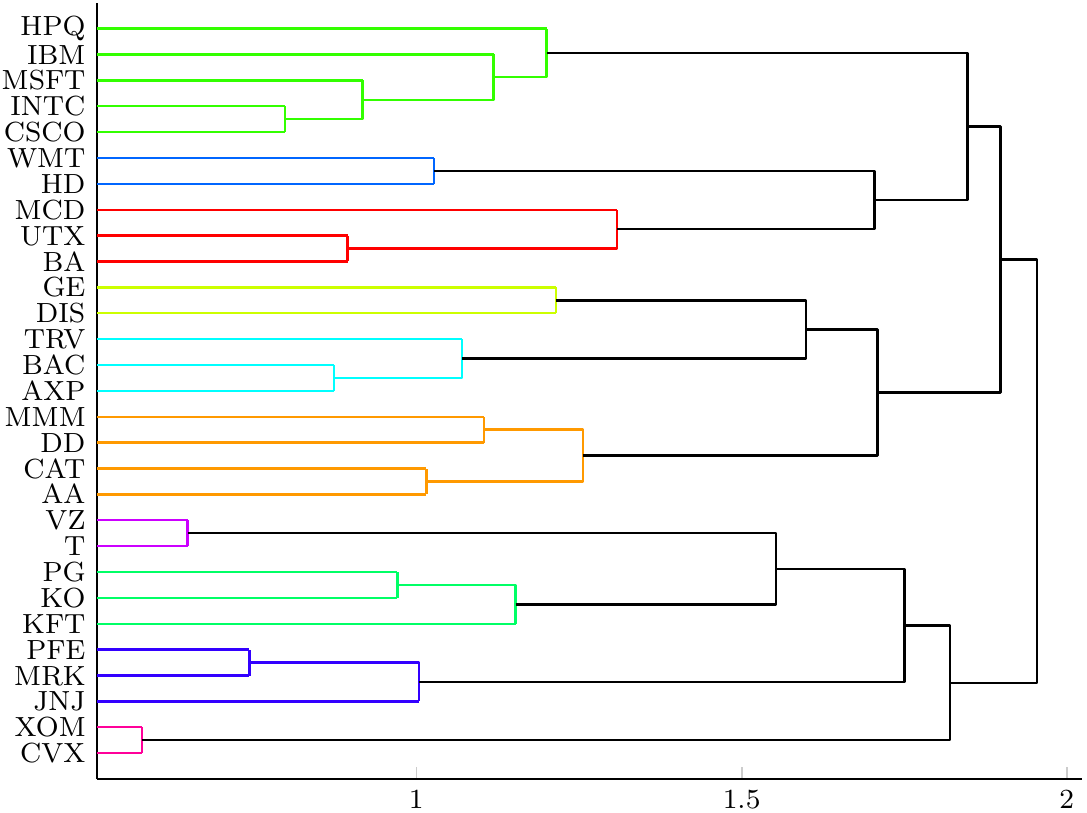}\\
 \includegraphics[scale=0.85]{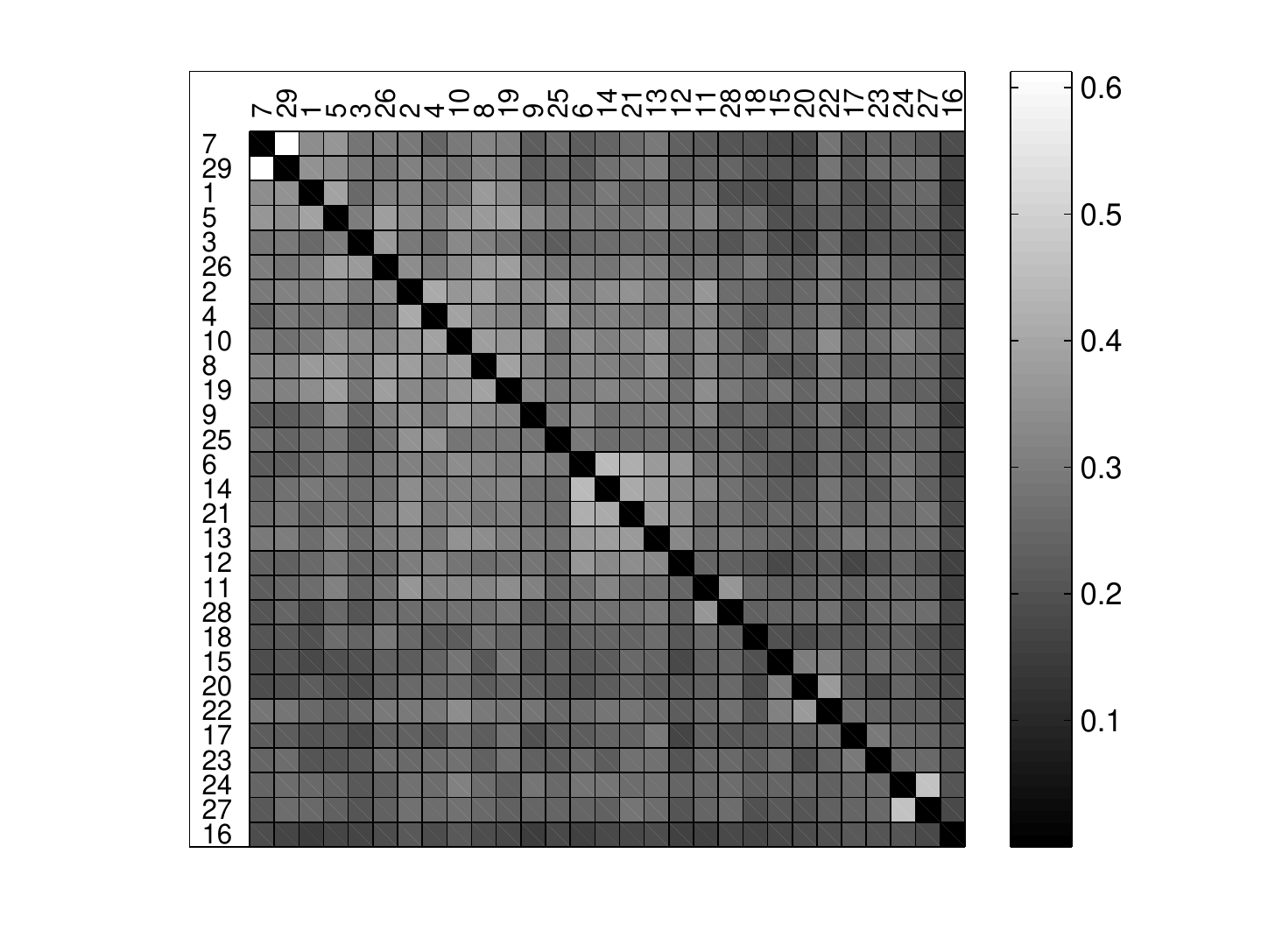}& \includegraphics[scale=0.85]{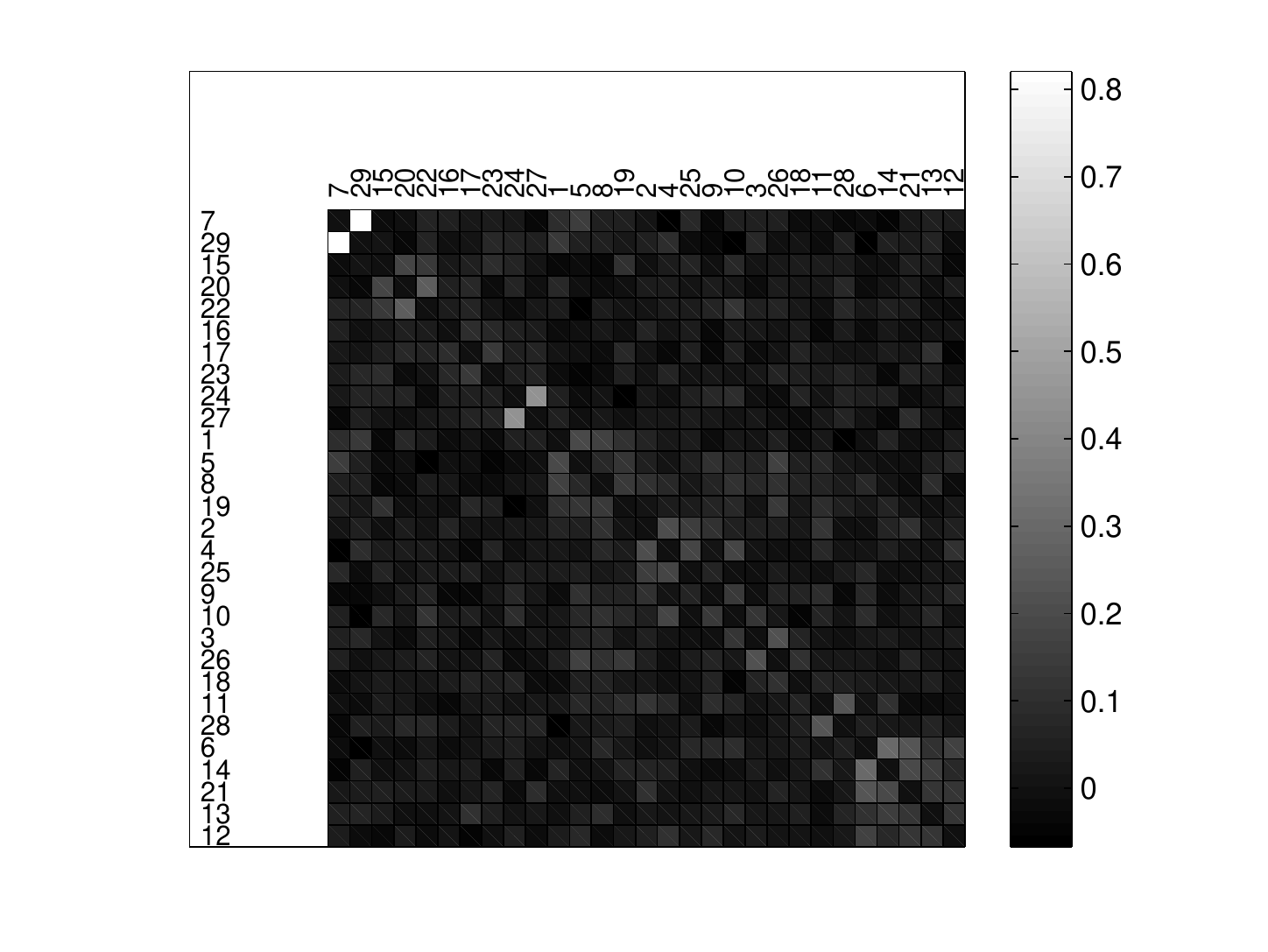}\\
\end{tabular}
}
\end{center}
\caption[Clusters of the Dow Jones]{Left: clusters and the matrix map obtained from the correlation matrix of the Dow Jones. Right: from the $\textbf{J}$ matrix. The matrices are reordered such that clusters stand on the diagonal (the numbers show the original ordering position).}
\label{fig:DJclusters2}
\end{figure}

Another method to visualize clusters is to plot the dendrogram (tree-like diagram). We illustrated the results obtained by using the correlation matrix and the mutual influence matrix ($\textbf{J}$) as dissimilarity matrices in Fig-\ref{fig:DJclusters2}. The identification is done by using the linkage function \footnote{see \url{http://www.mathworks.nl/help/stats/linkage.html}, for instance.} with complete standardized euclidian distance between clusters. A stock was removed due to lack of data.

The dendrogram obtained with the $\textbf{J}$-matrix returns the following sectors (the different clusters are shown in different colours): technology (IBM, HPQ, MSFT, INTC, CSCO), general distribution (WMT, HD), aviation industry (UTX, BA), TV broadcasting (DIS, GE), financial (TRV, BAC, AXP), chemicals (DD, MMM), heavy industry (CAT, AA), telecom (VZ, T), food and consumer goods (KFT, PG, KO), healthcare (PFE, MRK, JNJ) and oil industry (XOM, CVX). These sectors are consistent with those of the MST analysis and close to the companies profile.

\begin{figure}[!ht]
\begin{center}
\resizebox{1\textwidth}{!}{%
\begin{tabular}{c c}
 \includegraphics[height=8cm,width=8cm]{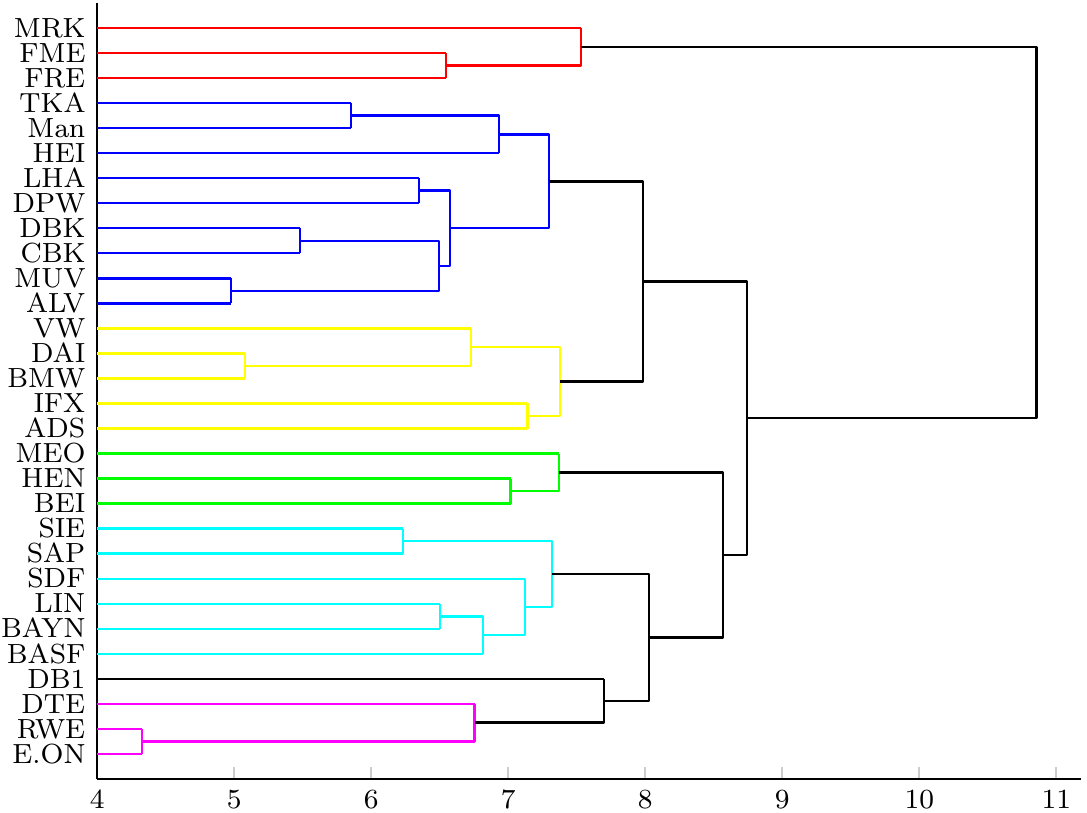}& \includegraphics[height=8cm,width=8cm]{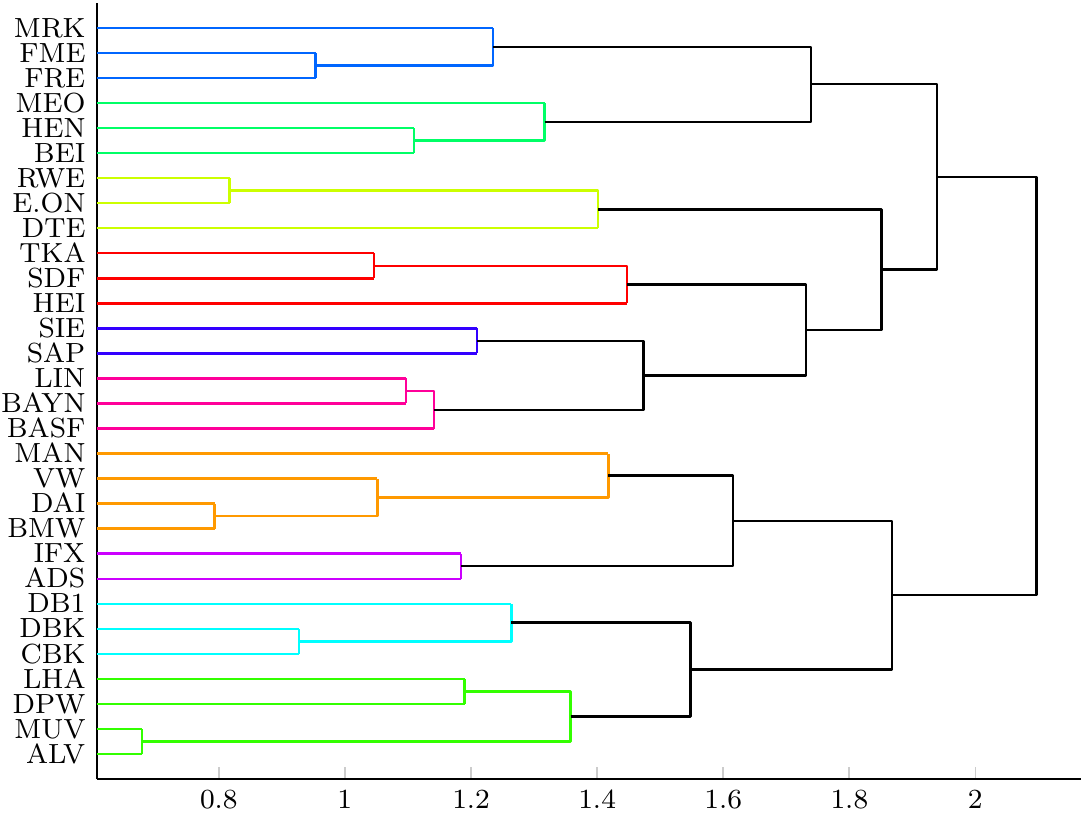}\\
\end{tabular}
}
\end{center}
\caption[Clusters of the DAX]{Left: clusters obtained from the correlation matrix of the DAX index. Right: from the $\textbf{J}$ matrix. The matrices are reordered such that clusters stand on the diagonal.}
\label{fig:DJclusters4}
\end{figure}

A more explicit example is the clustering of some stocks of the DAX  illustrated in Fig-\ref{fig:DJclusters4}. The clustering form the $\textbf{J}$ matrix seems more effective than with the correlation matrix.

\begin{figure}[!ht]
\begin{center}
\resizebox{\textwidth}{!}{%
\begin{tabular}{cc}
 \includegraphics{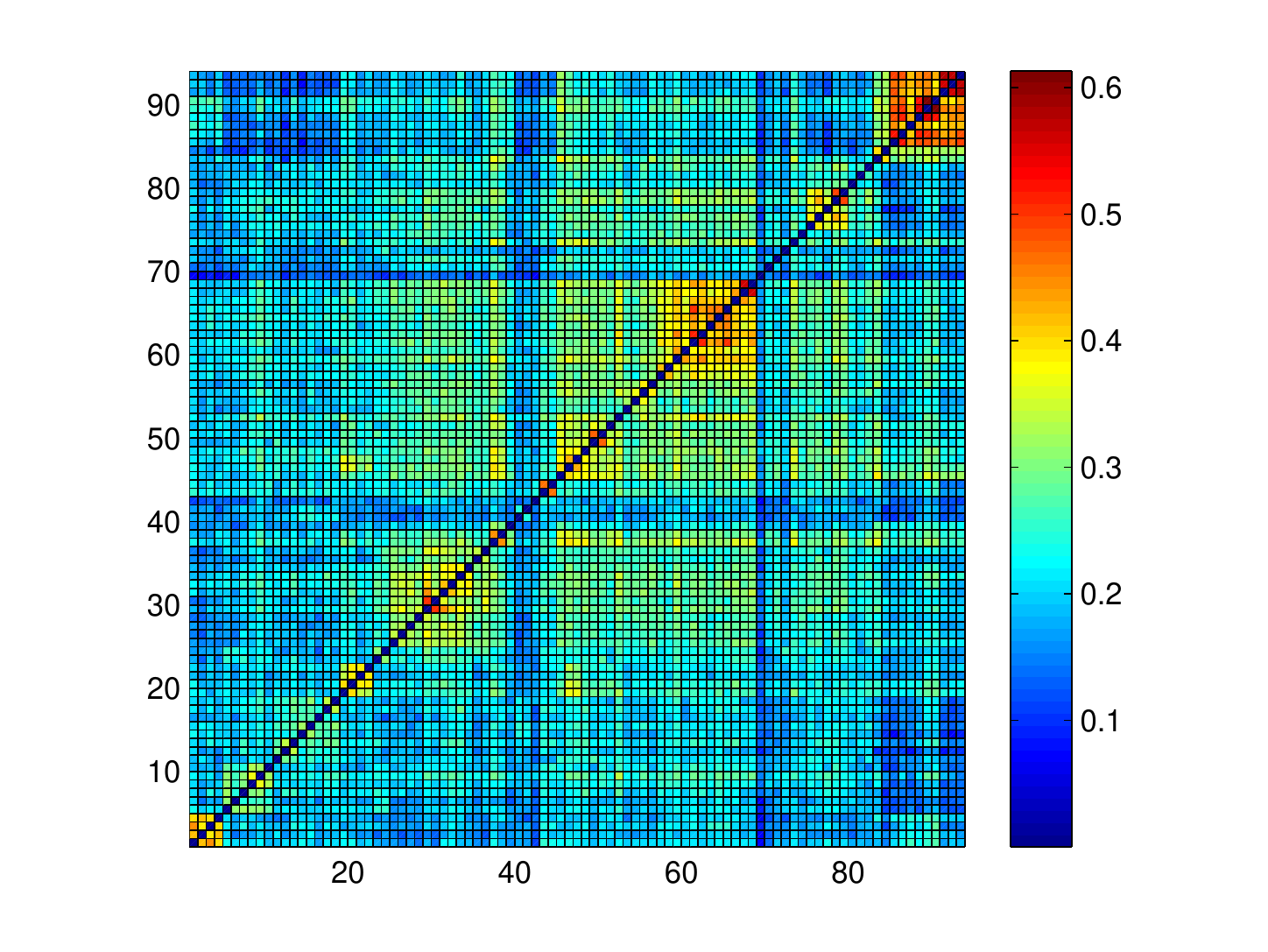} & \includegraphics{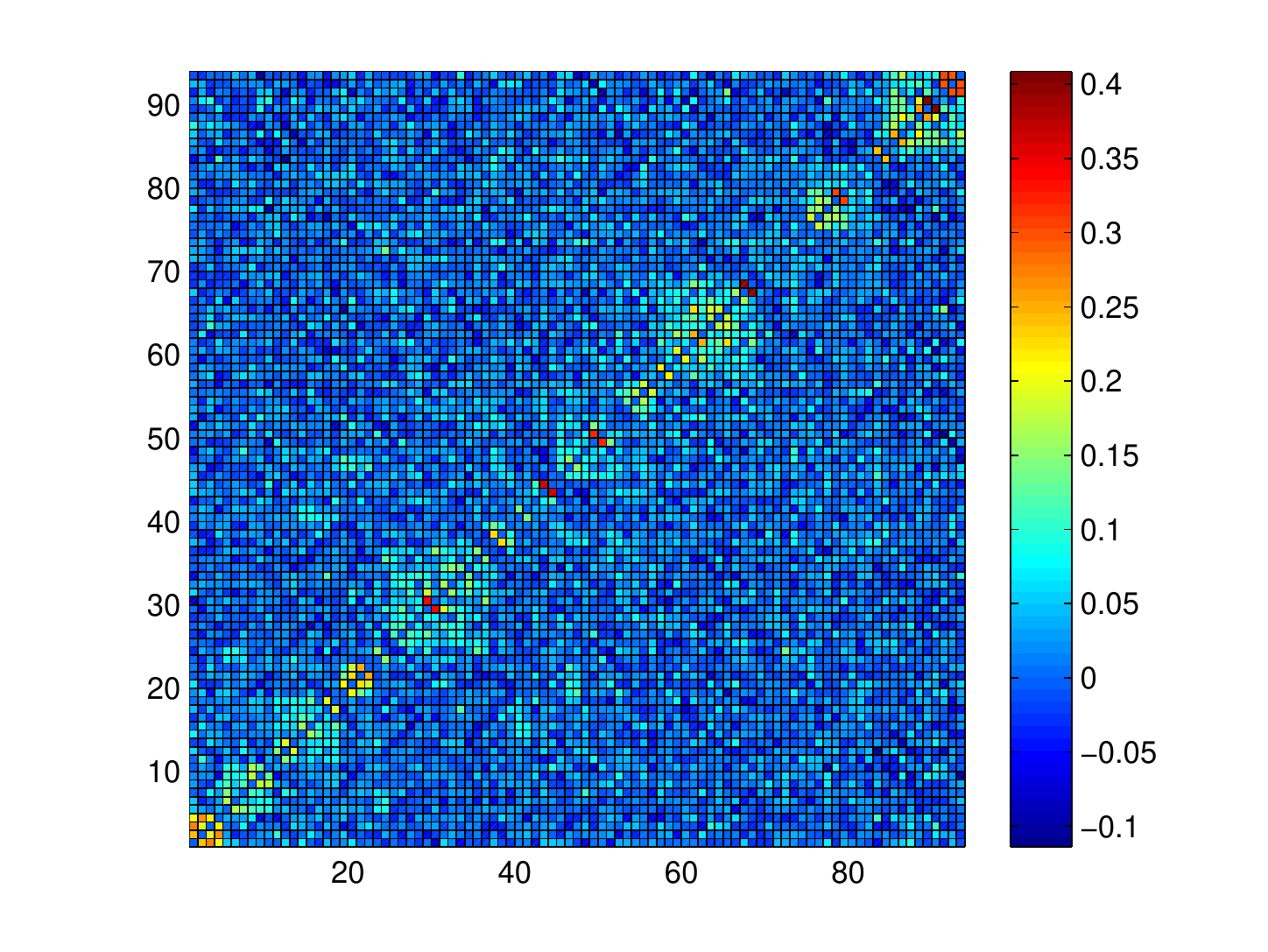}\\
\end{tabular}
}
\end{center}
\caption[Matrix maps]{Matrix map of the SP100 index. Left: map of the correlation matrix. Right: map of the $\mathbf{J}$-matrix. The matrices are reordered such that clusters stand on the diagonal.}
\label{fig:DJclusters5}
\end{figure}

We perform the same clustering method for a larger set, even with few data points ($T=2500$) and large number of entities ($N=95$), the different clusters seem to be close to the known profile of each company. The clusters are illustrated in Fig-\ref{fig:DJclusters3}. We also note that in general the $\textbf{J}$-matrix is more sparse than the correlation matrix. This feature may explain why the clustering using the $\textbf{J}$ matrix is more efficient than with the correlation matrix. The sparsity is illustrated in Fig-\ref{fig:DJclusters5}.


This clustering method may be useful in portfolio composition and capital allocation. Indeed, we may look after diversification and cluster identification is then a crucial feature.

\begin{figure}[!h]
\begin{center}
\resizebox{0.75\textwidth}{!}{%
\includegraphics{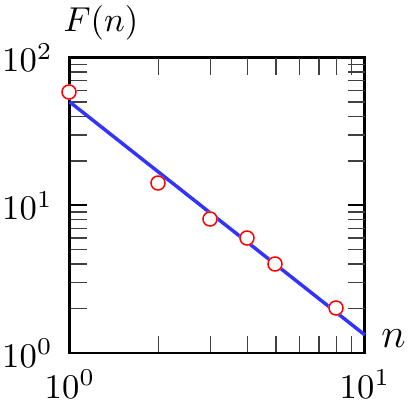}
\includegraphics{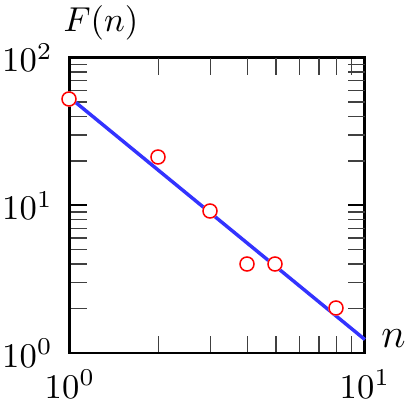}
}

\end{center}
\caption[Degree distribution]{The degree distributions during a growth period (left) and during a crash (right). The solid line is a power-law fit; the coefficients of determination are respectively $0.98$ and $0.93$.  }
\label{fig:SPdegree}
\end{figure}
\begin{figure}[!ht]
\begin{center}
\resizebox{0.95\textwidth}{!}{\rotatebox{90}{\includegraphics{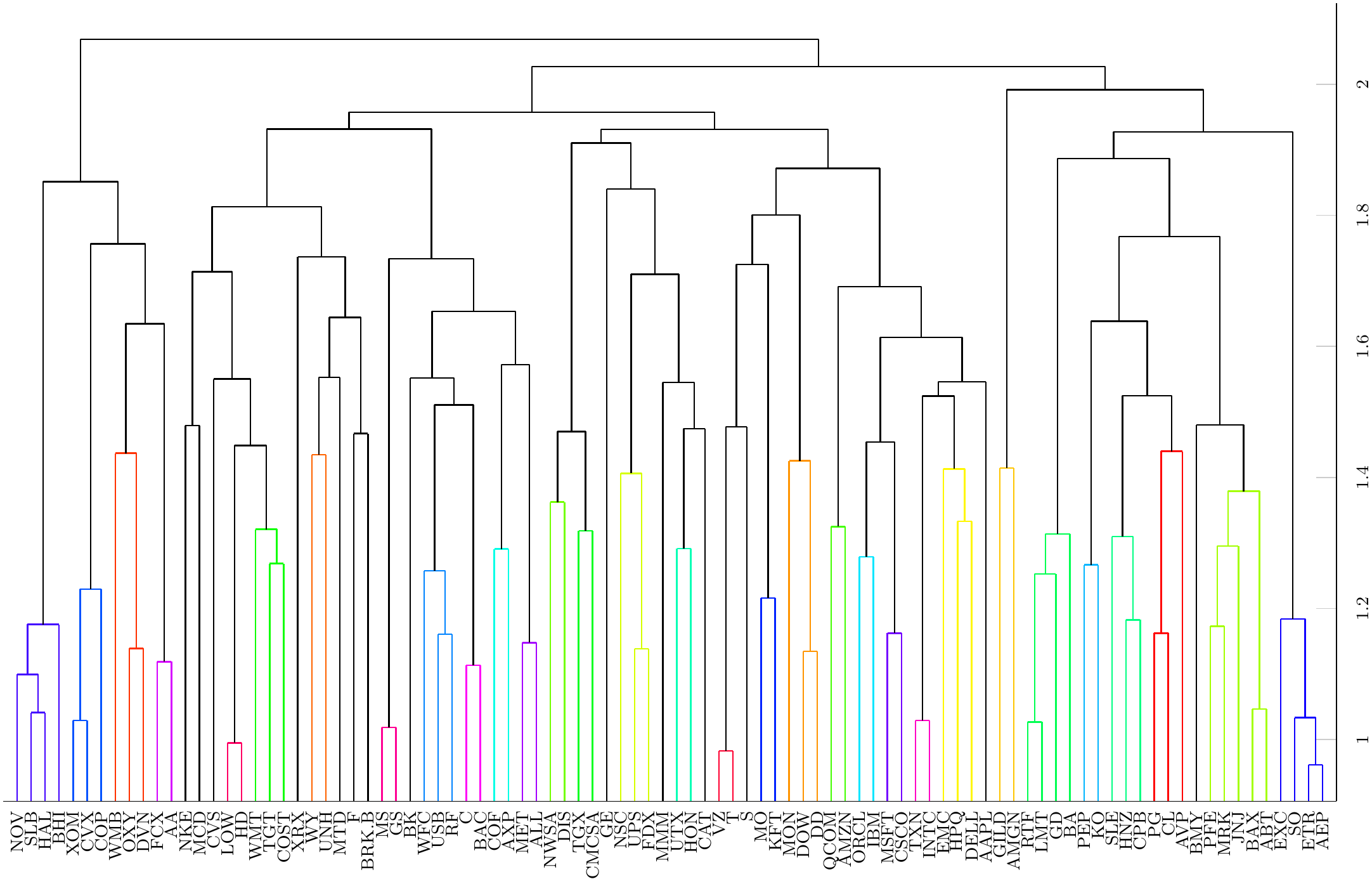}}}
\end{center}
\caption[Clusters of the SP100]{Clusters from the $\textbf{J}$ matrix of the SP100.}
\label{fig:DJclusters3}
\end{figure}

We can also study the topological structure of the remaining asset tree during a crash and a growth period. We will see that, as expected, the degree distribution follows a power law. We consider the stocks of the S\&P100 index on two intervals, from 1/10/2007 to 01/02/2009 (360 trading day crisis period) and from 1/02/2005 to 1/07/2007 (600 trading day growth period). The occurring frequencies of the vertex degrees are illustrated in Fig-\ref{fig:SPdegree}.

This study reveals that the degree distribution is a power law, $f(n)\sim n^{-\alpha}$, and the value of the exponent is similar for both periods. For the growth period, we obtain $\hat{\alpha}=1.64\pm 0.17$ and during a crash $\hat{\alpha}=1.58\pm 0.12$. They can be included in the confidence interval of each other, so they are similar.
The maximum degree is $n=8$ in the both periods. There are 58 vertices of degree $n=1$ during the crash. This value is slightly larger (about $10\%$) than the one corresponding to the growth period, 52 vertices of degree $n=1$. This explains the difference between both exponents. The asset tree topology is thus slightly different during a crash. The main change is the variation of the interaction strengths (the graph weights) rather than the variation of the vertex degrees.  In both regimes, the asset trees are thus scale-free networks. This implies that the edges are not drawn at random and the asset trees exhibit small-worldness (typical distance between two randomly chosen is proportional to the logarithm of total number of nodes in the network), as observed with another method in \cite{Petra}. Furthermore, the low value of this exponent implies that hubs (high-degree vertices)  represent a significant part of the total number of vertices. The market is thus sensitive to the failure of a hub (a highly connected company) whereas the failure of a leaf (terminal node) will only slightly affect the market. By example the hypothetic failure of the American Express Company (AXP) would leave a fragmented market whereas the bankruptcy of Kraft Food Inc. (KFT) would not change the topology of the asset tree significantly; see Fig-\ref{fig:DJclusters}. This could help in selecting the companies one has to save from an eventual bankruptcy in order to minimize the impact of such an event. This could also help to select which companies one has to monitor to prevent a hypothetical dramatic system failure.

\section{Conclusion}
We have seen that, without making assumptions on the market dynamics, the maximum entropy principle provides a rigourous pairwise model which is able to describe the data and the observed collective behaviours quantitatively. We showed that the collective phenomena emerge from simple pairwise interactions. The success of the pairwise model implies that markets exhibit some properties observed in magnetic materials and in neural networks. Indeed, we showed that an order-disorder transition occurs in such a system, as described by a pairwise model equivalent to the Ising model. We showed that the interaction strengths are time dependent meaning that an adaptive process occurs. Moreover, these Lagrange parameters are closely related to the orientation of the economic background. Furthermore, the $\textbf{J}$ matrix reveals itself a good measure of dissimilarity and allows cluster identification. This feature may be useful for capital allocation between different economic sectors (seeking for diversification) or to study the market structure.

The minimum spanning tree based on the $\textbf{J}$ matrix allows to determine the most connected economic nodes and may be used to determine which company is the most likely to overcome a systemic crash or which company may induce major impacts in case of bankruptcy.

In this view the system is more than the sum of its parts, is ruled by its entities pairs, exhibits collective behaviours and is quantitatively described by a pairwise model. It is surprising that such sophisticated collective behaviours, emergent structures and underlying complex trading rules are captured by a simple (a priori) scheme of interdependence involving only pairwise but no higher-order interactions.

\section*{Appendix}

\begin{figure}[!ht]
\begin{center}
\resizebox{0.8\textwidth}{!}{%
  \includegraphics{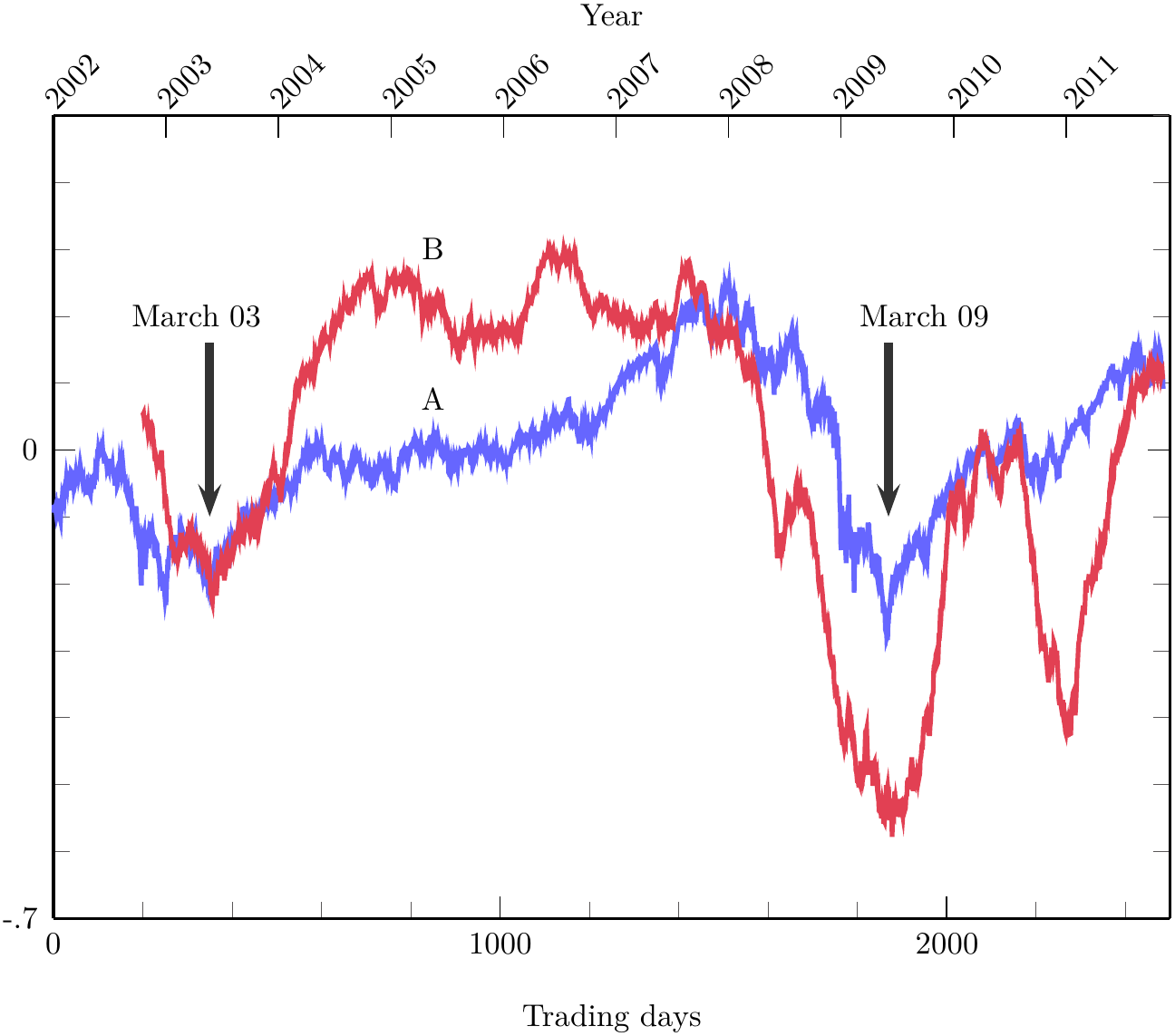}
}
\end{center}
\caption[Diagonal influences (large version)]{The normalized Dow Jones index is illustrated by the curve A; the trace minus its temporal average is the gray line (curve B). We used a sliding temporal window of width equal to 200 trading days translated each time by 1 trading day.}
\label{fig:DJspecHuge}
\end{figure}

\begin{figure}[!ht]
\begin{center}
\resizebox{0.8\textwidth}{!}{%
  \includegraphics{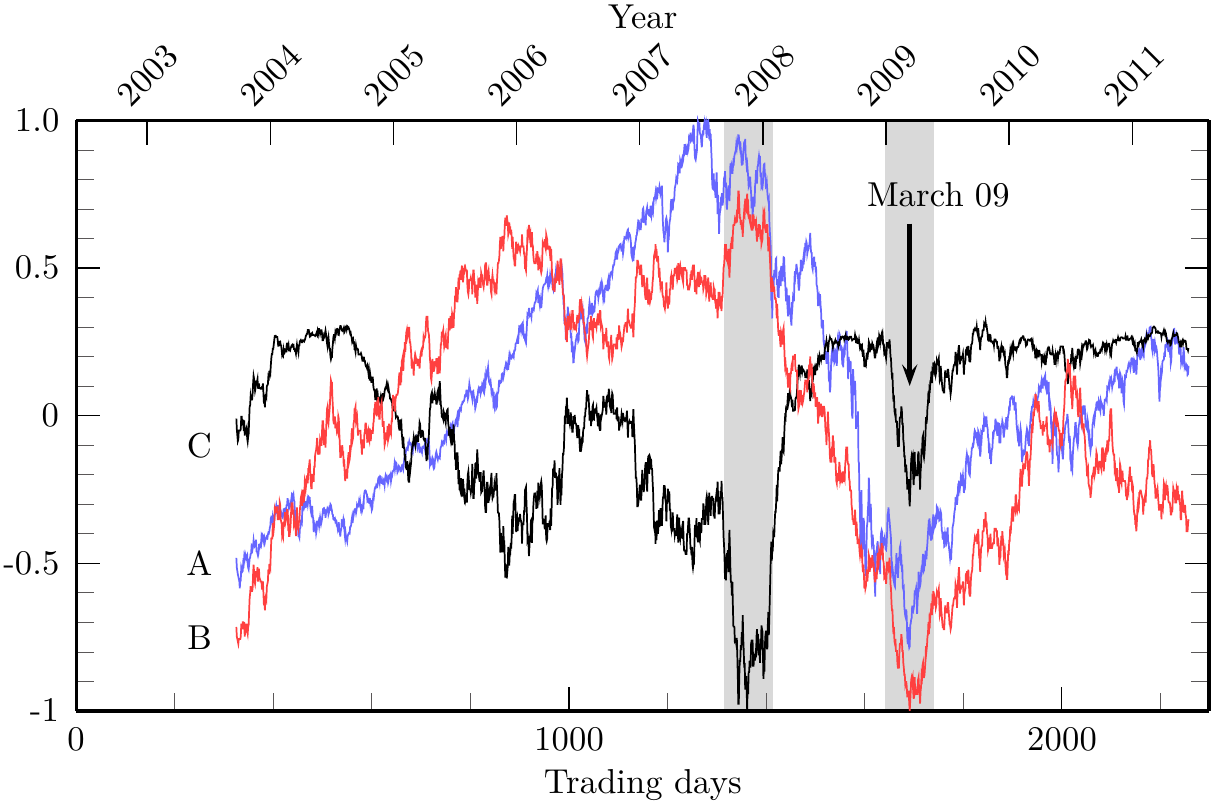}
}
\end{center}
\caption[Entropy during crises (large version)]{The normalized sum of indices (curve A), the normalized net orientation (curve B) and the normalized mean-field entropy (curve C). The last major crisis is pointed out by an arrow. The shaded portions show orientation extrema and entropy minima.}
\label{fig:EurEntr}
\end{figure}

\chapter{A statistical perspective on criticality in financial markets}\label{chap:crit}
\thispagestyle{empty}
\begin{summary}
Stock markets are complex systems exhibiting collective phenomena and particular features such as synchronization, fluctuations distributed as power-laws, non-random structures and similarity to neural networks. Such specific properties suggest that markets operate at a very special point. Financial markets are believed to be critical by analogy to physical systems but little statistically based evidence has been provided.
Through a data-based methodology and comparison to simulations inspired by statistical physics of complex systems, we show that the Dow Jones and indices sets are not rigorously critical. However, financial systems are significantly closer to criticality before a crisis.
\end{summary}

\newpage

\begin{figure}[!ht]
\begin{center}
\includegraphics[scale=0.8]{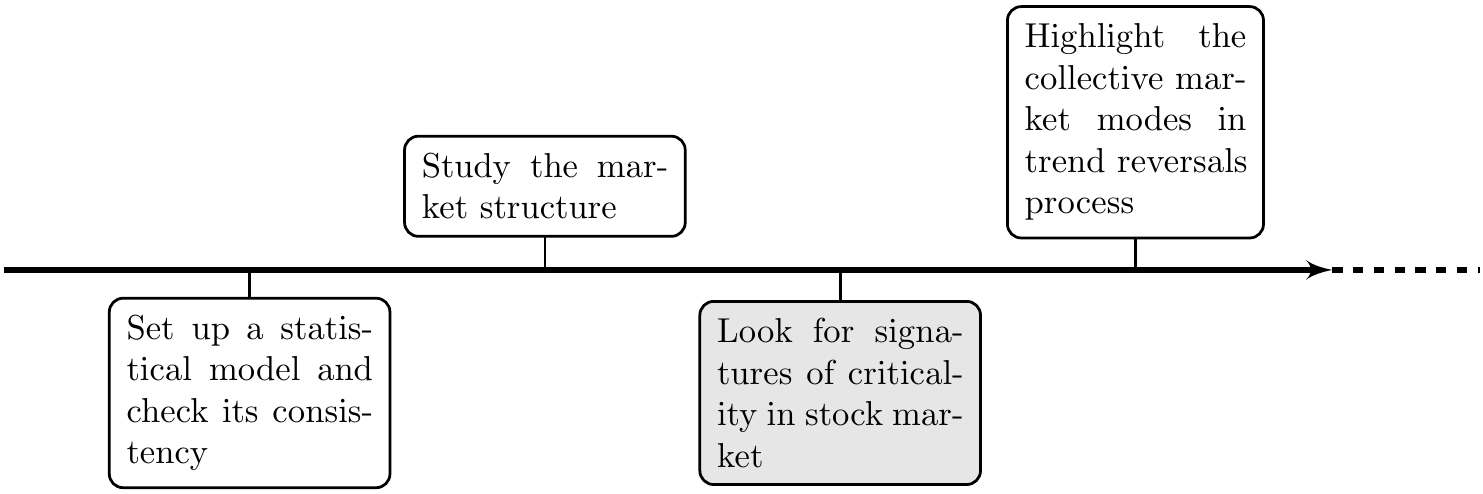}
\end{center}
\end{figure}

\newpage

\section{Introduction}
\label{06-intro}
The notion of scale invariance is widely used in finance and economics (fractional Brownian motion, detrended fluctuations analysis, volatility modelling, etc.). The scale invariance is crucial in finance because large absolute returns are power-law distributed \cite{Cont}. This lack of any characteristic scale is surprising at first glance but finds its foundation in the theory of complex systems. As complex systems composed of many correlated entities, financial markets exhibit collective behaviours like synchronization or non-random structure, propensity to self-arrange in large correlated structures as highlighted in \cite{Mant,OnnelaPRE,moi2}, large fluctuations \cite{Gabaix} and power-laws \cite{Cont}. Moreover it has been shown that financial networks share common properties with neural networks \cite{Petra}. One recovers those features in a class of models belonging to statistical physics, pairwise maximum entropy models which are particulary suited to capture collective behaviours. One knows that the market may exhibit some of the former features at a critical state, defined in a precise sense \cite{Huang} and that maximum entropy models may describe collective behaviours observed in neural networks \cite{Schneidman} and in financial markets \cite{moi2}. It is therefore tempting to think that financial markets are critical \cite{Sorn2} (in statistical physics sense) as it seems to be for neural networks \cite{fraiman,Egu,Tagl}.
It is not obvious how to validate empirically the presence of a critical state. Criticality was proposed for the approach of log-periodicity \cite{SorLog}. The phenomenological comparison to critical phenomena was done by substituting the temperature by the time which becomes therefore the control parameter \cite{Vande} but it is merely an analogy, log-periodicity should be understood as a dynamical feature rather than a second order phase transition. Indeed, several dynamical mechanisms generate log-periodicity \cite{SorDSI}. Criticality was also proposed for agent-based models exhibiting power-laws and volatility clustering at this particular state \cite{Lux,Zhou,Moro,Yeung,Cha,Sav}. However different rules and models lead to the same qualitative stylized facts. There is still ambiguity since there are non-critical mechanisms which generate stylized facts \cite{Newman}. Furthermore, detecting the criticality is not the same task as modelling complex systems, even if relations obviously exist. A rigourous approach of criticality detecting is the \emph{inverse} (or data-based) approach.  A transition between scale dependence and scale invariance is highlighted \cite{Kiyono} by this means. Here, we also follow an inverse (starting from the data without initial assumptions) procedure described in \cite{Mora} and applied in \cite{Step}. This procedure is also inspired by statistical physics and provides several statistical tests of criticality.

We find that the considered financial systems are not strictly critical even if some signatures are observed. It is more likely that financial systems do not stay in the same regime and are closer to the criticality when the system gets closer to a crash. The critical scaling parameter (see hereafter) reaches its maximal value in the vicinity of the beginning of the crash. Namely, the response function to a shock (a shock can be a modification of exogenous variables or of the level of stochasticity) has a peak and its position scales with system size towards the operating point (at which the probability distribution is the empirical one) for European market places. The operating point of the Dow Jones is far from the critical one but the criticality could be reached if the size of the index is large enough. The distribution of rank of configurations is not a power-law if the system is well sampled and the entropy is not a linear function of the log-likelihood. Moreover, we use a pairwise maximum entropy model \cite{moi1,moi2} to check that the variance of the log-likelihood and the variance of the overlap parameter reach their maximum at a value in line with the empirical ones. We compare empirical results to simulations of a multivariate GARCH process and a Monte Carlo Markov Chain. They corroborate the empirical findings. Last, we give an interpretation of criticality in financial markets. These findings can be important in a portfolio optimization which relies on the market structure (through the correlation matrix, for instance) and to figure out how market processes information which may eventually lead to a crash.

The chapter is organized as follows. In section \ref{sec6:crit}, the criticality is briefly presented. In section \ref{sec6:why}, we sketch the importance of criticality. In section \ref{sec6:recipe}, we give a practical recipe of the criticality test.  In section \ref{sec6:sig}, we recall the signatures of criticality. In section \ref{sec6:sampl}, we briefly discuss the sampling issues. In section \ref{sec6:res}, we present the empirical results. In section \ref{sec6:sim}, we present the outcomes of simulations.

\section{Criticality}\label{sec6:crit}
Criticality regroups phenomena occurring at a critical state which is a state delimiting the ordered and disordered phase, the order-disorder transition being continuous in our concern.  At the boundary between order and disorder, interesting phenomena occur such as power-law, increase of the correlation length, large fluctuations, slowing down, ergodicity breaking (some states can not be reached anymore). Strictly speaking, criticality only stands for infinite systems but for truly critical systems, the main features are qualitatively observed for finite sizes. We introduce some of these features through an example.

\begin{figure}[!ht]
\begin{center}
\includegraphics[width=\textwidth]{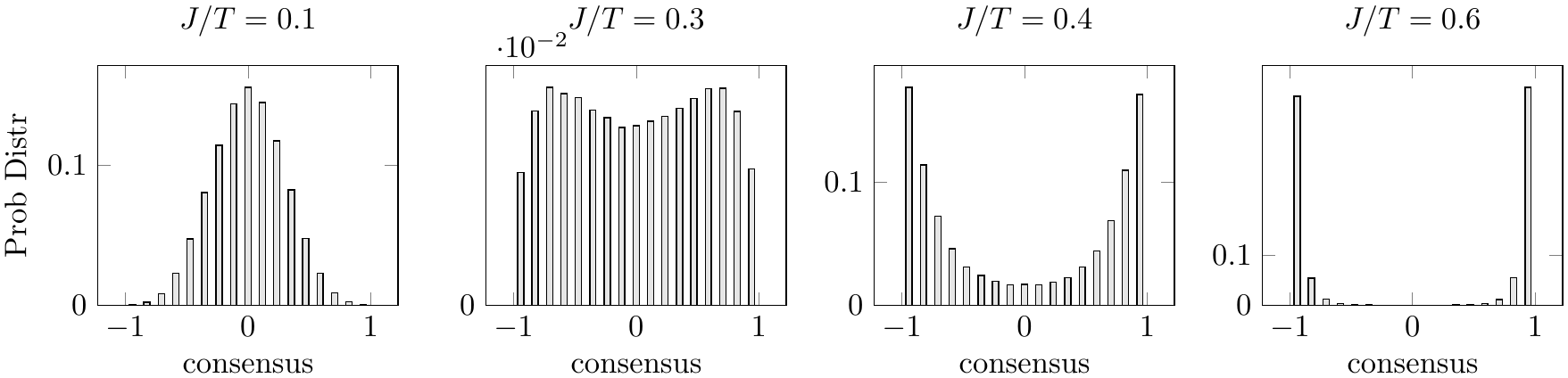}
\end{center}
\caption[Distributions of the net consensus for different interaction strengths]{\label{fig6:Erg} Distributions of the net consensus resulting from a variation of the social influence strength. From left to right: the social interaction strength increases, the system goes through a disordered state to an ordered state. Each distribution is estimated by a Monte Carlo Markov Chain ($1\times10^{4}$ equilibration steps and $1\times10^{6}$ recorded steps) for a system of size $N=16$. This example stands for the idealized city with nearest neighbours social interactions illustrated in Fig-\ref{fig:Lattice}.}
\end{figure}

We consider again an idealized city where each agent has exactly 4 neighbours as illustrated in Fig-\ref{fig:Lattice} which is related to the Schelling's model of segregation \cite{Stauff}. Each agent has to make a choice yes/no described by $s_{i}=\pm 1$. Lets take as utility function, the Brock and Durlauf's Social planner random utility \cite{Brock} $\mathcal{U}(\mathbf{s})+\epsilon(\mathbf{s})$ where $\mathcal{U}(\mathbf{s})=2^{-1}\sum_{ij}J_{ij}s_{i}s_{j}+\sum_{i}h_{i}s_{i}$ is the deterministic part and $\epsilon(\mathbf{s})$ the extreme-valued stochastic term. The stochasticity level can be handled by tuning a given parameter $T$ (thought as a common change of all the social influences $J_{ij}\rightarrow J_{ij}/T$). The resulting configuration distribution $P_{T}(\mathbf{s})$ is given by


\begin{equation}
P_{T}(\textbf{s})=\mathcal{Z}_{T}^{-1}\exp\left(\frac{1}{2T}\sum_{i, j}^{N}J_{ij}s_{i}s_{j}+\frac{1}{T}\sum_{i=1}^{N}h_{i}s_{i}\right)\equiv\frac {e^{ \mathcal{U}(\textbf{s})/T}}{\mathcal{Z}_{T}}\label{BDLagrange}
\end{equation}
which is equivalent to the pairwise maxent distribution (\ref{Lagrange}). As long as the system is finite, all the states can be reached and ergodicity is theoretically met but some issues may emerge following the kind of social influences \cite{Fischer}. As the variables describing the choices are bounded, interesting features arose from this model (multi equilibria, order-disorder transition, etc.). It is possible to show that this distribution is unimodal for the high levels of stochasticity and bimodal for the low levels of stochasticity as illustrated in Fig-{\ref{fig6:Erg}}.

The distribution goes continuously from unimodal to bimodal passing through a limiting case at a particular value of $T$ called the critical value of the scaling parameter $T_\mathrm{crit}$. At this value, the response functions to a shock in the idiosyncratic preferences (external inputs) $ \partial \langle s\rangle/\partial h$ and in the stochasticity level $-\partial\langle \mathcal{U} \rangle/\partial T$ reach their maximum values. For a Gibbs distribution, the response functions can be estimated by a Monte Carlo simulation using the relations

\begin{eqnarray}
  R_{m}(T)&=&\frac{\partial \langle s\rangle}{\partial h} = T^{-1} \VAR[\langle s\rangle] \\
  R_{\mathcal{U}}(T)&=&-\frac{\partial\langle \mathcal{U} \rangle}{\partial T}=T^{-2}\VAR[\mathcal{U}]
\end{eqnarray}

%

\begin{figure}[!ht]
\begin{center}
\includegraphics[width=\textwidth]{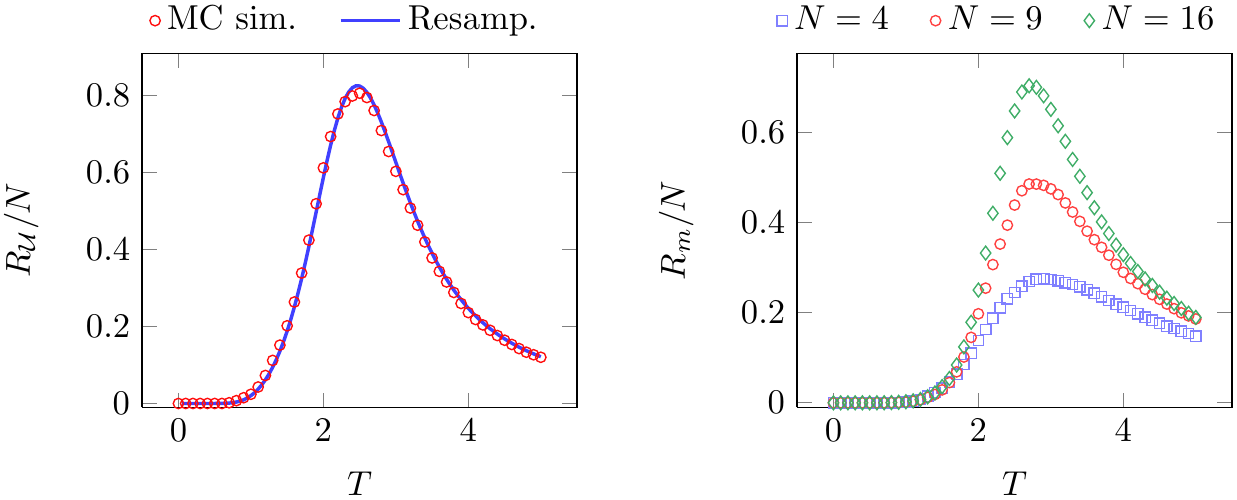}
\end{center}
\caption[The variances of the orientation and of the utility as a function of $T$]{\label{fig6:ResampMC} The response functions as a function of the scaling parameter. Left panel: the response to a shock in the stochasticity level obtained by a Monte Carlo simulation (circles) and by exact re-sampling (full line) for a system of size $N=16$. Right panel: the response function to a shock in idiosyncratic preference for different system sizes. }
\end{figure}

If the number of entities is not too large, one can perform an exact re-sampling of the probability distribution instead of simulations. For a Gibbs distribution $P(\mathbf{s})\propto \exp(\mathcal{U}(\mathbf{s}))$, the re-sampling is done by introducing the scaling parameter $T$ such that $P(\mathbf{s})\rightarrow P_{T}(\mathbf{s})\propto \exp(T^{-1}\mathcal{U}(\mathbf{s}))$. Both methods are illustrated in the left panel of Fig-\ref{fig6:ResampMC}. The maximal value of the response functions scales as the system size. The finite size scaling is illustrated in the right panel of Fig-\ref{fig6:ResampMC}.

\begin{figure}[!ht]
\begin{center}
\includegraphics[width=0.5\textwidth]{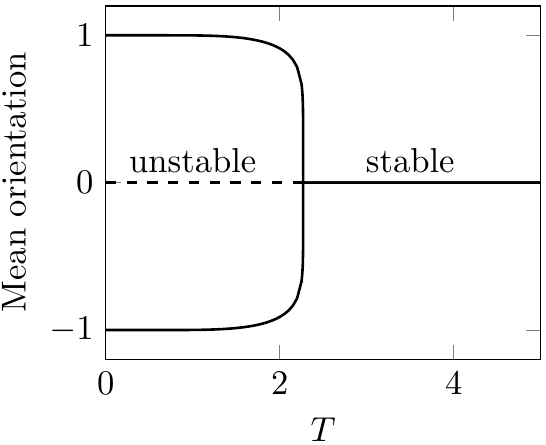}
\end{center}
\caption[The mean orientation as a function of the scaling parameter]{\label{fig6:IsingM} The mean orientation as a function of the scaling parameter. The stable solutions are illustrated by the full line. The stability is determined by the Hessian matrix of $\mathcal{F}=-T\ln \mathcal{Z}$ with respect to $\langle s_{i} \rangle$.}
\end{figure}

For the limiting case of infinite size $N\rightarrow\infty$, the response functions have an asymptote at a particular value of the scaling parameter $T_{\mathrm{crit}}=2/\ln(1+\sqrt{2})$ ($J_{ij}=1$ for the four nearest neighbours) and the mean orientation goes continuously from zero to a non-zero value as illustrated in Fig-\ref{fig6:IsingM}.

As we saw in Chap-\ref{chap:theory}, the stability of a solution is determined by the Hessian matrix $(\textbf{H}(\mathcal{F}))_{ij}=\partial^{2} \mathcal{F}/\partial q_{i}\partial q_{j}$ (where $\mathcal{F}=-T\ln \mathcal{Z}$ for the scaled distribution and $q_{i}=\langle s_{i} \rangle$) which is the inverse of the covariance matrix and should therefore be a positive semidefinite matrix.

\section{Why criticality is important}\label{sec6:why}

\begin{figure}[!ht]
\begin{center}
\includegraphics[width=0.5\textwidth]{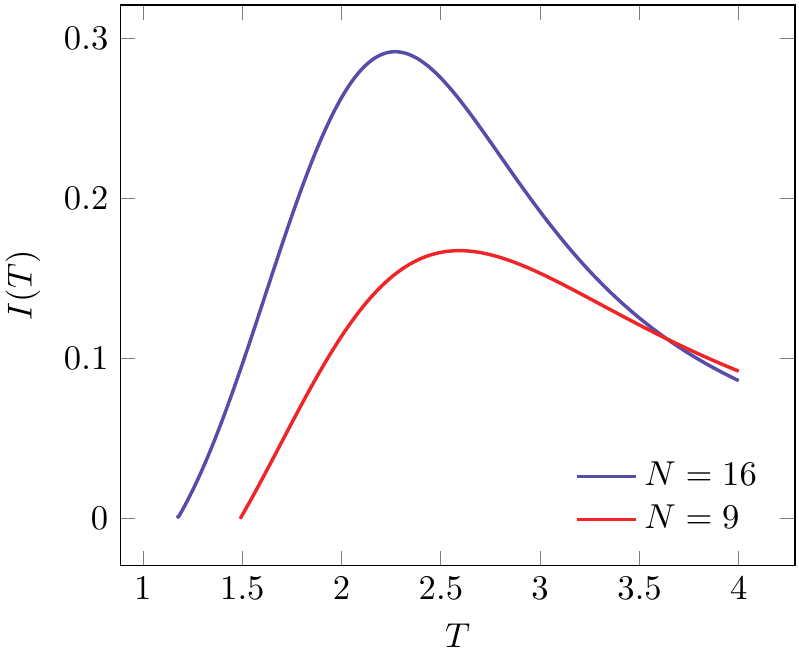}
\end{center}
\caption[Multi-information of an idealized city]{\label{fig6:IsingMI}The multi-information $I(T)$ as a function of the scaling parameter for the idealized city of sizes $N=9$ and N=$16$.}
\end{figure}

Criticality is a very special feature in several ways. As we saw, the response functions reach their maximum values which implies high reactivity and a global impact on the underlying network. It also implies a great structural malleability since the deviations to the mean likelihood (or equivalently, of the entropy $-\mathrm{E}[\ln p(\mathbf{s})]$) are the largest. An event can potentially affect the whole network. It has been shown that at criticality pairwise maxent models are prone to undertake avalanches triggered by external outputs \cite{Perkovic} (informally speaking, this feature is due to the suitable balance between fluctuations and co-movements). This idea is attractive  because its is exactly how the neural networks seems to process information \cite{Shew}.
Qualitatively, one can understand \emph{communication} through the network as follows. If the stochasticity level is to high, the noise is the dominant part and a message can not be passed from hub to hub. If the level of stochasticity is too low, the coupling is strong but the state of each hub varies slowly. A proper balance between coordination and fluctuation is met at criticality.
Formally, one can show that the multi-information peaks at the critical point for the former model of idealized city \cite{Erb}. This feature is illustrated in Fig-\ref{fig6:IsingMI}. A market close to the criticality is a market able to process information efficiently, quickly modify its structure but is also a system prone to crash.

Formally, it is also a state where the law of large numbers and the central limit theorem break down (the rate function of the large deviation principle becomes non-convex) \cite{Ellis}.

\section{Practical recipe}\label{sec6:recipe}

Before going into further detail, we give the practical recipe for testing the criticality of a complex system. All quantities will be defined in text.

\begin{enumerate}
\item Binarize the returns.
\item Test the statistical significance and determine the corresponding maximum size $N$.
      \begin{enumerate}
         \item Compute the empirical distribution of configurations $\hat{p}_{\mathbf{s}}$.
         \item Compute $m_{k}$ (the number of configurations sampled exactly $k$ times) and the empirical distribution of $K_{i}$ (the number of times the configuration $\mathbf{s}_{i}$ is observed in the sample).
         \item Deduce their entropies $H[\mathbf{s}]$ and $H[K]$.
         \item Locate the maximum of the relation  $H[\mathbf{s}]$ vs $H[K]$.
      \end{enumerate}
\item Get the response function $R_{\mathcal{U}}$ and find its maximum. Repeat for several ($\sim 100$) sets of $N$ randomly chosen entities. For each set:
      \begin{enumerate}
         \item Compute the empirical distribution of configurations $P(\mathbf{s})$.
         \item Rescale the empirical distribution as $P_{T}(\mathbf{s})=\frac{P(\mathbf{s})^{1/T}}{\sum_{\{\mathbf{s}\}}P(\mathbf{s})^{1/T}}$.
         \item Compute $R_{\mathcal{U}}=T \frac{\partial \mathcal{S}}{\partial T}$ where $\mathcal{S}(T)=-\sum_{\{\mathbf{s}\}}P_{T}(\mathbf{s})\ln P_{T}(\mathbf{s})$.
         \item Store the coordinates of its maximum.
      \end{enumerate}
\item Compare to a finite size version of a truly critical system.
       \begin{enumerate}
         \item Compute the relative difference $x=(T_{\mathrm{op}}-T_{\mathrm{max}})/T_{\mathrm{op}}$ where $T_{\mathrm{op}}=1$.
         \item Compute the Kullback-Leibler divergence (KLD) between $P_{T=T_{\mathrm{max}}}(\mathbf{s})$ and $P_{\mathrm{emp}}(\mathbf{s})$.
         \item Compare the obtained KLD value to the KLD between $P_{T=T_{\mathrm{crit}}}(\mathbf{s})$ and $P_{T=(1+x)T_{\mathrm{crit}}}(\mathbf{s})$ for the 2D nearest neighbours Ising model.
       \end{enumerate}
\item Perform a statistical test of Zipf's law as described in section \ref{sec3:DPL}.
\item Check the linearity of the relation $\mathcal{S}(\mathcal{U})$ vs $\mathcal{U}$ where $\mathcal{U}=\ln P(\mathbf{s})$.
\item Compare the empirical results to simulations.

   \begin{enumerate}
         \item Infer the Lagrange parameters (see section \ref{sec3:rpml}).
         \item Simulate data using a Monte Carlo Markov chain (see section \ref{sec3:MC}).
         \item Check if an order-disorder transition is allowed by computing the orientation distribution and by varying the scaling parameter $T$.
         \item Compute the variance of the log-likelihood and of the overlap parameter $q=N^{-1}\sum_{i}s_{i}^{(1)}s_{i}^{(2)}$ (two copies denoted by the superscript, linked with the covariance of the utility function $\mathcal{U}$). Compare empirical and simulated results for a common size. A large difference ($>10\%$) between the simulated value of $T_{\mathrm{max}}$ and the asymptotical value returned by fitting the empirical relation $T_{\mathrm{max}}(N)$ may reveal difficulties in the inference of Lagrange parameters \cite{Mastro} and therefore a poor fitting.
       \end{enumerate}
\end{enumerate}

\section{Signatures of criticality}\label{sec6:sig}
A critical state can be thought as a state where the system lies at the threshold between order and disorder. If there is no uncertainty, markets are perfectly ordered and thus homogeneous (either positive or negative). In the opposite situation where uncertainty is maximal, markets are completely random and uncorrelated; the probability to observe a positive or negative return is equal to $1/2$ whatever the returns (positive or negative) of other market exchanges. A critical state is halfway between these extreme states, letting markets on the edge of disorder and highly heterogenous.

Strictly, a critical state can be achieved only for infinite size systems. For finite systems, one will not observe divergences but we will still say \emph{critical} through abuse of language and we should compare the empirical results to a finite version of a model which  may actually reach the criticality (the nearest neighbour Ising model in two dimensions for instance) as proposed in \cite{Step}.


Statistical physics provides several tests of criticality. The signatures detailed in \cite{Mora} will be briefly recalled. First, we define a \emph{financial system} as a set of stocks (or indices). Relative stock returns of the $ith$ asset at period $t$ is taken as a random variable $r_{i,t}$ and can be rewritten as $r_{i,t}=\sgn(r_{i,t})|r_{i,t}|$. Signs of stock returns are sometimes considered as uncorrelated and attract less attention. However correlations may appear in complicated (non-linear) fashion as synchronization during crises \cite{Jr}. It is interesting to study orientation (sign of returns) changes since Ising-like models are suited to describe collective behaviours. Moreover the nature of the relative return sign is more subtle than the one of simple independent random variables and can render the particular structure of financial markets \cite{moi1,moi2}. The net orientation is defined as $m(t)=N^{-1}\sum_{i}s_{i,t}$, if $m(t)>0$ the market is bullish for the period $t$.
In order to study orientation changes, we consider a set of $N$ market indices or $N$ stocks described by binary variables $s_{i,t}\equiv \sgn(r_{i,t})$ ($s_{i,t}=\pm1$ for all $i=1,\cdots,N$ and for all periods $t=1,\cdots,M$). A system configuration will be described by a vector $\textbf{s}_{t}=(s_{1,t},\cdots,s_{N,t})$. The binary variable will be equal to one if the associated stock is bullish and equal to $-1$ if not. A configuration $(s_{1,t},\cdots,s_{N,t})$ may also be thought as a binary version of the returns.

One can formally write the probability $P(\textbf{s})$ of finding the system in state $\textbf{s}$ as a Gibbs distribution $P(\mathbf{s})=\mathcal{Z}^{-1}e^{ \mathcal{U}(\mathbf{s})}$ and without loss of generality set $\mathcal{Z}$ to $1$ which leads to the definition of the utility function (or \emph{energy} $\mathcal{H}=-\mathcal{U}$, \emph{potential}, etc.) as the log-likelihood: $\mathcal{U}(\textbf{s})=\ln P(\textbf{s})$. The rank $r(\mathrm{\textbf{s}})$ of a given configuration $\textbf{s}$ is defined as the number of configurations with a higher utility (more frequent) than the value associated to $\textbf{s}$.

\begin{figure}[ht!]
\begin{center}
\includegraphics[width=0.65\textwidth]{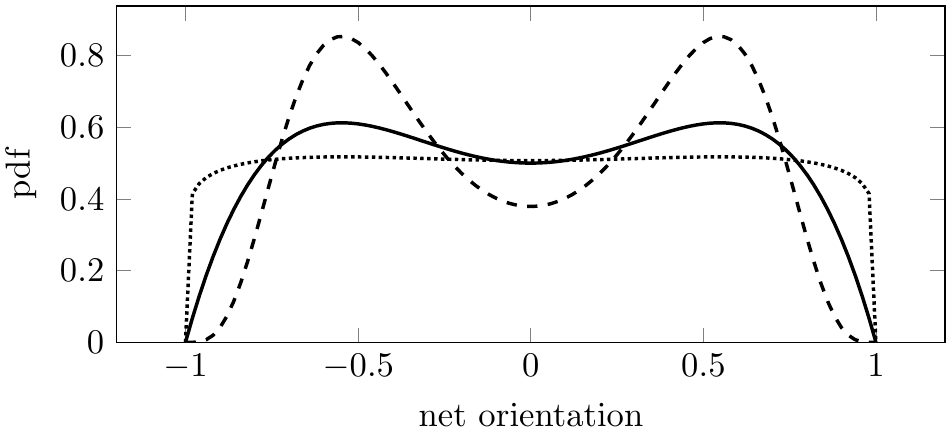}
\end{center}
\caption[Schematic illustration of the pdf rescaling]{Schematic illustration of the rescaling of a bimodal distribution as encountered in the Landau phenomenological theory of phase transition. The original probability density is illustrated by the full line at an arbitrary temperature $T^{*}$, the rescaled distributions are illustrated by the dashed line (at $T=0.25T^{*}$) and the dotted line (at $T=10T^{*}$).}
\label{fig:flat}
\end{figure}

A power-law $-\ln P(\textbf{s})=\alpha \ln r(\mathrm{\textbf{s}})+\mathrm{Cst}$ is a strong signature of criticality. In this framework, it is possible to obtain these quantities directly from a large enough sample and test the validity of Zipf's law. Another consequence of this law is the linearity of the Shannon entropy \cite{Step}, which measures the average \emph{surprise} or average \emph{log-likelihood}, expressed in term of an utility function \cite{Mora}. A weaker signature is the divergence of the variance of the likelihood at the operating point (in the limit of infinite number of entities). For finite systems, the variance of the likelihood should peak near the operating point if the system is in a critical state. This feature can also be checked directly from the data. The empirical relative frequencies are scaled as $P_{T}(\textbf{s})=P(\textbf{s})^{1/T}/\sum P(\textbf{s})^{1/T}$, the operating point corresponds to $T=1$. Noting that for such a Gibbs distribution we have the identity

\begin{equation}\label{FD}
  R_{\mathcal{U}}=-\frac{\partial\langle \mathcal{U} \rangle}{\partial T}=T^{-2}\left[\langle \mathcal{U}^{2}\rangle-\langle \mathcal{U}\rangle^{2}\right]
=T\frac{\partial \mathcal{S}}{\partial T}
\end{equation}
where $\mathcal{S}(T)$ is the Shannon entropy $-\sum_{\{\mathbf{s}\}}P_{T}(\textbf{s})\ln P_{T}(\textbf{s})$ of the rescaled distribution and the brackets stand for the average with respect to $P_{T}(\textbf{s})$.
In a statistical point of view, this extremum is the point where the deviation to equiprobability of events is the largest. Operating at this point involves that the variance of the log-likelihood reaches its largest value whereas for equiprobable events, the variance of the log-likelihood is equal to zero. A large variance of the log-likelihood also implies a large deviation from its mean value, the entropy, and thus large structural changes. The rescaling parameter $T$ can be thought as a randomness measure, changing this parameter leads to a reweighting of the empirical distribution. For $T>1$, the distribution will be flattened and closer to the uniform distribution as illustrated in Fig-\ref{fig:flat}. The entropy of the remaining distribution will thus be larger than the original one. The closer to the uniform distribution, the larger the entropy. We note that the expression $T\partial\mathcal{S}/\partial T$ is useful when direct sampling of probability distribution is feasible and the expression $T^{-2}\left[\langle \mathcal{U}^{2}\rangle-\langle \mathcal{U}\rangle^{2}\right]$ allows estimation through a Monte Carlo simulation even if direct sampling is unfeasible. When direct sampling is feasible, one can estimate the empirical distribution as $P(\mathbf{s})=M^{-1}\sum_{i=1}^{M}\delta_{\mathbf{s}_{t},\mathbf{s}}$ where $M$ is the sample length, compute the scaled distribution $P_{T}(\textbf{s})$ for any value of $T$ and then use the relation  $T\partial\mathcal{S}/\partial T$ for the empirical derivation of the response function.

\section{Sampling indices and stock exchanges}\label{sec6:sampl}
\begin{figure}[ht!]
\begin{center}
\includegraphics[width=\textwidth]{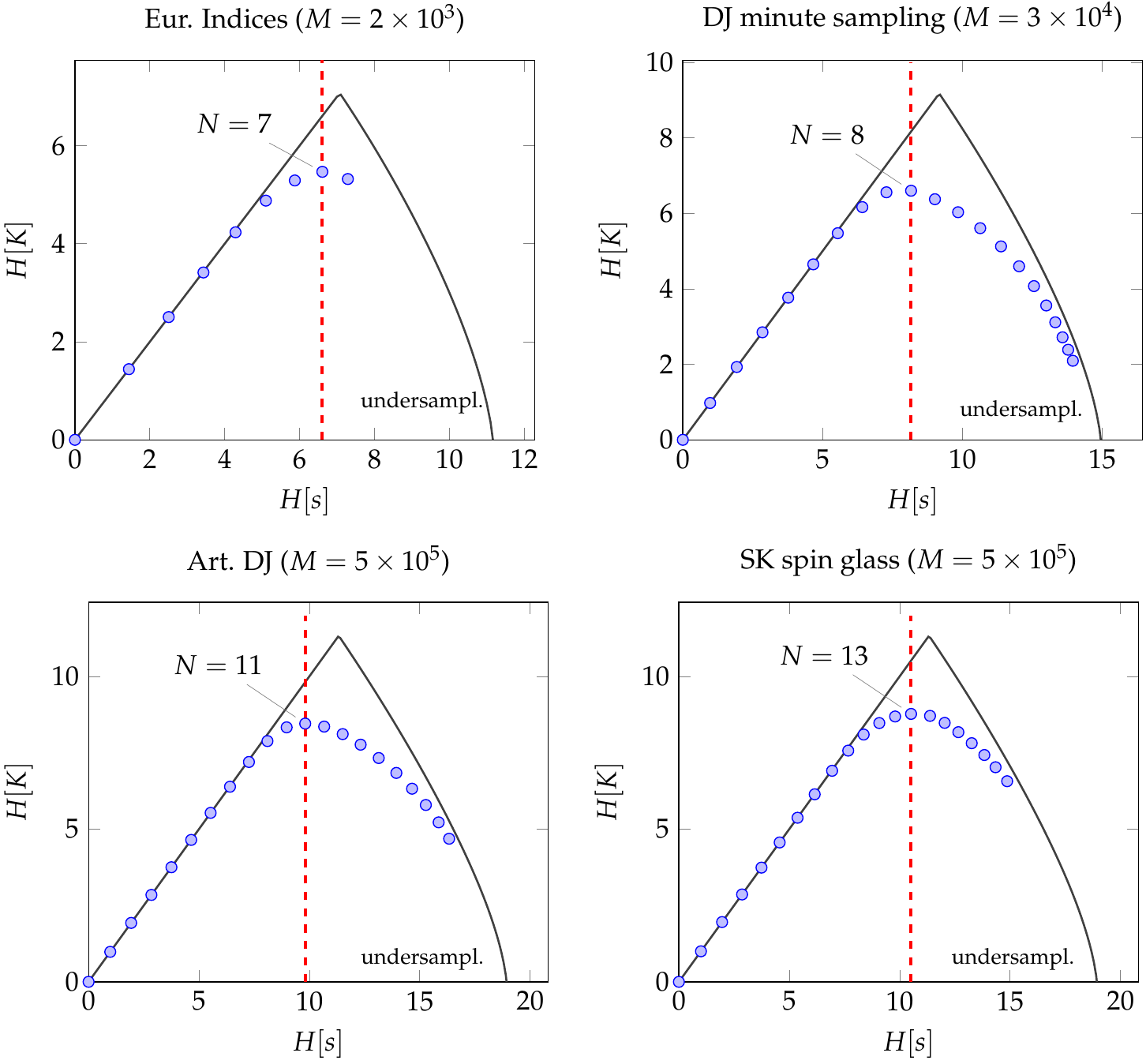}
\end{center}
\caption[Statistical significance of data sets]{Statistical significance of data sets. The configurations distribution $P[\mathbf{s}]$ is correctly sampled in the left part of the plane $\{H[K],H[\mathbf{s}]\}$, delimited by the dashed line. The full line stands for theoretical relation, $H[K]$ as a function of $H[\mathbf{s}]$. The dots stand for empirical values for each data set, as the system size increases (from left to right in the plane $\{H[K],H[\mathbf{s}]\}$). The right-bottom panel illustrates the results for a SK spin glass of size $N=25$ near the criticality.}
\label{fig:StatSig}
\end{figure}

We observed opening and closing prices of 8 European indices (AEX, BEL, CAC, DAX, EUROSTOXX, FTSE, IBEX, MIB), the sample length $M$ is 2300 trading days (approximatively nine trading years englobing two global crises, 2002-2011 period). We consider European stock exchanges because some issues (debt crisis, etc.) are specific to these market places and to ensure the simultaneity of time series. We also observed the stocks of the Dow Jones index during $3\times 10^{4}$ trading minutes and at daily sampling from 2002 to 2011. We consider two different time-scales to explore the differences when the correlations decrease. According to the Epps effect \cite{Epps}, we expect that systems sampled at low frequency (daily sampling) should be closer to the criticality than the systems sampled at larger frequencies (minute sampling, for instance). Positive returns are set to $1$ and negative returns to $-1$. The first sample is ten times larger than the number of possible configurations. Indeed, there are two possible values for each variable $s_{i}$, thus they are $2^{N=8}=256$ configurations. The second sample is not large enough for a satisfactory probability estimation (and thus a direct estimation of entropy).
Since entities may be strongly correlated, it is not obvious to know if the configurations are well sampled or not. In case of strongly correlated entities, the relevant region in the configurations space is narrow in comparison to independent entities. If the true configurations distribution is sharply peaked, there are only few relevant states. In this situation, a small ($M<2^N$) sample is enough to sample properly the configurations distribution. In the opposite case where entities are independent, every configuration has the same statistical weight and the sample size must be large ($M\gg 2^N$).
It is crucial to identify the maximum number entities one should consider to avoid undersampling of the configurations distribution $P(\mathbf{s})$ because power-laws occur spontaneously in the undersampling regime \cite{MMR}. In particular, Zipf's law is only a genuine feature if $P(\mathbf{s})$ is well sampled. To asses the maximum number of entities to consider in the analysis, we follow the procedure described in \cite{MMR}.
The limit between proper sampling and undersampling is defined by the coordinates of the maximum of $H[K]$ in the plane $\{H[K],H[\mathbf{s}]\}$ where $H[\mathbf{s}]$ is the entropy of the empirical configurations frequencies and $H[K]$ is the entropy of the random variable $K_{t}=K(\mathbf{s}_{t})$ which is the number of times the configuration $\mathbf{s}_{i}$ is observed in the sample. Beyond this point, $H[K]$ decreases when $H[\mathbf{s}]$ increases which means that configurations are sampled (approximatively) the same number of times.
Briefly, given a sample of $M$ independent configurations $(\mathbf{s}_{1},\cdots,\mathbf{s}_{M})$, the empirical distribution of the configurations is $\hat{p}_{\mathbf{s}}\equiv P(\mathbf{s}_{t}=\mathbf{s})=M^{-1}\sum_{t=1}^{M}\delta_{\mathbf{s}_{t},\mathbf{s}}$. The distribution of the random variable $K_{t}$, corresponding to the number of times the configuration $\mathbf{s}_{t}$ occurs in the sample, is written $P(K_{t}=k)=k\,m_{k}/M$ where $m_{k}=\sum_{\{\mathbf{s}\}}\delta_{k,M\hat{p}_{\mathbf{s}}}$ is the number of configurations that are sampled exactly $k$ times. Their entropies are
\begin{eqnarray}
  H[\mathbf{s}]&=& -\sum_{\{\mathbf{s}\}}\hat{p}_{\mathbf{s}}\ln \hat{p}_{\mathbf{s}}=-\sum_{k}\frac{k\,m_{k}}{M}\ln\frac{k}{M}\\
  H[K]&=& -\sum_{k}\frac{km_{k}}{M}\ln\frac{k\,m_{k}}{M}=H[\mathbf{s}]-\sum_{k}\frac{k\,m_{k}}{M}\ln m_{k}
\end{eqnarray}
These quantities can be evaluated to obtain the statistical significance of each data set. The points in Fig-\ref{fig:StatSig} have been obtained by considering increasing system size. Each point is obtained by this mean and by averaging over several sets of randomly chosen entities (see the caption). Moreover, the theoretical limit is given by the most informative samples (full lines in Fig-\ref{fig:StatSig}) which are those maximizing $H[K]$ with respect to $\{m_{k},k>0\}$ and satisfying the constraints $H[\mathbf{s}]\leq N$, $\sum_{k}k\,m_{k}=M$ and $H[K]\leq H[\mathbf{s}]$ since the random variable $K$ is a function of $\mathbf{s}$ (see \cite{MMR} for the complete discussion and derivation).

The statistical significance is illustrated in Fig-\ref{fig:StatSig} for each data set and for artificial data simulated by fitting a pairwise maximum entropy model (see hereafter). We simulate also a time series of a Sherrington-Kirkpatrick (SK) spin glass of size $N=25$ near the criticality.

The European indices set is correctly sampled up to 7 indices, the Dow Jones at minute up to 8 stocks. Increasing 15 times the sample length $M$, allows to consider up to $N=11$ entities.

A qualitative observation is that if entities are highly correlated (low stochasticity), almost all observed configurations (words) should be such that the mean orientation $m(t)=N^{-1}\sum_{i}s_{i,t}$ is non zero. One expects a $H[\mathbf{s}]$ significantly lower than the theoretical upper bound $N$ and $H[K]\simeq H[\mathbf{s}]$ since few different configurations are observed. On the other hand, nearly independent entities do not favour any value of $m(t)$. The configuration distribution $P[\mathbf{s}]$ should be approximatively uniform, $H[\mathbf{s}]$ should be close to $\mathrm{min}(N,\log_{2} M)$ for large system sizes and $H[K]$ should be small since each configuration is observed approximatively a same number of times.
From pairwise maximum entropy models \cite{Fischer}, one knows that criticality is a regime where no net orientation is observed but where fluctuations are the largest. We expect that the sampling of a truly critical regime should return a situation halfway between the two previous extreme cases, as illustrated in Fig-\ref{fig:StatSig} for the SK-model.

\begin{figure}[ht!]
\begin{center}
\includegraphics[width=0.75\textwidth]{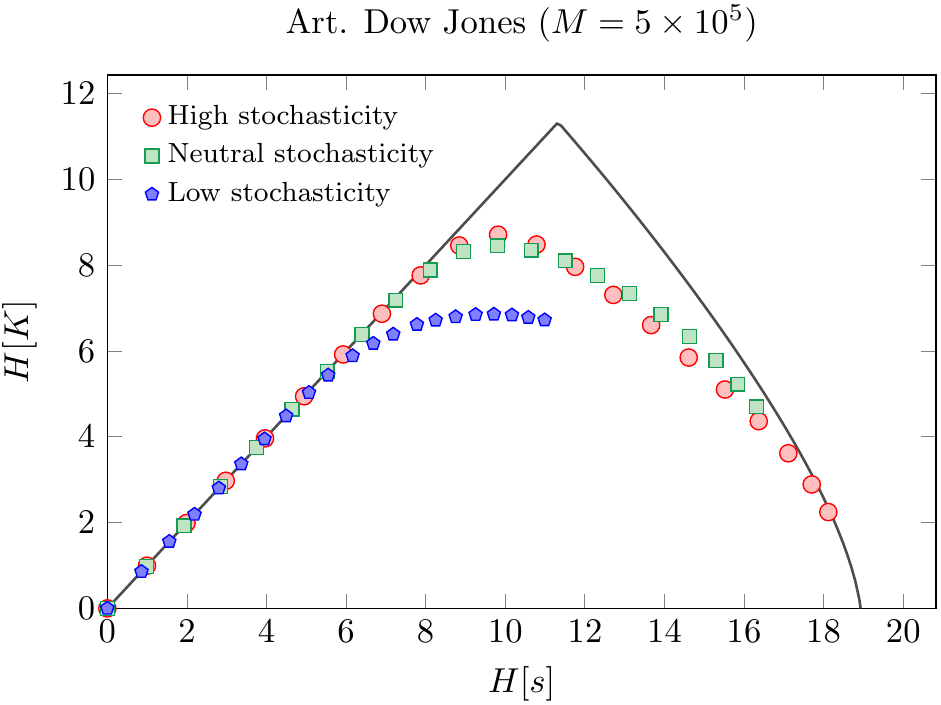}
\end{center}
\caption[Statistical significance of the Dow Jones data set]{Statistical significance of the Dow Jones data set for different levels of stochasticity. The full line stands for theoretical relation, $H[K]$ as a function of $H[\mathbf{s}]$. The dots stand for artificial data generated with a pairwise maxent model fitted on the Dow Jones data with 1 third larger stochasticity than actual one. The squares illustrate artificial data with the same level of stochasticity than the actual one and the pentagons illustrate data generated with 1 third lower stochasticity.}
\label{fig:StatSigDJ}
\end{figure}

After fitting a pairwise maxent model, we record artificial data for the Dow Jones varying the stochasticity by 1 third smaller and larger than the actual one. The results are illustrated in Fig-\ref{fig:StatSigDJ}.
It seems that the Dow Jones (minute sampling) is rather disordered, we will check this in detail hereafter.

\section{Results}\label{sec6:res}
\begin{figure}[ht!]
\begin{center}
\includegraphics[width=\textwidth]{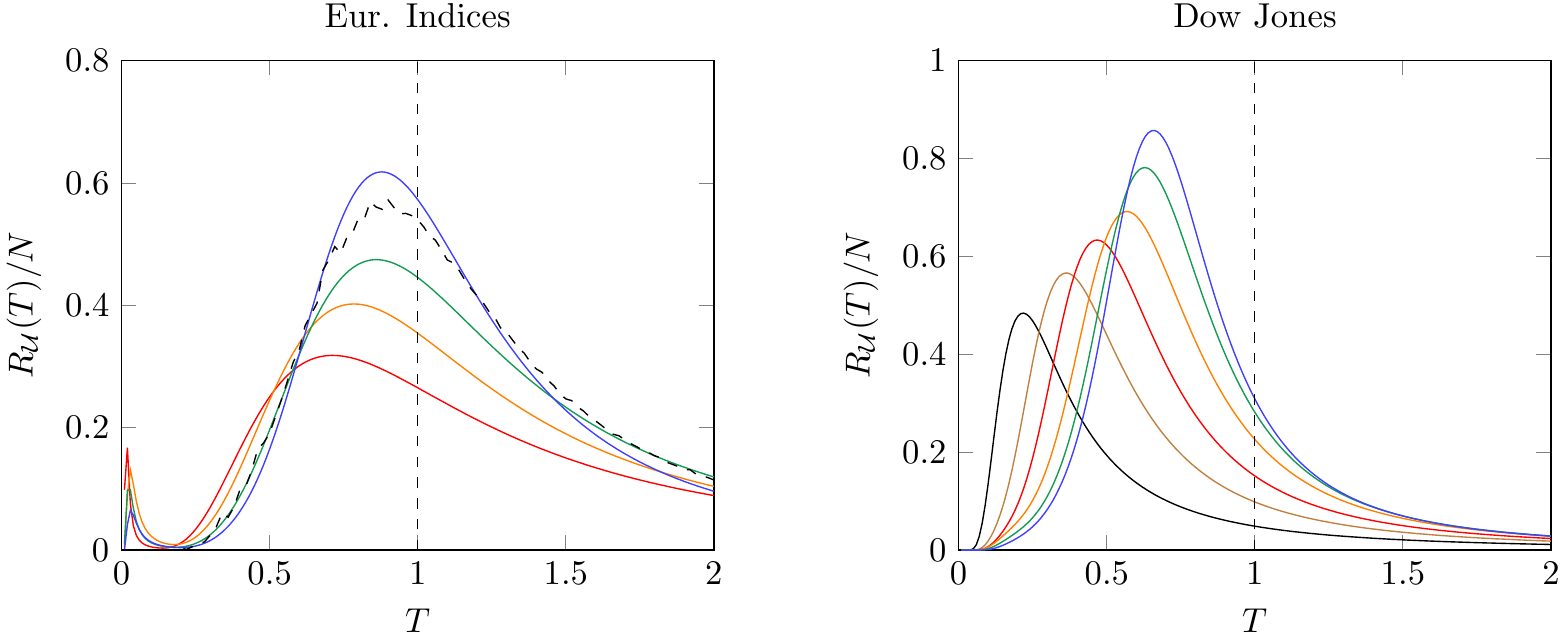}
\end{center}
\caption[Variance of the log-likelihood]{Variance of the log-likelihood for the European indices set (left) and for the Dow Jones at minute sampling (right) vs the rescaling parameter. The peak moves from left to right when we consider larger sets. For the European set, we plot the variance for sizes $N=$2,4,5,8. The dashed curve is a Monte Carlo simulation (see hereafter) for $N=8$. For the Dow Jones, we consider $N=2,\,4,\,6,\,8,\,10,\,12$, the last two values are not statistically significant. These curves have been obtained by direct sampling of the probability (and entropy) and by using the relation $T\partial\mathcal{S}/\partial T$.}
\label{fig:hc}
\end{figure}

In the following, we check if the signatures of criticality are observed in the considered data sets. The variance of the log-likelihood is illustrated in Fig-\ref{fig:hc}.


We can observe that the peak position scales with the system size, moving from left to right towards the operating point $T=1$ and that the maximum value of the variance becomes larger when the number of entities increases.

\begin{figure}[ht!]
\begin{center}
\includegraphics[width=\textwidth]{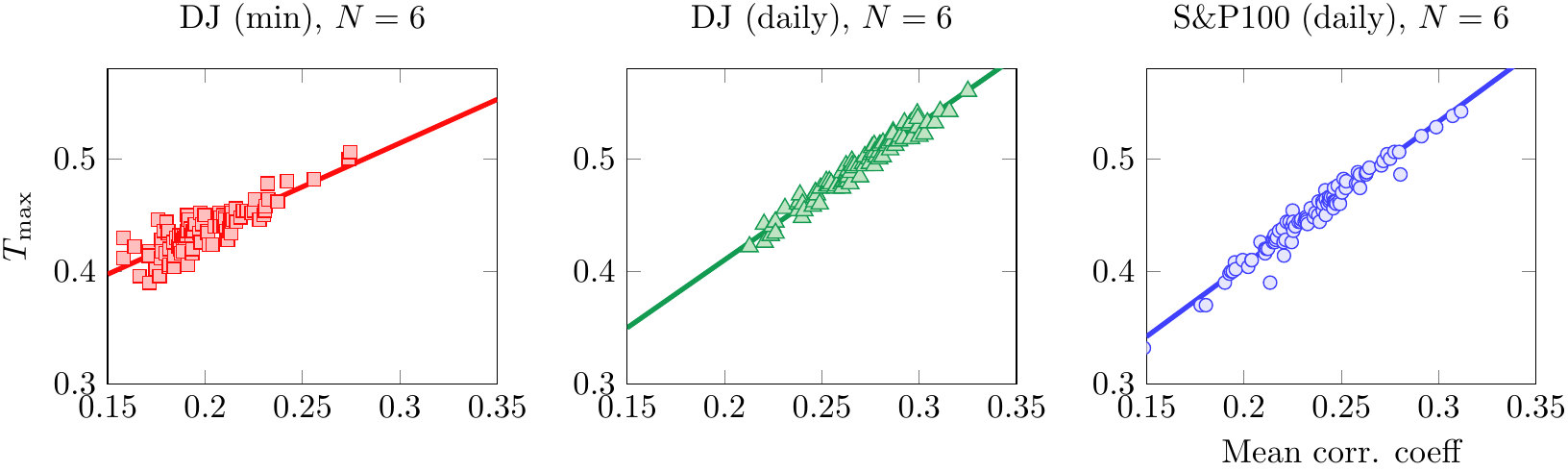}
\end{center}
\caption[The critical scaling parameter vs correlations]{The critical scaling parameter (x-axis coordinate of the maximum of the response function $R_{\mathcal{U}}$) versus the mean correlation coefficient of the considered set of $N=6$ randomly chosen entities. The results are illustrated for the Dow Jones at minute (squares, left panel) and daily samplings (triangles, center panel) and for the S$\&$P100 (circles, right panel). The size $N=6$ is chosen consistently with the latter analysis of statistical significance.}
\label{fig:TcriCorr}
\end{figure}

For a given and fixed size, one expects a larger value of the critical scaling parameter for sets (of $N$ randomly chosen entities) with a larger mean correlation coefficient. We consider 100 sets of $N=6$ randomly chosen entities for the Dow Jones (daily and minute samplings) and for the S$\&$P100. The results illustrated in Fig-\ref{fig:TcriCorr} suggest a roughly linear relation between the critical scaling parameter $T_{\mathrm{max}}$ and the mean correlation coefficient. Any further results will thus be averaged over several sets for each considered size.

\begin{figure}[ht!]
\begin{center}
\includegraphics[width=0.5\textwidth]{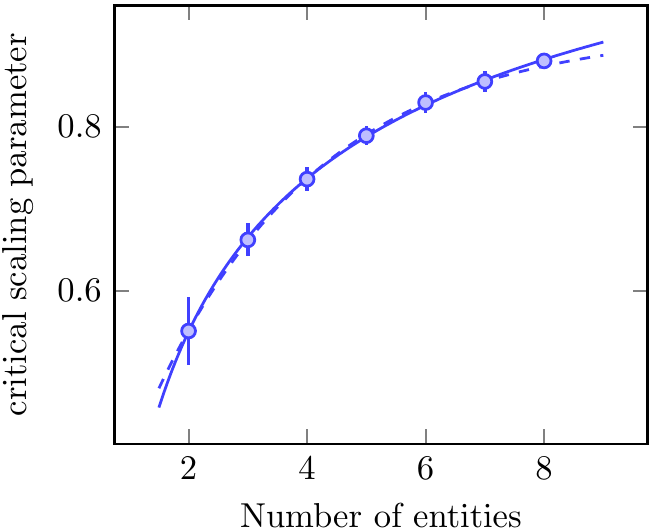}
\end{center}
\caption[Value of the critical scaling parameter (indices)]{Value of the scaling parameter at which response function $R_{\mathcal{U}}$ reaches its maximum value vs the number of entities $N$. Mean values $\bar{T}$ and error bars (1 standard deviation on $\bar{T}$) are computed over $\binom{8}{N}$ samples for European indices set. The full line stands for a power fit and the dashed line stands for an exponential fit on the 7 first values.}
\label{fig:Tcri}
\end{figure}

\begin{figure}[ht!]
\begin{center}
\includegraphics[width=0.5\textwidth]{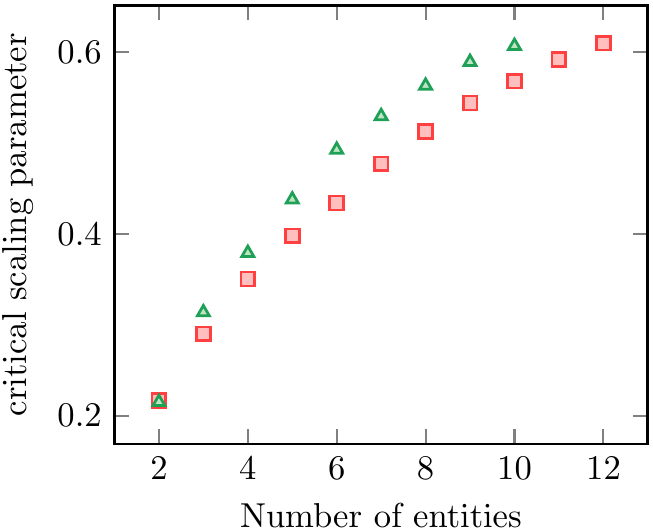}
\end{center}
\caption[Value of the critical scaling parameter (Dow Jones)]{Value of the scaling parameter at which response function $R_{\mathcal{U}}$ reaches its maximum value vs the number of entities $N$. Mean values are computed over 100 sets of $N$ randomly chosen stocks for the Dow Jones at daily (triangles) and minute (squares) samplings.}
\label{fig:Tcri2}
\end{figure}

To formalize the relation $T_{\mathrm{max}}=T_{\mathrm{max}}(N)$, we compute the value of the scaling parameter at which response function $R_{\mathcal{U}}$ reaches its maximum value for different sets of $N$ randomly chosen entities. Results are illustrated in Fig-\ref{fig:Tcri} for the European indices set and in Fig-\ref{fig:Tcri2} for the Dow Jones (daily and minute samplings).

The power and exponential fits return an asymptotic critical scaling parameter respectively equal to $1.38$ and $0.92$.

An exponential fit, on size up to $N=8$, of the DJ (min) returns an asymptotical critical parameter equal to $0.70$ and equal to $0.74$ if we fit up to $N=12$ (but the latter value is not trustful since the system is undersampled for $N>8$).
An exponential fit, on size up to $N=6$,  of the DJ (daily) returns an asymptotical critical parameter equal to $0.71$ and equal to $0.72$ if we fit up to $N=10$, (but the latter value is not trustful since the system is undersampled for $N>6$). Furthermore, even in the undersampled regime, we observe an increase of the critical scaling parameter.

\begin{figure}[ht!]
\begin{center}
\includegraphics[width=0.5\textwidth]{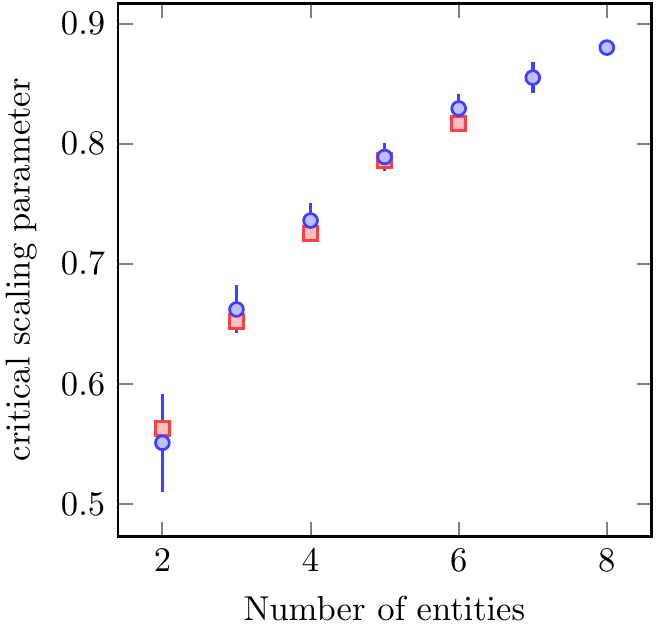}
\end{center}
\caption[Value of the critical scaling parameter (different time-windows)]{Value of the scaling parameter at which the response function $R_{\mathcal{U}}$ reaches its maximum value vs the number of entities $N$. Mean values and error bars (1 standard deviation) are computed over $\binom{8}{N}$ samples for European indices set (circles). The squares illustrate results for the same sets with scaled sample length $L(N)=2^{N+3}$ and averaged over 5 different time-windows.}
\label{fig:TcriScale}
\end{figure}

Larger correlations measured when size ($N$) increases may be a spurious effect due to the consideration of a particular time interval. One can perform to same study by changing size and scaling sample length simultaneously and considering different time-windows. For the set of European indices, we chose sample length $L(N)=2^{N+3}$ such that  $L(8)\simeq L_{\mathrm{max}}=2300$ and we average the results on 5 different time-windows. Results are illustrated in Fig-\ref{fig:TcriScale}. Each point (square) falls into the confidence interval of the constant size results excepted the last one ($N=6$). Larger correlations for increasing size is thus a genuine feature.

\begin{figure}[ht!]
\begin{center}
\includegraphics[width=0.5\textwidth]{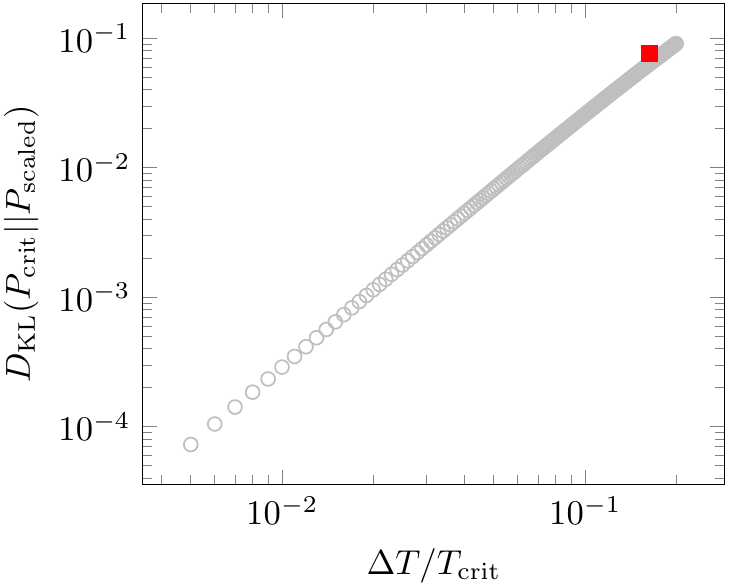}
\end{center}
\caption[KLD between the critical and the scaled distributions]{Kullback-Leibler divergence between the critical and the scaled distributions for the two dimensional square lattice nearest-neighbours Ising model $N=9$ (light grey circles) and for the set of 8 European indices (square).}
\label{fig6:DKL}
\end{figure}

As no inference method have been used, we expect that the Kullback-Leibler divergence (KLD) $D_{\mathrm{KL}}(P_\mathrm{crit}||P_\mathrm{emp})$ between the critical distribution $P[T=T_{\mathrm{max}}]$ (such that the maximum value of $R_{\mathcal{U}}$ is reached at $T_{\mathrm{max}}$) and the empirical distribution $P_\mathrm{emp}$ should be of the same order of magnitude  than for a truly critical system operating at $T_{\mathrm{crit}}+\Delta T$. The relative deviation $\Delta T/T_{\mathrm{crit}}$ and $(T_{\mathrm{op}}-T_{\mathrm{max}})/T_{\mathrm{max}}$ being equal (by definition $T_{\mathrm{op}}=1$). Following \cite{Step}, a reasonable benchmark is the two dimensional square lattice nearest-neighbours Ising model with periodic boundaries of size $N=9$. The response function $R_{\mathcal{U}}$ reaches its maximum value at $T_{\mathrm{crit}}=2.40$. We compute the exact distribution $P_\mathrm{crit}$ and the KLD with the scaled distribution $P_\mathrm{scaled}=P_\mathrm{crit}^{1/(1+x)}$ where $x=(T_{\mathrm{crit}}-T)/T_{\mathrm{crit}}$.
We found $T_{\mathrm{max}}=0.88$ and $D_{\mathrm{KL}}(P_\mathrm{crit}||P_\mathrm{emp})=0.070$ for empirical data (European indices). The results for the Ising model are illustrated in Fig-\ref{fig6:DKL}.

For both systems, the results are similar. Furthermore, we simulated (see hereafter) artificial binary returns with a Monte Carlo Markov Chain ($1\times 10^{4}$ equilibrations steps and $2.3\times 10^3$ recorded configurations for $N=8$) using a pairwise maximum entropy model fitted on the data. We obtained an absolute net orientation $\hat{|m|}=0.812\pm0.010$ (1 standard deviation). The empirical value is $\langle|m|\rangle=0.726$, not included in the confidence interval but near a critical state, a slight change in inferred parameters may leads to significant change of observables estimated by simulations \cite{Mastro}. To quantify the effect of a small reconstruction error on the estimated observable, we inferred Lagrange parameters with a regularized pseudo-maximum likelihood and we shifted slightly the parameters such that $\Delta=0.015$, consistently with \cite{Aurell}. The reconstruction error is $\Delta=\sqrt{N}\langle (J_{ij}-J_{ij}^{\mathrm{true}})^{2}\rangle^{1/2}$ and quantifies the ratio between the root mean square error of the reconstruction and a canonical standard deviation. We obtained $10\%$ of relative deviation between the two estimations of $|m|$. The empirical and critical values of $|m|$ are thus similar.

\begin{figure}[ht!]
\begin{center}
\includegraphics[width=0.75\textwidth]{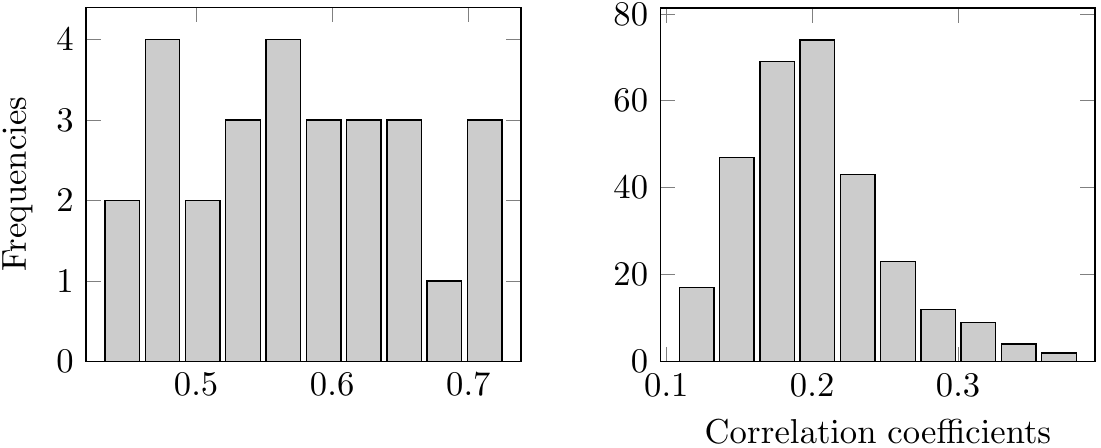}
\end{center}
\caption[Frequencies of correlation coefficients]{Frequencies of correlation coefficients between European indices (left) and stocks of the Dow Jones index at minute sampling (right).}
\label{fig:CorrCoef}
\end{figure}

The European market places seem to operate near the point corresponding to the maximum of the variance of the log-likelihood while for the Dow Jones (min and daily), the critical scaling parameter seems to be far away from the operating point $T_{\mathrm{op}}=1$ in the range of considered sizes. In Fig-\ref{fig:TcriArt}, we extend this plot for larger sizes by simulating artificial data (see hereafter). This may be explained by larger correlation coefficients between stock exchanges than between stocks of the Dow Jones as illustrated in Fig-\ref{fig:CorrCoef} and by the Epps effect (decreasing correlation magnitude with decreasing time-scale) \cite{Epps}.

Another observation is that the so-called critical exponent of the variance is equal to zero for each curve illustrated in the left panel of Fig-\ref{fig:hc} in agreement with the mean-field value of the Ising model at the critical temperature. The critical exponent can be obtained by taking the limit $\lim_{\epsilon\rightarrow0^{+}}\ln R_{\mathcal{U}}(\epsilon)/\ln \epsilon$ where $\epsilon=(T-T_{\mathrm{max}})/T_{\mathrm{max}}$ and $T_{\mathrm{max}}$ is such that $R_{\mathcal{U}}(T)$ reaches its maximum at this point \cite{Stanley}.

\begin{figure}[ht!]
\begin{center}
\includegraphics[width=\textwidth]{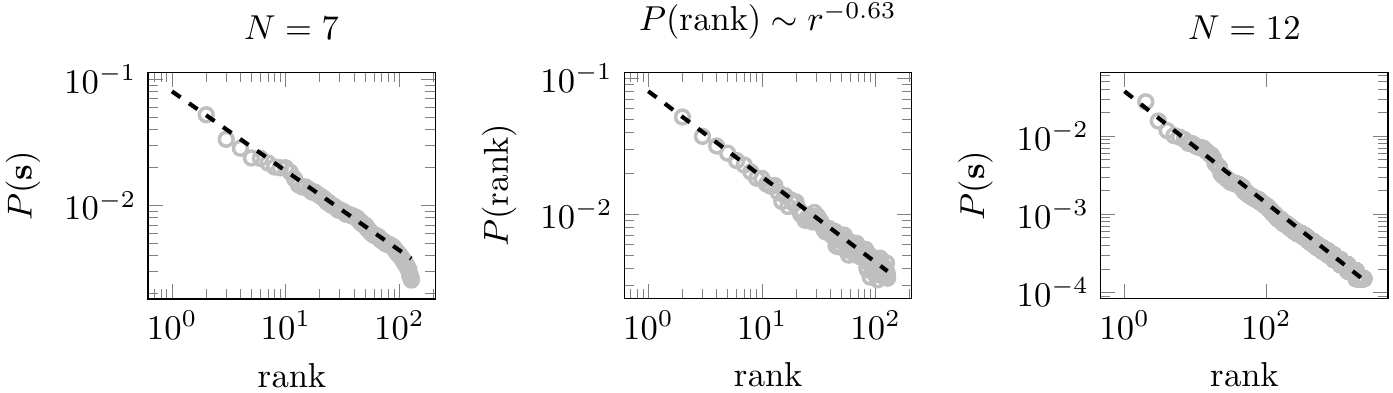}
\end{center}
\caption[Testing the power-law hypothesis]{From left to right: empirical relative frequencies of configurations vs the configurations rank of the observed time series for a set of 7 randomly chosen stocks of the Dow Jones, artificial rank distribution for a real power-law and empirical relative frequencies for a set of 12 randomly chosen stocks of the Dow Jones. The fit (dashed line) is obtained with the maximum likelihood estimator.}
\label{fig:zipf}
\end{figure}

We also study the distribution of the configuration rank. In order to know if we should reject or not Zipf's law, we perform a modified version (discrete power-law with a natural upper bound due to the finite number of configurations) of the statistical test described in \cite{Clauset}. If the p-value is smaller than $0.05$, the power-law hypothesis is ruled out and for p-value close to one, we can consider it as a good distribution \emph{candidate} (without guarantee that it is the \emph{correct} distribution). The empirical rank distribution is illustrated in Fig-\ref{fig:zipf}.

Test results for different sets are reported in Table-\ref{tab6:KS}. The considered size should not exceed $N=8$ for empirical data to have a good estimate of the distribution $P(\mathbf{s})$ by direct sampling. As expected, the power-law test outcomes depend on the system size. For the Dow Jones, the power-law is rejected when the system is properly sampled whereas in the undersampling regime the power-law is not rejected. As detailed in \cite{MMR}, the power-law is the most informative distribution when the distribution $P(\mathbf{s})$ is undersampled.

\begin{table}[!ht]
\caption[Statistical test of power-law hypothesis]{Statistical test of power-law hypothesis for sets of $N$ randomly chosen stocks of the Dow Jones ($3\times10^4$ points at minute sampling). We reported the maximum likelihood estimator $\hat{\alpha}$ of the power-law exponent $\alpha$ and its standard deviation $\sigma_{\alpha}$, the Kolmogorov-Smirnov statistic (D) and the p-value. One does not reject the power-law hypothesis if the p-value is larger than 0.05.}
\label{tab6:KS}
\begin{center}
\begin{tabular}{ccccc}
  \hline
  $\#$ of stocks & $\hat{\alpha}$ & $\sigma_{\alpha}$ & D      & p-val \\\hline
  6                  &                &                   & 0.0119 & 0.00 \\
  7                  &                &                   & 0.0117 & 0.00 \\
  8                  &                &                   & 0.0194 & 0.00 \\
  9                  & 0.6654         & 0.0038            & 0.0147 & 0.10 \\
  10                 & 0.6584         & 0.0035            & 0.0164 & 0.36 \\
  11                 & 0.7192         & 0.0027            & 0.0210 & 0.87 \\
  12                 & 0.7441         & 0.0025            & 0.0292 & 0.96 \\
  13                 & 0.7699         & 0.0024            & 0.0290 & 0.98 \\
  \hline
\end{tabular}
\end{center}
\end{table}

%
%
%
The maximum likelihood estimator (MLE) of the exponent is derived by the maximization of the log-likelihood

\begin{equation}
\ln L(\alpha)= -\alpha \sum_{i=1}^{N}\ln x_{i}-N \ln\left(\sum_{x=1}^{x_{\mathrm{max}}}x^{-\alpha}\right)
\end{equation}
where $x_{\mathrm{max}}$ is the upper bound. The standard deviation of this MLE is obtained by taking the expansion of the likelihood up to second order (Gaussian approximation). It reads

\begin{equation}
  \sigma_{\alpha_{\mathrm{MLE}}}=\frac{1}{\sqrt{N\left[\frac{\zeta^{''}(x_{\mathrm{max}},\alpha_{\mathrm{MLE}})}
  {\zeta(x_{\mathrm{max}},\alpha_{\mathrm{MLE}})}-\left(\frac{\zeta^{'}(x_{\mathrm{max}},\alpha_{\mathrm{MLE}})}
  {\zeta(x_{\mathrm{max}},\alpha_{\mathrm{MLE}})}\right)^{2}\right]}}
\end{equation}
where $\zeta(x_{\mathrm{max}},\alpha)=\sum_{x=1}^{x_{\mathrm{max}}}x^{-\alpha}$ and the prime stands for the derivative with respect to $\alpha$.

\begin{figure}[ht!]
\begin{center}
\includegraphics[width=0.5\textwidth]{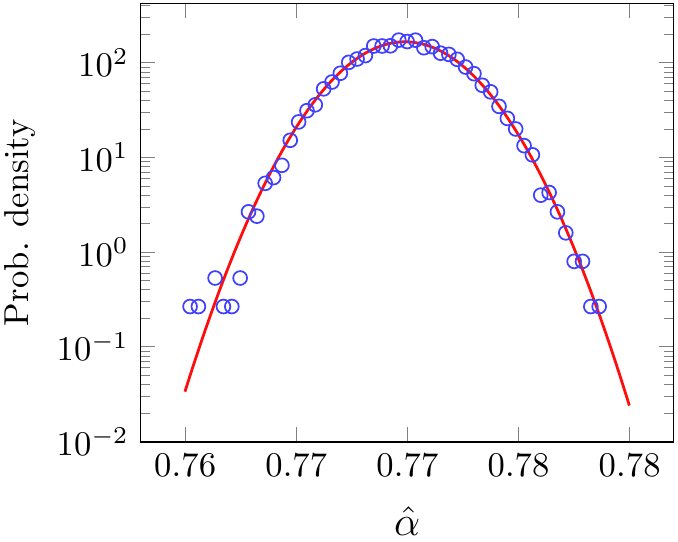}
\end{center}
\caption[Empirical pdf of the MLE estimator]{Empirical pdf of the MLE estimator for the size $N=13$ and $10^{4}$ tests (circles). The Gaussian approximation is illustrated by the full line.}
\label{fig:BetaSD}
\end{figure}

The empirical probability density function (pdf) of this estimator for  $N=13$ and $10^{4}$ tests and its Gaussian approximation are illustrated in Fig-\ref{fig:BetaSD}.

As a complement to the latter analyses, we study the linearity of the entropy expressed as a function of the utility.
The Zipf law induces a linear relation between entropy and the log-likelihood \cite{Mora}. The strict linearity can be achieved at a single value of the utility (as for the 2D nearest neighbour Ising model) or for \emph{any} value of the entropy if the distribution of the rank is a power-law \cite{Mora}. The expansion of the entropy around the mean utility $U$ is written (where $U$ is the notation for $\langle\mathcal{U}\rangle$)

\begin{equation}\label{LinEntr}
  S(\mathcal{U})\simeq S(U)-\frac{1}{T}(\mathcal{U}-U)+\frac{1}{2T^{2}R_{\mathcal{U}}} (\mathcal{U}-U)^{2}
\end{equation}
For ranks distributed following a power-law, the quadratic and higher order terms are sub-intensive; the entropy should be a linear function of the utility \cite{Step}.

%

\begin{figure}[ht!]
\begin{center}
\includegraphics[width=\textwidth]{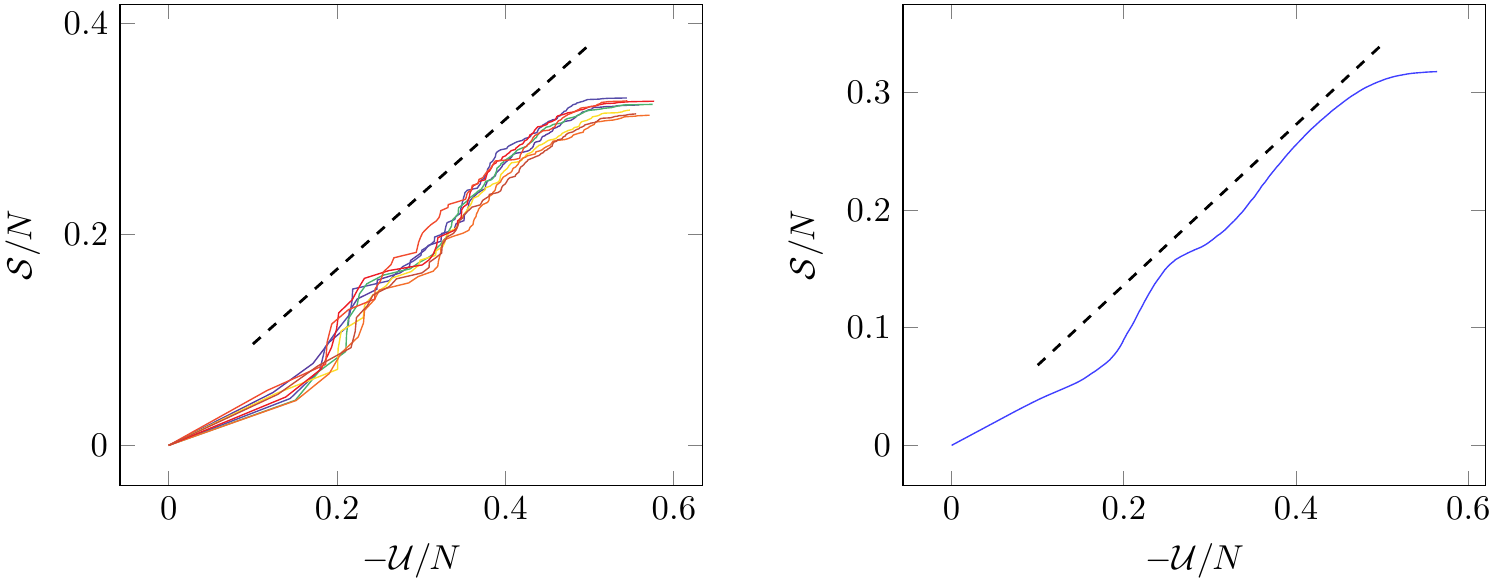}
\end{center}
\caption[Shannon entropy vs the opposite of the log-likelihood]{Left: Shannon entropy vs the opposite of the log-likelihood for several sets of 7 stocks (randomly chosen) of the Dow Jones index. Right: the average entropy-utility relation $\mathcal{S}(-\mathcal{U})$ for 100 sets of 7 randomly chosen stocks. The dashed line is the best linear fit with slope equal to 0.71 and 0.68 respectively.}
\label{fig:EntrLin}
\end{figure}

We check this property for several sets of 7 randomly chosen stocks of the Dow Jones Index. We compute the average entropy-utility relation $\mathcal{S}(-\mathcal{U})$ for 100 sets of 7 randomly chosen stocks, the results are illustrated in Fig-\ref{fig:EntrLin}.

\begin{figure}[!ht]
\begin{center}
\includegraphics[width=0.6\textwidth]{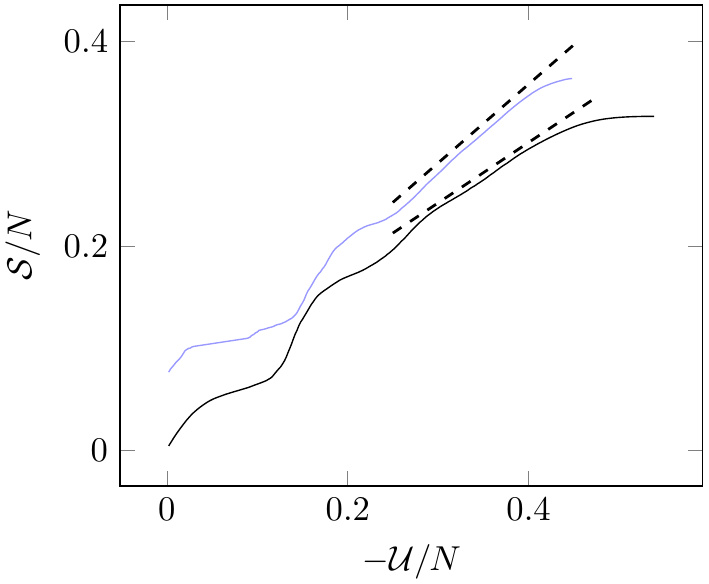}
\end{center}
\caption[Linearity of the entropy (simulation)]{Shannon entropy vs the opposite of the log-likelihood for 100 sets of 9 randomly chosen stocks. Artificial returns are simulated with a multivariate GARCH(2,2) process (light line) and with a pairwise maxent model (bold line). The dashed lines are a linear fit on a restricted range.}
\label{fig:EntrLinGARCH}
\end{figure}

We measured the relative non-linearity \cite{emancipator}, the typical value is $0.053$ (equal to zero if the function is exactly linear). The typical value of the slope is $0.71$. We also simulate $5\times 10^{5}$ artificial returns with a multivariate GARCH(2,2) and pairwise maxent processes fitted on the data. The entropy dependence on the log-likelihood is illustrated in Fig-\ref{fig:EntrLinGARCH}. The relative non-linearity is $0.032$ and $0.035$, the slope is equal to $0.77$ and $0.59$ respectively. For larger sample size the entropy is not linear either, however in a restricted utility range ($[0.3,0.4]$, about $10\%$ of the possible values of the utility, for instance) the entropy is almost linear (as measured by the relative non linearity).

As suggested by the Zip law check, the entropy is not a linear function of the log-likelihood. However, we can not reject the possibility of linearity in a restricted range or zero curvature in a single point as for the 2D nearest neighbor Ising model.

\begin{figure}[!ht]
\begin{center}
\includegraphics[width=0.75\textwidth]{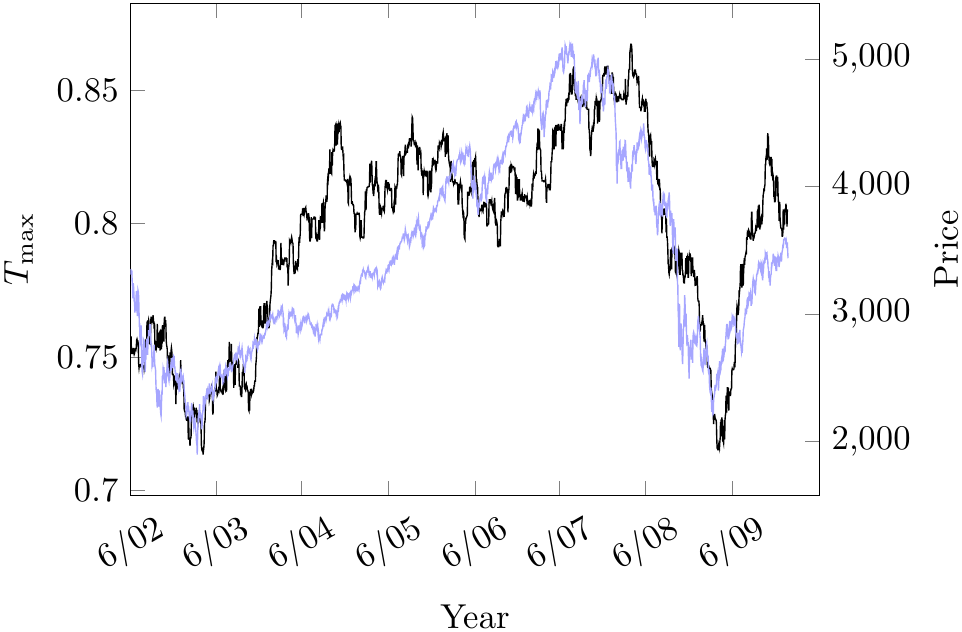}
\end{center}
\caption[Evolution of the critical scaling parameter]{Critical rescaling parameter $T_{\mathrm{max}}$ for 6 European indices (black curve, left ordinate) and the normalized sum of indices (light gray curve, right ordinate). The critical rescaling parameter is empirically estimated on a sliding window of $2^{N+2}$ trading days translated by $1$ trading day each step.}
\label{fig:TcritTime}
\end{figure}

Last, as the returns are believed to be non-stationary with volatility clustering (often modeled by a GARCH process), we study the evolution of the critical rescaling parameter $T_{\mathrm{max}}$ (at which the variance of the log-likelihood reaches its maximum value). As expected, for fixed size, $T_{\mathrm{max}}$ increases just before a crash (when fluctuations are the largest) as illustrated in Fig-\ref{fig:TcritTime} and gets closer to $T_{\mathrm{op}}$. Just before crises, financial markets undergo criticality outbursts.

\section{Link to maximum entropy models}\label{sec6:sim}
In the following, we use an inference procedure to check if the existence of a critical state is supported. One can show that the pairwise maximum entropy model is a consistent statistical model when the aim is to study collective behaviours rather than to give a precise (dynamical) model of the market \cite{moi1,moi2}. Rather than making specific assumptions of the underlying dynamics, we build a model which is consistent with the recorded data and the observed structure. This maxent model is directly linked to the former discussion since spin glasses and neural networks are also represented by pairwise maxent models which actually exhibit critical states. In this framework the configurations distribution $P(\mathbf{s})$ is rewritten as a Gibbs distribution

\begin{equation}
p_{2}(\textbf{s})=\mathcal{Z}^{-1}\exp\left(\frac{1}{2}\sum_{i, j}^{N}J_{ij}s_{i}s_{j}+\sum_{i=1}^{N}h_{i}s_{i}\right)\equiv\frac {e^{ \mathcal{U}(\textbf{s})}}{\mathcal{Z}}\label{Lagrange6}
\end{equation}
where $J_{ij}$ and $h_{i}$ are Lagrange multipliers (chosen to retrieve the first and second empirical moments). They can be thought as a measure of the pairwise mutual and individual influences.
Another well known application of the pairwise maxent model is the characterization of the neural network structure \cite{Schneidman} where the operating point seems to be a critical one \cite{Mora,fraiman}. One can show that this model is able to generate correlation matrices with non-Gaussian eigenvalues \cite{moi1} as observed in real financial time series \cite{Laloux} but also scale-free asset trees and order-disorder periods \cite{moi2}. This pairwise model gives more insights about the possibility of a critical operating point.
The rescaling of the Gibbs distribution is then viewed as a rescaling of all the parameters by a common factor $T^{-1}$. This rescaling is an investigation of a slice of the parameters space which corresponds to a stochasticity variation. A small value of $T$ favoris co-movements and a large value favoris the randomness. In this work, Lagrange multipliers are estimated with a regularized pseudo-maximum likelihood \cite{Aurell}. We note that close to $T=1$, many models are distinguishable and a slight change in parameters may lead to a significant change of the measured observables. One should compare artificial and empirical results.

%

\begin{figure}[!ht]
\begin{center}
\includegraphics[width=\textwidth]{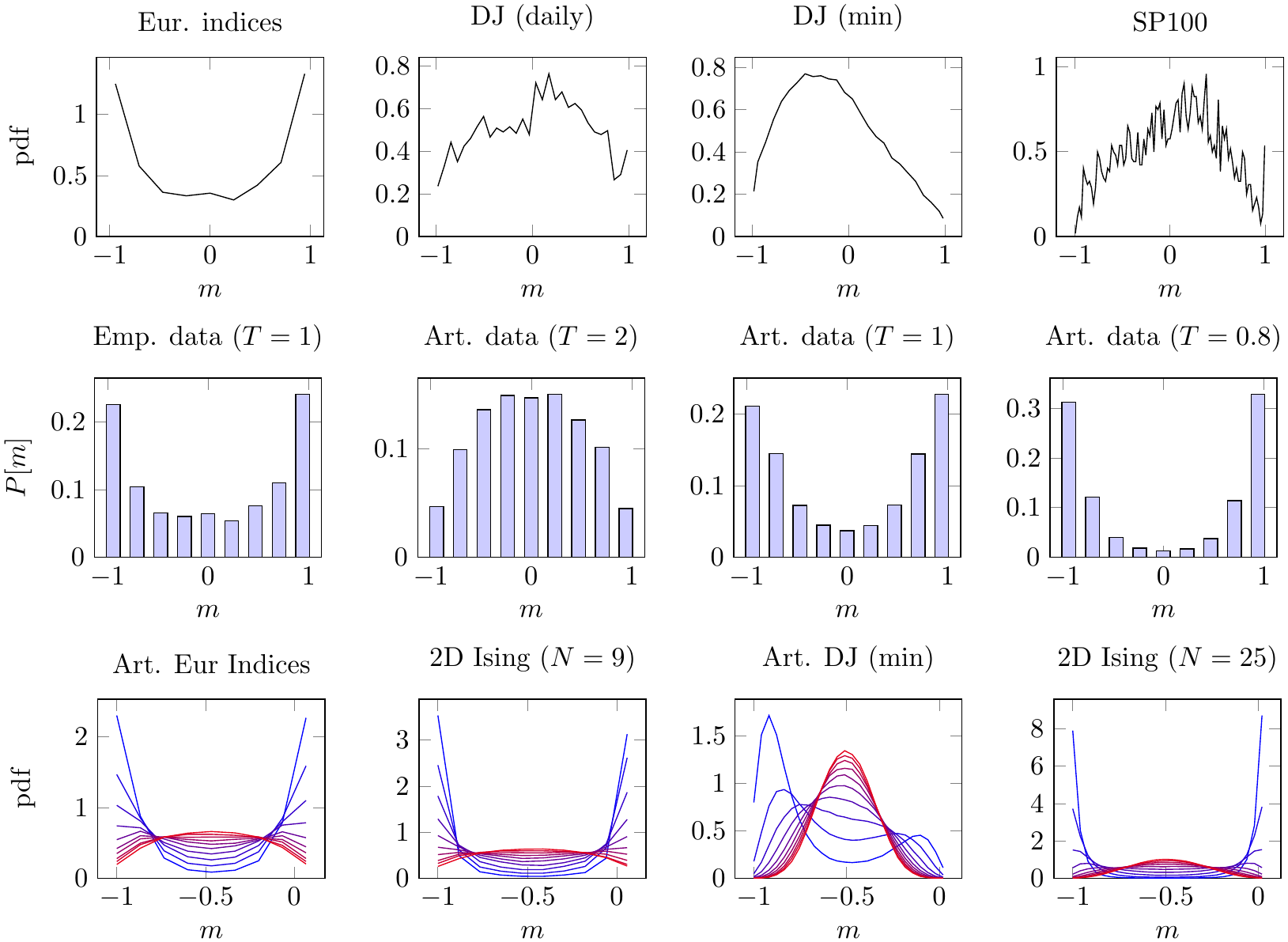}
\end{center}
\caption[Order-disorder transition?]{First row: the empirical probability density function (pdf) is illustrated for several data sets. Second row: comparison of the empirical probability mass function (pmf) of the net orientation to the artificial distributions resulting from simulations. Third row: 10 values of the stochasticity level $T$ (in the range $[0.8,2]$, blue to red respectively) are used to check if the pdf can go continuously from unimodal to bimodal, the results are compared to a 2D nearest neighbour Ising model without individual biases. The pdf and the pmf are estimated on $5\times 10^{5}$ Monte Carlo steps.}
\label{fig:Pm}
\end{figure}

First, we simulate artificial data with the estimated Lagrange parameters from the real time series. The Monte Carlo Markov chain (MCMC) is defined as follows (see Sec-\ref{sec3:MC} for details). A randomly chosen orientation is flipped if the conditional flipping probability $p(s_{i,t}=-s_{i,t-1}|s_{-i,t})$ is larger than a realization of a uniform law on the interval $[0,1]$, where $s_{-i,t}$ is the configuration excluding the $i$th entity. A configuration is recorded each $N$ flipping attempts, which defines a Monte Carlo step (MCS).
The result of the procedure applied to those artificial data is illustrated in Fig-\ref{fig:hc} by the dashed curve ($1\times 10^4$ equilibration MCS and $1\times10^5$ recorded MCS). This is consistent with the empirical variance, both peaks (blue and dashed curves) are located at the same value of the $T$-parameter.
If Lagrange parameters $\{J_{ij}\}$ are positive, the orientation distribution should be unimodal for large value of $T$ and bimodal for small value of $T$. As a qualitative test, we check if the empirical distributions are unimodal or bimodal and if they can become bimodal if we change the stochasticity level $T$, an order-disorder transition is then possible. As illustrated in the first row of Fig-\ref{fig:Pm}, the empirical distribution of the indices set is bimodal whereas the distributions of stock sets are unimodal as expected from the former empirical analyses. The second row of Fig-\ref{fig:Pm} illustrates the difference between the empirical orientation distribution and the simulated ones $P_{m}(T)$ at different stochasticity level. The indices set is clearly a rather ordered system, the probability mass peaks at the extremes values $-1,1$ of the net orientation. A disordered state exists for high level of stochasticity ($T=2$). The third row of Fig-\ref{fig:Pm}  illustrates the continuous deformation of the probability density function for a stochasticity varying from low level (blue) to high level (red). This deformation is compared to the one of the 2D nearest neighbour Ising model of corresponding size without individual biases.

\begin{figure}[ht!]
\begin{center}
\includegraphics[width=\textwidth]{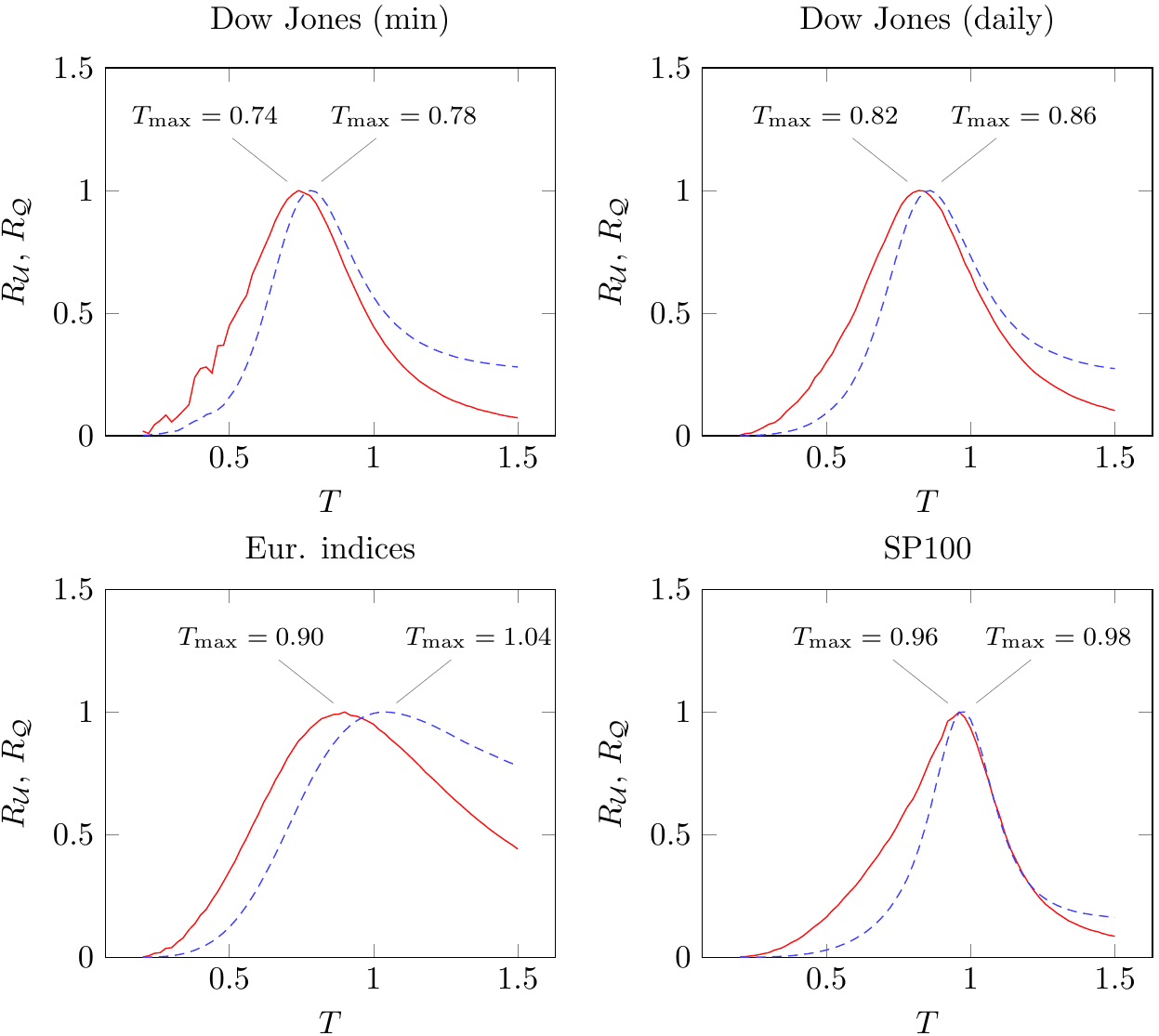}
\end{center}
\caption[The variances of the overlap parameter and of the log-likelihood]{The variances of the overlap parameter (dashed lines) and of the log-likelihood (full lines) for the 8 indices set, Dow Jones (daily and minute samplings) and SP100. Each point is computed over $5\times 10^5$ MCS after an equilibration period of $5\times 10^4$ MCS. The coordinate of the maximum is pinned (the coordinate on the left stands for the variance of the log-likelihood) .}
\label{fig:chi}
\end{figure}

The fitted maxent models allow an order-disorder transition which justifies their use in the criticality check. As mentioned in \cite{Mastro} such models are prone to accumulate in the vicinity of the critical point $T=1$ but are also highly distinguishable in this neighbourhood. Accordingly, we check if they return a $T_{\mathrm{max}}$ in line with the empirical results. One can estimate the variance $R_{\mathcal{Q}}$ of the overlap parameter $q=N^{-1}\sum_{i}s_{i}^{(1)}s_{i}^{(2)}$ and the variance of the log-likelihood. The overlap parameter measures the correlation between the configurations of two identical systems denoted by the superscript (1) and (2). The variances $R_{\mathcal{U}}$ and $R_{\mathcal{Q}}$ are known to peak at the critical value of the rescaling parameter \cite{Fischer}. If the operating point is indeed critical, we should find the peak near the value $T=1$. The results are illustrated in Fig-\ref{fig:chi}.

We note that the peaks are indeed located near the empirical values. For the indices set, the relative difference between empirical and simulated $T_{\mathrm{max}}$ is equal to $2\%$, slightly underestimated. For the Dow Jones (min), the relative difference is equal to $6\%$, slightly overestimated and for the Dow Jones (daily), $T_{\mathrm{max}}$ is overestimated of $14\%$.
The first two fitted models are consistent with the data and lead to the same conclusion: the indices set is close to the criticality ($1-T_{\mathrm{max}}\leq 10\%$) and the Dow Jones is far from criticality ($1-T_{\mathrm{max}}\geq 25\%$).
The larger deviation between empirical and simulated values for the Dow Jones (daily) may be due to inference errors in the Lagrange parameters estimation. The ratio $M/N$ (sample length on the number of entities) is too small, ten times smaller than for the Dow Jones (min). Consequently, one may expect the same relative error for the critical scaling parameter of the SP100 index.

\begin{figure}[ht!]
\begin{center}
\includegraphics[width=0.5\textwidth]{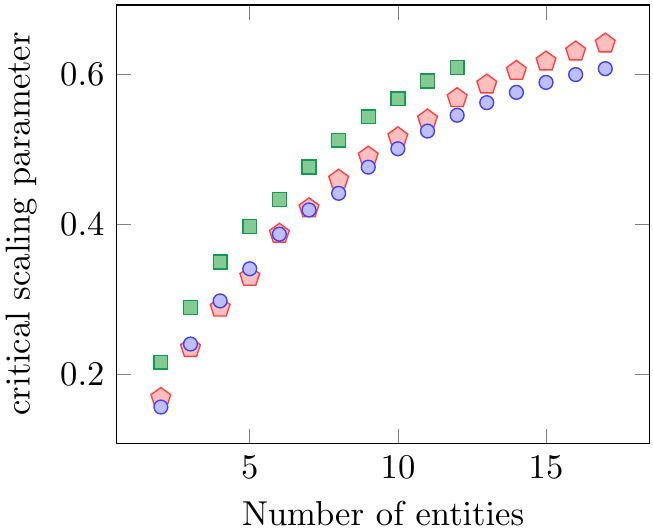}
\end{center}
\caption[Critical value of the scaling parameter (GARCH and MCMC)]{Value of $T$-parameter at which response function $R_{\mathcal{U}}$ reaches its maximum value vs the number of entities. The squares illustrate the real data, the pentagons stand for a multivariate $\mathrm{GARCH}(2,2)$ process and the dots for the MCMC.}
\label{fig:TcriArt}
\end{figure}

Since simulations are consistent with empirical results, we simulate data to complete Fig-\ref{fig:Tcri} for sizes larger than $N=8$. We simulate a binary sample of length $5\times10^6$ with the previous MCMC and also artificial returns with a multivariate $\mathrm{GARCH}$ process, known to capture the clustering of the volatility and the fat tail feature. We obtain results consistent with the empirical ones. The critical value of the rescaling parameter $T_{\mathrm{crit}}$ is illustrated in Fig-\ref{fig:TcriArt}. The critical value increases with size but is still far from $T=1$.

We note that in 2010, 12807 companies (excluding investment funds) have been listed in stock exchanges (see \url{http://www.world-exchanges.org/}). There is thus no obvious reason to consider the limit $N\rightarrow\infty$.
The market places system is significantly closer to the criticality despite its small size. It may be due to information aggregation of an index about the underlying stocks \cite{Shap}. A set of indices may operate as a system of larger size.

Furthermore, one can show that the financial network exhibits small-world organization \cite{Petra} and one knows that the Ising model on a complex network, among other, is a small-world one only at the critical temperature \cite{fraiman}.



\section{Discussion}

Stock markets are embedded in a non-uniform background. They should therefore be heterogeneous and go through regular periods interspersed with surprising events.  In a complex economic background, reactiveness is an expected behaviour. In the case of the Fukushima nuclear accident or the 2008 subprime crisis for instance, the market response was clear and prompt. All stocks fell quickly in an organized fashion. This behaviour can help to secure the profit made or prevent excessive losses if the situation goes even worse. Then, when the situation seems stabilized, or that stocks prices have fallen so dramatically that stocks became cheap and attractive, the market goes up again in an ordered fashion. These large positive-negative movements of the stock prices are encountered at any time scale \cite{Preis}. During such phases, the market exhibits large correlated structures and ordered state \cite{Dal,Jr,moi2} corresponding to an increase of the correlation strength. Such dramatic events impact globally the market (all economic sectors). On the other hand, some events (like the end of a state subsidy for eco-friendly goods, nuclear energy, etc.) have an impact on a single or few economic sectors. The criticality is then thought as a competition between global effects inducing homogeneity and local effects inducing heterogeneity in trades.

We have seen that Shannon entropy has an inflexion point near the operating point $T=1$ for the European indices set. We deduce that the micro-states number increases (or decreases) drastically following a variation of the stochasticity. The entropy is related to the logarithm of the averaged micro-states number and we can obtain this quantity by a simple integration of $R_{\mathcal{U}}(T)/T$. We observe that the largest slope stands approximatively at the actual operating point $T=1$ far from the saturation zones (where the slope is close to zero). In the neighbourhood of the operating point, the logarithm of the number of micro-states is almost linear with a large slope thus a variation of Lagrange parameters will induce a drastic (in an exponential fashion) change in the micro-structure.  It shows that the market network has a great structural malleability. The entropy also measures the degree of statistical dependency between stocks. If stocks did not influence each other, the system would be considered as a random one which implies small covariances and low reactiveness. Thus entropy would reach its largest value. In the opposite case, if stocks correlations are maximal (implying again low reactiveness), there would not be any incertitude anymore, the whole market state $\mathbf{s}$ would be predictable on the knowing of a individual state $s_{i}$ and the entropy would be zero. So if the slope of the entropy reaches its maximum value at the operating point, it means that the market is on the edge. Any variation can tip the market either towards a random (disordered, with independent trades) either towards a highly interactive (ordered, synchronized trades) state. We expect thus a large predictability exploiting instantaneous information: using the system configuration amputated of the $i$th entity $\mathbf{s}_{-i}$, one should be able to predict the state of this entity $s_{i}$ with high accuracy. This will be the subject of another work.

\begin{figure}[ht!]
\begin{center}
\includegraphics[width=\textwidth]{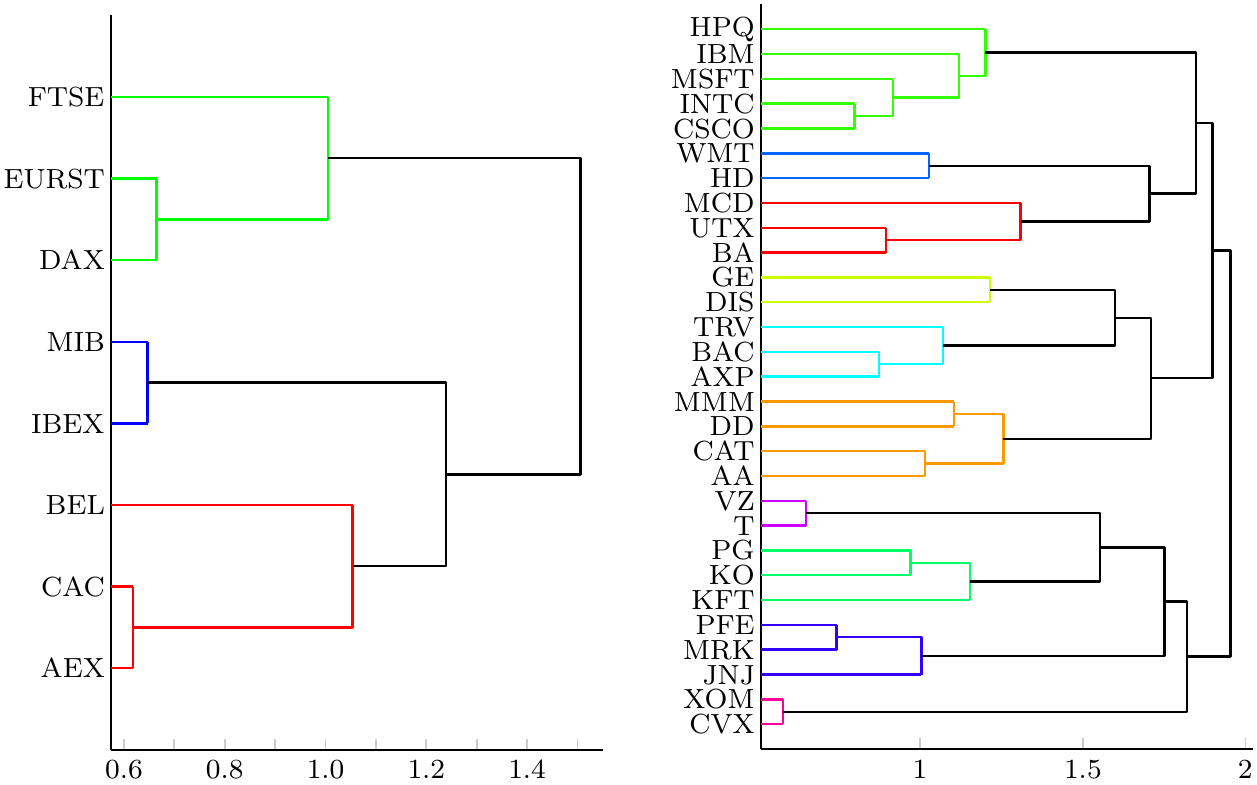}
\end{center}
\caption[The relative size of clusters]{Illustration of the clusters of each data sets. The clusters of the indices set (left) returns a partition of the European economy. The clustering of the Dow Jones (right) returns the different economic sectors (technologies, distribution, aircraft industry, TV broadcasting, finance, chemical and industrial companies, telecom, consumer goods, health care, oil).}
\label{fig:Clusters}
\end{figure}

Last, the fact that the European indices set is closer to the criticality than the Dow Jones may follow from information aggregation \cite{Shap}. A set of indices is a weighted average of stock prices. Considering the stocks as the fundamental hubs of the financial network, the indices represent super-hubs acting as a system of significantly larger size. The typical relative cluster size is also larger in the indices set where each cluster contains roughly $30\%$ of the total number of entities as illustrated in Fig-\ref{fig:Clusters} \cite{moi2}. For the Dow Jones, the cluster size is about $10\%$ of the index size. Correlated structures have thus a larger relative size in the indices set which may match the right balance between co-movements and fluctuations.

From the data analysis and simulations, we saw that the European market places seem to operate at a point where the variances of the log-likelihood is close to their largest values. An exponential empirical fit returns $T_{\mathrm{op}}=0.92$ as asymptotical value (thus maximum) for European indices and $T_{\mathrm{op}}=0.70$ for the Dow Jones at minute sampling. The entropy is not a linear function of the log-likelihood. The estimation of $T_{\mathrm{op}}$ with simulated data returns a value close to one but this value is suspected to be overestimated about $15\%$. For the Dow Jones, large simulated samples $M=5\times 10^{6}$ (using parameters obtained by fitting real data) return a consistent value $T_{\mathrm{op}}\simeq 0.65$.
Moreover, financial systems are closer to the criticality close to the beginning of a crash, meaning large fluctuation and large deviation from the uniform distribution of the configurations. This evolution also suggests a process of self-organization. The market is a highly adaptive system. By self-organization, the market reacts strongly to a change or unexpected events and by itself does not consider all possible events as equiprobable. However through the data analysis, the stock exchanges system is not exactly critical and the Dow Jones seems to be far from criticality. Furthermore, financial systems do not stay in the same regime and get closer to the criticality just before a crisis. An interesting finding because in such models, large avalanches occur more likely close to the criticality \cite{Perkovic}.

\chapter{Predicting trend reversals using market instantaneous state}\label{chap:flip}
\thispagestyle{empty}
\begin{summary}
Collective behaviours taking place in financial markets reveal strongly correlated states especially during a crisis period. A natural hypothesis is that trend reversals are also driven by mutual influences between the different stock exchanges. Using a maximum entropy approach, we find coordinated behaviour during trend reversals dominated by the pairwise component. In particular, these events are predicted with high significant accuracy by the ensemble's instantaneous state.
\end{summary}

\newpage

\begin{figure}[!ht]
\begin{center}
\includegraphics[scale=0.8]{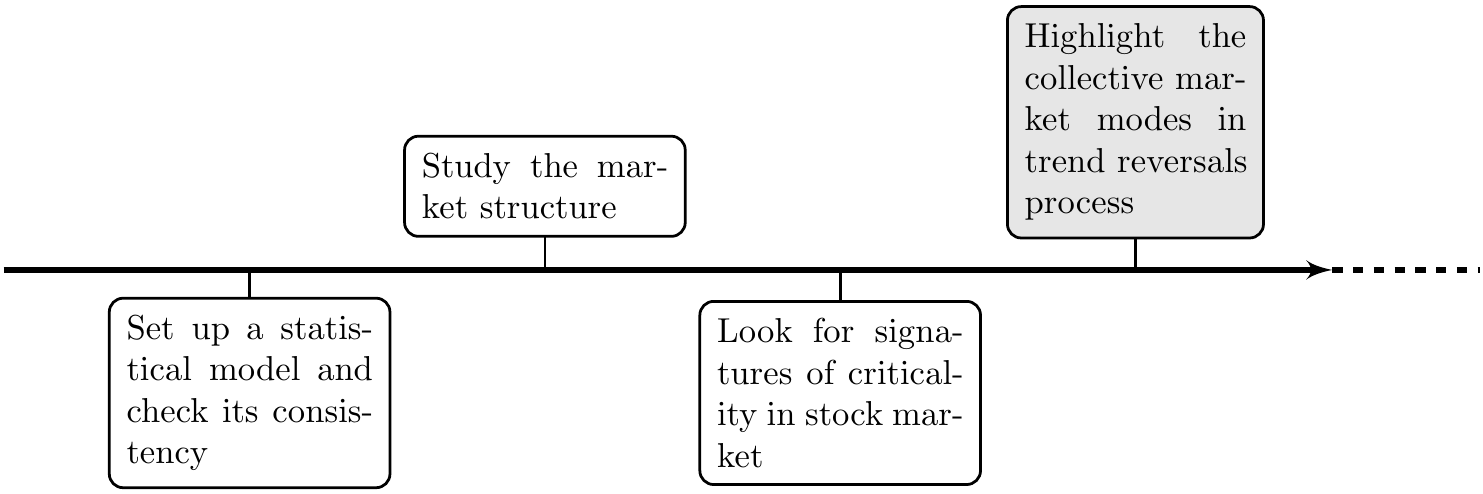}
\end{center}
\end{figure}

\newpage

\section{Introduction}
Despite abundant research focusing on estimating the level of stock returns, there are few studies examining the predictability of the sign of financial asset movements even though evidence of predictability of direction of excess return exists (the difference between returns and a defined benchmark) \cite{Chris,Moat,PreisGoogle,PreisComplex}. The herd behaviours of traders may explain this partial predictability \cite{Lux,Contherd,Sorn,Feng}. The orientation is an interesting quantity for capital allocation between different financial products but also because it allows the study of  collective behaviours as in neural networks and magnetic materials \cite{Fischer,Schneidman,moi2}.

Existing approaches of trend prediction are based on the connection between return volatility, skewness, kurtosis and return sign \cite{Chris}. Autologistic models (logistic models including past returns in a binary model) \cite{An} and a decomposition of the trade-to-trade price increments into three components (activity, direction and size) were considered as well as probit models with various commonly used financial variables as explanatory variables \cite{ny}. The problem with these models may be the use of a particular data generation process or the identification of relevant financial variables in the regression model. Moreover, observed collective behaviours in financial markets highlight the requirement of a multi-variate approach to capture co-movements that are a key feature to explain synchronization, order, non-random correlations and predictability \cite{Dal,Laloux,OnnelaPRE,mol,Plerou99}. We believe that any model intended to predict a financial quantity, like the sign of stock returns, should therefore be multivariate. Here we propose a statistical data-based model capturing almost all the correlation structure of a financial market, the so-called pairwise maximum entropy model \cite{moi1}. This qualitative model does not rely on a particular data generation dynamics, uses only a data-driven approach based on internal inputs (present and past returns) and takes into account co-movement.

The use of pairwise maximum entropy (maxent) models has led to a fruitful description of complex systems, particulary in phase transition and magnetic materials (Ising models and spin glasses) \cite{Fischer,Stanley}, but also in neuroscience \cite{Schneidman}.  They are related to graphical models, Boltzmann machines, error correcting codes, logistic regression, etc. \cite{Opper}. Maxent models are much more than models recovering moments from data, they are powerful effective models describing collective behaviours. However, one must pay attention to the scaling of parameters capturing co-movements (pairwise influences). In real neural networks, they seem to be size independent. Increasing the size is equivalent to lowering the temperature and freezing is prevented by the presence of negative pairwise couplings \cite{Schneidman} whereas in financial networks, couplings seem to scale as the inverse of the network size leading to a mean-field description \cite{moi2}.

The aim of the maxent approach is two-fold: use a statistical framework avoiding as much as possible any assumption and study the importance of co-movement (necessity of a multi-variate approach), especially in spatial predicting stock market orientation. We found that instantaneous conditional transitions (\emph{spatial} predictions) are able to predict in average $83\%$ of market place reversals which is far better than the individual model, thereby showing the importance of co-movements. Accuracy drops to $73\%$ for the components of the Dow Jones index. Such deviation may be induced by the lower correlations and the lack of large enough samples. Furthermore, we showed that history does not seem to improve the accuracy either by a genuine lack of memory or by a finite size effect in the parameter inference. These results suggest that some collective dynamics drives the global market trend \cite{Laloux,Plerou99}. They constitute another evidence of coordinated behaviours in financial markets. Moreover, they show that these collective modes are partially responsible for predictability of stock market orientation \cite{Mantegna,mol}.

We note that if a \emph{good} approximation of the collective dynamics was known together with dependencies between economic quantities, it would certainly lead to better predictions than those obtained by this simple autologistic model as it is the case in the related field of neural networks \cite{pill} and in econometric approaches \cite{An,ny}. We propose that this model serves as a benchmark with which to compare results of more sophisticated models embedding a real economic description.

The chapter is organized as follows. In section \ref{sec7:coll}, we derive instantaneous conditional probability of trend reversals. In section \ref{sec7:res}, we present the empirical results. In section \ref{sec7:noise}, we discuss the noise issues and the comparison to artificial networks. In section \ref{sec7:simult}, we compare different models of simultaneous trend reversals.

\section{Collective states}\label{sec7:coll}
We consider a set of 8 major European indices of the Eurozone (AEX, BEL, CAC, DAX, EUROSTOXX, FTSE, IBEX, MIB) observed during a ten year long daily time series including two large crises (2008 subprime and Euro-debt crises). The data were cleaned up to ensure simultaneity of the different time series (see appendix). An orientation reversal (or a \emph{flip}) is a trend reversal in two consecutive observed trading days. More precisely we consider daily returns (without the overnight period) defined as $r_{i,t}=(p^{\mathrm{c}}_{i,t}-p^{\mathrm{o}}_{i,t})/p^{\mathrm{o}}_{i,t}$, where $p^{\mathrm{c}}_{i,t}$ is the closing price of the $i$th stock of the  period $t$ and $p^{\mathrm{o}}_{i,t}$ the opening one. The index $i=1,\ldots,N$ labels assets ($N$ is the total number of assets). The index $t=1,\ldots,T$ labels time periods ($T$ is the total number observed periods). They can be rewritten as $r_{i,t}=s_{i,t}|r_{i,t}|$ where the binary variable $s_{i,t}\in\{-1,1\}$ is the sign or orientation of the index $i$ at period $t$. An orientation change occurs if $s_{i,t+1}=-s_{i,t}$. Such reversals are expressed as a binary variable $\mathbf{1}_{[s_{i,t+1}=-s_{i,t}]}$.  We consider the binary part of returns, $1$ for a positive return and $-1$ for a negative one. The resulting time series are strongly correlated, off-diagonal correlation coefficients lie between $0.43$ and $0.74$. We consider market orientation reversal as a multivariate stochastic process. This process can be decomposed in two main components, the instantaneous (influence within the defined time-bin unit or \emph{spatial} dependence) and the causal (temporal dependence) statistical dependencies among different market places.
The study of collective state and conditional flipping probability, causal and instantaneous, requires estimation of the probability distribution of a potentially high-dimensional system ($\sim N^{2}$ parameters) which is in general intractable without further constraints.
A way to tackle this problem is to use the maximum entropy principle \cite{Jaynes,Cover} restricted to second-order moments to infer a statistical model. One obtains a multivariate autologistic model (or Ising-like model). Pairwise statistical dependencies account for $95\%$ of all statistical dependencies as measured by the multi-information criterion \cite{Schneid_Multi,moi1} and this model is suitable for to description of collective behaviors. The resulting pairwise distribution is given by

\begin{equation}
p_{2}(s_{1,t};\cdots;s_{N,t})=\mathcal{Z}^{-1}\exp\left(\frac{1}{2}\sum_{i, j=1}^{N}J_{ij}s_{i,t}s_{j,t}+\sum_{i=1}^{N}h_{i}s_{i,t}\right)\label{7-Lagrange}
\end{equation}
where the binary variables $s_{i}\in \{-1,1\}$ describe the orientation of market places (respectively bearish or bullish), $\mathcal{Z}$ is a normalizing constant. The parameters $\{h_{i}\}$ and $\{J_{ij}\}$ are respectively Lagrange multipliers associated with first and second order constraints.
In this framework the instantaneous dependencies among indices (or stocks) are given in terms of conditional flipping probabilities of a given index. The flipping rate is given by

\begin{equation}\label{InstPr}
p(-s_{i,t-1}=s_{i,t}|\mathbf{s}_{-i,t})=\frac{\exp\left(-s_{i,t-1}\sum_{ j\neq i}J_{ij}s_{j,t}-h_{i}s_{i,t-1}\right)}{\exp\left(-\sum_{ j\neq i}J_{ij}s_{j,t}-h_{i}\right)+\exp\left(\sum_{j\neq i}J_{ij}s_{j,t}+h_{i}\right)}
\end{equation}
where $\mathbf{s}_{-i,t}$ is the observed market configuration at period $t$, excluding the $i$th entity.

\begin{figure}[ht!]
\begin{center}
\includegraphics[width=\textwidth]{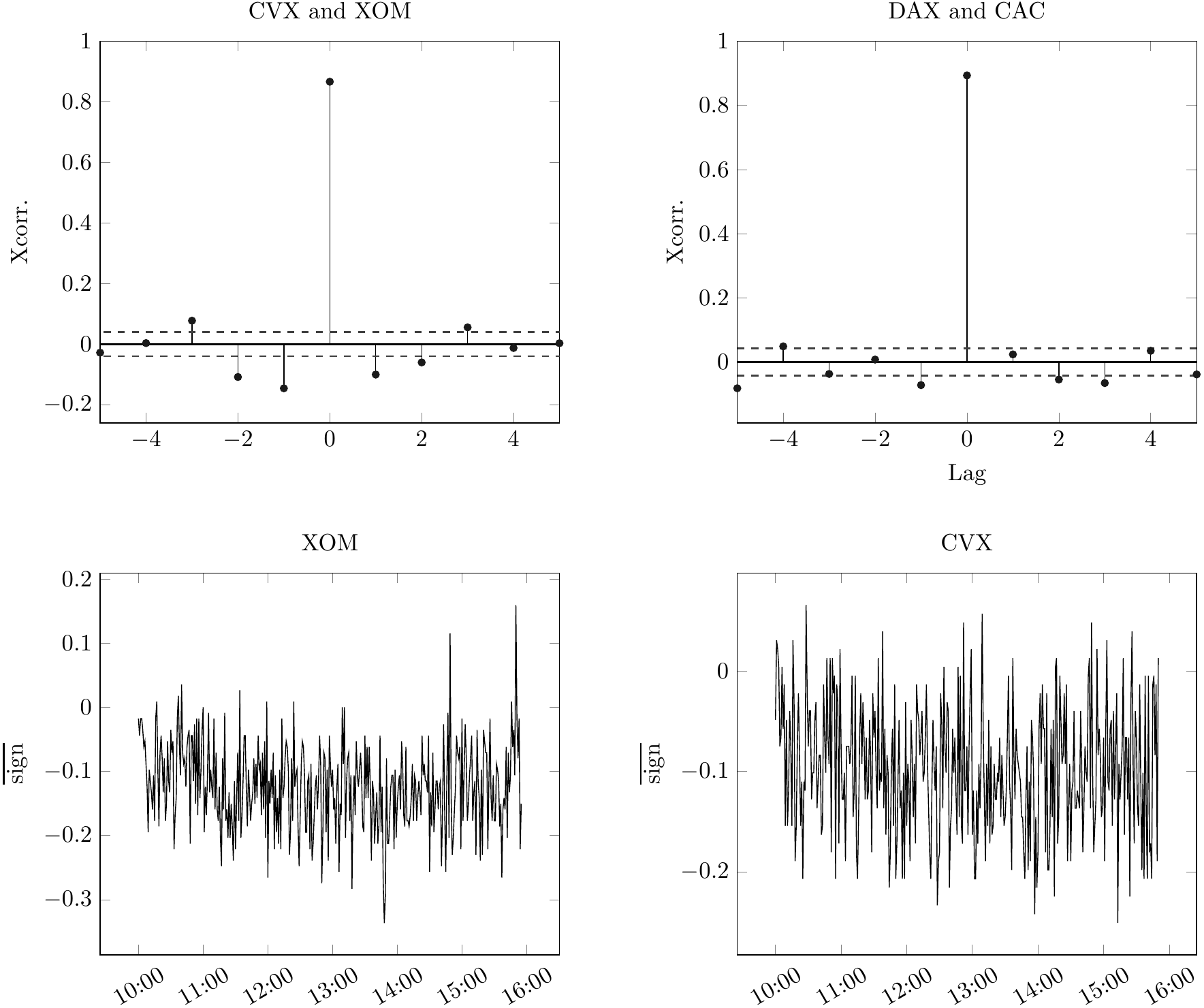}
\end{center}
\caption[Cross-correlogram]{\textbf{Top}: cross-correlogram between orientation of CVX and XOM (left) and between CAC and DAX indices. CVX and XOM are two main oil companies, 2500 daily returns have been used.
\textbf{Bottom}: The sign as a function of time for intraday data at minute sampling for XOM (left) and CVX (right). The bar stands for the temporal mean over 225 trading days (between March 2011 and May 2012).}
\label{fig:XCorr}
\end{figure}

One can enquire whether considering past states could help to predict flipping events. The conditional flipping probability (\ref{InstPr}) can be modified to include some memory and is given by

\begin{equation}\label{HistPr}
p(-s_{i,t-1}=s_{i,t}|\mathcal{H}_{t}^{T})=\frac{1}{2}
\left[1-s_{i,t-1}\tanh\left(\sum_{ j\neq i}J_{ij}s_{j,t}+h_{i}
+\sum_{\tau=1}^{T}\sum_{j}K_{ij}^{\tau}s_{j,t-\tau}\right)\right]
\end{equation}
where the history $\mathcal{H}_{t}^{T}$ denotes the sequence $(\mathbf{s}_{-i,t}; \mathbf{s}_{t-1};\ldots; \mathbf{s}_{t-T})$. We expect minor difference with the memoryless case since sign autocorrelations and pairwise cross-correlations are known to be insignificant for any lag (except the first one in some case) \cite{Cont,BouchaudBook} at the contrary of their absolute values \cite{Pod}; cross-correlations between CAC and DAX indices and between CVX and XOM stocks are illustrated in Fig-\ref{fig:XCorr}, for instance. However cross-correlations measure linear or monotonic dependencies. More sophisticated statistical relationships may exist. Maxent models are supposed to capture them as the entropy and related quantities provide a more general way to capture statistical dependencies \cite{Cover}.
Furthermore, we can check if our model is able to \emph{forecast} sign of returns by checking if the predictive power is significantly larger than $50\%$ when we consider only \emph{past} information (and so, make profit). We will see that it is not the case. This result is in line with the weak efficient market hypothesis (roughly: one can not forecast the \emph{sign} of excess returns using only past returns)\cite{Fama}.
In the following, we restrict ourself to two time-lags (since more lags means more parameters to estimate and decrease the prediction power). For higher sampling frequency (here, the minute timescale), specific features may influence the results. Firstly, prices move discretely (jumps) as they can only vary by $1$ cent increment. We have not considered this issue in the analysis but we considered highly capitalized and very liquid assets which can limit the impact of the so-called market structure noise. Secondly, the absolute intraday returns draw a concave curve with a minimum reached at lunch time (intraday seasonality). This deterministic pattern is observed throughout markets \cite{Andersen}. We looked for such seasonality in the sign of return. The mean over $225$ trading days of the intraday signs (between 10:00 am and 4:00 pm) is illustrated in bottom panels of Fig-\ref{fig:XCorr}. There is not a clear deterministic pattern neither in the time domain nor in the frequency domain (not illustrated here), meaning there is not a preferential direction of trades (sell or buy) at the opening and closing of a trading day.

Lagrange parameters were estimated by a regularized pseudo-maximum likelihood method (rPML) (see appendix and Sec \ref{sec3:rpml}) \cite{Aurell}. Once they were estimated, the flipping probability is obtained using (\ref{InstPr}) or (\ref{HistPr}).

However the distinction between statistical dependencies induced by \emph{correlated} common inputs $\{h_{i}\}$ and genuine pairwise ones should be done. In the pairwise maxent framework, if an input (says $h_{j}$) is dependent of another one (says $h_{i}$) this can lead to a non-diagonal covariance even if $J_{ij}$ are set to zero.

\section{Results}\label{sec7:res}

\subsection{Indices set}
\begin{figure}[ht!]
\begin{center}
\includegraphics[width=\textwidth]{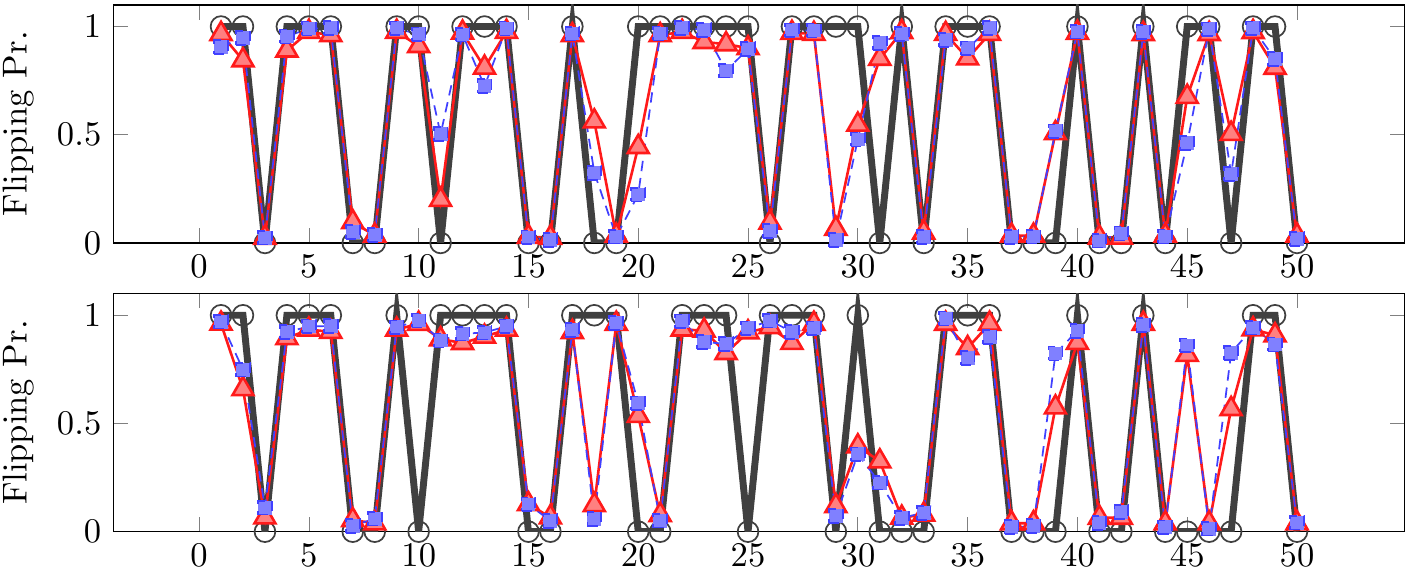}
\end{center}
\caption[Predicted series]{Predicted series for the CAC (top) and DAX (bottom) indices. The black circles represent the actual flipping time-series for 50 out-of-samples trading days. The red full line (triangles) illustrates the memoryless flipping probability and the blue dashed line (squares) the flipping probability including two time-lags.}
\label{fig:InstPr}
\end{figure}

First of all, we perform a preliminary test. We infer Lagrange parameters on a large time-window (more than 2000 trading days) and we compute flipping probabilities for 50 out-of-sample consecutive trading days using either instantaneous empirical data $\textbf{s}_{-i,t}$ in (\ref{InstPr}) or empirical sequence $\mathcal{H}_{t}^{T}$ in (\ref{HistPr}). The results for CAC and DAX indices are illustrated in Fig-\ref{fig:InstPr}.

\begin{figure}[ht!]
\begin{center}
\includegraphics[width=0.75\textwidth]{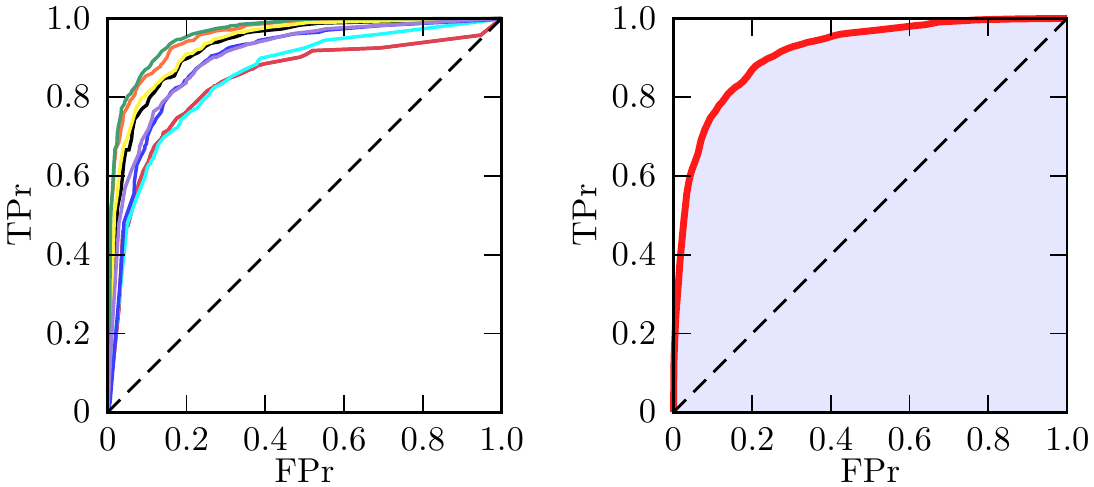}
\end{center}
\caption[ROC curves for European indices]{Prediction of a single market place trend reversal. The receiver operating characteristics (ROC) curves for 8 indices (left) and the resulting mean ROC curve (right). The ROC curve illustrates the true positive rate (TPr) as a function of the false positive rate (FPr). These curves were obtained with a ten-fold cross-validation scheme on the set of the 8 European indices. The shaded area below the mean ROC curve illustrates the \emph{area under the curve} (AUC).}
\label{fig:ROC}
\end{figure}

Both autologistic models give similar results close to the actual time series.
To assess the efficiency of instantaneous and historical models, we compare the true-positive (predicting a flip which actually occurs) rate to the false-positive (predicting a flip which does not occur) rate. Ideally, a good classifier is supposed to have a large accuracy, but also a large true-positive rate together with a low false-positive rate.
To evaluate these quantities, we consider the confusion matrix for varying detection level. The detection level $\alpha$ is the threshold value such that the flipping is considered as a true event if flipping probability is larger than $\alpha$. We used the so-called ROC (receiver operating characteristics) curves to illustrate the predictive power of the classifier \cite{fawcett}. We used a ten-fold cross-validation scheme to compare the performance of both methods on out-of-sample events because the fitting may lead to accurate predictions if predicted states are in the training set (in-sample) but poor predictions on the validation set (out-of-sample). The sample is divided in learning and testing blocks. Parameters are estimated on $90\%$ of the total amount of data (learning block). The prediction is performed on the validation sample ($10\%$ of the data set) using empirical orientations $\textbf{s}_{-i,t}$ (or $\mathcal{H}_{t}^{T}$) belonging to the testing block to infer $s_{i,t}$. The true-positive, false-positive and accuracy rates are measured for each validation fold and are averaged over the ten folds. The ROC curves are illustrated in Fig-\ref{fig:ROC}.

\begin{figure}[ht!]
\begin{center}
\includegraphics[width=0.75\textwidth]{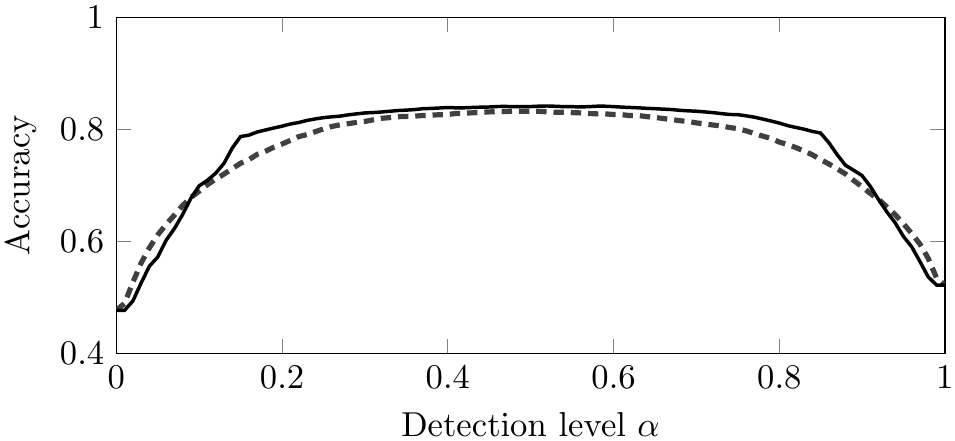}
\end{center}
\caption[Accuracy]{The mean accuracy as a function of the detection level for the set of 8 European indices. The accuracy of the memoryless case is illustrated by the full line and the causal model by the dashed line.}
\label{fig:Acc}
\end{figure}

For the memoryless model (\ref{InstPr}), the mean true-positive rate is about $76\%$ for less than $10\%$ false-positive rate. Another summary quantity is the \emph{area under the curve} (AUC). The random guessing produces the diagonal line and thus an AUC$=0.5$. A good classifier should have an AUC close to 1. The AUC may be interpreted as the probability that the model will assign a larger flipping probability to a randomly chosen sample containing a positive event. The AUC, illustrated by the shaded area in Fig-\ref{fig:ROC}, is equal to $0.914\pm0.042$ (mean $\pm$ s.d.). The lowest AUC for the set of 8 indices is equal to $0.849$ and the largest to $0.960$. We consider also the accuracy of the prediction as a function of the chosen detection level. The accuracy is  the number of true predictions divided by the total number of events. The mean accuracy versus the detection level is illustrated in Fig-\ref{fig:Acc}. The maximum mean accuracy is equal to $83\%$. In average $83\%$ of the total number of events were correctly predicted. The lowest value of these maximal rates is equal to $78\%$ and the largest maximal rate to $89\%$.

For the historical model (\ref{HistPr}), the mean true-positive rate is about $75\%$ for less than $10\%$ false-positive rate and the resulting AUC is $0.902\pm0.050$. This mean value is not included in the $96\%$ confidence interval of the memoryless AUC but the relative deviation between both AUC mean values is only $1.3\%$. The lowest AUC for the set of 8 indices is equal to $0.849$ and the largest to $0.960$ as for the memoryless model. The maximum mean accuracy is equal to $83\%$. In average $83\%$ of the total number of events were correctly predicted. The lowest value of these maximal rates is equal to $78\%$ and the largest maximal rate to $89\%$.

We note that the independent instantaneous model gives a very poor result and is nearly a random guessing (AUC $=0.51$). The independent model is defined by setting $\textbf{J}$ to zero in the instantaneous model. If we consider the historical model (\ref{HistPr}) without the instantaneous part, the maximal average accuracy is only $53\%$. Therefore, we conclude that the most important component is the one capturing instantaneous co-movements (here, the intra-day co-movements). We note that the econometrical model detailed in \cite{ny} correctly forecasts $59\%$ of out-of-sample events showing the importance of the knowledge, even partial, of the fundamental relationships between economic quantities.
Last, we note that a drawback of the historical model is the multiplication of parameters to be estimated. Each added time-step brings $(N^2-N)/2$ more parameters for each matrix $\mathbf{K}^{\tau}$. However the sum should be truncated at an optimal lag (the one where the accuracy reaches its maximum value, for instance).
We conclude that the most significant part of the prediction model is the one capturing instantaneous co-movements.

\subsection{Dow Jones}

The Dow Jones is an index regrouping highly capitalized US companies (AA, AXP, BA, BAC, CAT, CSCO, CVX,	DD,	DIS, GE, HD, HPQ, IBM, INTC, JNJ, KFT, KO, MCD, MMM, MRK, MSFT, PFE, PG, T, TRV, UTX, VZ, WMT, XOM).
We consider two different timescales: daily and 1 minute price sampling rates. The sample size for the daily sampling is about $2500$ trading days and $3\times10^4$ points for the minute timescale.
In this application, there are two main issues. For a satisfactory parameters estimation, we need large samples. A direct sampling would require a sample length several times larger than the total number of configurations $2^N$, which is huge for the Dow Jones ($\sim 10^9$ points which means 5 thousand trading years at this timescale). For the rPML method, the reconstruction may be done with fewer points, but still with large sample lengths $10^6$ to $10^8$ for a system size $N=64$ \cite{Aurell}. Secondly, the typical correlation coefficients between orientations are smaller than those of market places. The issues are thus twofold: parameter estimation may be flawed and low correlations may lead to intrinsically lower predictive power than in indices set analysis.

\begin{figure}[ht!]
\begin{center}
\includegraphics[width=\textwidth]{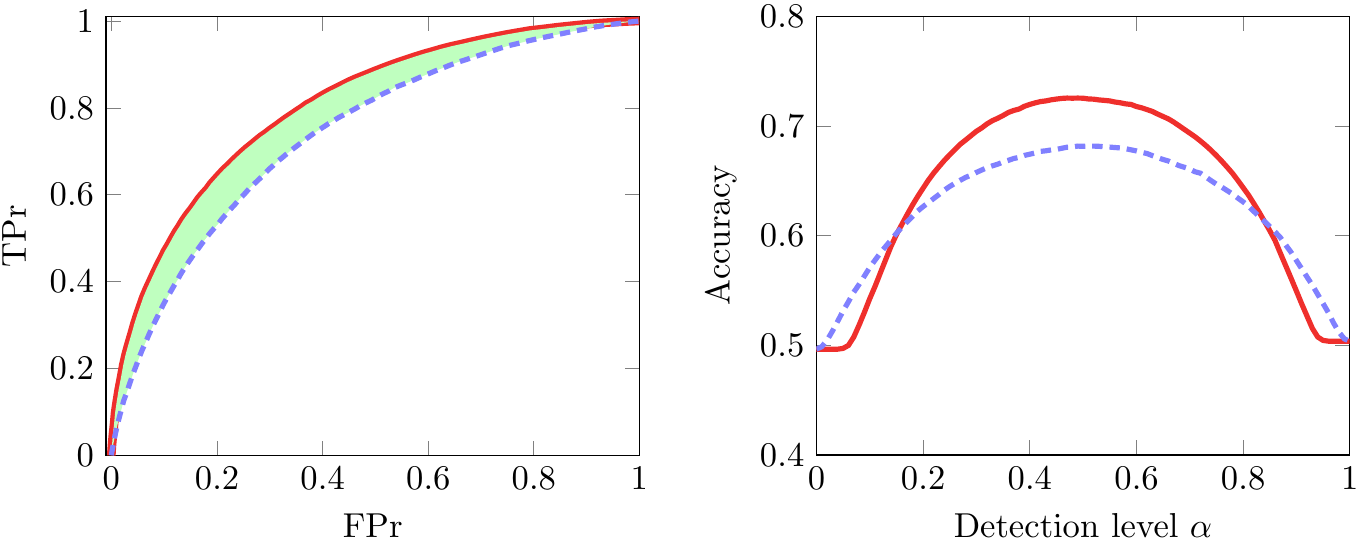}
\end{center}
\caption[Mean ROC curves for the Dow Jones (daily)]{Mean ROC curves for both models for the Dow Jones daily sampling (left) and the accuracy as a function of the detection level (right). The full line illustrates the memoryless model and the dashed line the historical one. These curves were obtained with a ten-fold cross-validation scheme. The shaded area illustrates the difference between AUC's.}
\label{fig:DJ}
\end{figure}

For the memoryless model, the AUC is equal to ($0.797\pm0.038$) and the mean maximum accuracy is equal to $73\%$.
For the historical model the AUC is equal to ($0.740\pm0.049$) and the mean maximum accuracy is equal to $68\%$. The difference between both AUC's is illustrated by the shaded area in the Fig-\ref{fig:DJ}. The predictive power is affected by the finite size estimation and the large number of parameters to be estimated (especially in the historical model).

To know if the timescale affects the predictive power, we performed the same analysis on a smaller timescale (3 orders of magnitude smaller).

\begin{figure}[ht!]
\begin{center}
\includegraphics[width=\textwidth]{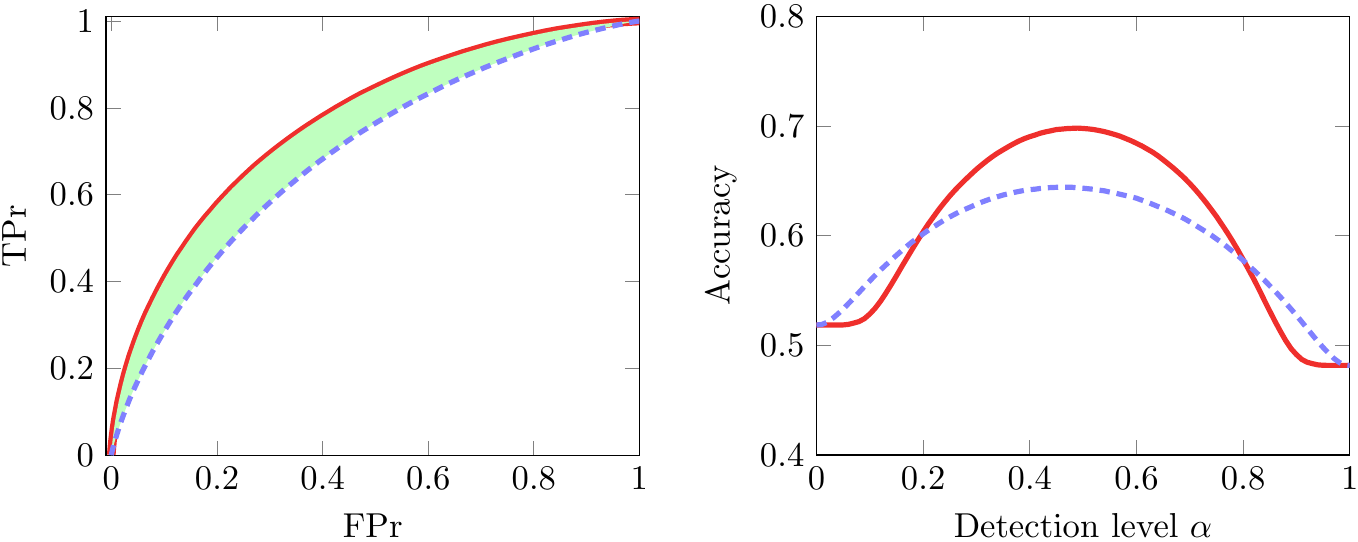}
\end{center}
\caption[Mean ROC curves for the Dow Jones (min)]{Mean ROC curves for the Dow Jones minute sampling (left) and the accuracy as a function of the detection level (right). The full line illustrates the memoryless model and the dashed line the historical one. These curves were obtained with a ten-fold cross-validation scheme. The shaded area illustrates the difference between AUC's.}
\label{fig:DJmin}
\end{figure}

For the memoryless model, the AUC is equal to ($0.763\pm0.029$) and the mean maximum accuracy is equal to $70\%$.
For the historical model the AUC is equal to ($0.695\pm0.037$) and the mean maximum accuracy is equal to $64\%$. The difference between both AUC's is illustrated by the shaded area in the Fig-\ref{fig:DJmin}.

These values are slightly lower than in the daily sampling analysis, the relative difference between accuracy of both timescales is equal to $4\%$. Moreover, the independent instantaneous model has an accuracy equal to $58\%$ significantly larger than for daily sampling results ($51\%$). These results are consistent with the observed lower correlation between returns at lower timescale (Epps effect) \cite{Epps}. The historical model is the least efficient. We conclude that the most significant part of the prediction model is the one capturing instantaneous co-movements.

Interestingly, the results are slightly improved if the pairwise influences $J_{ij}$ are set to their mean value (homogeneous influences) and if individual biases $h_{i}$ are set to zero. Given the relatively small width of the time-window, the reconstruction errors on these parameters induces biased results. However the improvement is slight, the relative difference with the heterogeneous case is about $2\%$. For the Dow Jones at minute sampling, the resulting accuracy is equal to $71\%$, the AUC is equal to $(0.786 \pm 0.026)$. For the Dow Jones at daily sampling, the accuracy is equal to $73\%$, the AUC is equal to $(0.810 \pm 0.030)$.

\subsection{Dependencies on number of units, sample length and distance}

\begin{figure}[ht!]
\begin{center}
\includegraphics[width=0.75\textwidth]{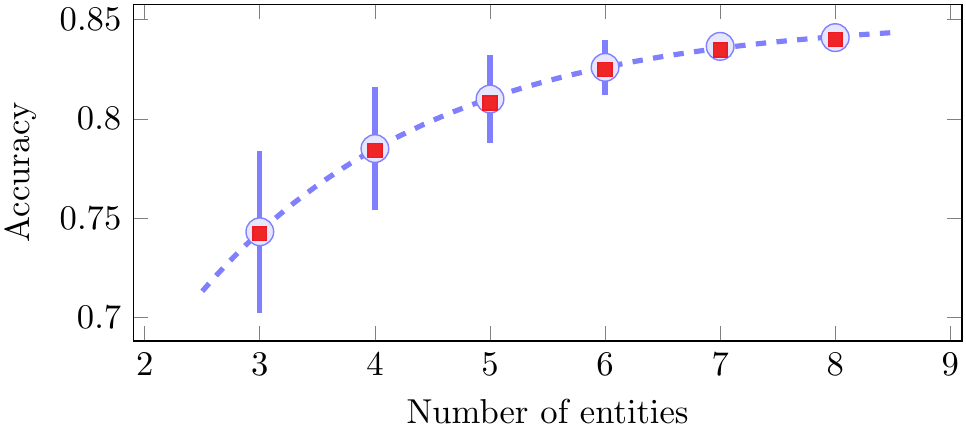}
\end{center}
\caption[Accuracy vs system size]{Accuracy as a function of number of indices. Dots illustrate the accuracy of the instantaneous model and squares the accuracy of the historical model. The dashed line is an exponential fit.}
\label{fig:AccN}
\end{figure}

The collective dynamics seems to be important for predicting flips. Adding more indices may improve the accuracy of the flipping detection. To study the dependency on system size, we let only $k$ indices visible among the $N=8$ European indices and we perform flipping prediction on the reduced system. For each value of $k$, we perform prediction on $N!/k!(N-k)!$ possible choices of indices set. Results are illustrated in Fig-\ref{fig:AccN}.

\begin{figure}[ht!]
\begin{center}
\includegraphics[width=0.75\textwidth]{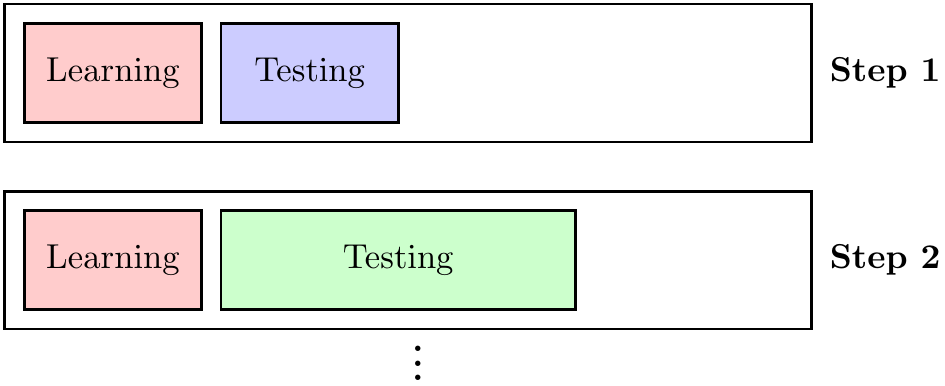}
\end{center}
\caption[Testing the dependence on the testing block length]{Schematic description of the method to check accuracy dependence on the length of the testing block. We divide the sample in blocks. Lagrange parameters are inferred on the learning block. We use these parameters and  empirical data of the testing block to perform flipping prediction.}
\label{fig:learning1}
\end{figure}

The accuracy may also depend on the length of the testing sample. To check this feature, we infer Lagrange parameters with the rPML method on a learning block and we perform prediction on a testing block of increasing length. This method is illustrated in Fig-\ref{fig:learning1}.

\begin{figure}[ht!]
\begin{center}
\includegraphics[width=0.75\textwidth]{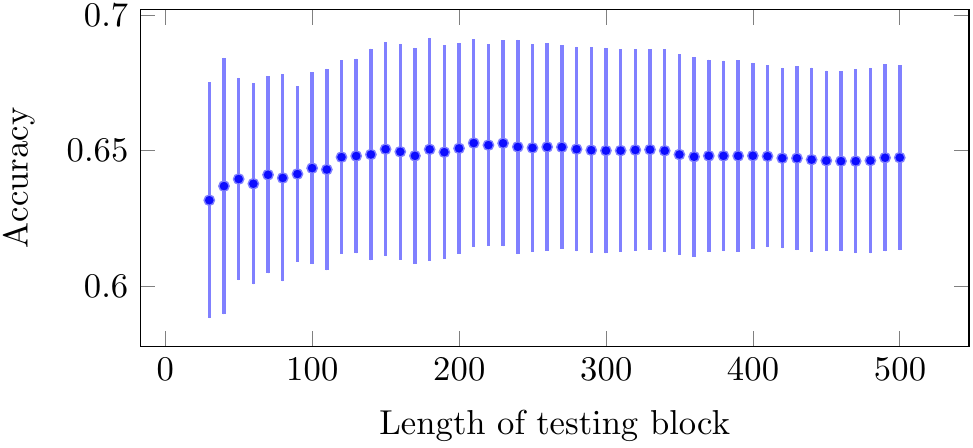}
\end{center}
\caption[Accuracy as a function of length of the testing block]{Accuracy as a function of length of the testing block. Error bars represent the standard deviation on 8 different testing blocks. Parameters are inferred on a learning block of 500 samples and accuracy is measured on 8 different testing blocks, each of length increasing from 30 to 500 points (2 trading years) by increment of 10 samples.}
\label{fig:AccLength}
\end{figure}

\begin{figure}[ht!]
\begin{center}
\includegraphics[width=0.75\textwidth]{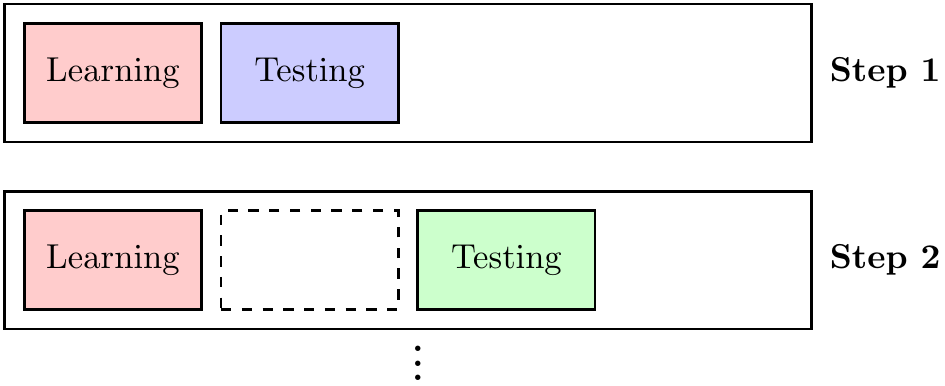}
\end{center}
\caption[Testing the dependence on the distance of the testing block]{Schematic description of the method to study the dependence on the distance between learning and testing blocks. We divide the sample in blocks. Lagrange parameters are inferred on the learning block. We use these parameters and  empirical data of the testing block the perform flipping prediction. Length of testing blocks is fixed.}
\label{fig:learning2}
\end{figure}

\begin{figure}[ht!]
\begin{center}
\includegraphics[width=0.75\textwidth]{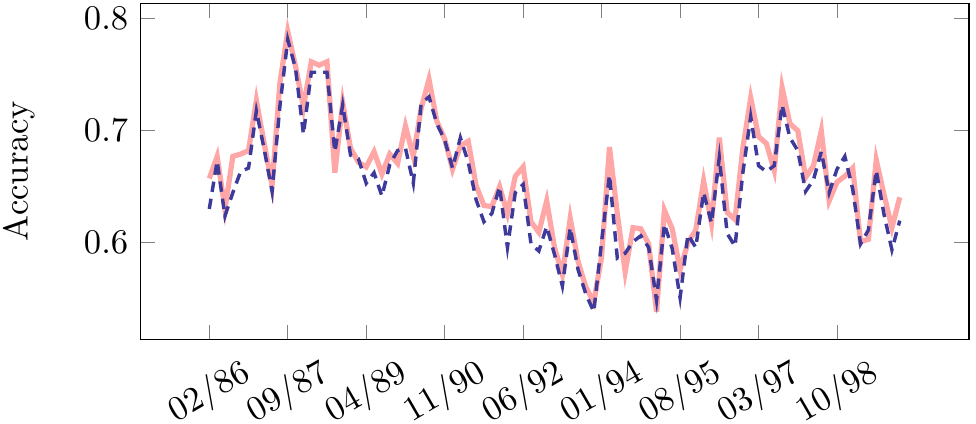}
\end{center}
\caption[Accuracy vs the distance between the learning and testing blocks]{Accuracy as a function of the distance between the learning and testing blocks for the Dow Jones index (1982-2000 period). Parameters inference is done on 1000 first points (1982-1985) and the accuracy is evaluated on 89 blocks of 40 points. The full line illustrates the instantaneous model and the dashed line the historical model.}
\label{fig:AccBlock}
\end{figure}

The accuracy seems to remain constant as the size of the testing block increases as illustrated in Fig-\ref{fig:AccLength}.
If the series was stationary, Lagrange parameters should be the same for the whole sample and we expect the accuracy to be constant. For a non-stationary time series, Lagrange parameters may vary through time and so the accuracy. However if significant deviations from their mean values only occur on small time-windows, the accuracy appears constant when computed on large time-windows.
To study this feature, we test the dependency of the accuracy on the distance between learning and testing blocks. Instead of taking larger and larger testing blocks, we consider testing blocks of fixed length but farther and farther from the learning block. This procedure allows to compare accuracy on these different time-windows of fixed length. This method is illustrated in Fig-\ref{fig:learning2} and results in Fig-\ref{fig:AccBlock}.

Returns exhibit volatility clustering, so we expect the accuracy will differ from its mean value only on small time windows and we should observe a nearly constant value on a large time-window for fixed Lagrange parameters. In Fig-\ref{fig:AccBlock}, we observe that accuracy reaches its maximum value in the testing block embedding Black Monday (October 19, 1987). A larger accuracy results from larger correlations during the crash. The difference between the maximum ($0.82$) and the minimum ($0.55$) accuracy is larger than the expected statistical error $40^{-1/2}\simeq0.16$, the increase of accuracy during crises is thus a genuine feature.

\begin{figure}[ht!]
\begin{center}
\includegraphics[width=0.75\textwidth]{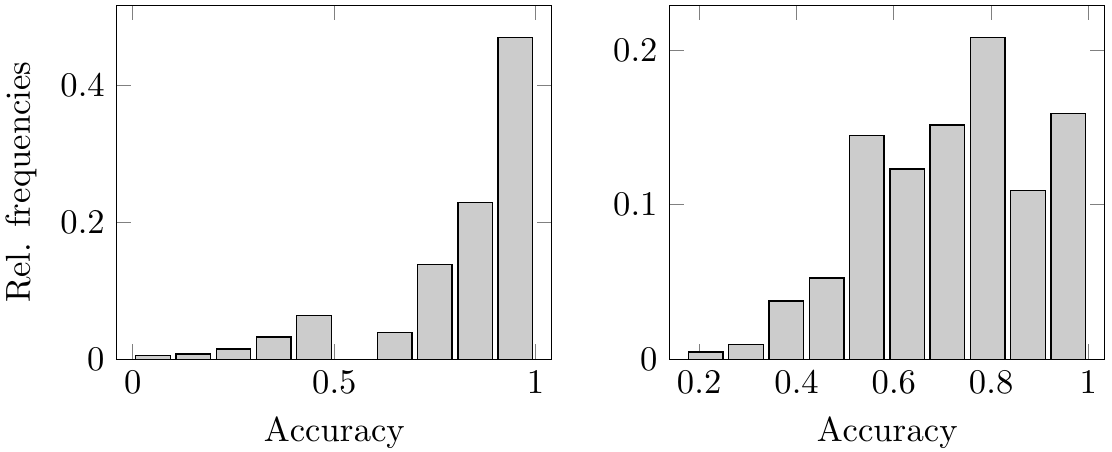}
\end{center}
\caption[Accuracy pmf]{Distribution of accuracy (averaged over the $N$ entities) over a time-window of 1000 trading days width for the European indices (left) and for the Dow Jones at daily sampling (right).}
\label{fig:AccFreq}
\end{figure}

Last, we note that over a time-window of 1000 trading days width, the averaged accuracy per trading day is rarely equal to zero as illustrated in Fig-\ref{fig:AccFreq}. For the European indices set, the averaged accuracy is equal to zero only for 6 trading days (31/08/2007, 18/10/2007, 22/04/2009, 14/04/2011, 06/02/2010, 23/02/2012). The first two occurrences happened just before the subprimes crisis, the third occurrence during the 2009 market rebound, the fourth at the end of the rebound following the Fukushima accident, the fifth and sixth happened during the recovery after the debt crisis (high risk periods). There is no obvious periodicity in the time series of accuracy (no fundamental frequency in the Fourier series). One could expect that Friday can be a day where accuracy decreases due to the expiration of securities but it is not observed in this analysis.

Another possibility is the one given in \cite{Livan11}: few driving forces can lead to a rich structure even in the bulk of the spectrum of the correlation matrix which is therefore not only due to noise. Such factors and clusters can also be thought as correlated structure appearing in the vicinity of the critical state of a pairwise maxent model. Global correlations (correlation length of the order of the network size) together with fluctuating clusters can coexist near the order-disorder boundary.

\section{Noise and comparison to artificial networks}\label{sec7:noise}
The estimation of the Lagrange parameters may introduce a bias in orientation prediction. Particularly because of noise due to finite size estimation and limitation of inference methods based on approximation scheme. Moreover, their values depend on the considered sample since they are inferred with a constrained regularized pseudo-likelihood, the constraints being the equality between empirical and theoretical first and second moments.

To quantify the bias, we estimate the noisy part of the standard deviation of the recovered $\textbf{J}$ matrix. We simulate binary time series (same sample length than the true data) with the maximum entropy conditional probability $p(s_{i,t}=-s_{i,t-1}|\mathbf{s}_{-i,t})$, known as the Glauber dynamics \cite{BinderMC}. A product is randomly chosen, a flipping attempt is accepted if the flipping probabilities $2^{-1}[1-s_{i}\tanh(\sum_{j}J_{ij}^{*}s_{j})]$ is larger than a randomly uniform number on the interval $[0,1]$. A configuration is recorded each Monte Carlo step (MCS). A MCS corresponds to $5N$ flipping attempts. In this data generation, the artificial $\textbf{J}^{*}$ matrix was taken homogeneous with all entries equal to the empirical mean of mutual influences. Then we estimate the influence matrix with the rPML method. Ideally, the standard deviation $\sigma_{\mathrm{noise}}$ of the estimated artificial influences should be much smaller than the one of real influences $\sigma_{J}$. Results are reported in Table-\ref{tab7:noise}. Depending on the sample length, the noise seems to be significant but not the dominant part of the estimation except for large system size.

\begin{figure}[ht!]
\begin{center}
\includegraphics[width=0.75\textwidth]{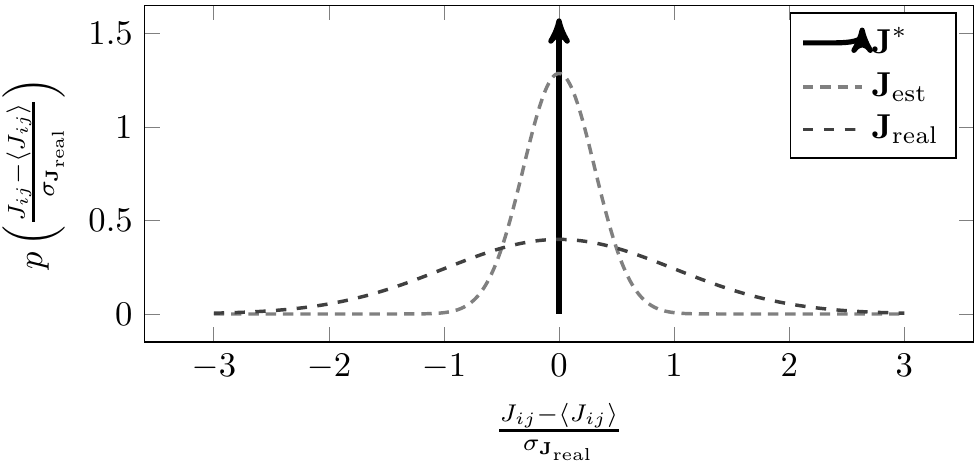}
\end{center}
\caption[Schematic representation of noise level estimation in parameters inference]{Schematic representation of noise level estimation in parameters inference. Artificial data are generated with homogeneous influences $\textbf{J}^{*}$ (probability density function illustrated by the green Dirac delta). Then we perform parameters estimation using these artificial data. Ideally the pdf of the estimated parameter $\textbf{J}_{\mathrm{est}}$ should be close to the pdf of $\textbf{J}^{*}$. Last, we compare the distribution of $\textbf{J}_{\mathrm{est}}$ to the variance of parameters resulting from real data $\textbf{J}_{\mathrm{real}}$ using their variance.}
\label{fig:noiseScheme}
\end{figure}

\begin{table}[!ht]
\caption[Quantification of the noise in influences inference]{Quantification of the noisy part of the standard deviation of the inferred mutual influences.}
\label{tab7:noise}
\begin{center}
\begin{tabular}{lcr}
\hline
Index/set    & sample length ($T$) & $\sigma_{\mathrm{noise}}/\sigma_{J}$\\ \hline
Eur. indices &  $2.5\times 10^{3}$ & 0.37                                \\
DJ(daily)    &  $2.5\times 10^{3}$ & 0.31                                \\
DJ(min)      &  $3.0\times 10^{4}$ & 0.24                                \\
\hline
\end{tabular}
\end{center}
\end{table}


We can also generate data with the estimated $\textbf{J}$ matrix from the data, infer the artificial $\textbf{J}^{*}$ matrix and compare $\textbf{J}^{*}$ to $\textbf{J}$. The reconstruction is satisfying if estimated Lagrange parameters $J_{ij}^{*}$ are close to their true values $J_{ij}$. To quantify deviation from the real network (defined by $\textbf{J}$), we use the reconstruction error $\Delta=\sqrt{N}\langle(J_{ij}^{*}-J_{ij})^{2}\rangle^{1/2}$ which represents the ratio between the root mean square error $\langle(J_{ij}^{*}-J_{ij})^{2}\rangle^{1/2}$ and a canonical standard deviation $1/\sqrt{N}$ \cite{Aurell}. This definition of the reconstruction error is believed to be consistent with financial networks \cite{moi2}. Results are reported in Table-\ref{tab7:rplm}. These results are consistent with those of \cite{Aurell} where the magnitude order of the reconstruction error is $10^{-2}$ for a complete network of size $N=64$ with $J_{ij}$ drawn from a Gaussian distribution $\mathcal{N}(0,N^{-1})$.

\begin{table}[!ht]
\caption[Quantification of the reconstruction error]{Quantification of the reconstruction error $\Delta$ with the regularized pseudo-likelihood. Artificial data are generated with the Glauber dynamics using $\textbf{J}$ inferred from real data as true influences matrix (a configuration was recorded each $5N$ flipping attempts).}
\label{tab7:rplm}
\begin{center}
\begin{tabular}{lcr}
\hline
Index/set       & sample length ($T$) & $\Delta$\\ \hline
Eur. indices    &  $2.5\times 10^{3}$ & 0.100                           \\
DJ(daily)       &  $2.5\times 10^{3}$ & 0.158                           \\
DJ(min)         &  $3.0\times 10^{4}$ & 0.035                           \\
DJ(min)         &  $1.0\times 10^{6}$ & 0.026                           \\
\hline
\end{tabular}
\end{center}
\end{table}


A useful benchmark to assess exactness of this autologistic model may be the predictive power computed from artificial data. We compute the mean accuracy and mean AUC for artificial data truly generated by a pairwise autologistic process and we compare them to the results obtained from financial data. These values are reported in Table-\ref{tab7:Art}.

\begin{table}[!ht]
\caption[Comparison of artificial accuracy and AUC to real accuracy and AUC]{Comparison of artificial accuracy and AUC to real accuracy and AUC. The artificial values are computed from data generated with a pairwise maximum entropy model (autologistic) and the real ones from financial data. Artificial samples are of the same length than the corresponding real samples.}
\label{tab7:Art}
\begin{center}
\begin{tabular}{lcccc}
\hline
Index/set            & Accuracy art. ($\%$)& Accuracy ($\%$)& AUC art.& AUC   \\
\hline
Eur. indices         & 87                  &  83            & 0.911   & 0.914 \\
DJ(daily)            & 75                  &  73            & 0.806   & 0.797 \\
DJ(min)              & 71                  &  70            & 0.769   & 0.763 \\

\hline
\end{tabular}
\end{center}
\end{table}

%

In general, the predictive power is slightly larger for artificial data. The relative difference between artificial and real data lies between $ 1\%$ and $5\%$. This benchmark reveals that sign of returns can be predicted with similar accuracy than finite size time-series truly generated by a pairwise instantaneous process. The artificial accuracy and AUC represent the maximum expected values that the model can return due to the finite size effects.

\section{Simultaneous trend reversals}\label{sec7:simult}
\begin{figure}[ht!]
\begin{center}
\includegraphics[width=\textwidth]{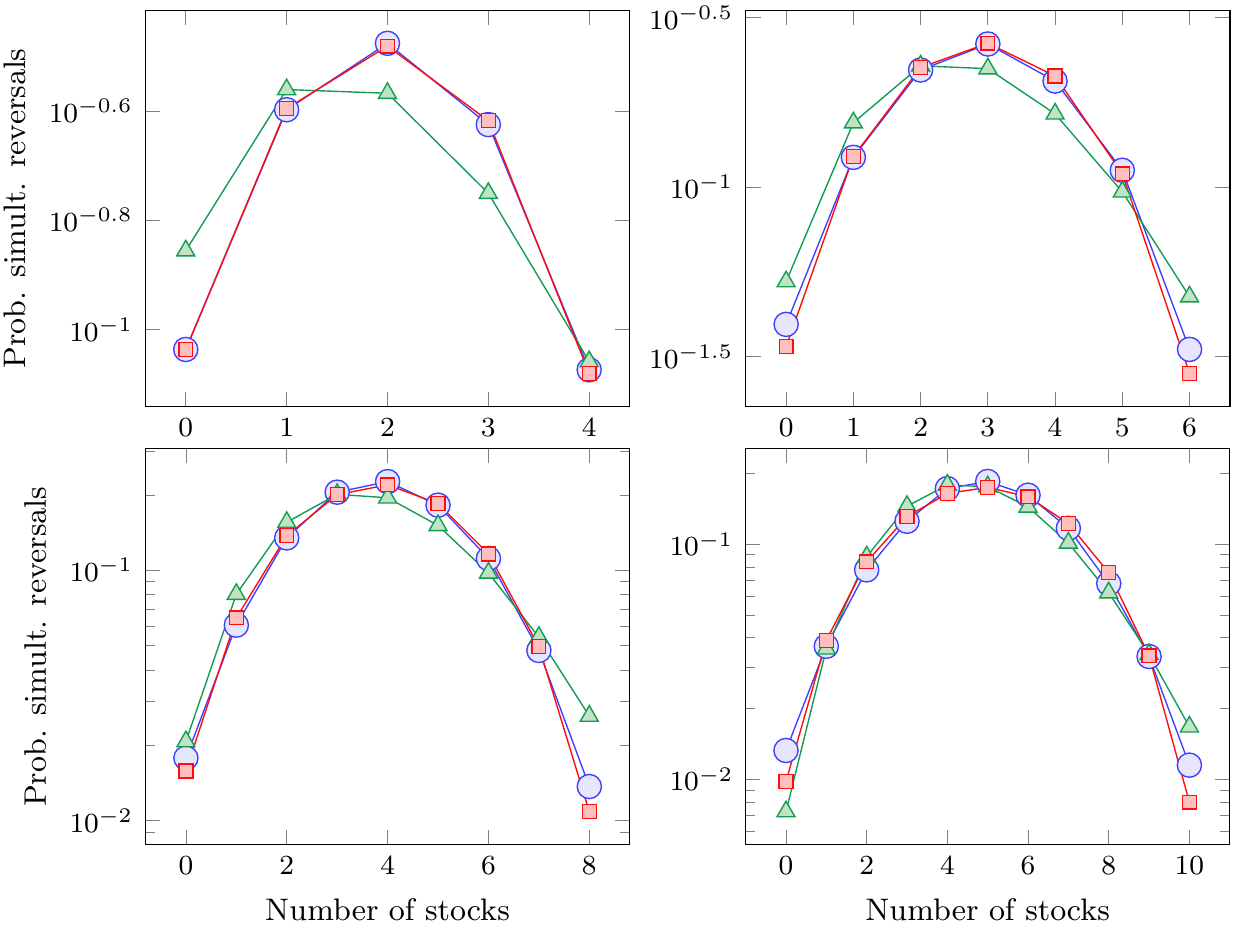}
\end{center}
\caption[The distributions of simultaneous trend reversals]{The distributions of simultaneous trend reversals. The empirical distribution is illustrated by dots, the pairwise distribution by squares and independent Poissonian model by triangles. The distribution is computed over 20 randomly chosen sets (for $N=4, 6, 8, 10$ stocks from top left to bottom right) of the Dow Jones at minute sampling.}
\label{fig:simDist}
\end{figure}

We also inquire if the pairwise autologistic model is able to estimate the distribution of simultaneous trend reversals. The occurrence of a trend reversal is expressed by a binary variable $x_{i,t}=\mathbf{1}_{[s_{i,t+1}=-s_{i,t}]}$. Using the maximum entropy principle, we get the following pairwise maxent model

\begin{equation}
p_{2}(x_{1,t};\cdots;x_{N,t})=\mathcal{Z}^{-1}\exp\left(\sum_{i, j=1}^{N}W_{ij}x_{i,t}x_{j,t}\right)\label{Lagrange2}
\end{equation}
where the matrix $\textbf{W}$ has a non null diagonal and can be estimated by the method detailed in \cite{Dick}. We also fit an independent trend reversal model (a Poisson distribution, using the maximum likelihood estimator). We compare the empirical, pairwise and independent distributions on 20 randomly chosen groups for different sizes (up to $N=12$ where direct sampling gives a good estimate of the distribution). Results are illustrated in Fig-\ref{fig:simDist}. The most frequent event is a reversal of approximatively half of the number of considered stocks.

\begin{figure}[ht!]
\begin{center}
\includegraphics[width=\textwidth]{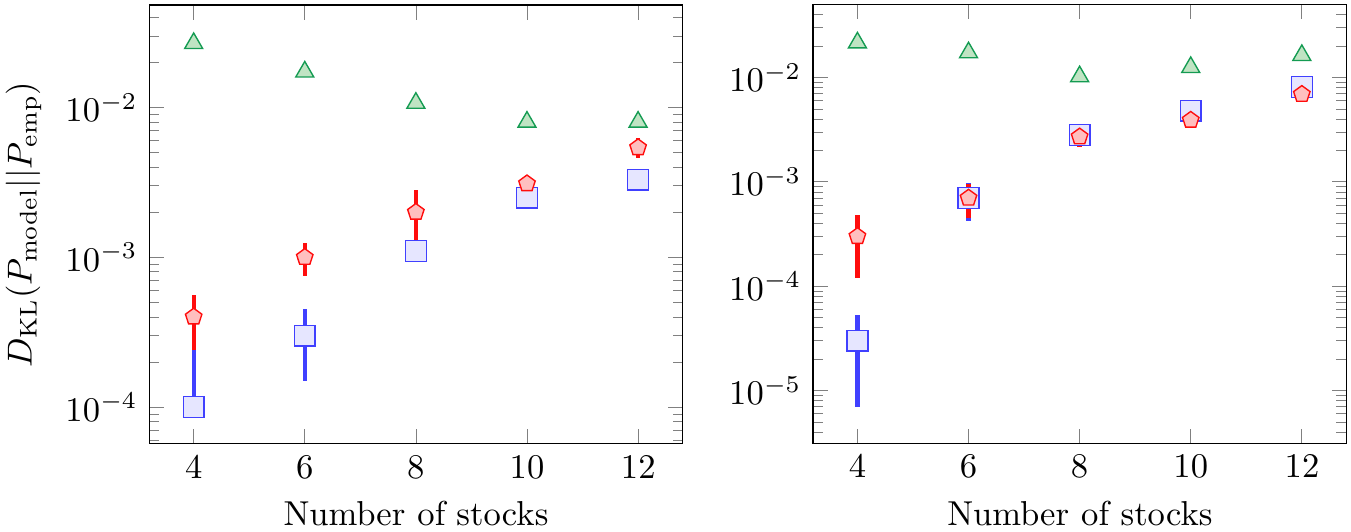}
\end{center}
\caption[Comparison of empirical and theoretical PMF of simultaneous reversals]{The average Kullback-Leibler divergence between the empirical distribution of simultaneous reversals and the pairwise (squares), independent (triangles) and dichotomized Gaussian (pentagons) models. The divergence is computed over 10 randomly chosen stock sets of the Dow Jones at daily sampling (left) and at minute sampling (right). Error bars represent the standard deviation over 10 randomly chosen stock sets.}
\label{fig:DKL}
\end{figure}

We computed the mean Kullback-Leibler divergence between the empirical distribution and the pairwise, independent and dichotomized Gaussian models. The results are illustrated in Fig-\ref{fig:DKL}. The pairwise model is the closest to the empirical distribution.

The dichotomized Gaussian (DG) model \cite{Amari,Macke09} is a threshold multivariate Gaussian model with mean and covariance matrix inferred to match the empirical first and second moments of the binary time series. It is an attractive alternative to the pairwise maxent model because the parameters are easier to infer and it can be used to characterized higher-order interactions \cite{Yu11}. As illustrated in Fig-\ref{fig:DKL}, its accuracy of simultaneous reversals prediction is similar to the one of the pairwise maxent model. Therefore, there is no reason to rule out the pairwise maxent. This result is consistent with the multi-information criterion which returns that pairwise statistical dependencies represent $95\%$ of statistical dependencies \cite{moi1}.


\section{Conclusion}
Our results suggest that trend reversals can be predicted using instantaneous collective states of other market places in the studied samples. This finding also reveals the strength of the collective dynamics underlying the flipping process since the individual instantaneous model is not able to make better than random predictions excepted at higher sampling frequency. Another advantage is that this pairwise maxent model satisfies \emph{all} the pairwise correlations simultaneously which can prevent the overcounting of dependencies using only the pairwise correlation when more than two entities are involved.
Including memory in this model does not improve the accuracy of prediction. This is a not very surprising result since the pairwise lagged cross-correlations are close to zero. Moreover, the sign of returns is poorly forecast ($53\%$ of accuracy) when we use only returns past information. This result is inline with the efficient market hypothesis and a profit can not be made using this model. However, the sample length is too small to estimate so many parameters.  The history may be important in more evolved models including a temporal filtering on the basis of a good approximation of the market dynamics (by analogy to the treatment of time series, especially in the neuroscience field) or modelling with exogenous economic variables. An interesting interpretation of the fine structure of the spectrum of the correlation matrix \cite{Livan11} is that such models allow global correlations (with characteristic length of the order of the network size) and fluctuating clusters coexist in the vicinity of the critical state \cite{Fischer}. This may account for the global collective mode, corresponding to the largest eigenvalue of the correlation matrix, and to the structure of the spectrum bulk which is not only due to noise but also accounts for clustering properties \cite{Livan11}.

It is interesting that such a minimal model returns an accuracy almost as good than the accuracy of pairwise autologistic models even if the market dynamics is undoubtedly much more complex than the model; this finding highlights the significant contribution of collective modes in trend prediction since individual biases are non relevant for the prediction excepted at higher sampling frequencies.

\begin{subappendices}

\section{Cleaning the data}
In this work, we consider instantaneous information (within the defined time bin). The timeseries should therefore be synchronous. The stock exchange closing days, pre-market and after hours trading exchanges are removed. If a time bin is missing for a particular asset, the same time bin should be deleted from the database. The latter case is marginal since we consider indices and highly capitalized companies.
\section{Regularized pseudo-maximum likelihood}

The rPML method is a powerful method for estimation of Lagrange parameters of pairwise maximum entropy model when common maximum likelihood is untractable \cite{Aurell}. This method can be thought as an autologistic regression in order to predict binary outcomes. The main idea is to factorize the distribution and to consider only conditional probabilities. For a N-dimensional sample of length $T$, the objective function to maximize is

\begin{equation}\label{PML2}
  \mathrm{PL}(\boldsymbol\theta)=\frac{1}{T}\sum_{t=1}^{T}\sum_{i=1}^{N}
  \log P(s_{i,t}|\mathbf{s}_{-i,t};\, \boldsymbol\theta)
\end{equation}
where conditional probabilities of the instantaneous model are

\begin{equation}
p(s_{i,t}|\mathbf{s}_{-i,t};\, \boldsymbol\theta)=
\frac{1}{2}
\left[1+s_{i,t}\tanh\left(\sum_{ j\neq i}J_{ij}s_{j,t}+h_{i}\right)\right]
\end{equation}
and

\begin{equation}
p(s_{i,t}|\mathcal{H}_{t}^{T};\, \boldsymbol\theta)=
\frac{1}{2}
\left[1+s_{i,t}\tanh\left(\sum_{ j\neq i}J_{ij}s_{j,t}+h_{i}
+\sum_{\tau=1}^{T}\sum_{j}K_{ij}^{\tau}s_{j,t-\tau}\right)\right]
\end{equation}
for the historical model.

A regularization term is added to the PL function to prevent overfitting which is a negative multiple of the $l_{2}$-norm of parameters to be estimated, for instance. The regularized PL (rPL) objective function is thus $\mathrm{PL}(\boldsymbol\theta)-\lambda\, \|\boldsymbol\theta\|_{2}^{2}$ with $\lambda>0$. If the network is believed to be sparse, a $l_{1}$ regularization term should be used \cite{Aurell} (small values of the parameters are projected on zero).

\section{Confusion matrix}

\begin{figure}[ht!]
\begin{center}
\includegraphics[width=0.75\textwidth]{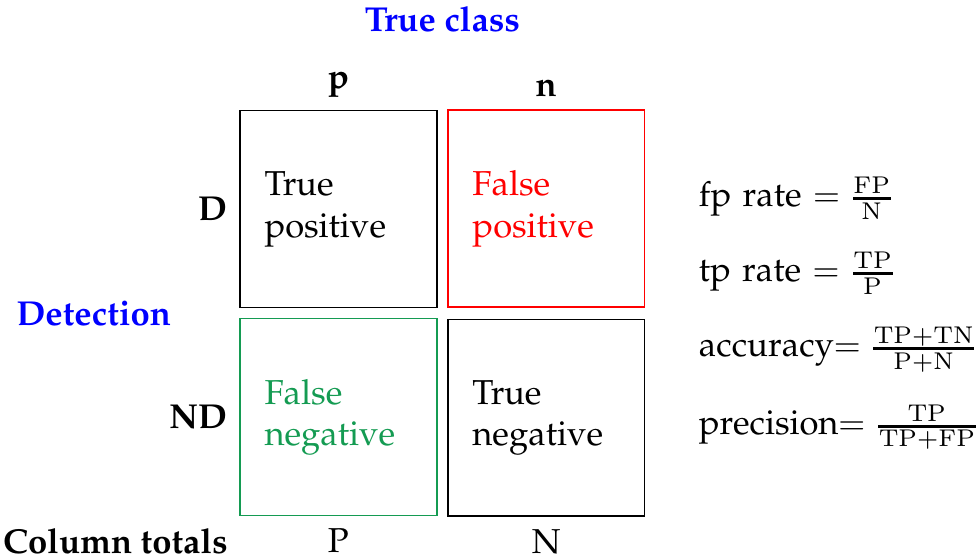}
\end{center}
\caption[Confusion matrix]{The confusion matrix defining the true positive rate, false positive rate and accuracy (among others).}
\label{fig:conf}
\end{figure}

The so-called confusion matrix is illustrated in Fig-\ref{fig:conf}.

Let's see how it works through a short example. Suppose that one observes and detects ten outcomes of a given process, as described in Tab-\ref{tab7:classifier}.

\begin{table}[!ht]
\caption[Confusion matrix, a short example]{Ten outcomes and corresponding detected events.}
\begin{center}
\begin{tabular}{lcccccccccccc}
\hline\label{tab7:classifier}

  Actual values (positive/negative)   & p & p & p & p & p & p & n & n & n & n \\
  Detected values (detected/not detected) & D & D & D & D & D & D & \textcolor{red}{D} & \textcolor{red}{D} & ND & ND \\

\hline
\end{tabular}
\end{center}
\end{table}

The quantities defined in Fig-\ref{fig:conf} are: P $=6$, N $=4$, TP $=6$, TN $=2$, FP $=2$, FN $=0$, fp rate $=2/4$, tp rate $=6/6$, accuracy $=8/10$, precision $=6/8$.

\section{Dichotomized Gaussian model}

A Dichotomized Gaussian model is a non-linear transformation (NLT) of a multi-variate random variable (random vector)  $U\sim\mathcal{N}(\gamma,\Lambda)$ with $\Lambda_{ii}=1$. The NLT is

\begin{equation}
 x_{i} = \begin{dcases*}
        1  & if $u_{i}>0$ \\
        0 & if $u_{i}\leq0$
        \end{dcases*}
\end{equation}

The NLT generates higher order dependencies between the new variables. Let $\boldsymbol{\mu}$ be the expected value of $\mathbf{x}$ and $\Sigma$ its covariance matrix.  The parameters $\{\gamma,\Lambda\}$ are fitted such that

\begin{eqnarray}
  \mu_{i} &=& \Phi(\gamma_{i}) \\
  \Sigma_{ii} &=& \Phi(\gamma_{i})\Phi(-\gamma_{i}) \\
  \Sigma_{ij} &=& \Psi(\gamma_{i},\gamma_{j},\Lambda_{ij})=\Phi_{2}(\gamma_{i},\gamma_{j},\Lambda_{ij})-\Phi(\gamma_{i})\Phi(\gamma_{j})
\end{eqnarray}
where $\Phi$ is the CDF of a $\mathcal{N}(0,1)$ and $\Phi_{2}$ the normal bivariate CDF with covariance $\Lambda_{ij}$. These equations can be solved to find the parameters $\{\gamma,\Lambda\}$.
\end{subappendices} 

\chapter{General conclusion}\label{chap:social}
\thispagestyle{empty}
\begin{summary}
We present socio-economic models in terms of maximum entropy models. We address briefly the significance of deviations from the most probable state and the notion of equilibrium in heterogeneous systems. Lastly, the final conclusion and perspectives are given.
\end{summary}

\section{Introduction}
\label{8-intro}
So far, we have focused on financial networks using the maximum entropy principle which has led to simple but rich and complex effective models. Actually, the MEP can be used in several economic settings. Axelrod and Bennett have introduced a theory of aggregation \cite{Axelrod} which is equivalent to a deterministic version of the pairwise maxent model. Each actor has to choose one of the two possible actions: cooperate or fight. Each pair of actors $(i,j)$ has a propensity $J_{ij}$ to work together. The coalition is found by minimizing the so-called "energy" $E(\mathbf{s})=-\sum_{i<j}J_{ij}s_{i}s_{j}$. Such kind of clustering and combinatorial problems were formulated in terms of maximum entropy models ten years before Axelrod-Bennett \cite{Fischer}. It is interesting that they formulated an equivalent model as they come from an (a priori) unrelated discipline. This sheds light on the convergence of models dealing with complexity. These models fall in the branch of mathematics dealing with combinatorial optimization problems. Is it surprising? Not really since entropy itself can be thought as a combinatorial object. For the interested reader, an extensive discussion of the use of the combinatorial and information theoretic entropy approach in economics can be found in \cite{Aoki}. Several agent-based models use also a binary pairwise component, most often they are built following the \emph{direct} approach (starting from plausible economic assumptions to derive a price dynamics) \cite{Born,Zhou}. It is also possible to show that the Schelling segregation Model can be rewritten as a maximum entropy model \cite{Stauff}.

The main breakthrough done by Brock and Durlauf \cite{Brock} is the link between the underlying optimization process and the emergence of the Gibbs distribution. A recent paper \cite{MMR} details the general derivation of the functional dependence of the configuration probability on the (a priori unknown) optimized utility function. Moreover, sampling such a complex system actually brings information about the utility function. If the configurations $\{\mathbf{s}\}$ are the outcome of an optimization of an unknown utility function $\mathcal{U}(\mathbf{s})$ and if the configuration probability distribution is accurately sampled, then the utility is proportional to the log-likelihood $\mathcal{U}(\mathbf{s}) \propto \ln \mathrm{P_{emp}}(\mathbf{s})$, where $\mathrm{P_{emp}}(\mathbf{s})$ is the empirical configurations distribution. One can thus extract information about the underlying maximization process by sampling a complex system.

Hereafter, we shortly present how to derive the Brock-Durlauf model \cite{Brock} (noted: BD model) in terms of statistical inference over some data set. The task is to model the influence of social interactions when agents make a binary choice, the agents are not supposed to be rational and are allowed to make mistakes.
Deriving the utility function including a social component on economic considerations requires several assumptions. The application of the maximum entropy principle provides a useful statistical inverse formulation which can be interpreted and linked to the optimization process. Furthermore, we will show that the maxent formulation also provides a convenient framework to discuss the equilibria and their stability. In presence of heterogeneity, the decision making problem is significantly harder. The time to reach the equilibrium (relaxation time) can be large in comparison to the characteristic time scale of the decision process. Lastly, we draw the conclusion of this thesis.


The chapter is organized as follows. In section \ref{sec8:BD}, the Brock-Durlauf model is reviewed, we give a maxent derivation and we consider static collective behaviours. In section \ref{sec8:ccl}, the final conclusion is drawn. In section \ref{sec8:perspec}, we give some perspectives.

\section{The Brock-Durlauf model}\label{sec8:BD}

The Brock-Durlauf model is a random utility (or payoff) model taking into account the role of social interactions when economic agents face a binary choice. First, we give a review of this model \cite{Brock}. Each individual in a population of N agents must choose a binary action (yes/no, buy/sell, etc.). Each of these actions are denoted by a binary variable $s_{i}\in\{-1,1\}$. The population configuration will be described by a vector $\mathbf{s}=(s_{1},\cdots,s_{N})$ and the choices of all the agents other than $i$ will be denoted by  $\mathbf{s}_{-i}=(s_{1},\cdots,s_{i-1},s_{i+1},\cdots,s_{N})$. Individual utility $V(s_{i})$ is assumed to consist in a private utility (payoff) $u_{s_{i}}$, a social component $S(s_{i},\mu_{i}^{e}(\mathbf{s}_{-i}))$ where $\mu_{i}^{e}(\mathbf{s}_{-i})$ denotes the conditional probability measure that agent $i$ places on the choices of others at the time of making his own decision. Last, a random utility term $\epsilon(s_{i})$ independently and identically distributed (IID) across agents. Regrouping all these term, one gets

\begin{equation}\label{BDutility}
  V(s_{i})=u(s_{i})+S(s_{i},\mu_{i}^{e}(\mathbf{s}_{-i}))+\epsilon(s_{i})
\end{equation}

This model is then restricted to parametric representations of the social utility and of the probability density function of the random utility. The social utility is assumed to exhibit a constant and totalistic strategic complementarity with intensity $J>0$. This assumption leads to the form

\begin{equation}\label{Sutility}
  S(s_{i},\mu_{i}^{e}(\mathbf{s}_{-i}))=Js_{i} \mu_{i}^{e}(\mathbf{s}_{-i})
\end{equation}
where $\mu_{i}^{e}(\mathbf{s}_{-i})$ may be replaced by $(N-1)^{-1}\sum_{j\neq i}\mathrm{E}[s_{j}]$ if one imposes rational expectations (agents assess rationally the other choices).

The second assumption is that the errors are extreme-value distributed such that the differences $\epsilon(-1)-\epsilon(1)$ are logistically distributed

\begin{equation}\label{Error}
  \Pr[\epsilon(-1)-\epsilon(1)\leq x]=\frac{1}{1+\exp(-\beta x)}
\end{equation}

Under these assumptions, the resulting noncooperative (no communication, an agent makes his choice based on beliefs on the mean choice, here the rational expectation) probability of a configuration $\mathbf{s}$ is

\begin{equation}\label{Prob}
  p(\mathbf{s})=\mathcal{Z}^{-1}\exp\left(\beta u(s_{i})+\beta\frac{J}{N-1} s_{i}\sum_{j\neq i}\mathrm{E}[s_{j}]\right)
\end{equation}
Using linear individual utility $u(s_{i})=h_{i}s_{i}$, one obtains self-consistent relations for the equilibrium mean choices

\begin{equation}\label{Equil}
  \mathrm{E}[s_{i}]=\tanh\left(\beta h_{i}+\beta J \frac{\sum_{j\neq i} \mathrm{E}[s_{j}]}{N-1}\right)
\end{equation}
Stating the invariance $\mathrm{E}[s_{i}]=\mathrm{E}[s_{j}]$ and homogeneous individual preferences $h_{i}=h$, it comes

\begin{equation}\label{8-MF}
  \mathrm{E}[s]=\tanh\left(\beta h+\beta J  \mathrm{E}[s]\right)
\end{equation}

Last, if a social planner sets the choices accordingly to the individual utility then the deterministic component of the social planner's utility is the sum of the individual deterministic utilities. The random component is chosen extreme-value distributed over configurations $\epsilon(\mathbf{s})$ for mathematical tractability.

\subsection{Maximum entropy formulation}

An alternative derivation based only on statistical consideration is done using the maximum entropy principle (MEP) \cite{Jaynes}. The MEP allows to derive the less structured model consistent with some knowledge of the system (inverse model). It selects the distribution which leaves us with the largest remaining uncertainty consistent with our knowledge (constraints) of the system. In this way, we have not introduced any additional assumptions. Maximizing the entropy can be viewed as a maximization of the likelihood of the distribution $p$ closest to the uniform distribution $U$ without range restriction since $D_{\mathrm{KL}}(p||U)=-\mathcal{S}[p]+\ln N$ where $D_{\mathrm{KL}}$ is the Kullback-Leibler divergence and $\mathcal{S}[p]$ the entropy of the probability distribution $p$.

Moreover, maximum entropy models are particularly suited to the description of collective behaviours as appearing in the BD model. They are used to model collective modes in neuroscience, physics, finance, information processing for instance \cite{Schneidman,Fischer,Nish,moi1}. The reason for such eclecticism is that the description of macroscopic collective behaviours does not require necessarily the exact knowledge of the individual dynamics. The key features are the topology of the network, the order of influences  (pairwise or higher), their range and their boundedness.

Assume that one observes first and second moments over a population of $N$ agents. The MEP reads

\begin{eqnarray}\label{8-maxent}
  & &\max_{\substack{\{p(\textbf{s})\}}} \mathcal{S}[p(\textbf{s})]= \max_{\substack{p(\textbf{s})}} \left\{-\sum_{\{\textbf{s}\}}p(\textbf{s}) \,\ln p(\textbf{s})\right\}  \\ \nonumber
   &\mathrm{s.t}&  \sum_{\{\textbf{s}\}}p(\textbf{s})=1,\quad \sum_{\{\textbf{s}\}}p(\textbf{s})s_{i}=m_{i},
   \quad \sum_{\{\textbf{s}\}}p(\textbf{s})s_{i}s_{j}=q_{ij} \nonumber
\end{eqnarray}
leading to the two-agent probability distribution

\begin{equation}
p_{2}(\textbf{s})=\frac{1}{\mathcal{Z}}\exp\left(\frac{1}{2}\sum_{i, j}^{N}J_{ij}s_{i}s_{j}+\sum_{i=1}^{N}h_{i}s_{i}\right)\label{8-Lagrange}
\end{equation}

The self-consistent equation of the BD model is exactly the self-consistent equation for a fully connected social network with homogenous interactions. Indeed, setting $J_{ij}=J/N$ for $i\neq j$ and $h_{i}=h$, we get the social planner's problem of the BD model:

\begin{equation}
p_{2}(\textbf{s})=\frac{1}{\mathcal{Z}}\exp\left(\mathcal{U}(\mathbf{s})\right)
=\frac{1}{\mathcal{Z}}\exp\left(\frac{J}{2N}(\sum_{i}s_{i})^{2}+h\sum_{i}s_{i}\right)
\label{SP}
\end{equation}
where $\mathcal{U}(\mathbf{s})$ is the social planner's utility corresponding to the sum of the deterministic parts of individual utilities. Using individual or social planner's utilities, we get the same consistent equation (\ref{8-MF}) than in BD model if the the social influences are globally rescaled by a control parameter ($\beta$): $J\rightarrow \beta J$. This control parameter can be thought as the propensity for making error. For $\beta\rightarrow\infty$, the decision making is deterministic.

The description of non-cooperative decision making is thus the same than in the BD model. The decision making is non-cooperative because each agent evaluates rationally the social pressure acting on himself $h_{i}^{\mathrm{eff}}=N^{-1}J\sum_{j\neq i}s_{j}+h$. As long as the expected consensus $M_{\backslash i}=\sum_{j\neq i}s_{j}$ takes the same value, the equilibrium is unchanged.
We conclude that the maxent potential $\mathcal{U}(\mathbf{s})$ is the sum of the deterministic individual utilities. Lagrange parameters $J_{ij}$ and $h_{i}$ are respectively interpreted as social influences and idiosyncratic preferences. As the social influence matrix $\mathbf{J}$ is conjugated to the covariance matrix, it is a symmetric matrix.

Last, we saw in Sec-\ref{sec3:Relation} that the maximum entropy principle can be thought as a way to approximated the social network. The pairwise maxent model takes into account only unary and binary social interactions. The statistical significance of each order can be tested using the multi-information criterion (see Sec-\ref{sec3:TestMEP}).

\subsection{Static collective behaviours}

As the decision making is characterized by the Gibbs distribution (\ref{Lagrange}), interesting known results (multiple equilibria, hierarchical structure, large fluctuations, etc.) follow under the assumption of rational expectation \cite{Fischer,Brock,Cont}. Multiple Nash equilibria may exist in the non-cooperative decision making (agents do not communicate). The resulting collective behaviours, according to \cite{Cont}, are denoted by \emph{static} ones as they are described by an equilibrium distribution of a Markov process. In this case, Lagrange parameters $\{J_{ij},h_{i}\}$ are time independent. We emphasize those equilibria may be derived with a variational method since $\ln \mathcal{Z}$ is the cumulant generating function

\begin{equation}\label{8-CumAve}
\langle s_{i_{1}}\ldots s_{i_{N}}\rangle_{\mathrm{c}}=\beta^{-N}\partial^{N}\ln \mathcal{Z}/\partial h_{i_{1}}\ldots \partial h_{i_{N}}
\end{equation}

where the cumulants are denoted by $\langle\cdot\rangle_{\mathrm{c}}$. However for large $N$, says $N>20$, and heterogenous influences the partition function $\mathcal{Z}$ can not be computed exactly since it involves $2^{N}$ terms. Several approximations exist. Suppose that $p$ is the true and untractable Gibbs distribution $p(\textbf{s})=\mathcal{Z}_{p}^{-1}\exp\left(-\beta\mathcal{H}(\textbf{s})\right)$ and $q$ a tractable but misspecified Gibbs distribution $q(\textbf{s})=\mathcal{Z}^{-1}_{q}\exp\left(-\beta\mathcal{H}_{q}(\textbf{s})\right)$. We minimize the dissimilarity, measured by the Kullback-Leibler divergence, between $q$ and $p$ (see Sec-\ref{sec3:Var} for details)

\begin{equation}
D_{\mathrm{KL}}(q||p)=\mathcal{F}_{\beta}[q]+\ln \mathcal{Z}_{p} \geq 0
\end{equation}

The Kullback-Leibler divergence minimization is thus equivalent to the minimization of the $\mathcal{F}_{\beta}$-functional\footnote{$\mathcal{F}_{\beta}\equiv
\beta^{-1}\mathcal{F}=-\beta^{-1}\ln \mathcal{Z}$} over the space of the $q$-distributions. Doing so for heterogeneous social influences \cite{Barb}, one gets at the first order

\begin{equation}\label{MFeq}
  \mathrm{E}[s_{i}]=\tanh\left(\sum_{j}\beta J_{ij}\mathrm{E}[s_{j}]+\beta h_{i}\right)
\end{equation}
This first order equilibrium self-consistent equation is equivalent to (\ref{8-MF}) of the BD model if the invariance $\mathrm{E}[s_{i}]=\mathrm{E}[s_{j}]$ is stated and if the idiosyncratic preferences are homogeneous. This equation is exact if social interactions are homogeneous and properly scaled ($\sim N^{-1}$) and if the idiosyncratic preferences are homogeneous. For more general social networks, the fluctuations of the mean choice should be considered (see Sec-\ref{sec3:eq}). To include a part of fluctuations (one agent fluctuations), we consider the second order variational approximation (one can also consider an expansion in term of cumulants as in Sec-\ref{sec3:eq}.)

\begin{equation}\label{TAPeq}
  \mathrm{E}[s_{i}]=\tanh\left(\sum_{j}\beta J_{ij}\mathrm{E}[s_{j}]-
  \sum_{j}\beta^{2}J_{ij}^{2}\mathrm{E}[s_{i}](1-\mathrm{E}[s_{j}]^{2})
  +\beta h_{i}\right)
\end{equation}

\begin{figure}[ht!]
\begin{center}
\includegraphics[width=0.75\textwidth]{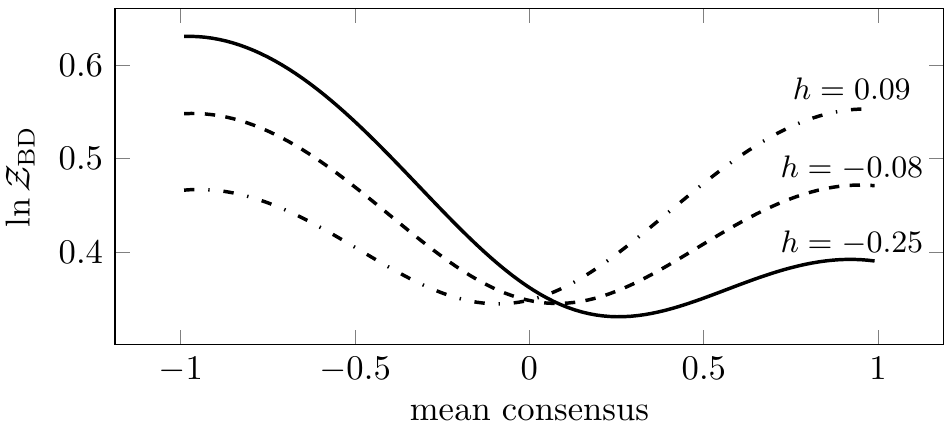}
\end{center}
\caption[BD partition function]{The logarithm of the BD partition function illustrated for different values of idiosyncratic preferences $h$ and for $\beta J=2$.}
\label{fig:Z}
\end{figure}

The stability of these equilibria directly results from (\ref{VarPrinc}). For the BD model, the cumulant generating functional has a global maximum for the equilibrium of the same sign of the individual preference $h$ meaning a higher expected utility for each agent and thus a larger welfare as illustrated in Fig-\ref{fig:Z}.

\begin{figure}[ht!]
\begin{center}
\includegraphics[width=\textwidth]{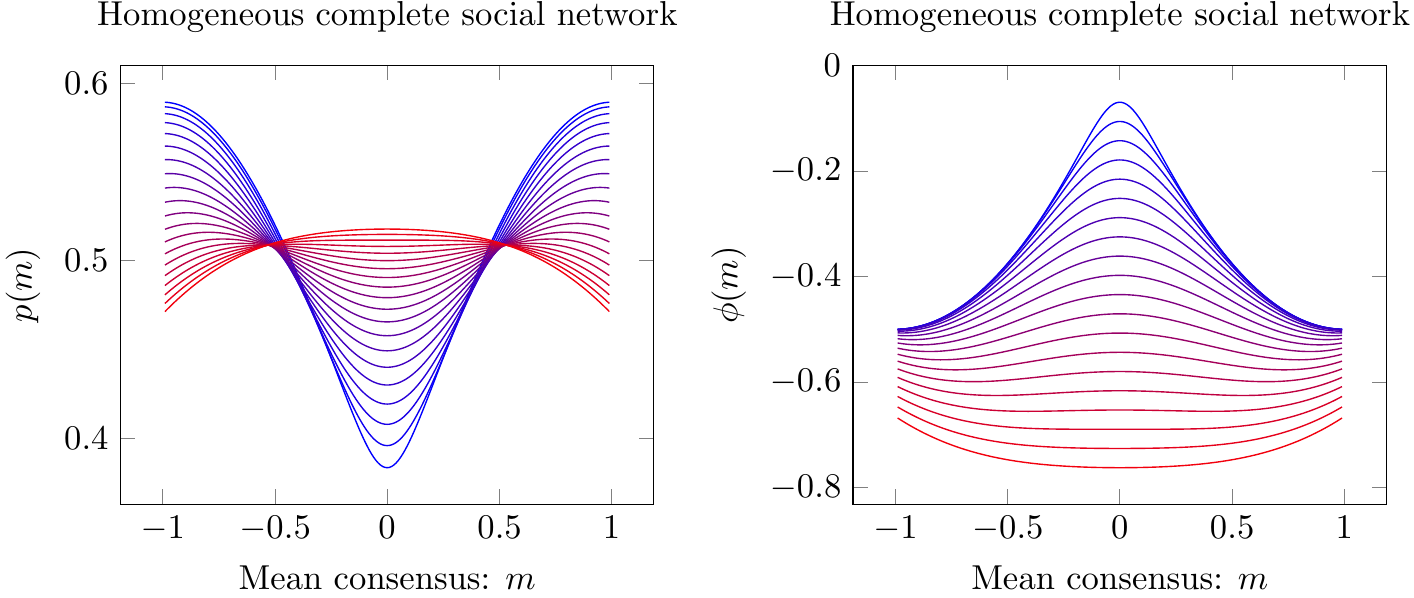}
\end{center}
\caption[Mean consensus pdf for different values of $\beta$]{The mean consensus pdf and $\phi$ as a function of $m$ illustrated for different values of the stochasticity level $\beta$. From red to blue, $\beta$ increases.}
\label{fig:Zb}
\end{figure}

Moreover the cumulant generating function does not have the same curvature for all value of the stochasticity level $\beta$. For a complete social network with influences scaling as $N^{-1}$, the exact relation between the net consensus $m$ and $N^{-1}\ln \mathcal{Z}$ is known \cite{Baxter}. The partition function is


\begin{eqnarray}
  \mathcal{Z} &=& \sum_{\{\mathbf{s}\}}\exp\left( \frac{\beta J}{2N}N^{2}(\sum_{i}s_{i})^{2}+\beta h N \sum_{i}s_{i}\right) \\
    &=& \sqrt{\frac{\beta J N}{2\pi}}\int_{-\infty}^{\infty}\ud m \sum_{\{\mathbf{s}\}}
    \exp\left( \frac{\beta J}{2N}m^{2}+\beta J m (\sum_{i}s_{i})+\beta h  \sum_{i}s_{i}\right)
\end{eqnarray}
The Gaussian (or Hubbard-Stratonovich) transformation leaves us with decoupled agents but introduces a fluctuating component $m$ \cite{Opper}. The equilibrium condition $\partial \ln\mathcal{Z}/\partial m =0$ provides the identification $m=N^{-1}\sum_{i}\langle s_{i}\rangle$.  For large systems ($N\gg1$) the later integral is well approximated by the saddle point method. Rearranging the terms, the partition function is given by

\begin{equation}\label{8-Zmf2}
\mathcal{Z}=\sqrt{\frac{\beta J N}{2\pi}}\int_{-\infty}^{\infty}\ud m \exp\left(-\beta J N \phi(m)\right)
\end{equation}
where $\phi(m)=2^{-1}m^{2}-TJ^{-1}\ln\left(2\cosh T^{-1}(J m+h)\right)$, illustrated in the right panel of Fig-\ref{fig:Zb}.

The probability density function (pdf) to observe a mean consensus equal to $m$ in the BD model is thus

\begin{equation}\label{8-Pm}
p(m)=\frac{\exp\left(-\beta JN \phi(m)\right)}{\mathcal{Z}}
\end{equation}
this pdf is illustrated in Fig-\ref{fig:Zb}. For high stochasticity, there is only one stable equilibrium ($m=0$) and for low stochasticity level (large $\beta$ value), there are two stable equilibria (if the idiosyncratic preferences are set to zero) as proven in the BD model \cite{Brock}. One notes the continuous deformation from a single peak curve to a bimodal curve as $\beta$ increases.

Finally, the large deviation theory tells us that the probability of a large deviation from the equilibrium consensus $m_{\mathrm{eq}}$ is given by $\mathrm{P}[m]\approx\exp\left(N \beta J[\phi(m)-\phi(m_{\mathrm{eq}})]\right)$. For large systems ($N\gg 1$), large deviations are statistically unlikely.

\begin{figure}[ht!]
\begin{center}
\includegraphics[width=0.75\textwidth]{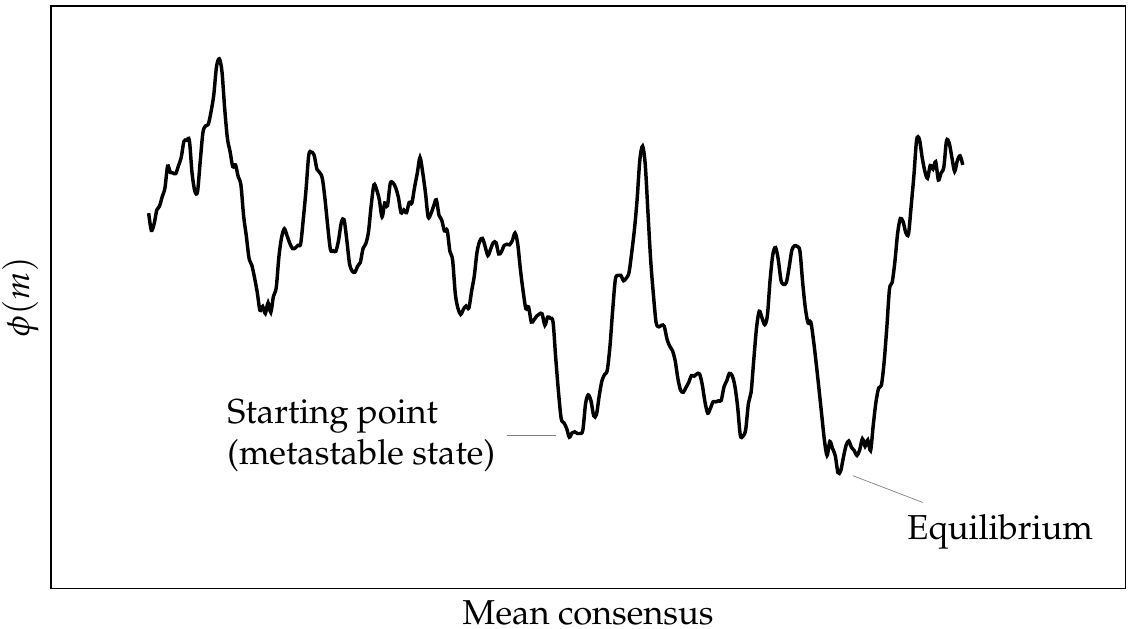}
\end{center}
\caption[$\phi(m)$ for heterogenous social networks]{Schematic plots of the $\phi$-function for low stochasticity level. Reaching the equilibrium from the metastable state (quasi-equilibrium) is very unlikely since the system must firstly pass through a high barrier (which is highly unfavourable).}
\label{fig8:Z2}
\end{figure}

For heterogeneous social networks, the profile of the cumulant generating function $\ln\mathcal{Z}$ and of the $\phi$-function can be much more complex as schematically illustrated in Fig-{\ref{fig8:Z2}}. This feature implies that for a low stochasticity level, the system dynamics could be very slow depending on the system size and on the variance of the social influences \cite{Fischer}.

\subsection{Dynamic collective behaviours}

The landscape of the partition function can be complex in presence of heterogenous conformity $J_{ij}>0$ and dissimilarity $J_{ij}<0$. The valleys between maxima may be deep. In a dynamic framework, passing from a local maximum to the global maximum can be slow especially for particular values of the social influences \cite{Fischer}.

The dynamic evolution of the mean choice is given by the master equation (the variation of the probability is equal to the inward minus outward flows)

\begin{equation}
\frac{\mathrm{d}}{\mathrm{d}t}p(\textbf{s}; t)=
\sum_{i = 1}^{N} \omega(s_{i}|\, -s_{i})\;p(\Flip{i}\textbf{s}; t)-\omega(-s_{i}|\, s_{i})\;p(\textbf{s}; t)\label{MEC}
\end{equation}
where $\Flip{i}\textbf{s}=\Flip{i}(s_{1},\ldots,s_{i},\ldots,s_{N})=(s_{1},\ldots,-s_{i},\ldots,s_{N})$ and the transition rates are related to the transition probability $W(-s_{i},\epsilon|\, s_{i},0)$ by the golden rule (where $\epsilon$ is an infinitesimal time)

\begin{equation}
W(-s_{i},\epsilon|\, s_{i},0)=\omega(-s_{i}|\, s_{i})\,\epsilon+o(\epsilon)
\end{equation}
In the Markov chain theory, the sojourn times are exponentially distributed \cite{Lawler}. A state survives at least a time equal to $\epsilon$ if there is no transition during this interval: $\mathrm{P}[T_{i}>\epsilon]=W(s_{i},\epsilon|\, s_{i},0)= e^{\mu_{i}\epsilon}= 1-\mu_{i}\epsilon+o(\epsilon)$ which implies that the sojourn time is exponentially distributed with a characteristic time scale equal to the opposite of the rate transition.

The detailed balance condition

\begin{equation}
  \omega(s_{i}|\, -s_{i})\;p^{s}(\Flip{i}\textbf{s})-\omega(-s_{i},|\, s_{i})\;p^{s}(\textbf{s})=0
\end{equation}
ensures the convergence to equilibrium. Several updating schemes satisfying the detailed balance can be chosen. The dynamic updating defined in \cite{Cont} is the following. An agent updates his choice at random time intervals such that there is almost surely no simultaneous updating. The agent is free to observe other agents in his neighbourhood (agents socially connected to him) meaning exchange of information (as opposed to synchronous updating). Using the rates

\begin{equation}\label{rates}
   \omega(-s_{i}|s_{i})= \frac{1}{2\tau}\left[1-s_{i,t}\tanh\left(\beta\sum_{j}J_{ij}s_{j,t}+\beta h_{i}\right)\right]
\end{equation}
one gets the following relaxation evolution

\begin{equation}\label{EMO}
   \frac{\mathrm{d}m_{i}(t)}{\mathrm{d}t}=
   -\frac{1}{\tau}\left( m_{i}(t)-\mathrm{E}\left[\tanh(\beta \sum_{j}J_{ij}s_{j,t}+\beta h_{i})\right]\right)
\end{equation}
where $m_{i}\equiv\mathrm{E}[s_{i}]$ and $\tau$ is the characteristic time scale of the transition.

\subsubsection{Homogeneous network}

\begin{figure}[!ht]
\begin{center}
\includegraphics[width=\textwidth]{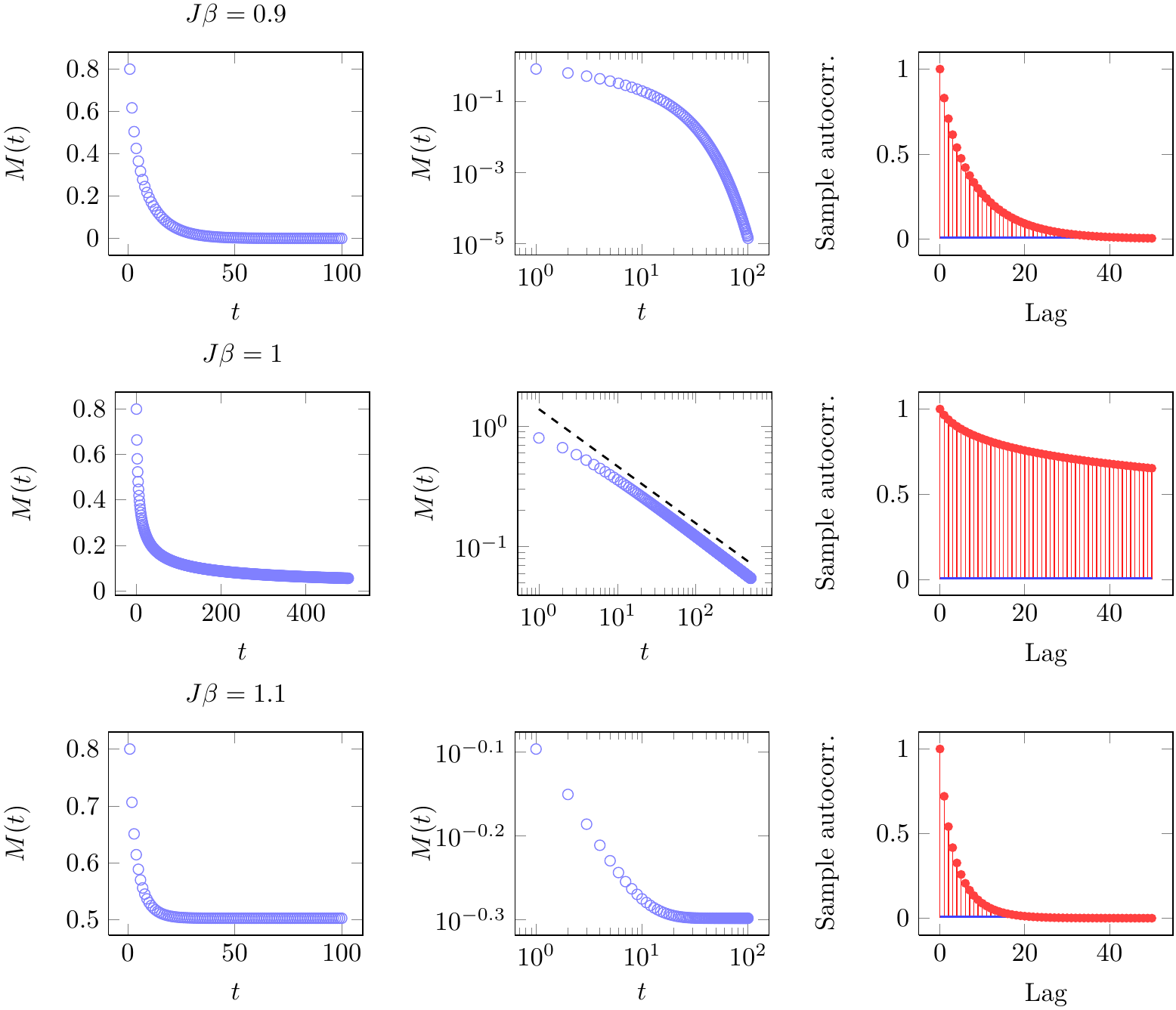}
\end{center}
\caption[The evolution of the mean consensus]{The evolution of the mean consensus $M(t)=N^{-1}\sum_{i}s_{i}$ in the Brock-Durlauf model for three cases: $J\beta<1$ (zero consensus), $J\beta=1$ (the bifurcation point) and $J\beta>1$ (non zero consensus).}
\label{fig:BDconsensus}
\end{figure}

For a complete and homogeneous social network ($J_{ij}=J/N$ for $i\neq j$ as in the BD model), the exact evolution of the mean choice is given by

\begin{equation}\label{8-EMOMFdis}
   \frac{\mathrm{d}m(t)}{\mathrm{d}t}=
   -\frac{1}{\tau}\left[ m(t)-\tanh(\beta Jm(t)+\beta h)\right]
\end{equation}
The discrete time version is

\begin{equation}\label{8-EMOMF}
   m(t+1)=\tanh(\beta Jm(t)+\beta h)
\end{equation}
By construction, this dynamics describes the relaxation towards the equilibrium consensus. The equilibrium consensus is generally quickly reached excepted for the bifurcation case $J\beta=1$. It is possible to show that the autocorrelation time reaches its maximum value at the bifurcation point \cite{BinderMC}. The autocorrelation time $\tau$ is the characteristic decay time of the autocorrelation function $R(t)\propto \exp(-t/\tau)$. The dynamics is illustrated in Fig-{\ref{fig:BDconsensus}}.

We note that the dynamics is significantly different at the bifurcation point $J\beta=1$. The Consensus decreases to zero as $\sim t^{-0.5}$ instead of in an exponential fashion. The autocorrelation decreases slower meaning an increase of the autocorrelation time $\tau$ and a persistent effect of the initial condition (memory).
\subsubsection{General social networks}

For heterogeneous social influences, the exact evolution is not tractable due to the averaging of the hyperbolic tangent. The dynamical counterpart of the variational approximation exposed in Sec-\ref{sec3:Var} can be found in \cite{RoudiTAP}. At the first order, it comes

\begin{equation}
 m_{i}(t+1) = \tanh\left(h_{i}(t)+\sum_{j}J_{ij}m_{j}(t) \right)
\end{equation}
the second order reads

\begin{equation}
 m_{i}(t+1)= \tanh\left(h_{i}(t)+\sum_{j}J_{ij}m_{j}(t)-m_{i}(t+1)
   \sum_{j}J_{ij}^2(1-m_{j}^{2}(t))\right)
\end{equation}
The continuous time version is obtained by substituting $m_{i}(t+1)$ by $m_{i}(t)+ \mathrm{d}m_{i}(t)/\mathrm{d}t$.
The former discussion shows that the welfare analysis is a difficult task in case of heterogeneity and that the optimal state (maximizing the average utility and the entropy) can be reached very slowly. This feature has potentially an important consequence, the optimal state (or equilibrium) can never be reached if the lifetime of a metastable state is larger than the order of magnitude of the characteristic time scale of the decision making. If it is so, a \emph{local} approach is better than a global maximization. In place of searching the \emph{global} optimum state, one should restrict oneself to the range of states which can be reached in a reasonable time (compared to the time scale of the process).

\section{Conclusion}\label{sec8:ccl}

Through this thesis, we explored the market structure by means of statistical methods avoiding as much as possible analogies and rules design.

Namely, we showed that pairwise maximum entropy models are good candidates to describe the market structure. Their statistical formulation is simple but they are rich and complex effective models. They capture collective modes, they do not depend on the nature of the constituting entities, their order is statistically testable and they allow to perform simulations. Furthermore they are stated with a minimal set of assumptions and are thought as statistical (inference) models rather than \emph{physical} models.

Using these models together with data analysis, we shed light on the relation between collective phenomena and structural changes of financial networks. In particular, we emphasized that the stock market does not stand in a given regime but goes back and forth through order and disorder and exhibit a great malleability. These studies also showed that markets do not stand rigourously at a critical state but get closer to it before a crash. Criticality is an important concept since complex systems process efficiently information in this regime and it is the state where the deviation to the equiprobability of events is the largest.

Other applications are found in the economic modelling of social systems. They can be addressed as such statistical models and several methods, coming from economics and physics, can be straightforwardly applied. In particular, the link between the underlying optimization process and the emergence of the Gibbs distribution is another evidence of the convergence of fields in the background of complex systems.

\section{Perspectives}\label{sec8:perspec}
\begin{figure}[!ht]
\begin{center}
\includegraphics[scale=0.95]{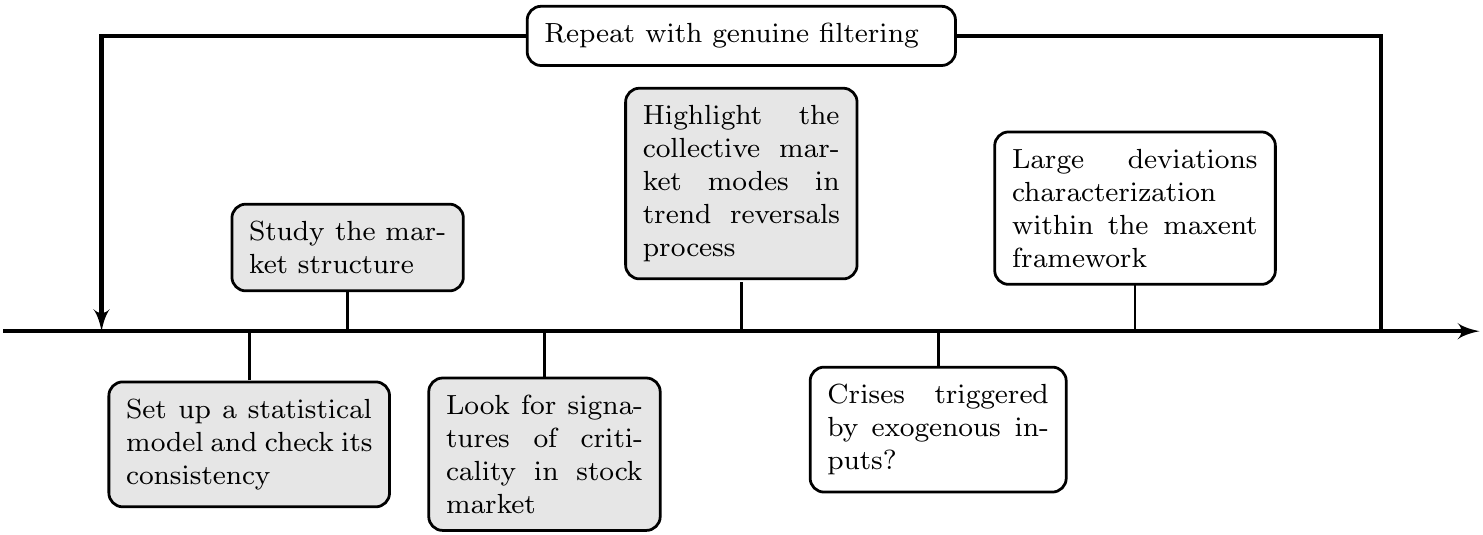}
\end{center}
\caption[Thought-line, step $\infty$]{The logical ordering of the series extensions.}
\label{fig:TLfinal}
\end{figure}

These results lead to perspectives, not yet launched. We conclude this thesis with a few clues for further work. Some of them are in line with the present approach and are straightforward extensions (\emph{series} extensions), the others are linked to this work but involve a complete revision and thus much efforts (\emph{parallel} extensions), see Fig-\ref{fig:TLfinal}.

\begin{itemize}
  \item \textbf{Large deviations} \\
  Perhaps the most interesting issue is the large deviation analysis of stock market. We could consider the large deviations to the mean orientation in the maxent framework and check the consistency with empirical results. A starting point could be (\ref{8-Pm}), check if the large deviations are well described or not by such a rate function. This idea is supported by a recent paper which emphasizes the possibility of emergence of power-laws without fine tuning (depending on the distribution of external information) \cite{Schwab2}.

  \item \textbf{Correlation matrix filtering}\\
  A better characterization of the financial network could be obtained if a genuine filtering of the correlation matrix is preliminary applied. However, as recently shown in \cite{Livan11}, the fine structure of the correlation matrix involves a tricky filtering. The simple removing of eigenvalues of the bulk (sometimes thought as "noise") could be too rough a filtering. The first step will be to consider the naive filtering: diagonalize the covariance matrix (which is possible since it is a real symmetric matrix) $\boldsymbol{\Lambda}=\textbf{O}^{\mathrm{T}}\textbf{C}\textbf{O}$ (where $\boldsymbol{\Lambda}$ is the diagonal matrix whose entries are the eigenvalues and the $\textbf{O}^{\mathrm{T}}$ stands for the transpose matrix), remove the eigenvalues corresponding to noise (following the semi-circle law), inverse the change of basis $\textbf{C}_{\mathrm{filt}}=\textbf{O}\boldsymbol{\Lambda}_{\mathrm{filt}}\textbf{O}^{\mathrm{T}}$, infer the influence matrix $\textbf{J}$ based on $\textbf{C}_{\mathrm{filt}}$, redo the analyses and compare with the results without filtration.

  \item \textbf{Non-extensive formulation}\\
  Long range interactions lead to the non-additivity of the entropy, alternative approaches (as the Tsallis entropy, for instance) could shed light on relaxation, fat tail and other topics. The additivity is a feature which is a non necessary axiom for a measure of uncertainty. The Shannon entropy is a particular case of the Tsallis entropy. Therefore, we could consider the Tsallis entropy in place of the Shannon entropy in the maximum entropy principle, see \cite{Tsallis} and references within. Then we could redo and extend the analyses with the new two-agent distributions. We note that a new (or a modification of an existing) inversion method is needed to infer the Lagrange parameters.

  \item \textbf{Generalized linear models, point processes and magnitude of returns}\\
   The non-stationarity of the financial markets can results from adaptive or learning process (temporal evolution of the Lagrange parameters). An attractive class of models to tackle the non-stationarity issue is the class of generalized linear models (not to be confused with general linear models) and point processes used in neuroscience. A possibility could be the study of high frequency data or threshold the data such that the occurrence time is itself stochastic. We could binarize the data using a proper threshold method (a naive method could be: $0$ if the absolute return is smaller than $\alpha\%$ and $1$ if the return is larger or equal to $\alpha\%$, with $\alpha>0$) and infer the transition probability within the point process framework, see \cite{Truc} for instance. Then, we could redo this analysis for different threshold level $\alpha$ going from moderate to large absolute returns. A further extension could be the three-state $\{-1,0,1\}$ model (also called the \emph{Potts model} in the literature). The three-state variable being set to $-1$ if the return is negative and smaller than $-\alpha\%$, to $0$ if the return is between $-\alpha\%$ and $\alpha\%$ and to $1$ if the return is larger than $\alpha\%$. We note that the Lagrange parameters can be inferred with a pseudo-maximum likelihood, as in the binary case.

  \item \textbf{Triggered crises}\\
   The Barkhausen effect (avalanches in critical systems triggered by exogenous inputs) could be thought as a crises formation process in the maxent framework. Depending on the randomness of the external information, avalanches could be observed or not, as explained in \cite{Perkovic} for instance.

  \item \textbf{Portfolio composition}\\
   One could compare portfolios (return and risk) created using the correlation matrix as usual and the influence matrix. In modern portfolio theory, the portfolio volatility is defined using the correlation coefficients. We saw that the influence matrix $\mathbf{J}$ is a better quantification of the statistical dependencies than correlation coefficients. A new definition of the \emph{risk} could be derived in this framework. Then, we could compare the performance of both kinds of portfolios.
\end{itemize}


\backmatter


\printbibliography

\end{document}